\documentclass[twocolumn,
 showpacs,  
 preprintnumbers,  
 aps,  
 prd,  
 a4paper,  
 superscriptaddress,  
 nofootinbib,  
 tightenlines,  
 floats,floatfix  
 ]{revtex4}

\usepackage{graphicx}
\usepackage{subfigure}
\usepackage{latexsym}
\usepackage{amsmath,amssymb}  
\usepackage{dcolumn}

\def\tX{\tilde{X}}

\def\rd{{\rm d}}

\def\be{\begin{equation}}
\def\ee{\end{equation}}
\def\bea{\begin{eqnarray}}
\def\eea{\end{eqnarray}}

\def\nn{\nonumber \\}

\def\d{{\rm d}}

\def\5{\overline 5}
\def\vp{\varphi}

\def \wt {\widetilde}

\newcommand{\prt}{\partial}

\newcommand{\om}{\Omega_{m}^{(0)}}

\def\be{\begin{equation}}
\def\ee{\end{equation}}
\def\ba{\begin{eqnarray}}
\def\ea{\end{eqnarray}}

\begin{document}

\title{Dynamics of dark energy}
\author {Edmund J.~Copeland}
\affiliation{School of Physics and Astronomy, University of
Nottingham, University Park, Nottingham NG7 2RD, United Kingdom \\
{\tt Email:ed.copeland@nottingham.ac.uk}}

\author{M.~Sami}
\affiliation{Centre for Theoretical Physics, Jamia Millia Islamia, New Delhi,
India}

\affiliation{Department of Physics, Jamia Millia Islamia, New Delhi, 
India\\
{\tt Email:sami@iucaa.ernet.in; sami@jamia-physics.net}}

\author{Shinji Tsujikawa}
\affiliation{Department of Physics, Gunma National College of
Technology, Gunma 371-8530, Japan \\
{\tt Email:shinji@nat.gunma-ct.ac.jp}}

\date{\today}

\vskip 1pc
\begin{abstract}

In this paper we review in detail a number of approaches that have been 
adopted to try and explain the remarkable observation of our accelerating Universe. 
In particular we discuss the arguments for and recent progress made towards 
understanding the nature of dark energy.
We review the observational evidence for the current accelerated expansion 
of the universe and present a number of dark energy models 
in addition to the conventional cosmological constant, 
paying particular attention to scalar field models such as quintessence, 
K-essence, tachyon, phantom and dilatonic models.
The importance of cosmological scaling solutions is emphasized when  
studying the dynamical system of scalar fields including coupled dark 
energy. We study the evolution of cosmological perturbations allowing 
us to confront them 
with the observation of the Cosmic Microwave Background and 
Large Scale Structure and demonstrate how  
it is possible in principle to reconstruct the equation of state of dark energy
by also using Supernovae Ia observational data. 
We also discuss in detail the nature of tracking solutions in cosmology, 
particle physics and braneworld models of dark energy, the nature of 
possible future singularities, the effect of higher order curvature 
terms to avoid a Big Rip singularity, and approaches to modifying gravity 
which leads to a late-time accelerated expansion without recourse to 
a new form of dark energy.

\end{abstract}


\maketitle
\vskip 1pc

\begin{widetext} 
\tableofcontents{}
\end{widetext} 

\newpage

\section{Introduction}
\label{intro}

Over the course of the past decade, evidence for the most striking result in modern 
cosmology has been steadily growing, namely the existence of 
a cosmological constant which is driving the current acceleration 
of the Universe as first observed in Refs.~\cite{perlmutter,riess}. 
Although it may not have come as such a surprise to a few theorists 
who were at that time considering the interplay between a number 
of different types of observations  \cite{Krauss}, for the majority 
it came as something of a bombshell. The Universe is not only expanding, 
it is accelerating. The results first published in 
Refs.~\cite{perlmutter,riess} 
have caused a sea change in the way we have started thinking about the universe. 

Conventionally, the world of particle physics and cosmology has been seen as 
overlapping in the early universe, particle physics providing much needed 
sources of energy density during that period, leading to processes like inflation, 
baryogenesis, phase transitions etc... Now though we need to understand 
the impact particle physics has on cosmology today, how else 
can we explain the nature of this apparent cosmological constant? 
Theorists never short of ideas, have come up with a number of particle physics
related suggestions (as well as a number completely unrelated to particle physics) 
to help us understand the nature of the acceleration.

There is a key problem that we have to explain, 
and it is fair to say it has yet to be understood. 
The value of the energy density stored in the cosmological constant today, 
which rather paradoxically is called dark energy and has nothing to do 
with dark matter, this value has to be of order the critical density, 
namely $\rho_{\Lambda} \sim 10^{-3}\,{\rm eV}^4$. 
Unfortunately, no sensible explanation exists as to why a true 
cosmological constant should be at this scale, it should naturally 
be much larger. Typically, since it is conventionally associated 
with the energy of the vacuum in quantum theory we expect it to 
have a size of order the typical scale of early Universe phase transitions. 
Even at the QCD scale it would imply a value 
$\rho_{\Lambda} \sim 10^{-3}\,{\rm GeV}^4$. 
The question then remains, why has $\Lambda$ got the value it has today? 

Rather than dealing directly with the cosmological constant a number of alternative 
routes have been proposed which skirt around this thorny 
issue \cite{Varunreview,Carrollreview,Paddyreview,PR03,Norbert}. 
They come in a a number of flavors. 
An incomplete list includes: Quintessence 
models \cite{Wetterich88,peebles} (see also Refs.~\cite{earlyqu,Carroll:1998zi})
which invoke an evolving canonical scalar 
field  with a potential (effectively providing an inflaton for today) and 
makes use of the scaling properties \cite{Ferreira97,CLW}  and 
tracker nature \cite{caldwell98,Paul99} of such scalar fields evolving in the presence 
of other background matter fields; scalar field models where 
the small mass of the quintessence field is protected by an approximate 
global symmetry by making the field a pseudo-Nambu-Goldstone 
boson \cite{Frieman:1991tu}; Chameleon fields in which the scalar field 
couples to the baryon energy density and is homogeneous being allowed to 
vary across space from solar system to cosmological scales \cite{Khoury04,Brax04};  
a scalar field with a  non-canonical kinetic term, known as K-essence 
\cite{COY,AMS1,AMS2} 
based on earlier work of K-inflation \cite{kinf}; modified gravity arising out of both string 
motivated \cite{DGP} or more generally General Relativity 
modified \cite{Capo,Carroll:2003wy,odintsovl} actions which
both have the effect of introducing large length scale corrections and 
modifying the late time evolution of the Universe; the feedback of 
non-linearities into the evolution equations which can significantly 
change the background evolution and lead to acceleration at late 
times without introducing any new matter \cite{Ellis87}; 
Chaplygin gases which attempt to unify dark energy and 
dark matter under one umbrella by allowing for a fluid with an equation 
of state which evolves between the two \cite{Kamenshchik:2001cp,Bilic02,Bento:2002ps}; 
tachyons \cite{Padmanabhan:2002cp,Bagla03} arising in string theory \cite{Asen}; 
the same scalar field responsible for both inflation in the early Universe 
and again today, known as Quintessential inflation \cite{Peebles:1998qn}; 
the possibility of a network of frustrated topological defects forcing the universe 
into a period of accelerated expansion today \cite{Battye:1999eq}; 
Phantom Dark Energy \cite{Caldwell02} and Ghost Condensates \cite{Arkani-hamad,PT}; 
de-Sitter vacua with the flux compactifications in string theory \cite{KKLT};
the String Landscape arising from the multiple numbers of vacua that exist when the string
moduli are made stable as non-abelian fluxes are turned on \cite{Susskind:2003kw}; 
the Cyclic Universe \cite{cyclic}; causal sets in the context of Quantum Gravity \cite{sorkin};
direct anthropic arguments 
\cite{Linde:1984ir,Weinberg:1988cp,Efstathiou:1990xe,Garriga:1999bf}, 
all of these are more or less  exotic solutions to the dark energy question. 

These possibilities and more, have been discussed in the literature and many of them 
will be discussed in detail in this review.  Given the strength of the data which are all effectively indicating the presence of a cosmological 
constant type term today, then any dynamically evolving contribution must resemble a cosmological 
constant today. If we are to see evidence of dynamics in the dark energy equation of state, 
we have to probe back in time. A number of routes in that direction have been suggested 
and plans are underway to extend this even further. For example by looking at the detailed 
patterns of the anisotropies in the cosmic microwave background (CMB), 
we are seeing when and under what conditions the photons left the surface of
last scattering. As they propagated towards us today, they will have traveled 
through gravitational potentials determined by the nature of the 
dark matter and dark energy, and so different forms of dark energy could 
in principle have led to different contributions to quantities such as the separation of 
CMB Peaks \cite{Doran:2000jt,Doran:2001ty,Corasaniti:2001mf}, 
the integrated Sachs Wolfe effect \cite{Corasaniti:2004sz},
the nature of galaxy formation \cite{Alcaniz:2001uy}, the clustering of large scale structure (LSS) 
as measured through quantities such as $\sigma_8$ \cite{Doran:2001rw,Kunz03}, 
the propagation of light through weak and strong gravitational 
lenses \cite{Glensing,Weinberg:2002rd}, 
and simply through the evolution of the Hubble expansion rate itself which is 
a function of the energy contributions to the Friedmann equation \cite{statefinder}. 

On the other hand, what if the data is misleading us and we do not 
require an effective cosmological constant \cite{Blanchard:2003du} ? 
A minority of cosmologists have argued forcefully that the majority of 
the data as it presently stands can be  interpreted without recourse to a 
cosmological constant, rather we can explain it through other physical
processes, for example by relaxing the hypothesis that the fluctuation 
spectrum can be described by a single power law \cite{Blanchard:2003du}. 
On the other hand perhaps we do not yet fully understand how 
Type Ia supernova evolve and we may have to eventually think of 
alternative explanations. Although this might well be the case, 
there is a growing body of evidence for the presence of a cosmological 
constant which does not rely on the supernova data to support it 
(in relation to this and the comment above see Ref.~\cite{Blanchard:2005ev}). 

In the same vein Plaga recently discussed observations of a cluster of galaxies 
``Abell 194" and has argued that the distribution of galaxy redshifts 
is fitted better with an Einstein-Straus vacuole region of space time
as opposed to the cosmological concordance model 
with a $\Lambda$ \cite{Plaga:2005pa}. Of course, this is based on 
limited data, but we should remember the need to always be 
prepared to test the standard model against observation. 

However, the more accepted interpretation of the data is that it is 
becoming clear that consistency between the anisotropies in the 
CMB \cite{WMAP3,WMAP} and LSS \cite{Tegmark-sdss} 
observations imply we live in a Universe where the energy 
density is dominated by a cosmological constant type contribution. 
An impressive aspect of this consistency check is the fact that the 
physics associated with each epoch is completely different and of 
course it occurs on different time scales. It appears that consistency 
is obtained for a spatially flat universe with the fractional energy 
density in matter contributing today with 
$\Omega_{m}^{(0)} \sim 0.3$ 
whereas for the cosmological constant we have 
$\Omega_{\Lambda}^{(0)} \sim 0.7$ \cite{Seljak:2004xh}. 

In this review we assume that the dark energy is really there in some form, either dominating 
the energy density or through some form of modified gravity, in both cases 
driving our Universe into a second period  of accelerated 
expansion around a redshift of $z= {\cal O}(1)$. Most of the observational results are based on the years of analysing the first year WMAP data \cite{WMAP}, and has not yet reached the stage of analysing the beautiful new data published around the same time as this review was completed \cite{WMAP3}. We have attempted to include the new results where possible and where appropriate. Fortunately for us, many of the key results of WMAP1 have stood the test of time and statistics and appear to be holding true in the three year data as well (with some notable exceptions of course). 

Our goal is to introduce the reader to some of the theoretical model 
building that has gone into understanding the nature of dark energy. 
We will include string inspired models, uninspired models, 
phenomenological models, modified gravity models, etc.
We will look into the observational implications associated 
with dynamical dark energy, and investigate the ways we may 
determine whether or not there may be a $\Lambda$ term out
there governing our Universe today. 

Now a word of caution. The reader is about to spend a great deal 
of time learning (we hope!) about models of dark energy. The fact remains that 
although many of us believe some sort of dynamics is responsible 
for the dark energy, such is the sensitivity of current observations, 
there is no evidence of an evolving dark energy component, 
everything remains perfectly consistent with the simplest model 
(not from the particle physics point of view)  of a time 
independent cosmological constant \cite{Corasaniti:2004sz}. 
Indeed if we include the number of required extra parameters needed to 
allow for dynamical dark energy as a part of the selection criteria and 
apply Bayesian information criteria to carry out cosmological model selection, 
then there is no need at present to allow anything other than the 
cosmological constant \cite{Liddle:2004nh,Mukherjee:2005tr}.
Nevertheless this may change in the future as observations improve even more, 
and it remains  important to pursue alternative models of dark 
energy to distinguish them from the cosmological constant observationally.

Before we set off, it is worth mentioning here the approach we are 
adopting with regard the way we are classifying models, 
because to some, having a list of apparently unrelated possibilities 
may not seem the best way forward. We are treating all of these 
possibilities separately, whereas in principle a number of them 
can be related to each other as variants of theories carrying 
the same sort of signature -- see for example 
Refs.~\cite{Uzan:2004my,Uzan:2004qr,Uzan:2006mf}. 
Our reason for doing this is that we believe the models 
themselves have now become accepted in their own right 
and have had so much work done on them that they 
are better being treated separately without trying 
in this review to discuss the conformal transformations 
which link them - although we take on board 
the fact that some of them can be related. 

This paper is organized as follows. 
In Sec.~\ref{FRW} we introduce Einstein's equations in a homogeneous
and isotropic background and provide the basic tools to study the 
dynamics of dark energy.
In Sec.~\ref{obser} we discuss the observational evidence for 
dark energy coming from supernova constraints.
Sec.~\ref{Cconstant} is devoted to the discussion of the cosmological 
constant, whereas 
in Sec.~\ref{scalarmodel} we introduce a number of 
scalar-field dark energy models which can act as alternatives to the 
cosmological constant.
This is followed in Sec.~\ref{cdynamics} where the cosmological dynamics of 
scalar-field dark energy models in the presence of 
a barotropic fluid is presented.
In Sec.~\ref{scalingsec} we derive the condition for the 
existence of scaling solutions for more general scalar-field Lagrangians.
In Sec.~\ref{QK} we turn to discuss a number of  aspects of 
quintessence scenarios, paying particular attention to 
particle physics models of Quintessence.
In Sec.~\ref{cdenergy} we present coupled dark energy scenarios showing 
how accelerated expansion can be realized for a class of scaling solutions.
Sec.~\ref{valpha} is devoted to a discussion of varying fine structure 
constant ($\alpha$) models which although somewhat controversial 
opens up an important avenue, allowing us in principle to  distinguish 
between quintessence and a cosmological constant observationally.
In Sec.~\ref{perturbations} we study the evolution of 
cosmological perturbations in a dark energy universe and 
show several situations in which 
analytic solutions for perturbations can be obtained.
This is followed in Sec.~\ref{reconstruct} where we provide 
reconstruction equations
for a general scalar-field Lagrangian including a coupling to dark matter.
Sec.~\ref{eosobser} is devoted to a number of approaches to reconstructing
the equation of state of the dark energy by parameterizing it 
in terms of the redshift $z$. In Sec.~\ref{fate} we investigate a possibility 
that there may be future singularities 
in a dark energy scenario, and classify these into five classes.
In Sec.~\ref{hcorrections} we study the effect of higher-order
curvature terms to the cosmological evolution around the 
singularities discussed in Sec.~\ref{fate} and in  Sec.~\ref{modified} 
we discuss modified gravity theories in  which an accelerated 
expansion can be realized without recourse to dark energy. 
We conclude in the final section.

Throughout the review we adopt  natural units $c=\hbar=1$ 
and have a metric signature $(-,+,+,+)$.
We denote the Planck mass as 
$m_{\rm pl}=G^{-1/2}=1.22 \times 10^{19}\,{\rm GeV}$
and the reduced Planck mass as 
$M_{\rm pl}=(8\pi G)^{-1/2}=2.44 \times 10^{18}\,{\rm GeV}$.
Here $G$ is Newton's gravitational constant.
We define $\kappa^2=8\pi G=8\pi m_{\rm pl}^{-2}=M_{\rm pl}^{-2}$
and will use the unit $\kappa^2=1$ in some sections (but will make it clear when we are doing so).

Finally we would like to a provide guide lines for approaching this review.
Some of the sections/subsections are of specific interest and may be
skipped over in the first reading. For many, it may be preferable first time round to skip over the details of the  KKLT scenario described 
in Sec.~\ref{Cconstant}. Similarly a  brief look at sections \ref{fate} and \ref{hcorrections} 
may be sufficient for a first reading of the review.

\section{Elements of FRW cosmology}
\label{FRW}

The dynamics of the universe is described by the Einstein equations which are in general
complicated non-linear equations. However they exhibit simple analytical 
solutions in the presence of generic symmetries. 
The Friedmann-Robertson-Walker (FRW) metric is based
upon the assumption of homogeneity and isotropy of the universe 
which is approximately true on large scales. 
The small deviation from homogeneity at early epochs played a very 
important role in the dynamical history of our universe. 
Small initial density perturbations grew
via gravitational instability into the structure we see today in the universe.
The temperature anisotropies observed in the Cosmic Microwave 
Background (CMB) are believed to have originated from quantum 
fluctuations generated during an inflationary stage in the early universe. 
See Refs.~\cite{LLbook,Lidsey95,Riotto02,BTW,scott,Padmanabhan:2006kz,Trodden:2004st} 
for details on density perturbations
predicted by inflationary cosmology.
In this section we shall review the main features of 
the homogeneous and isotropic cosmology
necessary for the subsequent sections.

The FRW metric is given 
by \cite{Weinbergbook,Rocky,Paddybook,LLbook}
\begin{eqnarray}
\rd s^2=-\rd t^2+a^2(t)\left[\frac{\rd r^2}{1-Kr^2}
+r^2(\rd\theta^2+\sin^2\theta\rd\phi^2)\right] 
\nonumber \,, \\
\label{frwmet}
\end{eqnarray}
where $a(t)$ is scale factor with cosmic time $t$. 
The coordinates $r$, $\theta$ and $\phi$ are known as {\it
comoving} coordinates. 
A freely moving particle comes to rest in these coordinates.

Equation (\ref{frwmet}) is a purely kinematic statement. In this problem the dynamics
is associated with the scale factor-- $a(t)$. Einstein equations allow us to determine
the scale factor provided the matter content of the universe is specified. The 
constant $K$ in the metric (\ref{frwmet}) describes the geometry
of the spatial section of space time, with closed, flat and open universes
corresponding to $K=+1, 0, -1$, respectively.

It may be convenient to write the metric (\ref{frwmet}) in the 
following form:
\begin{eqnarray}
\rd s^2=-\rd t^2+a^2(t)\left[\rd \chi^2+f_K^2 (\chi) 
(\rd\theta^2+\sin^2\theta\rd\phi^2)\right]\,,
\label{frwmet2}
\end{eqnarray}
where 
\begin{eqnarray}
f_K (\chi) =  \left\{\begin{array}{lll}
{\rm sin} \chi\,, \quad & K=+1\,, \\
\label{fK0}
\chi\,, \quad & K=0\,, \\
{\rm sin h} \chi\,, \quad & K=-1\,.
\end{array} \right. 
\end{eqnarray}

\subsection{Evolution equations}

The differential equations for the scale factor and the matter density follow from
Einstein's equations \cite{Weinbergbook}
\begin{equation}
G^{\mu}_{\nu} \equiv R^{\mu}_{\nu}-\frac{1}{2}
\delta^{\mu}_{\nu}R=8 \pi G T^{\mu}_{\nu}\,,
\label{Einsteineq}
\end{equation}
where $G^{\mu}_{\nu}$ is the Einstein tensor, and 
$R^{\mu}_{\nu}$ is the Ricci tensor
which depends on the metric and its derivatives, $R$ is the Ricci scalar
and $T^{\mu}_{\nu}$ is the energy momentum tensor.
In the FRW background (\ref{frwmet}) the curvature terms 
are given by \cite{Rocky}
\begin{eqnarray}
R_{0}^0 &=& \frac{3\ddot{a}}{a}\,, \\
R^{i}_j&=& \left(\frac{\ddot{a}}{a}+\frac{2\dot{a}^2}{a^2}+
\frac{2K}{a^2} \right) \delta^{i}_j\,, \\
R&=&6 \left(\frac{\ddot{a}}{a}+\frac{\dot{a}^2}{a^2}
+\frac{K}{a^2}\right)\,,
\end{eqnarray}
where a dot denotes a derivative with respect to $t$.

Let us consider an ideal perfect fluid as the source of 
the energy momentum tensor $T^{\mu}_{\nu}$.
In this case we have
\begin{equation}
T^{\mu}_{\nu}={\rm Diag} \left(-\rho,p,p,p \right)\,,
\end{equation}
where $\rho$ and $p$ are the energy density and the pressure 
density of the fluid, respectively.
Then Eq.~(\ref{Einsteineq}) gives the two independent equations
\begin{eqnarray}
 \label{HubbleeqI}
&& H^2 \equiv \left(\frac{\dot{a}}{a}\right)^2
=\frac {8\pi G \rho}{3}-\frac {K}{a^2}\,, \\
\label{dotHeq}
&&\dot{H}=-4\pi G(p+\rho)+\frac{K}{a^2}\,,
\end{eqnarray}
where $H$ is the Hubble parameter, $\rho$ and $p$ denote the
total energy density and pressure of all the species present
in the universe at a given epoch.

The energy momentum tensor is conserved by virtue of the Bianchi 
identities, leading to the continuity equation
\begin{equation}
\dot{\rho}+3 H(\rho+p)=0\,.
\label{conteq}
\end{equation}
Equation (\ref{conteq}) can be derived from 
Eqs.~(\ref{HubbleeqI}) and (\ref{dotHeq}), 
which means that two of Eqs.~(\ref{HubbleeqI}), (\ref{dotHeq}) and 
(\ref{conteq}) are independent. 
Eliminating the $K/a^2$ term 
from Eqs.~(\ref{HubbleeqI}) and (\ref{dotHeq}), we obtain
\begin{eqnarray}
\label{acceleq}
\frac{\ddot{a}}{a}=
-\frac {4 \pi G}{3} \left(\rho+3p\right)\,.
\end{eqnarray}
Hence the accelerated expansion occurs for $\rho+3p<0$.

One can rewrite Eq.~(\ref{HubbleeqI}) in the form:
\begin{equation}
\Omega(t)-1=
\frac{K}{(aH)^2}\,,
\end{equation}
where $\Omega(t) \equiv \rho(t)/\rho_c(t)$ is the dimensionless density parameter 
and $\rho_c(t)=3H^2(t)/8 \pi G$ is
the critical density. 
The matter distribution clearly determines the spatial geometry of 
our universe, i.e., 
\begin{eqnarray}
&& \Omega >1~~{\rm or}~~\rho>\rho_c~~\to K=+1\,, \\
&& \Omega =1~~{\rm or}~~\rho=\rho_c~~\to K=0\,, \\
&&\Omega <1~~{\rm or}~~\rho<\rho_c~~\to K=-1\,.
\end{eqnarray}
Observations have shown that the current universe is very close to 
a spatially flat geometry ($\Omega \simeq 1$) \cite{WMAP3}.
This is actually a natural result from inflation in the early 
universe \cite{LLbook}. 
Hence we will therefore consider a flat universe ($K=0$)
in the rest of this section.

\subsection{The evolution of the universe filled with a perfect fluid}

Let us consider the evolution of the universe filled with a barotropic 
perfect fluid with an equation of state 
\begin{equation}
w=p/\rho\,,
\end{equation}
where $w$ is assumed to be constant.
Then by solving the Einstein equations given in Eqs.~(\ref{HubbleeqI})
and (\ref{dotHeq}) with $K=0$, we obtain
\begin{eqnarray}
 \label{sol1}   
&& H =\frac{2}{3(1+w)(t-t_{0})}\,, \\
\label{sol2} 
&& a(t) \propto (t-t_{0})^{\frac{2}{3(1+w)}} \,, \\
\label{sol3} 
&& \rho \propto a^{-3(1+w)}\,,
\end{eqnarray}
where $t_{0}$ is constant.
We note that the above solution is valid for $w \neq -1$.
The radiation dominated universe corresponds to $w=1/3$, whereas
the dust dominated universe to $w=0$. 
In these cases we have 
\begin{eqnarray}
{\rm Radiation}:~~a(t) \propto (t-t_{0})^{1/2}\,,~~~
\rho \propto a^{-4}\,, \\
{\rm Dust}:~~a(t) \propto (t-t_{0})^{2/3}\,,
~~~\rho \propto a^{-3}\,.
\end{eqnarray}
Both cases correspond to a decelerated 
expansion of the universe.

{}From Eq.~(\ref{acceleq}) an accelerated expansion 
($\ddot{a}(t)>0$) occurs for the equation of state given by 
\begin{eqnarray}
w<-1/3\,.
\label{dcondition}
\end{eqnarray}
In order to explain the current acceleration of the universe, 
we require an exotic energy dubbed ``dark energy'' 
with equation of state satisfying Eq.~(\ref{dcondition}).
We note that Newton gravity can not account for the accelerated 
expansion. 
Let us consider a homogeneous sphere whose radius and energy density 
are $a$ and $\rho$, respectively.
The Newton's equation of motion for a point particle with mass $m$
on this sphere is give by 
\begin{eqnarray}
m \ddot{a} &=&
-\frac{Gm}{a^2} \left(\frac{4\pi a^3 \rho}
{3} \right)\,, \nonumber \\
\to \quad
\frac{\ddot{a}}{a} &=&
-\frac{4\pi G}{3}\rho\,.
\label{Newtoneq}
\end{eqnarray}
The difference compared to the Einstein equation (\ref{acceleq})
is the absence of the pressure term, $p$. 
This appears in Einstein
equations by virtue of relativistic effects.
The condition (\ref{dcondition}) means that we essentially require 
a large negative pressure in order to give rise to an accelerated expansion.
We stress here that Newton gravity only leads to a decelerated expansion 
of the universe.

{}From Eq.~(\ref{conteq}) the energy density $\rho$ is constant 
for $w=-1$. In this case the Hubble rate is also 
constant from Eq.~(\ref{HubbleeqI}), 
giving the evolution of the scale factor:
\begin{eqnarray}
a \propto e^{Ht}\,,
\end{eqnarray}
which is the de-Sitter universe.
As we will see in the Sec.~\ref{Cconstant}, this exponential expansion 
also arises by including a cosmological 
constant, $\Lambda$, in the Einstein equations.

So far we have restricted our attention to the equation of state: $w\ge -1$. 
Recent observations suggest that the equation of state which is less than 
$-1$ can be also allowed \cite{ASSS}. 
This specific equation of state corresponds to a 
{\it phantom} (ghost) dark energy \cite{Caldwell02} component 
and requires a separate consideration
(see also Ref.~\cite{staphan}).
We first note that Eq.~(\ref{sol2}) describes
a contracting universe for $w<-1$.
There is another expanding solution given by 
\begin{equation}
a(t)=(t_s-t)^{\frac{2}{3(1+w)}}\,,
\end{equation}
where $t_s$ is constant.
This corresponds to a super-inflationary solution 
where the Hubble rate and the scalar curvature grow:
\begin{eqnarray}
& &H=\frac{n}{t_s-t}\,,~~~n=-\frac{2}{3(1+w)}>0\,, \\
& &R=6\left(2H^2+\dot{H}\right)
=\frac{6n(2n+1)}{(t_s-t)^{2}}\,.
\label{phantomevol}
\end{eqnarray}

The Hubble rate diverges as $t\to t_s$, which corresponds to
an infinitely large energy density at a finite time in the future. The
curvature also grows to infinity as $t\to t_s$. 
Such a situation is referred to as a Big Rip singularity \cite{CKW}. 
This cataclysmic conclusion is not inevitable in these models, 
and can be avoided in specific models of phantom fields 
with a top-hat potential \cite{Carroll03,SSN}.
It should also be emphasized that we expect quantum 
effects to become important
in a situation when the curvature of the universe becomes large. 
In that case we should take into account higher-order 
curvature corrections to the Einstein Hilbert action which 
crucially modifies the structure of
the singularity, as we will see in Sec.~\ref{fate}.

\section{Observational evidence for dark energy}
\label{obser}

In this section we briefly review the observational evidence 
for dark energy, concentrating on the types of observation 
that have been introduced. 
Later, in Sec.~\ref{eosobser} we will return to discuss 
in more detail the observational constraints on 
the dark energy equation of state. 

\subsection{Luminosity distance}

In 1998 the accelerated expansion of 
the universe was pointed out by two groups 
from the observations of Type Ia Supernova
(SN Ia)  \cite{perlmutter,riess}.
We often use a redshift to describe the evolution of the universe. 
This is related
to the fact that light emitted by a stellar object becomes red-shifted due 
to the expansion of the universe. 
The wavelength $\lambda$ increases proportionally to the scale 
factor $a$, whose effect can be quantified by the redshift $z$, as
\begin{equation}
\label{redshift}    
1+z=\frac{\lambda_0}{\lambda} 
=\frac{a_{0}}{a}\,,
\end{equation}
where the subscript zero denotes the quantities given at the present epoch.

Another important concept related to observational tools 
in an expanding background is
associated to the definition of a distance. 
In fact there are several ways of measuring distances 
in the expanding universe.
For instance one often deals with the comoving distance 
which remains unchanged during the evolution
and the physical distance which scales proportionally to the scale factor.
An alternative way of defining a distance is through the luminosity 
of a stellar object.
The distance $d_L$ known as the luminosity distance, plays 
a very important role in astronomy
including  the Supernova observations.

In Minkowski space time the absolute luminosity $L_{s}$ 
of the source and the energy flux ${\cal F}$ at a distance 
$d$ is related through ${\cal F}=L_s/(4\pi d^2)$.
By generalizing this to an expanding universe, the luminosity 
distance, $d_{L}$,  is defined as
\begin{equation}
 \label{dldef}
d_{L}^2 \equiv \frac{L_s}{4\pi {\cal F}}\,.
\end{equation}
Let us consider an object with absolute luminosity $L_{s}$
located at a coordinate distance $\chi_{s}$ from an observer at $\chi=0$
[see the metric (\ref{frwmet2})].
The energy of light emitted from the object with 
time interval $\Delta t_1$ is denoted as $\Delta E_1$, whereas 
the energy which reaches at the sphere with radius $\chi_{s}$ is 
written as $\Delta E_0$. We note that $\Delta E_1$
and $\Delta E_0$ are proportional to the frequencies of 
light at $\chi=\chi_s$ and $\chi=0$, respectively, 
i.e., $\Delta E_{1} \propto \nu_1$ and $\Delta E_{0} \propto \nu_{0}$.
The luminosities $L_{s}$ and $L_0$ are given by 
\begin{equation}
 \label{LSL0}
 L_{s}=\frac{\Delta E_1}{\Delta t_{1}}\,, \quad 
 L_0=\frac{\Delta E_0}{\Delta t_{0}}\,.
\end{equation}
The speed of light is given by $c=\nu_ 1\lambda_1
=\nu_0 \lambda_0$, where $\lambda_1$ and $\lambda_0$
are the wavelengths at $\chi=\chi_{s}$ and $\chi=0$.
Then from Eq.~(\ref{redshift}) we find 
\begin{equation}
 \label{lamre}
\frac{\lambda_{0}}{\lambda_{1}}=\frac{\nu_{1}}{\nu_{0}}
=\frac{\Delta t_{0}}{\Delta t_{1}}=\frac{\Delta E_{1}}{\Delta E_{0}}
=1+z\,,
\end{equation}
where we have also used $\nu_{0} \Delta t_0=\nu_1 \Delta t_{1}$.
Combining Eq.~(\ref{LSL0}) with Eq.~(\ref{lamre}), 
we obtain 
\begin{equation}
L_{s}=L_0 (1+z)^2\,.
\end{equation}

The light traveling along the $\chi$
direction satisfies the geodesic equation 
$\rd s^2=-\rd t^2+a^2(t)\rd \chi^2=0$. 
We then obtain
\begin{equation}
 \label{chis}
\chi_{s}=\int_{0}^{\chi_{s}} \rd \chi=
\int_{t_1}^{t_{0}} \frac{\rd t}{a(t)}
=\frac{1}{a_0H_0} \int_{0}^z
\frac{\rd z'}{h(z')}\,,
\end{equation}
where $h(z)=H(z)/H_0$. Note that we have used the relation $\dot{z}=-H(1+z)$
coming from Eq.~(\ref{redshift}). From the metric (\ref{frwmet2}) we find that 
the area of the sphere at $t=t_0$
is given by $S=4\pi (a_{0}f_K(\chi_{s}))^2$. 
Hence the observed energy flux is 
\begin{equation}
\label{flux}    
{\cal F}=\frac{L_0}{4\pi (a_{0}f_K(\chi_{s}))^2}\,.
\end{equation}
Substituting Eqs.~(\ref{chis}) and (\ref{flux}) for Eq.~(\ref{dldef}), 
we obtain the luminosity distance in an expanding universe:
\begin{equation}
d_{L}=a_0 f_K(\chi_{s}) (1+z)\,.
\end{equation}

In the flat FRW background with $f_{K}(\chi)=\chi$
we find 
\begin{equation}
 \label{dLform}
d_{L}=\frac{1+z}{H_{0}}
\int_{0}^z \frac{\rd z'}{h(z')}\,,
\end{equation}
where we have used Eq.~(\ref{chis}).
Then the Hubble rate $H(z)$ can be expressed 
in terms of $d_{L}(z)$:
\begin{equation}
H(z)=\left\{ \frac{\rd}{\rd z}
\left(\frac{d_{L}(z)}{1+z}\right)\right\}^{-1}\,.
\label{HdLz}
\end{equation}
If we measure the luminosity distance observationally, 
we can determine the expansion rate of the universe.

The energy density $\rho$ on the right hand side 
of Eq.~(\ref{HubbleeqI}) includes all components 
present in the universe, namely,
non-relativistic particles, relativistic particles, 
cosmological constant, etc:
\begin{equation}
\rho=\sum_i \rho^{(0)}_i (a/a_{0})^{-3(1+w_{i})}
=\sum_i{\rho_i^{(0)} (1+z)^{3(1+w_i)}}\,,
\label{rhoz}
\end{equation}
where we have used Eq.~(\ref{redshift}).
Here $w_{i}$ and $\rho^{(0)}_i$ correspond to 
the equation of state and the present energy density of each 
component, respectively.

\begin{figure}
\includegraphics[height=3.2in,width=3.3in]{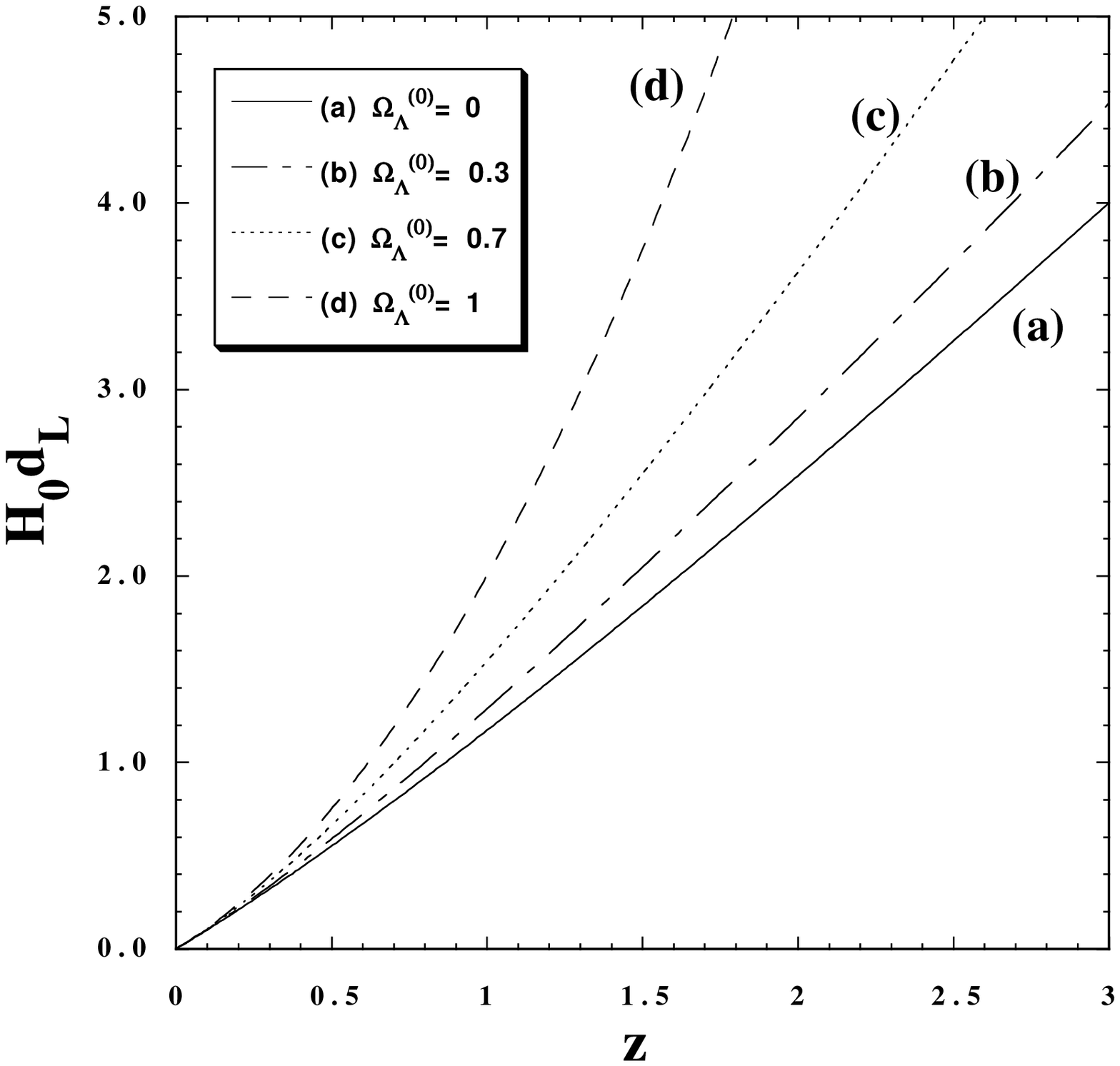}
\caption{Luminosity distance $d_{L}$ in the units of 
$H_{0}^{-1}$ for a two component flat 
universe with a non-relativistic fluid ($w_{m}=0$)
and a cosmological constant ($w_{\Lambda}=-1$).
We plot $H_{0}d_{L}$ for various values of
$\Omega_{\Lambda}^{(0)}$.}
\label{distancefig}
\end{figure}

Then from Eq.~(\ref{HubbleeqI}) 
the Hubble parameter takes the convenient form
\begin{equation}
H^2=H^2_0 
\sum_i{\Omega^{(0)}_i(1+z)^{3(1+w_i)}} \,,
\label{Hz}
\end{equation}
where $\Omega^{(0)}_i \equiv 
8\pi G\rho_i^{(0)}/(3H_0^2)=\rho_i^{(0)}/\rho_c^{(0)}$ 
is the density parameter 
for an individual component at the present epoch.
Hence the luminosity distance in a flat geometry is given by 
\begin{equation}
d_L=\frac{(1+z)}{H_0}\int^z_0{\frac {\rd z'}
{\sqrt{\sum_i{\Omega_i^{(0)} (1+z')^{3(1+w_i)}}}}}\,.
\label{DLF}
\end{equation}
In Fig.~\ref{distancefig} we plot the luminosity distance (\ref{DLF})
for a two component flat universe (non-relativistic fluid with $w_m=0$
and cosmological constant with $w_{\Lambda}=-1$) satisfying 
$\Omega_{m}^{(0)}+\Omega_{\Lambda}^{(0)}=1$.
Notice that $d_L \simeq z/H_0$ for small values of $z$.
The luminosity distance becomes  larger when the cosmological 
constant is present.

%
\subsection{Constraints from Supernovae Ia}

The direct evidence for the current acceleration of the universe is
related to the observation of luminosity distances of
high redshift supernovae \cite{perlmutter,riess}.
The apparent magnitude $m$ of the source with an 
absolute magnitude $M$ is related to the luminosity 
distance $d_{L}$ via the 
relation \cite{Varunreview,Paddyreview}
\begin{equation}
m-M=5 \log_{10}
\left(\frac{d_L}{{\rm Mpc}}\right)+25\,.
\label{magnitude}
\end{equation}
This comes from taking the logarithm of Eq.~(\ref{dldef})
by noting that $m$ and $M$ are related to the 
logarithms of ${\cal F}$ and $L_s$, respectively.
The numerical factors arise because of 
conventional definitions of $m$ and $M$ in astronomy.

The Type Ia supernova (SN Ia) can be observed when white dwarf
stars exceed the mass of the Chandrasekhar limit and explode.
The belief is that SN Ia are formed in the same way irrespective of 
where they are in the universe, which means that they have a 
common absolute magnitude $M$ independent of 
the redshift $z$. Thus they can be treated as an ideal standard  
candle. We can measure the apparent magnitude $m$ and the redshift 
$z$ observationally, which of course depends upon the objects we observe.

In order to get a feeling of the phenomenon let us consider
two supernovae 1992P at low-redshift $z=0.026$ with 
$m=16.08$ and 1997ap at high-redshift redshift $z=0.83$ 
with $m=24.32$ \cite{perlmutter}.
As we have already mentioned, the luminosity distance is approximately 
given by $d_L(z) \simeq z/H_0$ for $z \ll 1$.
Using the apparent magnitude $m=16.08$ of 1992P at $z=0.026$, 
we find that the absolute
magnitude is estimated by $M=-19.09$ from Eq.~(\ref{magnitude}). 
Here we adopted the value $H_{0}^{-1}=2998h^{-1}\,{\rm Mpc}$ 
with $h=0.72$.
Then the luminosity distance of 1997ap is obtained
by substituting $m=24.32$ and $M=-19.09$ for Eq.~(\ref{magnitude}):
\begin{equation}
H_0d_L \simeq 1.16\,,~~~{\rm for}~~~
z=0.83\,.
\end{equation}
{}From Eq.~(\ref{DLF}) the theoretical estimate for the luminosity distance
in a two component flat universe is
\begin{eqnarray}
&& H_{0}d_{L} \simeq 0.95,~~~~\Omega_m^{(0)} \simeq 1 \,, \\
&& H_{0}d_{L} \simeq 1.23,~~~~\Omega_m^{(0)} 
\simeq 0.3,~\Omega_{\Lambda}^{(0)} \simeq 0.7\,.
\end{eqnarray}
This estimation is clearly consistent 
with that required for a dark energy dominated universe as can be seen also in  
Fig.~\ref{distancefig}.

Of course, from a statistical point of view, one can not strongly claim that 
that our universe is really accelerating
by just picking up a single data set.
Up to 1998 Perlmutter {\it et al.} [supernova cosmology project (SCP)] 
had discovered 42 SN Ia in the 
redshift range $z=0.18$-$0.83$ \cite{perlmutter}, whereas 
Riess {\it et al.} [high-$z$ supernova team (HSST)]
had found 14 SN Ia in the range $z=0.16$-$0.62$ and 34 
nearby SN Ia \cite{riess}.
Assuming a flat universe
($\Omega_m^{(0)}+\Omega_\Lambda^{(0)}=1$), 
Perlmutter {\it et al.} found 
$\Omega_m^{(0)}=0.28^{+0.09}_{-0.08}$ (1$\sigma$ statistical) 
${}^{+0.05}_{-0.04}$ (identified systematics), thus 
showing that about 70 \%
of the energy density of the present 
universe consists of dark energy.

\begin{figure}
\includegraphics[height=3.2in,width=3.2in]{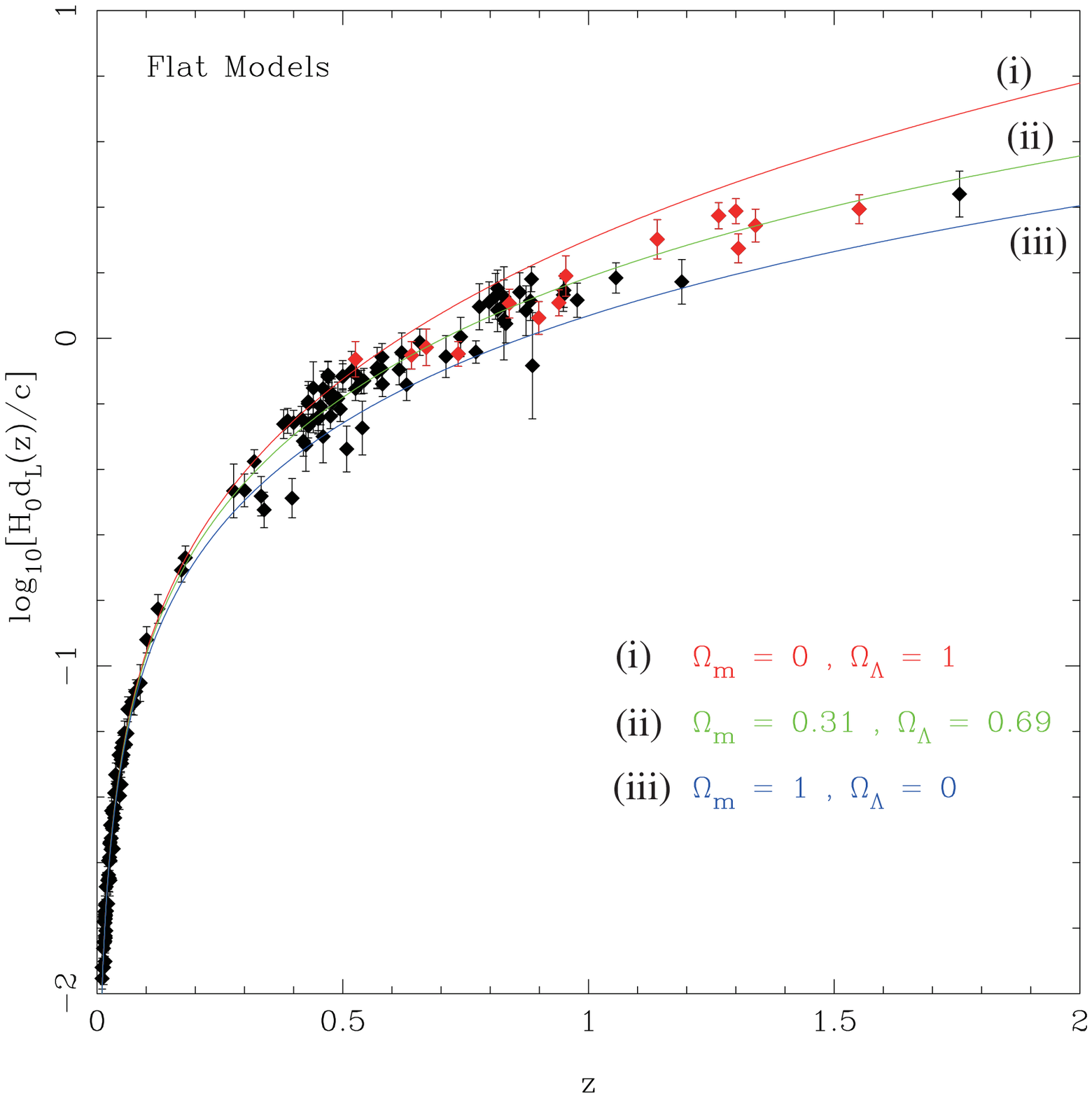}
\caption{The luminosity distance $H_{0}d_{L}$ (log plot)
versus the redshift $z$ for a flat cosmological model.
The black points come from the ``Gold'' data sets by Riess 
{\it et al.} \cite{riess2}, whereas the red points show the recent
data from HST.
Three curves show the theoretical values 
of $H_{0}d_{L}$ for (i) $\Omega_{m}^{(0)}=0$, 
$\Omega_{\Lambda}^{(0)}=1$, 
(ii) $\Omega_{m}^{(0)}=0.31$, 
$\Omega_{\Lambda}^{(0)}=0.69$ and 
(iii) $\Omega_{m}^{(0)}=1$, 
$\Omega_{\Lambda}^{(0)}=0$.
{}From Ref.~\cite{CP05}.}
\label{fitting}
\end{figure}

In 2004 Riess {\it et al.} \cite{riess2} reported the measurement of 16 
high-redshift SN Ia with redshift $z>1.25$ with the Hubble Space 
Telescope (HST). By including 170 previously known SN Ia data points, they 
showed that the universe exhibited a transition from deceleration to 
acceleration at $>99$ \% confidence level.
A best-fit value of $\Omega_m^{(0)}$ was found to be 
$\Omega_m^{(0)}=0.29^{+0.05}_{-0.03}$ (the error bar is $1\sigma$).
In Ref.~\cite{CP05} a likelihood analysis was performed by including the 
SN data set by Tonry {\it et al.} \cite{Tonry} together with the one by 
Riess {\it et al.} \cite{riess2}.
Figure \ref{fitting} illustrates the observational values of 
the luminosity distance $d_{L}$ versus redshift $z$
together with the theoretical curves derived from 
Eq.~(\ref{DLF}).
This shows that a matter dominated universe without a cosmological 
constant ($\om=1$) does not fit to the data.
A best-fit value of $\om$ obtained in a joint analysis 
of Ref.~\cite{CP05} is 
$\om=0.31^{+0.08}_{-0.08}$, which is consistent with the result by 
Riess {\it et al.} \cite{riess2}.
See also Refs.~\cite{SNobser} for 
recent papers about the SN Ia data analysis.

In Ref.~\cite{Jassal05}, a comparison is made of the constraints on models of dark energy from
supernova and CMB observations. The authors argue that models preferred by these
observations lie in distinct parts of the parameter space but there is no
overlap of regions allowed at the $68\%$ confidence level. They go on to
 suggest that this may indicate unresolved systematic errors in one of the
observations, with supernova observations being more likely to suffer from
this problem due to the very heterogeneous nature of the data sets available
at the time. 
Recently observations of high redshift supernovae from the SuperNova Legacy
Survey have been released \cite{Astier}. The survey has aimed to reduce systematic errors
by using only high quality observations based on using a single instrument to
observe the fields.  The claim is that through a rolling search technique the sources 
are not lost and data is of superior quality. 
Jassal {\it et al.} claim that the data set is in 
better agreement with WMAP \cite{Jassal:2006gf}. 
In other words the high redshift supernova data from 
the SNLS (SuperNova Legacy Survey) project is 
in excellent agreement with CMB observations. 
It leaves open the current state of supernova observations 
and their analysis, as compared to that of the CMB. 
The former is still in a state of flux and any conclusions 
reached using them need to be understood 
giving due regard to underlying assumptions.

It should be emphasized that the accelerated expansion is by cosmological standards really a 
late-time phenomenon, starting at a redshift $z \sim 1$. 
{}From Eq.~(\ref{Hz}) the deceleration parameter, 
$q \equiv -a\ddot{a}/\dot{a}^2$, is given by 
\begin{equation}
q(z)=\frac32 \frac{\sum_{i} \Omega_{i}^{(0)}
(1+w_i)(1+z)^{3(1+w_i)}}
{\sum_{i}  \Omega_{i}^{(0)} (1+z)^{3(1+w_i)}}-1\,.
\label{qeq}
\end{equation}
For the two component flat cosmology, the universe enters an 
accelerating phase ($q<0$) for 
\begin{equation}
\label{zccon}
z<z_{c} \equiv 
\left(\frac{2\Omega_{\Lambda}^{(0)}}
{\Omega_m^{(0)}}\right)^{1/3}-1\,.
\end{equation}
When $\Omega_m^{(0)}=0.3$ and 
$\Omega_\Lambda^{(0)}=0.7$, we have $z_c=0.67$.
The problem of why an accelerated expansion should occur now in the 
long history of the universe is called the ``coincidence problem''.

We have concentrated in this section on the use of 
SN Ia as standard candles. There are other possible candles 
that have been proposed and are actively being investigated. 
One such approach has been to use FRIIb radio 
galaxies \cite{Podariu:2002jj,Daly:2002kn}. 
{}From the corresponding redshift-angular size data 
it is possible to constrain cosmological parameters in 
a dark energy scalar field model. 
The derived constraints are found to be consistent with 
but generally weaker than those determined using 
Type Ia supernova redshift-magnitude data. 

However, in Ref.~\cite{Daly:2003iy}, the authors have gone further and 
developed a model-independent approach (i.e. independent of 
assumptions about the form of the dark energy) using a set of 
20 radio galaxies out to a redshift $z \sim 1.8$, 
which is further than the SN Ia data can reach. 
They conclude that the current observations indicate the universe 
transits from acceleration to deceleration at a redshift greater than 0.3, 
with a best fit estimate of about 0.45, and have best fit values 
for the matter and dark energy contributions to $\Omega$ in broad 
agreement with the SN Ia estimates. 

Another suggested standard candle is that of Gamma Ray Bursts 
(GRB), which may enable the expansion 
rate of our Universe to be measured out to 
very high redshifts ($z > 5$). 
Hooper and Dodelson \cite{Hooper:2005xx} have explored 
this possibility and found that GRB have the potential to 
detect dark energy at high statistical significance, 
but in the short term are unlikely 
to be competitive with future supernovae missions, 
such as SNAP, in measuring the properties of the dark energy. 
If however, it turns out there is appreciable dark energy 
at early times, GRB's will provide an excellent probe of 
that regime, and will be a real complement for the SN Ia data. 
This is a rapidly evolving field and there has recently been 
announced tentative evidence for a dynamical equation of 
state for dark energy, based on GRB data out to redshifts 
of order 5 \cite{G-R-B}. 
It is far too early to say whether this is the correct 
interpretation, or whether GRB are good standard candles, 
but the very fact they can be seen out to such large redshifts, 
means that if they do turn out to be standard candles, 
they will be very significant complements 
to the SN Ia data sets, and potentially more significant.  

\subsection{The age of the universe and the cosmological constant}

Another interesting piece of evidence for the existence of a
cosmological constant emerges when we compare 
the age of the universe ($t_{0}$) to the age of the oldest stellar populations ($t_{s}$).
For consistency we of course require $t_{0}>t_{s}$, but it is difficult to 
satisfy this condition for a flat cosmological model  
with a normal form of matter as we will see below.
Remarkably, the presence of cosmological constant 
can resolve this age problem.

First we briefly mention the age of the oldest stellar objects
have been constrained by a number of groups.
For example, Jimenez {\it et al.} \cite{Jimenez} determined
the age of Globular clusters in the Milky Way
as $t_{1}=13.5 \pm 2$\,Gyr by using a distance-independent 
method.
Using the white dwarfs cooling sequence method, 
Richer {\it et al.} \cite{Richer02} 
and Hansen {\it et al.} \cite{Hansen02} 
constrained the age of the globular cluster M4
to be $t_{1}=12.7 \pm 0.7$\,Gyr.
Then the age of the universe needs to satisfy the lower bound:
$t_0>11$-12\,Gyr. Assuming a $\Lambda$CDM model, 
the most recent WMAP3 data produces 
a best fit value of $t_0= 13.73^{+0.13}_{-0.17}$ 
Gyrs for the age of the universe \cite{WMAP3}.

Let us calculate the age of the universe from the Friedmann equation 
(\ref{HubbleeqI}) with $\rho$ given by (\ref{rhoz}).
We shall consider three contributions: 
radiation ($w_r=1/3$), pressureless dust ($w_{m}=0$)
and cosmological constant ($w_{\Lambda}=-1$).
Then Eq.~(\ref{HubbleeqI}) is written as 
\ba
H^2 &=& 
H_0^2 [\Omega_r^{(0)} (a/a_{0})^{-4}
+\Omega_m^{(0)} (a/a_0)^{-3} \nonumber \\
& & +\Omega_{\Lambda}^{(0)}-
\Omega_{K}^{(0)} (a/a_0)^{-2}]\,,
\label{Hubblere}
\ea
where $\Omega_{K}^{(0)} \equiv K/(a_0^2H_0^2)$.
Then by using Eq.~(\ref{redshift}) one can express $H$ in terms 
of $z$. The age of the universe is given by 
\ba
t_0 &=& \int_{0}^{t_0} \rd t=\int_{0}^{\infty}
\frac{\rd z}{H(1+z)} \nonumber \\
&=& \int_{0}^{\infty}
\frac{\rd z}{H_0x [\Omega_r^{(0)} x^4
+\Omega_{m}^{(0)}
x^3+\Omega_{\Lambda}^{(0)}
-\Omega_{K}^{(0)}x^2]^{1/2}}\,, \nonumber \\
\label{agecal}
\ea
where $x(z) \equiv 1+z$.
It is a good approximation 
to neglect the contribution of the radiation term 
in Eq.~(\ref{agecal}) since the radiation 
dominated period is much shorter than the total 
age of the universe. In other words the integral 
coming from the region $z \gtrsim 1000$ hardly affects the 
total integral (\ref{agecal}).
Hence we set $\Omega_r^{(0)}=0$ when we evaluate $t_{0}$.

We shall first study the case in which the cosmological constant 
is absent ($\Omega_{\Lambda}^{(0)}=0$).
Since $\Omega_{K}^{(0)}=\Omega_{m}^{(0)}-1$ 
from Eq.~(\ref{Hubblere}), 
the age of the universe is given by 
\ba
t_0 =\frac{1}{H_{0}} \int_{0}^{\infty}
\frac{\rd z}{(1+z)^2\sqrt{1+
\Omega_{m}^{(0)} z}}\,.
\label{tzero}
\ea
For a flat universe 
($\Omega_{K}^{(0)}=0$ and $\Omega_{m}^{(0)}=1$), 
we obtain 
\ba
t_{0}=\frac{2}{3H_{0}}\,.
\label{ageflat}
\ea
{}From the observations of the Hubble Space Telescope 
Key project \cite{Freedman01} the present Hubble parameter is 
constrained to be 
\begin{eqnarray}
& &H_0^{-1}=9.776 h^{-1}\, {\rm Gyr}\,,\quad
0.64<h<0.80\,.
\end{eqnarray}
This is consistent with the conclusions arising from observations of the 
CMB \cite{WMAP3} and large
scale structure \cite{Tegmark-sdss,Seljak:2004xh}.
Then Eq.~(\ref{ageflat}) gives $t_{0}=8$-10\,Gyr, 
which does not satisfy the stellar age bound:
$t_0>11$-12\,Gyr.
Hence a flat universe without a cosmological constant 
suffers from a serious age problem.

\begin{figure}
\includegraphics[height=3.1in,width=3.3in]{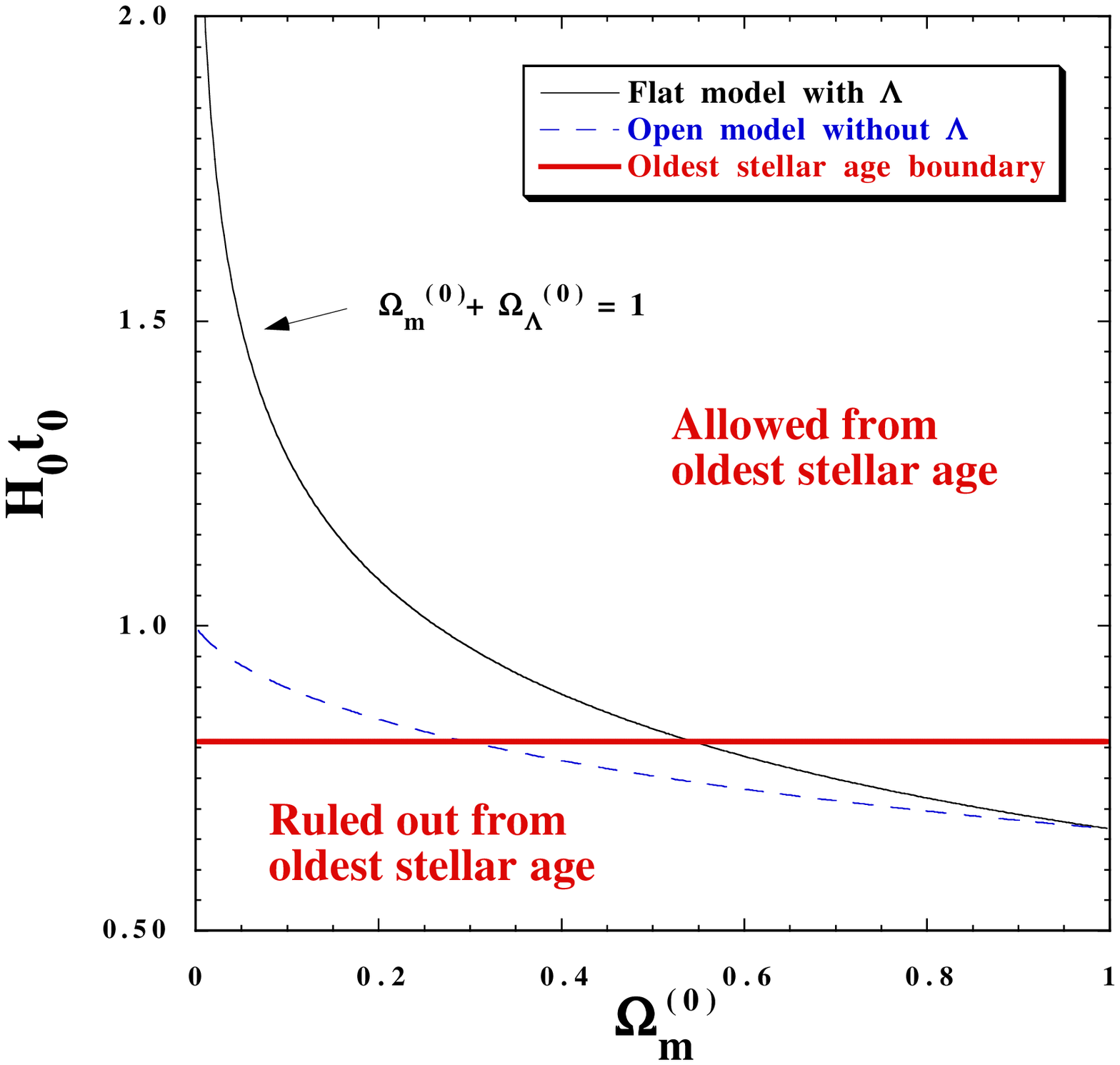}
\caption{The age of the universe (in units of $H_0^{-1}$) is 
plotted against $\Omega_m^{(0)}$
for (i) a flat model with $\Omega_m^{(0)}+
\Omega_{\Lambda}^{(0)}=1$ 
(solid curve) and (ii) a open model (dashed curve).
We also show the border $t_0=11$ Gyr coming from the 
bound of the oldest stellar ages. The region 
above this border is allowed for consistency.
This constraint strongly supports the evidence 
of dark energy.}
\label{agefig} 
\end{figure}

In an open universe model ($\Omega_m^{(0)}<1$), Eq.~(\ref{tzero})
shows that the age of the universe is larger than the flat model explained above.
This is understandable, as the amount of
matter decreases, it would take longer for gravitational interactions to slow
down the expansion rate to its present value. In this case 
Eq.~(\ref{tzero}) is integrated to give
\begin{equation}
H_0t_0=\frac{1}{1-\Omega^{(0)}_m}-\frac{\Omega^{(0)}_m}
{2(1-\Omega^{(0)}_m)^{3/2}}
\ln \left( \frac{1-\sqrt{1-\Omega_{m}^{(0)}}}
{1+\sqrt{1-\Omega_{m}^{(0)}}} \right)\,,
\end{equation}
{}from which we have $H_0t_0 \to 1$
for $\Omega_m^{(0)} \to 0$ and  
$H_0t_0 \to 2/3$ for $\Omega_m^{(0)} \to 1$.
As illustrated in Fig.~\ref{agefig}, $t_{0}$ monotonically increases
toward $t_{0}=H_{0}^{-1}$ with the decrease of $\Omega_m^{(0)}$.
The observations of the CMB \cite{WMAP3} constrain 
the curvature of the universe to be very close to flat, 
i.e., $|\Omega_{K}^{(0)}|=|\Omega_m^{(0)}-1| \ll 1$.
However, since $\Omega_m^{(0)} \simeq 1$ in this case, the age of the universe 
does not become larger than the oldest stellar age (see Fig.~\ref{agefig}).

The age problem can easily  be solved in a flat
universe ($K_{0}=0$) with a cosmological constant 
($\Omega_\Lambda^{(0)} \neq 0$).
In this case Eq.~(\ref{agecal})  gives 
\ba
H_0t_0 &=& \int_{0}^{\infty}
\frac{\rd z}{(1+z) \sqrt{\Omega_m^{(0)} (1+z)^3+
\Omega_\Lambda^{(0)}}} \nonumber \\
&=& \frac{2}{3\sqrt{\Omega_\Lambda^{(0)}}}
\ln \left( \frac{1+\sqrt{\Omega_\Lambda^{(0)}}}
{\sqrt{\Omega_m^{(0)}}} \right)\,,
\ea
where $\Omega_m^{(0)}+\Omega_{\Lambda}^{(0)}=1$.
The asymptotic values are 
$H_0t_0 \to \infty$ for $\Omega_m^{(0)}\to 0$ and  
$H_0t_0 \to 2/3$ for $\Omega_m^{(0)} \to 1$.
In Fig.~\ref{agefig} we plot the age $t_0$ versus 
$\Omega_m^{(0)}$. 
The age of the universe increases as
$\Omega_m^{(0)}$ decreases.
When $\Omega_m^{(0)} = 0.3$ and 
$\Omega_{\Lambda}^{(0)} = 0.7$ one has $t_{0}=0.964H_{0}^{-1}$, 
which corresponds to $t_0=13.1$\,Gyr for $h=0.72$.
Hence this easily satisfies the constraint $t_0>11$-12\,Gyr coming from 
the oldest stellar populations.
Thus the presence of $\Lambda$ elegantly solves 
the age-crisis problem. In \cite{Gruppuso:2005xy}, 
the authors  manage to go further and find the solution 
for the scale factor in a flat Universe driven by dust plus 
a component characterized by a constant parameter of state 
which dominates in the asymptotic future. 

\subsection{Constraints from the CMB and LSS}

The observations related to the CMB \cite{WMAP3} and large-scale 
structure (LSS) \cite{Tegmark-sdss,Seljak:2004xh}
independently support the ideas of a dark energy dominated universe. 
The CMB anisotropies observed by COBE in 1992 and by WMAP 
in 2003 exhibited a nearly scale-invariant spectra 
of primordial perturbations, which 
agree very well with the prediction of inflationary cosmology. However, note that the best fit power-law flat $\Lambda$CDM model obtained from using only the WMAP data now gives a scalar spectral tilt of $n_s = 0.951^{+0.015}_{-0.019}$, significantly less than scale invariant! \cite{WMAP3}.
The position of the first acoustic peak around $l =200$
constrains the curvature of the universe to be 
$|1-\Omega_{\rm total}|=0.030^{+0.026}_{-0.025} \ll 1$ \cite{LPage}
as predicted by the inflationary paradigm. It is worth pointing out that Weinberg in Ref.~\cite{Weinberg:2000ts} 
provides an analytic expression for the position of the first peak 
showing how it depends on the background distribution 
of energy densities between matter and a cosmological constant.  

Using the most recent WMAP data \cite{WMAP3} with an assumption of 
constant equation of state $w_{\rm DE}=-1$ for dark energy, then combining WMAP and 
the Supernova legacy Survey implies $\Omega_{K}^{(0)}= -0.015^{+0.02}_{-0,016}$, 
consistent with a  flat universe. Combining with the HST key project constraint 
on $H_0$ provides a tighter constraint, 
$\Omega_{K}^{(0)} 
= -0.010^{+0.016}_{-0,009}$ and $\Omega_{\Lambda}^{(0)} = 
0.72\pm 0.04$ 
(to be compared with earlier pre WMAP3 results 
$\Omega_\Lambda^{(0)}=0.69^{+0.03}_{-0.06}$, 
which assumed a flat universe with a prior for the 
Hubble constant $h=0.71 \pm 0.076$ \cite{Sievers}) . 

\begin{figure}
\includegraphics[height=3.6in,width=3.2in]{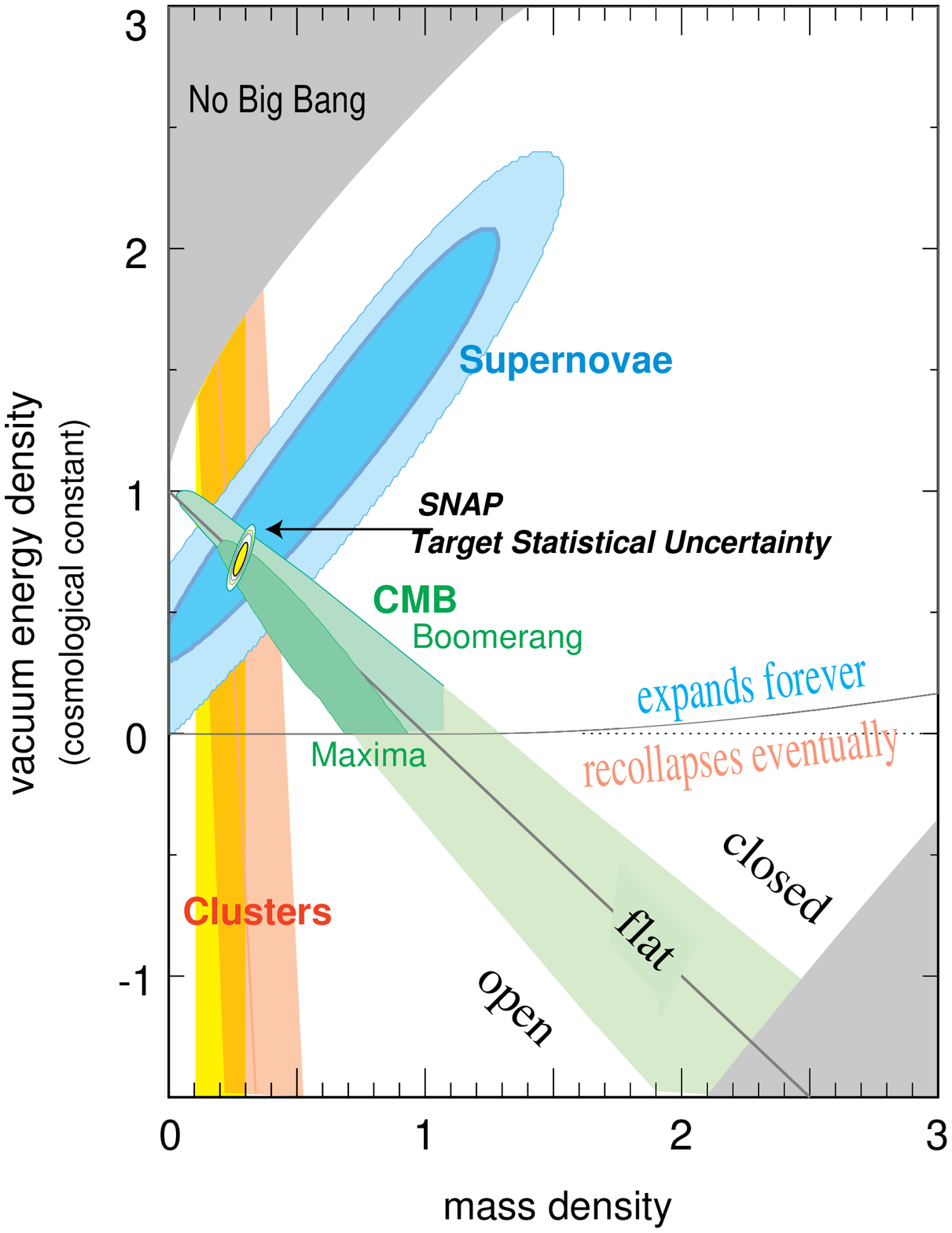}
\caption{\label{figsnap}
The $\Omega_{m}^{(0)}$-$\Omega_{\Lambda}^{(0)}$ 
confidence regions constrained  from the observations of SN Ia, 
CMB and galaxy clustering.
We also show the expected confidence region 
from a SNAP satellite for a flat universe with 
$\Omega_{m}^{(0)}=0.28$.
{}From Ref.~\cite{aldering}.}
\label{cmbfig}
\end{figure}

In Fig.~\ref{cmbfig} we plot the confidence regions coming from 
SN Ia, CMB(WMAP1) and large-scale galaxy clustering \cite{aldering}
(see Ref.~\cite{Bahcall} for an earlier work introducing the ``cosmic triangle'').
Clearly the flat universe without a cosmological constant is ruled out.
The compilation of three different 
cosmological data sets strongly reinforces the need for  a dark energy dominated universe
with $\Omega_{\Lambda}^{(0)} \simeq 0.7$ and 
$\Omega_m^{(0)} \simeq 0.3$.
Amongst the matter content of the universe, 
baryonic matter amounts to 
only 4 \%. The rest of the matter (27 \%)
is believed to be in the form of a non-luminous component of non-baryonic nature with
a dust like equation of state ($w=0$) known as Cold Dark Matter (CDM). 
Dark energy is distinguished from dark matter in the sense that its
equation of state is different ($w<-1/3$), allowing it to give rise to an 
accelerated expansion.

The discussion in this section has been based on the assumption that the equation of state
of dark energy is constant ($w_{\Lambda}=-1$).
This scenario, the so called $\Lambda$CDM model, 
has  become the standard model for  modern cosmology.
However, it may be that this is not the true origin of dark energy. 
If scalar fields turn out to be responsible for it, then the equation 
of state of dark energy can be dynamical.
In order to understand the origin of dark energy it is 
important to distinguish between the cosmological constant 
and dynamical dark energy models.
The observations of SN Ia alone are still not sufficient to establish  
evidence of a dynamically changing equation of state, 
but this situation could well improve through future observations.
In a dark energy dominated universe the gravitational potential 
varies unlike the case of matter dominated universe, which 
leads to an imprint on the CMB power spectrum \cite{CT96}.
This phenomenon, the so called Integrated Sachs-Wolfe (ISW)
effect \cite{Sciama}, could also be important in helping to distinguish 
the cosmological constant and dynamical dark energy models, 
since the evolution of the gravitational potential 
strongly depends upon the dynamical property 
of the equation of state of dark energy.

At present the observations of WMAP are perfectly consistent with
a non varying dark energy contributed by a cosmological constant. 
Tensions which appeared to exist between the  WMAP and the
Gold SN data set \cite{Jassal05} appear to have disappeared in the more recent SNLS 
data \cite{Astier,Jassal:2006gf}, although it is still early days in 
the search for the true nature of the dark energy. 
However given the consistency of a true cosmological constant, 
we shall first discuss the  problem and highlight recent progress 
that has been made in determining the existence of a pure 
cosmological constant, before  proceeding  to discuss dynamical
dark energy models in subsequent sections.

\section{Cosmological constant}
\label{Cconstant}

As mentioned earlier, the cosmological constant $\Lambda$,  was originally 
introduced by Einstein in 1917 to achieve a static universe. 
After Hubble's discovery of the expansion of the universe
in 1929, it was dropped by Einstein as it was no longer required.  
{}From the  point of view of particle physics, however, the cosmological 
constant naturally arises as an energy density of the vacuum.
Moreover, the energy scale of $\Lambda$ 
should be much larger than that of the present Hubble 
constant $H_{0}$, if it originates from the vacuum energy density.  
This is the ``cosmological constant problem'' \cite{Weinberg:1988cp}
and was well known to exist long before the discovery of 
the accelerated expansion of the universe in 1998.

There have been a number of attempts to solve this problem.
An incomplete list includes: adjustment 
mechanisms \cite{Dolgov82,Brown:1987dd}, anthropic 
considerations \cite{Linde:1984ir,Anth,Garriga:1999bf,Anth2,Anth3,Anth4,MartelS}, 
changing gravity \cite{Van82},
quantum gravity \cite{Coleman88}, 
degenerate vacua \cite{Yokoyama},
higher-dimensional gravity \cite{highdimen1,highdimen2},
supergravity \cite{supergravity,burgess-6d}, string 
theory \cite{Bousso00,Silverstein,Feng01,KKLT,BKQ03}, 
space-time foam approach \cite{foam} and
vacuum fluctuations of the energy density \cite{paddyvac} (see also \cite{Gurzadyan:2000ku}).
In this section we shall first address the fine-tuning problem 
associated with the cosmological constant 
$\Lambda$. We will then discuss recent progress to construct 
de-Sitter vacua in the context of string theory \cite{KKLT} and 
proceed to discuss several attempts to explain the origin 
of $\Lambda$.

\subsection{Introduction of $\Lambda$}

The Einstein tensor $G^{\mu \nu}$ and the energy momentum tensor 
$T^{\mu \nu}$ satisfy the Bianchi identities
$\nabla_{\nu}G^{\mu \nu}=0$ and energy conservation 
$\nabla_{\nu} T^{\mu \nu}=0$.
Since the metric $g^{\mu \nu}$ is constant with respect to 
covariant derivatives ($\nabla_{\alpha} g^{\mu \nu}=0$), 
there is a freedom to add a term $\Lambda g_{\mu \nu}$
in the Einstein equations (see Refs.\,\cite{nareshphilosophy} for
a nice discussion on the related theme).
Then the modified Einstein equations are given by 
\begin{eqnarray}
R_{\mu \nu}-\frac12 g_{\mu \nu}R+\Lambda g_{\mu \nu}
=8\pi G T_{\mu \nu}\,.
\label{mEinstein}
\end{eqnarray}
By taking a trace of this equation, we find that 
$-R+4\Lambda=8\pi G T$.
Combining this relation with Eq.~(\ref{mEinstein}), we obtain 
\begin{eqnarray}
R_{\mu \nu}-\Lambda g_{\mu \nu}=8 \pi G
\left( T_{\mu \nu}- \frac12 Tg_{\mu \nu}\right)\,.
\label{mEinstein2}
\end{eqnarray}

Let us consider Newtonian gravity with metric $g_{\mu \nu}=
\eta_{\mu \nu}+h_{\mu \nu}$, where $h_{\mu \nu}$
is the perturbation around the Minkowski metric $\eta_{\mu \nu}$.
If we neglect the time-variation and rotational effect 
of the metric, $R_{00}$ can be written by a gravitational potential 
$\Phi$, as $R_{00} \simeq -(1/2)\Delta h_{00}=\Delta \Phi$.
Note that $g_{00}$ is given by $g_{00}=-1-2\Phi$. 
In the relativistic limit with $|p| \ll \rho$, we have 
$T_{00} \simeq -T \simeq \rho$.
Then the 00 component of Eq.~(\ref{mEinstein2}) gives 
\begin{eqnarray}
\Delta \Phi=4\pi G \rho-\Lambda\,.
\label{Poisson}
\end{eqnarray}

In order to reproduce the Poisson equation in Newtonian gravity, 
we require that $\Lambda=0$ or $\Lambda$ is sufficiently 
small relative to the $4\pi G \rho$ term in Eq.~(\ref{Poisson}).
Since $\Lambda$ has dimensions of $[{\rm Length}]^{-2}$,
the scale corresponding to the cosmological constant needs to be
much larger than the scale of stellar objects on which 
Newtonian gravity works well.
In other words the cosmological constant becomes 
important on very large scales.

In the FRW background given by (\ref{frwmet}) 
the modified Einstein equations (\ref{mEinstein}) give
\begin{eqnarray}
\label{accelerationL0}
&&H^2=\frac{8\pi G}{3}\rho-\frac{K}{a^2}+\frac{\Lambda}{3} \,, \\
&&\frac{\ddot{a}}{a}=-\frac{4\pi G}{3}\left(\rho+3p \right)
+\frac{\Lambda}{3}\,.
\label{accelerationL}
\end{eqnarray}
This clearly demonstrates that the cosmological constant contributes 
negatively to the pressure term
and hence exhibits a repulsive effect.

Let us consider a static universe ($a={\rm const}$)
in the absence of $\Lambda$.
Setting $H=0$ and $\ddot{a}/a=0$
in Eqs.~(\ref{HubbleeqI}) and (\ref{acceleq}), we find 
\ba
\label{Einin}
\rho=-3p=\frac{3K}{8\pi G a^2}\,.
\ea
Equation (\ref{Einin}) shows that either $\rho$ or
$p$ needs to be negative.
When Einstein first tried to construct a static universe,
he considered that the above solution is not 
physical\footnote{We note however that the negative 
pressure can be realized by scalar fields.} and so added the  
cosmological constant to the original field equations
(\ref{Einsteineq}).

Using the modified field equations 
(\ref{accelerationL0}) and (\ref{accelerationL})
in a dust-dominated universe ($p=0$), 
we find that the static universe 
obtained by Einstein corresponds to
\ba
\label{Einin2}
\rho=\frac{\Lambda}{4\pi G}\,, \quad
\frac{K}{a^2}=\Lambda\,.
\ea
Since $\rho>0$ we require that $\Lambda$
is positive. This means that the static universe is
a closed one ($K=+1$) with a radius $a=1/\sqrt{\Lambda}$.
Equation (\ref{Einin2}) shows that the energy density $\rho$
is determined by $\Lambda$.

The requirement of a cosmological constant to achieve a static 
universe can be understood by having a look at the 
Newton's equation of motion (\ref{Newtoneq}).
Since gravity pulls the point particle toward the center of
the sphere, we need a repulsive force to realize a situation 
in which $a$ is constant.
This corresponds to adding a cosmological constant term 
$\Lambda/3$ on the right hand side of Eq.~(\ref{Newtoneq}).

The above description of the static universe was abandoned
with the discovery of the redshift of distant stars, but it is 
intriguing that such a cosmological constant should return
in the 1990's to explain the observed acceleration of the universe.

Introducing the modified energy density and pressure 
\ba
\wt{\rho}=\rho+\frac{\Lambda}{8\pi G}\,, \quad
\wt{p}=p-\frac{\Lambda}{8\pi G}\,,
\ea
we find that Eqs.~(\ref{accelerationL0}) and (\ref{accelerationL})
reduce to Eqs.~(\ref{HubbleeqI}) and (\ref{acceleq}).
In the subsequent sections we shall use the field equations 
(\ref{HubbleeqI}) and (\ref{acceleq}) when we study the 
dynamics of dark energy.

\subsection{Fine tuning problem}

If the cosmological constant originates from a vacuum
energy density, then this suffers from a severe fine-tuning problem.
Observationally we know that $\Lambda$ is of order the 
present value of the Hubble parameter $H_{0}$, that is 
\ba
\label{lamobser}
\Lambda \approx H_{0}^2=
(2.13h \times 10^{-42}\, {\rm GeV})^2\,.
\ea
This corresponds to a critical density $\rho_{\Lambda}$,
\ba
\label{rhoLam}
\rho_{\Lambda}=\frac{\Lambda m_{{\rm pl}}^2}{8\pi}
\approx 10^{-47}\,{\rm GeV}^4\,.
\ea

Meanwhile the vacuum energy density evaluated by the sum of  
zero-point  energies of quantum fields with mass $m$ is given by 
\ba
\label{cutint}
\rho_{\rm vac} &=&\frac12 \int_{0}^\infty \frac{\rd^3 {\bf k}}{(2\pi)^3}
 \sqrt{k^2+m^2} \nonumber \\
&=& \frac{1}{4\pi^2} \int_{0}^\infty
\rd k\, k^2\sqrt{k^2+m^2}\,.
\ea
This exhibits an ultraviolet divergence: $\rho_{\rm vac} \propto k^4$.
However we expect that quantum field theory is valid up to some cut-off
scale $k_{\rm max}$ in which case the integral (\ref{cutint}) is finite:
\ba
\rho_{\rm vac} \approx \frac{k_{\rm max}^4}{16\pi^2}\,.
\ea
For the extreme case of General Relativity we expect it to be valid to just below 
the Planck scale: $m_{\rm pl}=1.22 \times 10^{19}\,{\rm GeV}$.
Hence if we pick up $k_{\rm max}=m_{\rm pl}$, we find that the 
vacuum energy density in this case is estimated as 
\ba
\rho_{\rm vac} \approx 10^{74}\,{\rm GeV}^4\,,
\ea
which is about $10^{121}$ orders of magnitude larger than 
the observed value given by Eq.~(\ref{rhoLam}).
Even if we take an energy scale of QCD for $k_{\rm max}$, 
we obtain $\rho_{\rm vac} \approx 10^{-3}\,{\rm GeV}^4$
which is still much larger than $\rho_{\Lambda}$.

We note that this contribution is related to the ordering 
ambiguity of fields and disappears when normal ordering is adopted.
Since this procedure of throwing away the vacuum energy is
ad hoc, one may try to cancel it by introducing counter terms.
However this requires a fine-tuning to adjust 
$\rho_{\Lambda}$ to the present energy density of the universe.
Whether or not the zero point energy in field theory 
is realistic is still a debatable question.

A nice resolution of the zero point energy is provided by supersymmetry.
In supersymmetric theories 
every bosonic degree of freedom has its Fermi counter part 
which contributes to the zero point energy 
with an opposite sign compared to the
bosonic degree of freedom thereby canceling 
the vacuum energy. Indeed, for a field with spin $j>0$, 
the expression (\ref{cutint}) for the vacuum 
energy generalizes to
\ba
\label{cutints}
\rho_{\rm vac} &=&\frac{1}{2}(-1)^{2j}(2j+1) 
\int_{0}^\infty \frac{\rd^3 {\bf k}}{(2\pi)^3}
 \sqrt{k^2+m^2} \nonumber \\
&=& \frac{(-1)^{2j}(2j+1)}{4\pi^2} \int_{0}^\infty
\rd k\, k^2\sqrt{k^2+m^2}\,.
\ea

Exact supersymmetry implies an equal number of fermionic and 
bosonic degrees of freedom  for a given value of the mass $m$ such that 
the net contribution to the vacuum energy vanishes.
It is in this sense that 
supersymmetric theories do not admit 
a non-zero cosmological constant.
However, we know that we do not live in a 
supersymmetric vacuum state and hence it should be broken today. 
For a viable supersymmetric scenario, for instance if it is to be 
relevant to the hierarchy problem,
 the supersymmetry breaking scale should be around
$M_{{\rm SUSY}} \sim 10^{3}\,{\rm GeV}$.
Indeed, the presence of a scalar field (Higgs field)
in the standard model of particle physics (SM) is necessary to ensure the
possibility of a spontaneous breakdown of the gauge symmetry.

However, the same scalar field creates what has come to be
known as the ``hierarchy problem''. 
The origin of this problem lies in the quadratic nature of 
the divergence of the scalar self-energy arising out of scalar loops. 
A way out of this is supersymmetry
(SUSY) which as we have mentioned demands a fermionic partner for every boson
and vice versa with the two having the same mass \cite{collins1,collins2}. 
Since fermionic loops come
with an overall negative sign, the divergence in the scalar
self energy due to the scalar loop and its SUSY partner cancel
out. However, particles in nature do not come with degenerate
partners as demanded by SUSY and hence SUSY must be
broken. With a broken SUSY, one of course wants to ensure that
no new scales are introducted between the electroweak scale
of about ~246 GeV and the Planck scale. The superpartners of
the Standard Model particles thus are expected to have masses of
the order of TeV. 
Masses much lower than this are ruled
out from null experimental results in present day accelerators and
specific bounds for the masses for the various superpartners of
SM particles are available from analysis of experimental data. 
Theoretically, a consistent scheme of spontaneous breakdown of
SUSY is technically far more complicated than in the non SUSY
version. Nevertheless several approaches are available 
where this can be achieved.

With supersymmetry breaking around $10^{3}\,{\rm GeV}$,
we are still far away from the observed value of 
$\Lambda$ by many orders of magnitudes.
At present we do not know how the Planck scale or 
SUSY breaking scales are really  
related to the observed vacuum scale.

The above cosmological constant problem has led many 
many authors to try a different approach to the dark energy issue. 
Instead of assuming we have a small cosmological constant, 
we ignore it, presume it is zero due to some as yet unknown mechanism, 
and investigate the possibility that the dark energy is caused 
by the dynamics of a light scalar field. 
It does not solve the cosmological constant problem, 
but it does open up another avenue of attack as we will shortly see.  

\subsection{$\Lambda$ from string theory}

Recently there has been much progress in constructing 
de-Sitter vacua in string theory or supergravity.
According to the no-go theorem in Refs.~\cite{gibbons-nogo,maldacena-nunez}
it is not possible to find de-Sitter solutions 
only in the presence of the lowest order terms
in the 10 or 11 dimensional supergravity action.
However this situation is improved when $\alpha'$ or quantum 
corrections to the tree-level action are taken into account or extended
objects like D-branes are present.
In fact Kachru, Kallosh, Linde and Trivedi (KKLT) \cite{KKLT}
constructed de-Sitter vacua by incorporating nonperturbative corrections
to a superpotential
in the context of type IIB string theory compactified on a Calabi-Yau
manifold in the presence of flux. 
The importance of flux to insolving the cosmological constant problem was 
originally realized in Ref.~\cite{Bousso00}.
In what follows we shall briefly discuss the effect of a four-form gauge
flux to construct de-Sitter vacua \cite{Bousso00}
and then proceed to the review of the KKLT scenario using  flux compactification.

\subsubsection{Four-form fluxes and quantization}

Let us consider a four-form flux field 
$F_4^2=F_{\mu \nu \rho \sigma}F^{\mu \nu \rho \sigma}$ 
which appears in M theory. 
The starting point in Ref.~\cite{Bousso00} is a four
dimensional gravity action in the presence of 
a negative bare cosmological constant $-\Lambda_b$
and the four-form flux field:
\ba
S=\int {\rm d}^4 x \sqrt{-g} \left(
\frac{1}{2\kappa^2}R+\Lambda_{b}-
\frac{1}{2 \cdot 4!} F_4^2 \right)\,,
\ea
which arises as an effective action arising, e.g., from 
a $M^4 \times S^7$ compactification.
The bare cosmological constant is should be negative 
at the perturbative regime of string theory if it exists.

The four-form equation of motion, 
$\nabla_{\mu}(\sqrt{-g}F^{\mu \nu \rho \sigma})=0$, 
gives the solution 
$F^{\mu \nu \rho \sigma}=c\epsilon^{\mu \nu \rho \sigma}$,
where $F^{\mu \nu \rho \sigma}$ is an antisymmetric tensor 
with $c$ being constant. 
Since $F_{4}^2=-24c^2$, we find that the effective cosmological 
constant is given by 
\ba
\Lambda=-\Lambda_b-\frac{1}{48}F_{4}^2
=-\Lambda_b+\frac{c^2}{2}\,.
\ea
This shows that it is possible to explain a small value of $\Lambda$
provided that the bare cosmological constant is nearly canceled by the 
term coming from the four-form flux.
However as long as the contribution of the flux is continuous, one can
not naturally obtain the observed value of $\Lambda$.

Bousso and Polchinski tackled this problem by quantizing 
the value of $c$ \cite{Bousso00}. This implies that $c$ is discontinuous as 
$c=nq$, where $n$ is an integer.
Although a single flux is not sufficient to explain the small
values of $\Lambda$ because of the large steps involved, this situation is 
improved by considering $J$ multiple fluxes.
In this case an effective cosmological constant is given by 
\ba
\Lambda=-\Lambda_b+\frac12 \sum_{i=1}^J
n_{i}^2q_{i}^2\,.
\ea
It was shown in Ref.~\cite{Bousso00} that one can 
explain the observed value of $\Lambda$ with 
$J$ of order  $100$, which is not unrealistic.
The work of Bousso and Polchinski did not address the problem 
of the stabilization of the modulus fields, but 
it opened up a new possibility for 
constructing large numbers of de-Sitter 
vacua using fluxes--and this has been called the ``string landscape'' \cite{Susskind:2003kw}.

\subsubsection{The KKLT scenario}

KKLT \cite{KKLT} provided a mechanism to construct
de-Sitter vacua of type IIB string theory based on flux compactifications
on a Calabi-Yau manifold.
They first of all fixed all the moduli associated with the compactification 
in an anti de-Sitter vacua by preserving supersymmetry.
Then they incorporated nonperturbative corrections to the 
superpotential to obtain de-Sitter vacua.

The low energy effective action of string/M-theory 
in four dimensions
is described by $N=1$ supergravity \cite{collins2}
\begin{eqnarray}
\hspace*{-2.0em}
S&=&\int
\rd^4x\,\sqrt{-g}\Biggl[\frac{M_{\rm pl}^2}{2}R+
g^{\mu\nu}K_{\alpha\bar{\beta}}\partial_{\mu}
\varphi^{\alpha} \partial_{\nu}\bar{\varphi}^{\beta}
\nonumber\\
\hspace*{-2.0em}
&&-e^{K/M_{\rm pl}^2}\left(K^{\alpha\bar{\beta}}
D_{\alpha}WD_{\bar{\beta}}\bar{W}-\frac{3}{M_{\rm pl}^2}|W|^2
\right)\Biggr],
\label{stringaction}
\end{eqnarray}
where $\alpha,\beta$ run over all moduli fields $\vp$. 
Here $W(\varphi^\alpha)$ and $K(\varphi^\alpha,\bar{\varphi}^\beta)$
are the superpotential and the K\"ahler
potential, respectively, and 
\begin{eqnarray}
K_{\alpha\bar{\beta}} \equiv \frac{\partial^2K}
{\partial\varphi^\alpha \partial\bar{\varphi}^\beta}\,, \quad
D_\alpha W \equiv \frac{\partial
W}{\partial\varphi^\alpha}+\frac{W}
{M_{\rm pl}^2}\frac{\partial
K}{\partial\varphi^\alpha}\,.
\end{eqnarray}
The supersymmetry is unbroken only
for the vacua in which $D_\alpha W=0$ for all $\alpha$, which means that 
the effective cosmological constant is not positive from 
the action (\ref{stringaction}). 
We use the units $M_{\rm pl}^2=1$ for the rest of this section.

The authors in Ref.~\cite{sgref}
adopted the following tree level
functions for $K$ and $W$ in the flux compactification 
of Type IIB string theory \cite{pmref,sgref}:
\begin{eqnarray}
K&=&-3\ln[-i(\rho-\bar{\rho})]-\ln[-i(\tau-\bar{\tau})] 
\nonumber \\
& &-\ln[-i\int_{\cal M}\Omega\wedge\bar{\Omega}]\,, \\
W&=&\int_{\cal M} G_3\wedge\Omega\,,
\end{eqnarray}
where $\rho$ is the volume modulus 
which includes the volume of the 
Calabi-Yau space and an axion coming from the R-R 4-form $C_{(4)}$,
and $\tau=C_{(0)}+ie^{-\Phi}$ is the axion-dilaton
modulus. $\Omega$ is the holomorphic
three-form on the Calabi-Yau space and 
$G_3$ is defined by $G_3=F_3-\tau H_3$ where
$F_3$ and $H_3$ are the R-R flux and the  NS-NS flux,
respectively, on the 3-cycles of the internal Calabi-Yau
manifold ${\cal M}$.

Since $W$ is not a function of $\rho$, we obtain
$K^{\rho\bar{\rho}}D_{\rho}WD_{\bar{\rho}}\bar{W}=3|W|^2$,
Then Eq.~(\ref{stringaction}) gives the supergravity potential 
\begin{eqnarray}
V&=& e^{K}\left(K^{i\bar{j}}
D_iWD_{\bar{j}}\bar{W}\right)\,,
\end{eqnarray}
where $i,j$ run over all moduli fields except for $\rho$.
The condition $D_iW=0$ fixes all
complex moduli except for $\rho$ \cite{sgref}, 
which gives a zero effective cosmological constant. 
On the other hand, the supersymmetric
vacua satisfying $D_{\rho}W=0$ gives $W=0$, whereas,
the non-supersymmetric vacua yield $W=W_0 \neq 0$.

To fix the volume modulus $\rho$ as well, 
KKLT \cite{KKLT} added a
non-perturbative correction \cite{ew96} to
the superpotential, which is given by 
\begin{eqnarray}
W=W_0+Ae^{ia\rho}\,,
\end{eqnarray}
where $A$ and $a$ are constants and $W_0 \equiv 
\int G_3\wedge\Omega$ is the tree level contribution.
This correction is actually related to the effect of 
brane instantons. Note that the conditions
$D_iW=0$ are automatically satisfied.
For simplicity we set the axion-dilaton
modulus to be zero and take $\rho=i \sigma$.
Taking real values of $A, a$ and $W_{0}$, we find that 
the supersymmetric condition $D_{\rho}W=0$ gives
\begin{eqnarray}
W_{0}=-Ae^{-a \sigma_{c}}
\left(1+\frac23 a \sigma_{c} \right)\,,
\end{eqnarray}
which  fixes the volume modulus $\rho$ in terms of $W_0$. 
This produces the anti de-Sitter vacua, that is
\begin{eqnarray}
V_{\rm AdS}=-3e^{K}|W|^2
=-\frac{a^2A^2e^{-2a\sigma_{c}}}{6\sigma_{c}}\,.
\label{VKKLT}
\end{eqnarray}
Hence all the moduli are stabilized while
preserving supersymmetry with a negative cosmological constant.

In order to obtain a de-Sitter vacuum, KKLT introduced an anti-D3
brane in a warped background.
Since fluxes $F_3$ and $H_3$ are also the sources 
for a warp factor \cite{sgref,pmref},
models with fluxes generically correspond to a warped
compactification, whose metric is given by 
\begin{eqnarray}
\rd s^2_{10}=
e^{2B(y)}g_{\mu\nu}(x)\rd x^{\mu}
\rd x^{\nu}+e^{-2B(y)}\tilde{g}_{mn}(y)
\rd y^m \rd y^n, \nonumber \\
\label{warp}
\end{eqnarray}
where the factor, $e^B$, can be computed in the regions closed to a conifold
singularity of the Calabi-Yau manifold. 
This warp factor is exponentially suppressed at the tip of the throat,
depending on the fluxes as 
\begin{eqnarray}
e^{B_{\rm min}}\sim
\exp\left(-\frac{2\pi N}{3g_sM}\right)\,,
\label{min}
\end{eqnarray}
where $g_s$ is the string coupling, integers $M$ and $N$ are 
the R-R and NS-NS three-form flux, respectively. 
While the warp factor is of order
one at generic points in the $y$-space, its minimum value can be
extremely small for a suitable choice of fluxes.

The background fluxes generate a potential for the
world-volume scalars of  the anti-D3 brane, which means that they 
do not introduce additional moduli \cite{skref}. 
The anti-D3 brane, however,
provides an additional energy to the supergravity potential
\cite{skref,KKLT}:
\begin{eqnarray}
\delta V=\frac{2b_0^4T_3}{g_s^4}\frac{1}{({\rm Im}\rho)^3}\,,
\label{DVKKLT}
\end{eqnarray}
where $T_{3}$ is the brane tension and 
$b_0$ is the warp factor at the location of the anti-D3 brane.
The anti-D3 brane energetically prefers to sit at the tip of the throat,
giving $b_{0}=e^{B_{\rm min}}$.
The total potential is the sum of Eqs.~(\ref{VKKLT}) and (\ref{DVKKLT}), 
that is
\begin{eqnarray}
V=\frac{2b_0^4T_3}{g_s^4}\frac{1}{({\rm Im}\rho)^3}
-\frac{a^2A^2e^{-2a\sigma_{c}}}{6\sigma_{c}}\,.
\end{eqnarray}
Then one can obtain positive cosmological constant by tuning 
the flux integers $M$ and $N$.

The life time of the vacua was found to be larger than 
the age of the universe and hence these solutions can be considered
as stable for practical purposes \cite{KKLT}. 
Although a fine-tuning problem of $\Lambda$ still remains in this 
scenario, it is interesting that string theory in principle gives rise to a stable 
de-Sitter vacua with all moduli fixed.
A remarkable and somewhat controversial argument about 
the nature of the cosmological constant problem has developed 
recently out of this realisation that there are many possible de-Sitter vacua. 
The fact that there are a vast number of different choices of fluxes
leads in principle to a complicated string landscape with more 
than $10^{100}$ vacuum \cite{Susskind:2003kw}. 
Surely, the argument goes, it should be possible to find 
a vaccua which is identical to the one we live in! 
In an interesting paper, Liddle and Urena-Lopez have 
examined the conditions needed to unify the description of 
dark matter, dark energy and inflation within the context of the 
string landscape \cite{Liddle:2006qz}. 
They claim that incomplete decay of the inflaton field offers 
the possibility that a single field might be responsible for 
all of inflation, dark matter and dark energy, whereas, 
unifying dark matter and dark energy into a single field 
which is separate from the inflaton appears very difficult.

\subsubsection{Relaxation of $\Lambda$ in string theory}

In Ref.~\cite{Feng01}, the authors developed an earlier approach in 
Refs.~\cite{Brown:1987dd} to relax the effective 
cosmological constant through the nucleation of branes coupled to 
a three-index gauge potential. 
The influence of string theory in the new approach is important, 
the brane depends on the compactification of the extra dimensions 
which in turn can provide the required very small quantized unit 
for jumps in the effective cosmological term.
As well as this feature, when considering multiple 
coincident branes, in Ref.~\cite{Feng01}, the authors show 
that the  internal degrees of freedom for such a configuration 
can dramatically enhance tunneling rates by exponentially large density 
of states factors. 

For consistency, the dynamics of the system must be such that 
the cosmological constant relaxes quickly enough from high energy 
scales, but today remains stable on a time scale of the universe,
a constraint which leads to a non-trivial relation 
between the scale of supersymmetry breaking and the value of
the cosmological constant. In particular the constraint becomes
\begin{equation}
M^2_{\rm SUSY} \leq (10^{-3} {\rm eV}) (M_{\rm Planck}),
\end{equation}
which rules out large supersymmetry breaking scales 
for these relaxation models, with the largest possible 
scale still viable in nature. 
Time will tell whether the relaxation mechanism is 
sufficiently versatile to uniquely pick out the actual 
vacuum we live in, but it is certainly a novel 
approach to determining it.

\subsubsection{$\Lambda$ from a self-tuning universe}

In \cite{highdimen1}, both sets of authors develop an approach 
to the cosmological constant problem which relies on the presence 
of an extra dimension. Rather than making the vacuum energy 
small, this approach proceeds by removing the gravitational effect 
of vacuum energy on the expansion of the universe. 
Considering Poincare invariant domain wall (``3-brane'') 
solutions to some 5-dimensional effective theories which can arise 
naturally in string theory, the basic idea behind the models is that 
the Standard Model vacuum energy ``warps'' the higher-dimensional 
spacetime while preserving 4D flatness. 
In the strong curvature region the size of the extra dimension is 
effectively cut off (under certain assumptions about the nature of 
the singularity in the strong curvature regime), giving rise to 
macroscopic 4D gravity without a cosmological constant. 
Although the higher-dimensional gravity dynamics is treated 
classically, the Standard Model is fully quantum field-theoretic, 
leading the authors to argue that 4D flatness of their solutions is 
stable against Standard Model quantum loops and changes to 
Standard Model couplings. 

In \cite{Carroll:2001zy}, the authors point out how such a self 
tuning scenario requires changing of the Friedmann equation 
of conventional cosmology, and investigate in the context of 
specific toy models of self tuning the difficulties that arise 
in obtaining cosmological evolution compatible with 
observation in this context. It remains to be seen whether 
this mechanism will eventually work, but the idea that 
by making the metric insensitive to the value of the 
cosmological constant as opposed to trying to make the 
vacuum energy small itself is intriguing.

\subsubsection{$\Lambda$ through mixing of degenerate vacua}

In Refs.~\cite{Kane:2003qh}, the authors suggest a mechanism 
in string theory, where the large number $N$ of connected degenerate 
vacua that could exist, can lead to a ground state with much lower 
energy than that of any individual vacuum. 
This is because of the effect of level repulsion in quantum theory 
for the wavefunction describing the Universe. 
To make it more quantitative, they consider a scenario 
where initial quantum fluctuations give an energy density 
$\sim m_{\rm SUSY}^2m_{{\rm pl}}^2$, 
but the universe quickly cascades to an energy 
density $\sim m_{\rm SUSY}^2m_{{\rm pl}}^2/N$. 
The argument then proceeds, as the universe expands and 
undergoes a series of phase transitions there are large 
contributions to the energy density and consequent 
rearrangement of levels, each time followed by a rapid 
cascade to the ground state or near it. 
The ground state which eventually describes 
our world is then  a superposition of a large number 
of connected string vacua, with shared superselection 
sets of properties such as three families etc.. 
The observed value of the cosmological constant 
is given in terms of the Planck mass, the scale of 
supersymmetry breaking and the number of 
connected string vacua, 
and they argue can quite easily be very small.

\subsection{Causal sets and $\Lambda$}

String theory is not the only candidate for a quantum theory of gravity. 
There are a number of others, and one in particular is worthy of mention 
in that it makes a prediction for the order of magnitude expected of 
the cosmological constant. In the context of Causal sets 
(for a review of Causal sets see \cite{Sorkin:2003bx}), 
Sorkin \cite{sorkin}, back in the early 1990's, 
predicted that a fluctuating cosmological term $\Lambda(x)$ 
would arise under the specific modification of General Relativity 
motivated by causal sets. 
The predicted fluctuations arise as a residual (and non-local) 
quantum effect from the underlying space-time discreteness. 

Roughly speaking, the space-time discreteness leads to 
a finite number $N$ of elements, and the space-time volume 
${\cal V}$ is a direct reflection of $N$. Now $\Lambda$ is conjugate 
to ${\cal V}$, and  fluctuations in ${\cal V}$ arise from the Poisson 
fluctuations in $N$ (which have a typical scale $\sqrt{N}$), 
implying there will be ever decreasing fluctuations in 
$\Lambda$ given by \cite{sorkin}
\be
\Delta \Lambda \sim {1 \over \Delta {\cal V}} 
\sim {1 \over \sqrt{{\cal V}}}.
\label{causet-fluc}
\ee
This could be used to explain why $\Lambda$ is not exactly zero 
today, but why is it so near to zero? Sorkin addresses this issue 
by pointing out that the space-time volume ${\cal V}$ is roughly 
equal to the fourth power of the Hubble radius $H^{-1}$. 
It follows that at all times we expect the energy density in 
the cosmological constant to be of order the critical density 
$\rho_c$, i.e., 
\be 
\rho_\Lambda \sim {\cal V}^{-1/2} \sim 
H^2 \sim \rho_{\rm crit}\,.
\label{causet-lambda}
\ee

Therefore, the prediction for today's $\Lambda$ has the 
right order of magnitude that agrees with current observations 
of the dark energy, and it fluctuates about zero due to the 
non-discrete nature of space-time. Interestingly another prediction 
is that this agreement is true for all times, implying a kind of 
scaling (or tracking) behaviour arising out of causal sets. 
In \cite{Ahmed:2002mj}, this basic paradigm is put to the 
test against observations, and appears to have survived the
first set of tests showing evidence of ``tracking" 
behaviour with no need for fine tuning and consistency 
with nucleosynthesis and structure formation constraints. 
This is a fascinating idea and certainly deserves further 
attention to compare it with more detailed observations, 
although of course the actual mechanism to generate the 
cosmological constant based on causal sets remains to be solved. 

\subsection{Anthropic selection of $\Lambda$}

The use of the anthropic principle has generated much debate
in the cosmology community. It has been used in physics 
on many occasions to explain some of the observed features 
of our Universe, without necessarily explaining the features 
from an underlying theory. For many, it is the solution you 
introduce when you have given up 
on finding any physical route to a solution. For others, 
it is a perfectly plausible weapon in the physicists armoury 
and can be brought out and used when the need arises. 

In a cosmological context, it could be argued that discussions 
related to the use of the anthropic principle were meaningless 
without an underlying cosmological model, to place it 
in context with. The Inflationary Universe provided such 
a paradigm and Linde discussed the anthropic principle 
in this context in the famous Proceedings of the Nuffield 
Symposium in 1982 \cite{Linde:1982gg}. 
In \cite{Linde:1984ir}, he proposed a possible anthropic 
solution to the cosmological constant problem. 
Assuming unsupressed quantum creation of the universe 
at the Planck energy density, he noted that vacuum energy 
density could be written as a sum of contributions from 
the effective potential of the scalar field $V(\phi)$ 
and that of fluxes $V(F)$. The condition for the universe 
to form was that the sum of these two terms matched the 
Planck energy density, i.e., $V(\phi) + V(F) =1$ 
in suitable units. However as the universe inflates, 
the field slowly rolls to its minimum at some different value 
$\phi_0$, leaving a different vacuum energy density 
$\Lambda = V(\phi_0) + V(F)$. Since $V(\phi)$ can take 
any value subject to the initial constraint $V(\phi) + V(F) =1$, 
it leads to a flat probability distribution for the final value 
of the cosmological constant $\Lambda = V(\phi_0) + V(F)$ 
a condition which is required for the anthropic solution of
the cosmological constant problem. 

If ever a problem  required an anthropic argument to explain 
it, then it could well be that the cosmological 
constant is that problem. There has been considerable 
work in this area over the past twenty or 
so years \cite{Weinberg:1988cp,Anth}.  

In \cite{Anth4}, the authors extended an idea first explored in \cite{Linde:1986dq},  
the possibility that the dark energy is due to a potential of 
a scalar field and that the magnitude and the slope of 
this potential in our part of the universe are largely 
determined by anthropic selection effects. 
A class of models are consistent with observations in that 
the most probable values of the slope are very small, 
implying that the dark energy density stays constant to 
very high accuracy throughout cosmological evolution. 
However, in other models, the most probable values of 
the slope make it hard to have sufficient slow-roll condition, 
leading to a re-collapse of the local universe on a time-scale 
comparable to the lifetime of the sun. Such a situation leads 
to a rapidly varying effective equation of state with the 
redshift, leading to a number of testable predictions 
(see also \cite{Dimopoulos:2003iy} for a related model).

According to the anthropic principle,
only specific values of the fundamental constants of nature 
can have lead to intelligent life in our universe. 
Weinberg \cite{Weinberg:1988cp} 
was the first to point out that once the cosmological constant comes 
to dominate the dynamics of the universe, then
structure formation stops because density perturbations cease to grow.
Thus structure formation should be completed before the domination of
vacuum energy, otherwise there could be no observers now.
This leads to the following bound arising out of 
an anthropic argument \cite{Weinberg:1988cp}
\bea
\rho_{\Lambda}  < 500 \rho_m^{(0)}\,,
\eea
which is two orders of magnitude away from the observed 
value of the vacuum energy density. 

The situation can  change if the vacuum energy differs in different regions 
of the universe. In this case one should define a conditional probability density
to observe a given value of $\rho_{\Lambda}$ \cite{Efstathiou:1990xe,Garriga:1999bf}
\bea
\rd {\cal P}(\rho_{\Lambda})={\cal P}_* (\rho_{\Lambda})
n_G (\rho_{\Lambda}) \rd\rho_{\Lambda}\,,
\eea
where $n_G(\rho_{\Lambda})$ is the average number of galaxies that can form 
per unit volume
for a given value of the vacuum energy density and 
${\cal P}_* (\rho_{\Lambda})$ is the {\it a priori} probability 
density distribution.
For a flat distribution of ${\cal P}_* (\rho_{\Lambda})$, 
it was shown in Ref.~\cite{MartelS} that ${\cal P}(\rho_{\Lambda})$ 
peaks around $\rho_{\rm vac} \sim 8 \rho_m^{(0)}$. 

There are two important aspects of the anthropic selection, one is related to 
the prediction of the a priori probability and the other to
the possibility of $\Lambda$ assuming different values in different regions of the universe.
The existence of a vast landscape of de-Sitter vacua  in string theory makes the
anthropic approach especially interesting.  On the other hand, the prediction of a priori probability
arising out of  fundamental theory is of course non-trivial (perhaps impossible!) 
and this could perhaps be more difficult than 
the derivation of an observed value of $\Lambda$ itself. The anthropic
arguments can not tell us how the present observed scale of 
$\Lambda$ is related to the scales arising in particle physics, e.g., 
SUSY breaking scale, but many believe it is important  to carry on 
the investigation of whether or not the anthropic principle
has real predictive power in the context of the cosmological constant.

\subsection{A Dynamical Approach to the Cosmological Constant}

In \cite{Mukohyama:2003nw} a dynamical approach to the cosmological 
constant is investigated. The novel feature is that a scalar field exists which 
has non-standard kinetic terms whose coefficient diverges at zero curvature. 
Moreover, as well as having the standard kinetic term, the field has a  
potential whose minimum occurs at a generic, but negative, value 
for the vacuum energy.  The divergent coefficient of the kinetic term
means that the lowest energy state is never achieved. 
Instead, the cosmological constant automatically stalls 
at or near zero. The authors argue that the model is stable under 
radiative corrections, leads to stable dynamics, despite the singular 
kinetic term, and can reduce the required fine-tuning by at least 60 
orders of magnitude. They also point out that the model could provide 
a new mechanism for sampling possible cosmological constants 
and implementing the anthropic principle.

\subsection{Observing dark energy in the laboratory ?}

At present we do not really know how quantum field theory could {\it naturally} 
lead to the present observed scale of cosmological constant. 
Assuming that we have solved this problem let us ask whether it is possible 
to observe the cosmological constant directly through laboratory experiments? It is a question that is fascinating and has generated quite a bit of debate. There is no consensus yet as to the answer.

We remind the reader that so far all the evidence for the presence of dark energy 
has been astrophysical in nature.
In fact there is little doubt that vacuum fluctuations are found in nature. 
One example of their role, is that they are responsible for the quantum noise 
found in dissipative systems, noise which has been 
detected experimentally. Quantum noise should emerge in a dissipative 
system due to uncertainty principal for a simple reason. 
Classically, the stable state of a dissipative system corresponds
to a zero momentum state which is not permissible quantum mechanically. 
Thus quantum noise should be present in the system which would keep it going 
\cite{gardiner}.

For simplicity let us assume that the vacuum fluctuations are  
electromagnetic in nature \cite{Beck:2005pr}. These
fluctuations are then represented by an ideal gas of harmonic 
oscillators. Quantum statistical mechanics tells us that the 
spectral energy density of the fluctuations with a frequency 
$\nu$ and a temperature $T$ is given by
\begin{eqnarray}
\rho(\nu, T)=\rho_0(\nu)+\rho_{{\rm rad}}(\nu,T)\,,
\label{noise}
\end{eqnarray} 
where 
\begin{eqnarray}
\rho_0(\nu)=\frac{4\pi h \nu^3}{c^3}
\end{eqnarray} 
corresponds to the zero-point fluctuation, and
\begin{eqnarray}
\rho_{{\rm rad}}(\nu,T)=\frac{8\pi h \nu^3}{c^3}
\frac{1}{\exp (h \nu/k_{B}T)-1}
\end{eqnarray} 
describes the thermal fluctuations of a Planck spectrum.
Note that we have explicitly written Planck's constant $h$,
the speed of light $c$ and the Boltzmann constant $k_{B}$ 
following standard convention.

The energy density $\rho_0(\nu)$ is formally
infinite, so, as before, we introduce a cut-off $\nu =\nu_{\Lambda}$
to handle it \cite{Beck:2005pr}
\begin{equation}
\rho_v \equiv \int_{0}^{\nu_{\Lambda}}{\rho_0(\nu)d\nu}
=\frac{\pi h}{c^3} \nu_{\Lambda}^4\,.
\end{equation}
Identifying the vacuum energy density with the observed 
value of the dark energy we obtain an estimate 
for the cut-off frequency
\begin{equation}
\nu =\nu_{\Lambda} \simeq 1.7 \times 10^{12}\,\,{\rm Hz}\,.
\label{nucut}
\end{equation}
If the vacuum fluctuations are responsible for dark energy, we should observe 
a cut-off (\ref{nucut}) in the spectrum of fluctuations. 

Let us now briefly describe an experimental
set up to investigate the nature of vacuum fluctuations. 
Over two decades ago, Koch {\it et al.} carried out experiments
with devices based upon Josephson junctions \cite{Koch1,Koch2}. 
They were interested in obtaining the spectrum of quantum noise
present in their particular experiment
that could remove the thermal part of the noise
because it ran at  low temperatures.
The results of this experiment are in agreement with Eq.~(\ref{noise}) up to
the maximum frequency of $ \nu_{\rm max}=6 \times 10^{11}\,{\rm Hz}$ 
they could reach in their experiment. 

The results of Koch {\it et al.} demonstrate the existence 
of vacuum fluctuations in the spectrum through the linear part of the spectrum. 
However, on the basis of these findings, we can say nothing
about the inter-relation of vacuum fluctuations to dark energy. 
We still need to investigate the spectrum up to frequencies three times larger than 
$\nu_{\rm max}$ to beat the threshold. 
And if a cut-off is observed in the spectrum around $\nu_{\Lambda}$,
it will be suggestive that vacuum fluctuations could be responsible for dark energy. 
In the next few years it would be possible to cross the threshold frequency as suggested 
in Ref.~\cite{Beck:2004fh} (see also \cite{Beck:2006pv}). 
The outcome of such an experiment may be dramatic not only 
for cosmology but also for string theory \cite{Frampton:2005fb}. 
However, we should remind the reader that there is some debate 
as to whether this technique can actually produce evidence of a $\Lambda$ in the laboratory.
In \cite{Jetzer:2004vz}, Jetzer and Straumann claim 
that Dark Energy contributions can not be determined 
from noise measurements of Josephson junctions as 
assumed in \cite{Beck:2004fh}.
This claim is then rebutted by Beck and Mackey 
in \cite{Beck:2006cv}, with Jetzer and Straumann 
arguing against that conclusion in \cite{Jetzer:2006pt} (see also Ref.\cite{Doran:2006ki} on the related theme). 
Time will tell who (if either) are correct. 

{}From now on we assume we have solved the underlying 
$\Lambda$ problem.
It is zero for some reason and dark energy 
is to be explained by some other mechanism. 
Readers only interested in a constant $\Lambda$, 
may want to skip to Sec.~\ref{eosobser} on 
the observational features of dark energy 
as a way of testing for $\Lambda$.

\section{Scalar-field models of dark energy}
\label{scalarmodel}

The cosmological constant corresponds to a fluid with a constant  equation of state
$w=-1$.  Now, the observations which constrain the value of $w$ today 
to be close to that of the cosmological constant, these observations actually 
say relatively little about the time evolution of $w$, and so we can broaden our horizons 
and consider a situation in which 
the equation of state of dark energy changes with time, such as in inflationary cosmology.
Scalar fields naturally arise in particle physics including string theory
and these can act as candidates for dark energy. 
So far a wide variety of scalar-field dark energy models have been 
proposed. These include quintessence, phantoms, K-essence, tachyon, ghost 
condensates and dilatonic dark energy 
amongst many. We shall briefly describe these models in this section.
We will also mention the Chaplygin gas model, although it
is different from scalar-field models of dark energy.
We have to keep in mind that the contribution of the dark matter 
component needs to be taken into account for 
a complete analysis.
Their dynamics will be dealt with in detail 
in Sec.~\ref{cdynamics}.
In the rest of the paper we shall study a flat FRW universe ($K=0$)
unless otherwise specified. 

\subsection{Quintessence}

Quintessence is described by an ordinary scalar field $\phi$ 
minimally coupled to gravity, but as we will see with particular 
potentials that lead to late time inflation. 
The action for Quintessence is given by 
\begin{eqnarray}
S=\int {\rm d}^4x \sqrt{-g} \left[ -\frac12 
(\nabla \phi)^2-V(\phi) \right]\,, 
\label{NG}
\end{eqnarray}
where $(\nabla \phi)^2=g^{\mu \nu} 
\partial_{\mu}\phi \partial_{\nu} \phi$
and $V(\phi)$ is the potential of the field.
In a flat FRW spacetime the variation of the action (\ref{NG}) 
with respect to $\phi$ gives 
\begin{eqnarray}
\ddot{\phi}+3H \dot{\phi}+
\frac{\rd V}{\rd \phi}=0\,.
\label{phieqF}
\end{eqnarray}

The energy momentum tensor of the field is 
derived by varying the action (\ref{NG})
in terms of $g^{\mu \nu}$:
\begin{equation}
T_{\mu\nu}=-\frac{2}{\sqrt{-g}}\frac{\delta S}{\delta
g^{\mu\nu}}\,.
\end{equation}
Taking note that $\delta \sqrt{-g}=-(1/2)\sqrt{-g} g_{\mu \nu}
\delta g^{\mu \nu}$, we find
\begin{equation}
T_{\mu \nu}=\partial_{\mu} \phi \partial_{\nu} \phi-g_{\mu \nu} 
\left[{1 \over 2}g^{\alpha \beta} \partial_{\alpha} 
\phi \partial_{\beta} \phi+V(\phi)\right]\,.
\end{equation}
In the flat Friedmann background we obtain the energy density and 
pressure density of the scalar field:
\begin{equation}
\rho=-T_0^0=\frac12 \dot{\phi}^2+V(\phi)\,,~~~
p=T_i^i=\frac12 \dot{\phi}^2-V(\phi)\,.
\end{equation}
Then Eqs.~(\ref{HubbleeqI}) and (\ref{acceleq}) yield
\begin{eqnarray}
\label{H2}
& &H^2=\frac{8\pi G}{3} \left[\frac12 \dot{\phi}^2
+V(\phi) \right]\,, \\
\label{ddota}
& &\frac{\ddot{a}}{a}=-\frac{8\pi G}{3} 
\left[\dot{\phi}^2-V(\phi) \right]\,.
\end{eqnarray}
We recall that the continuity equation (\ref{conteq}) is
derived by combining these equations. 

{}From Eq.~(\ref{ddota})  we find that the universe accelerates
for $\dot{\phi}^2<V(\phi)$. This means that one requires a
flat potential to give rise to an accelerated 
expansion. In the context of inflation the slow-roll parameters 
\begin{eqnarray}
\epsilon=\frac{m_{\rm pl}^2}{16\pi}
\left(\frac{1}{V}\frac{\rd V}{\rd \phi} \right)^2\,,~~~
\eta=\frac{m_{\rm pl}^2}{8\pi}
\frac{1}{V}\frac{\rd^2 V}{\rd \phi^2}\,,
\end{eqnarray}
are often used to check the existence of an inflationary 
solution for the model (\ref{NG}) \cite{LLbook}.
Inflation occurs if the slow-roll conditions, $\epsilon \ll 1$
and $|\eta| \ll 1$, are satisfied.
In the context of dark energy these slow-roll conditions 
are not completely trustworthy, since there exists dark matter 
as well as dark energy.
However they still provide a good measure 
to check the existence of a solution 
with an accelerated expansion.
If we define slow-roll parameters in terms of the time-derivatives 
of $H$ such as $\epsilon=-\dot{H}/H^2$, this is a good measure 
to check the existence of an accelerated expansion since they 
implement the contributions of both dark energy and dark matter.

The equation of state for the field $\phi$ is given by 
\begin{eqnarray}
w_{\phi}=\frac{p}{\rho}=
\frac{\dot{\phi}^2-2V(\phi)}
{\dot{\phi}^2+2V(\phi)}\,.
\end{eqnarray}
In this case the continuity equation (\ref{conteq}) can be 
written in an integrated form:
\begin{equation}
\label{rhoint}
\rho=\rho_{0} \exp \left[ -\int 3(1+w_{\phi})
\frac{{\rm d}a}{a} \right]\,,
\end{equation}
where $\rho_{0}$ is an integration constant.
We note that the equation of state for the field $\phi$ ranges in 
the region $-1 \le w_{\phi} \le 1$.
The slow-roll limit, $\dot{\phi}^2 \ll V(\phi)$, corresponds to 
$w_{\phi}=-1$, thus giving $\rho={\rm const}$ from Eq.~(\ref{rhoint}).
In the case of a stiff matter characterized by $\dot{\phi}^2 \gg 
V(\phi)$ we have $w_{\phi}=1$, in which case 
the energy density evolves as 
$\rho \propto a^{-6}$ from Eq.~(\ref{rhoint}).
In other cases the energy density behaves as 
\begin{equation}
\rho \propto a^{-m}\,,~~~0< m<6\,.
\end{equation}
Since $w_{\phi}=-1/3$ is the border of acceleration and deceleration,
the universe exhibits an accelerated expansion for $0 \le m<2$ 
[see Eq.~(\ref{sol3})].

It is of interest to derive a scalar-field potential that gives rise
to a power-law expansion:
\begin{equation}
\label{plaw}
a(t) \propto t^p\,.
\end{equation}
The accelerated expansion occurs for $p>1$.
{}From Eq.~(\ref{dotHeq}) we obtain the relation
$\dot{H}=-4\pi G \dot{\phi}^2$.
Then we find that $V(\phi)$ and $\dot{\phi}$ can be expressed
in terms of $H$ and $\dot{H}$:
\begin{eqnarray}
& &V=\frac{3H^2}{8\pi G} \left(1+
\frac{\dot{H}}{3H^2}\right)\,, \\
& & \phi=\int {\rm d}t
\left[-\frac{\dot{H}}{4\pi G}\right]^{1/2}\,.
\end{eqnarray}
Here we chose the positive sign of $\dot{\phi}$.
Hence the potential giving the power-law expansion (\ref{plaw}) 
corresponds to 
\begin{eqnarray}
\label{expo}
V(\phi)=V_{0} \exp \left(-\sqrt{\frac{16\pi}{p}}
\frac{\phi}{m_{\rm pl}}\right)\,,
\end{eqnarray}
where $V_{0}$ is a constant. The field 
evolves as $\phi \propto {\rm ln}\,t$.
The above result shows that the exponential potential
may be used for dark energy provided that $p>1$.

In addition to the fact that exponential potentials can give
rise to an accelerated expansion, they possess cosmological 
{\it scaling solutions} \cite{CLW,BCN99} 
in which the field energy density ($\rho_{\phi}$)
is proportional to the fluid energy density ($\rho_{m}$). 
Exponential potentials were used in one of the earliest  
models which could accommodate a period of acceleration today within it, the loitering universe 
\cite{Sahni:1991ks} (and see \cite{Sahni:2004fb} for 
an example of a loitering universe in the
braneworld context).

In Sec.\,\ref{cdynamics} we shall 
carry out a detailed analysis of the cosmological 
dynamics of an exponential potential in the presence of a barotropic fluid.

The above discussion shows that scalar-field potentials
which are not steep compared to exponential potentials
can lead to an accelerated expansion.
In fact the original quintessence models \cite{peebles,caldwell98} 
are described by the power-law type potential
\begin{equation}
\label{Qpo}    
V(\phi) = \frac{M^{4+\alpha}}{\phi^{\alpha}} \,,
\end{equation}
where $\alpha$ is a positive number 
(it could actually also be negative \cite{LS99})
and $M$ is constant.
Where does the fine tuning arise in these models?
Recall that we need to match the energy density 
in the quintessence field to the current critical 
energy density, that is
\begin{equation}
\rho_{\phi}^{(0)} \approx 
m_{\rm pl}^2 H_0^2 \approx 10^{-47}~{\rm GeV}^4\,.
\end{equation}

The mass squared of the field $\phi$ is given by 
$m_{\phi}^2 =\dfrac{\rd^2 V}{\rd \phi^2} 
\approx \rho_{\phi}/\phi^2$, whereas 
the Hubble expansion rate is given by 
$H^2 \approx \rho_{\phi}/m_{\rm pl}^2$.
The universe enters a tracking regime in which 
the energy density of the field $\phi$ catches
up that of the background fluid
when $m_{\phi}^2$ decreases to of 
order $H^2$ \cite{peebles,caldwell98}.
This shows that the field value at present is of 
order the Planck mass ($\phi_{0} \sim m_{\rm pl}$), 
which is typical of most of the quintessence models.
Since $\rho_{\phi}^{(0)} \approx V(\phi_{0})$,
we obtain the mass scale
\begin{equation}
M=\left( \rho_\phi^{(0)} 
m_{\rm pl}^\alpha \right)^{\frac{1}{4+\alpha}}\,.
\end{equation}

This then constrains the
allowed combination of $\alpha$ and $M$. For example 
the constraint implies $M= 1\,{\rm GeV}$ for 
$\alpha =2$ \cite{Paul99}.
This energy scale can be compatible with the one 
in particle physics, which means that the severe fine-tuning 
problem of the cosmological constant is alleviated. 
Nevertheless a general problem we always 
have to tackle is finding such quintessence 
potentials in particle physics. 
One of the problems is highlighted in Ref.~\cite{Carroll:1998zi}. 
The Quintessence field must couple to ordinary matter, which even if 
suppressed by the Planck scale, 
will lead to long range forces and time dependence of the constants of nature. 
There are tight constraints on such forces and variations and any successful 
model must satisfy them. 
In Sec.~\ref{QK} we shall present a number of quintessence models 
motivated by particle physics.

\subsection{K-essence} \label{formalism} 

Quintessence relies on the potential energy of scalar fields
to lead to the late time acceleration of the universe.
It is possible to have a situation where the accelerated expansion
arises out of modifications to the kinetic energy of the scalar fields.
Originally kinetic energy driven inflation, called K-inflation, was
proposed by Armendariz-Picon {\it et al.} \cite{kinf} to explain 
early universe inflation at high energies.
This scenario was first applied to dark energy by 
Chiba {\it et al.} \cite{COY}. 
The analysis was extended to a more general Lagrangian 
by Armendariz-Picon {\it et al.} \cite{AMS1,AMS2}
and this scenario was called ``K-essence''.

K-essence is characterized by a scalar field with a non-canonical 
kinetic energy. The most general scalar-field action which is a function of
$\phi$ and $X \equiv -(1/2)(\nabla \phi)^2$ is given by 
\begin{eqnarray}
\label{actionK}
S=\int {\rm d}^4x \sqrt{-g}\,p(\phi, X)\,,
\end{eqnarray}
where the Lagrangian density $p(\phi, X)$ corresponds to 
a pressure density.
We note that the action (\ref{actionK}) includes quintessence models.
Usually K-essence models are restricted to the Lagrangian density
of the form \cite{COY,AMS1,AMS2}:
\begin{eqnarray}
\label{Kesse}
p(\phi, X)=f(\phi) \hat{p}(X)\,.
\end{eqnarray}

One of the motivations to consider this type of Lagrangian originates from 
string theory \cite{kinf}.
The low-energy effective string theory generates  
higher-order derivative terms coming from 
$\alpha'$ and loop corrections
(here $\alpha'$ is related to the string length scale $\lambda_{s}$
via the relation $\alpha'=\lambda_{s}/2\pi$).
The four-dimensional effective string action is generally given by 
\begin{eqnarray} 
\label{staction}
S &=&  \int {\rm d}^4x \sqrt{-\tilde{g}} \{B_g(\phi) \wt{R}  + 
B_\phi^{(0)}(\phi) (\wt{\nabla}\phi)^2 \nonumber \\
& &-\alpha'[c_1^{(1)} B_\phi^{(1)}(\phi) (\wt{\nabla}\phi)^4+\cdots]
+{\cal O}(\alpha'^2) \}\,,
\end{eqnarray}
where $\phi$ is the dilaton field that controls the strength 
of the string coupling $g_s^2$ via the relation 
$g_s^2=e^\phi$ \cite{Gas}.
Here we set $\kappa^2=8\pi G=1$.
In the weak coupling regime ($e^{\phi} \ll 1$) the coupling 
functions have the dependence
$B_g \simeq B_\phi^{(0)} \simeq B_\phi^{(1)} \simeq e^{-\phi}$.
As the string coupling becomes of order unity, the form of the 
couplings should take more complicated forms.
If we make a conformal transformation 
$g_{\mu \nu}=B_{g}(\phi) \wt{g}_{\mu \nu}$, the string-frame 
action (\ref{staction}) is transformed to the 
Einstein-frame action \cite{Maeda89,Gas,kinf}:
\begin{eqnarray} 
S_E=\int {\rm d}^4x \sqrt{-g}
\left[\frac12 R+K(\phi)X+L(\phi)X^2+\cdots \right],
\nonumber \\
\end{eqnarray}
where 
\begin{eqnarray} 
\label{Kphi} 
& & K(\phi)=\frac{3}{2}
\left(\frac{1}{B_g} \frac{\rd B_g}{\rd \phi}
\right)^2 - \frac{B_\phi^{(0)}}{B_g}\,, \\
\label{Lphi}
& & L(\phi)=2c_1^{(1)}\alpha' B_{\phi}^{(1)}(\phi)\,.
\end{eqnarray}
Hence this induces a Lagrangian with noncanonical 
kinetic terms:
\begin{eqnarray} 
\label{KL}
p(\phi, X)=K(\phi)X+L(\phi)X^2\,.
\end{eqnarray}

If we make the field redefinition
\begin{eqnarray}
\phi_{\rm new}=\int^{\phi_{\rm old}} {\rm d}\phi \sqrt{\frac{L}{|K|}}\,,
\end{eqnarray}
the Lagrangian (\ref{KL}) transforms into \cite{COY}
\begin{eqnarray}
\label{pX2}    
p(\phi, X)=f(\phi) (-X+X^2)\,,
\end{eqnarray}
where $\phi \equiv \phi_{\rm new}$, 
$X \equiv X_{\rm new}=(L/|K|) X_{\rm old}$
and $f(\phi)=K^2(\phi_{\rm old})/L(\phi_{\rm old})$.
This shows that the model given by (\ref{KL})
falls into the category of K-essence 
(\ref{Kesse}) with a choice $\hat{p}(X)=-X+X^2$
after an appropriate field definition.

For the pressure density (\ref{pX2}) we find that the 
energy density of the field $\phi$ is given by 
\begin{eqnarray}
\label{rhoK}    
\rho=2X \frac{\partial p}{\partial X}-p
=f(\phi)(-X+3X^2)\,.
\end{eqnarray}
Then the equation of state of the field is given by 
\begin{eqnarray}
\label{eosK}
w_{\phi}=\frac{p}{\rho}
=\frac{1-X}{1-3X}\,.
\end{eqnarray}
This shows that $w_\phi$ does not vary for constant $X$.
For example we obtain the equation of state of a cosmological 
constant ($w_{\phi}=-1$) for $X=1/2$.
The equation of state giving rise to an accelerated expansion is 
$w_{\phi}<-1/3$, which translates into the condition 
$X<2/3$.

We recall that the energy density $\rho$ satisfies 
the continuity equation (\ref{conteq}).
During the radiation or matter dominant era
in which the equation of state of the background fluid 
is $w_{m}$, the evolution of the 
Hubble rate is given by $H=2/[3(1+w_{m})(t-t_{0})]$
from Eq.~(\ref{sol1}).
Then the energy density $\rho$ of the field 
$\phi$ satisfies
\begin{eqnarray}
\label{rhodot}
\dot{\rho}=-\frac{2(1+w_{\phi})}
{(1+w_m)(t-t_{0})}\rho\,.
\end{eqnarray}
For constant $X$ (i.e., constant $w_{\phi}$) 
the form of $f(\phi)$ is constrained to be  
\begin{eqnarray}
f(\phi) \propto (\phi-\phi_{0})^{-\alpha}\,,\quad
\alpha=\frac{2(1+w_{\phi})}{1+w_{m}}\,,
\end{eqnarray}
where we used Eqs.~(\ref{rhoK}) and (\ref{rhodot}).

When $w_{\phi}=w_{m}$ the function $f(\phi)$
behaves as $f(\phi) \propto (\phi-\phi_{0})^{-2}$
in the radiation or matter dominant era.
This corresponds to the scaling solutions, 
as we will see in Sec.~\ref{scalingsec}.
In the case of $w_{\phi}=-1$ we find that 
$f(\phi)={\rm const}$ with $X=1/2$.
This corresponds to the ghost condensate scenario
proposed in Ref.~\cite{Arkani-hamad}.
In order to apply this to dark energy we need to 
fine-tune $f(\phi)$ to be of order 
the present energy density of the universe.
We caution that the above function $f(\phi)$
is obtained by assuming that the energy density of 
the field is much smaller than that of the background
fluid ($\rho \ll \rho_{m}$). Hence this is no longer 
applicable for a dark energy dominated universe.
For example even for $f(\phi) \propto (\phi-\phi_{0})^{-2}$
there exists another solution giving an accelerated expansion
other than the scaling solutions at late times. In fact this case marks the 
border between acceleration and deceleration.
We will clarify these issues in Sec.~\ref{cdynamics}.

Equation (\ref{eosK}) shows that the kinetic term $X$
plays a crucial role in determining the equation of state of $\phi$.
As long as $X$ belongs in the range $1/2<X<2/3$,
the field $\phi$ behaves as dark energy for
$0 \le \alpha \le 2$.
The model (\ref{pX2}) describes one of 
the examples of K-essence.
In fact Armendariz-Picon {\it et al.} \cite{AMS1,AMS2}
extended the analysis to more general forms of 
$\hat{p}(X)$ in Eq.~(\ref{Kesse}) 
to solve the coincident problem of dark energy.
See Refs.~\cite{Kessencepapers} for various aspects 
of K-essence.

\subsection{Tachyon field}

Recently it has been suggested that rolling tachyon condensates, in
a class of string theories, may have interesting cosmological consequences.
Sen \cite{Sen} showed that the decay of D-branes 
produces a pressureless gas with finite energy density 
that resembles classical dust (see also Refs.~\cite{tac}). 
A rolling tachyon has
an interesting equation of state whose parameter smoothly
interpolates between $-1$ and $0$ \cite{Gibbons}. This has led to a flurry of 
attempts being made to construct viable
cosmological models using the tachyon as a suitable candidate 
for the inflaton at high energy \cite{tacinflation}. 
However tachyon inflation in open string models is typically 
plagued by several difficulties \cite{KL} associated with 
density perturbations and 
reheating\footnote{We note that these 
problems are alleviated in D-branes in 
a warped metric \cite{GST}
or in the case of the geometrical tachyon \cite{geo}.}.
Meanwhile the tachyon can also act as a source of dark energy 
depending upon the form of the tachyon
potential \cite{Paddy,Bagla,AF,AL,GZ,CGST}. 
In what follows we shall consider the tachyon as a field from which 
it is possible to obtain viable models of dark energy.

The effective Lagrangian for the tachyon on a non-BPS 
D3-brane is described by 
\begin{eqnarray}
S=-\int {\rm d}^{4}x\,V(\phi)\sqrt{-\det(g_{ab}+\prt_{a}
\phi\prt_{b}\phi)}\,, 
\label{taction}
\end{eqnarray}
where $V(\phi)$ is the tachyon potential.
The effective potential
obtained in open string theory has the form \cite{Kutasov:2003er}
\begin{equation}
V(\phi)=\frac{V_0}{\cosh \left(\phi/\phi_0 \right)}\,,
\label{tpot}
\end{equation}
where $\phi_0=\sqrt{2}$ for the non-BPS D-brane in the superstring 
and  $\phi_0={2}$ for the bosonic string.
Note that the tachyon field has a ground state at $\phi \to \infty$.
There exists another type of tachyon potential which appears as 
the excitation of massive scalar fields on the anti-D branes \cite{GST}. 
In this case the potential is given by 
$V(\phi)=V_0 e^{\frac12 m^2\phi^2}$ and it has a minimum at $\phi=0$.
In this review we keep the tachyon potential as general as possible 
and will carry out a detailed analysis of the associated dynamics in Sec.~\ref{cdynamics}.

The energy momentum tensor which follows 
from the action (\ref{taction}) has the form
\begin{equation}
T_{\mu \nu}=\frac{V(\phi)\partial_{\mu }\phi \partial_{\nu }\phi}
{\sqrt{1+g^{\alpha \beta} \partial_{\alpha}\phi
\partial_{\beta}\phi}} 
-g_{\mu \nu} V(\phi)
\sqrt{1+g^{\alpha \beta}\partial_{\alpha}\phi
\partial_{\beta}\phi}\,.
\end{equation}
In a flat FRW background
the energy density $\rho$ and the pressure density 
$p$  are given by
\begin{eqnarray}
\label{rhotach}
& &\rho=-T^0_0 = {V(\phi) \over {\sqrt{1-\dot{\phi}^2}}}\,,\\
\label{ptach}
& &p=T_i^i = -V(\phi)\sqrt{1-\dot{\phi}^2}\,. 
\end{eqnarray}

{}From Eqs.~(\ref{HubbleeqI}) and (\ref{conteq}) 
we obtain the following equations of motion:
\begin{eqnarray}
\label{tachH}
& &H^2=\frac{8\pi GV(\phi)}{3\sqrt{1-\dot{\phi}^2}}\,,\\
\label{tachevo}
& &{\ddot{\phi} \over {1-\dot{\phi}^2}}+3H \dot{\phi}+
\frac{1}{V} \frac{{\rm d}V}{{\rm d}\phi}=0\,.
\end{eqnarray}
Combining these equations gives
\begin{eqnarray}
\label{tachddota}
\frac{\ddot{a}}{a}=\frac{8\pi G V(\phi)}
{3\sqrt{1-\dot{\phi}^2}}
\left(1-\frac32 \dot{\phi}^2 \right)\,.
\end{eqnarray}
Hence an accelerated expansion occurs for 
$\dot{\phi}^2<2/3$.

The equation of state of the tachyon is given by 
\begin{eqnarray}
\label{EOStac}    
w_{\phi}=\frac{p}{\rho}=\dot{\phi}^2-1\,.
\end{eqnarray}
Now the tachyon dynamics is very different from the standard field case. 
Irrespective of the steepness of the tachyon potential, the equation of state 
varies between 0 and $-1$, in which case the tachyon energy density 
behaves as $\rho \propto a^{-m}$ with $0<m<3$ from 
Eq.~(\ref{rhoint}).

One can express $V(\phi)$ and $\phi$ in terms of 
$H$ and $\dot{H}$, as we did in the case of 
Quintessence\footnote{Note that a ``first-order formalism'' 
which relates the potential to the Hubble parameter is given in 
Ref.~\cite{Bazeia}}.
{}From Eqs.~(\ref{tachH}) and (\ref{tachddota}) we find 
$\dot{H}/H^2=-(3/2)\dot{\phi}^2$. Then together with 
Eq.~(\ref{tachH}) we obtain \cite{Paddy}
\begin{eqnarray}
& &V=\frac{3H^2}{8\pi G} \left(1+
\frac{2\dot{H}}{3H^2}\right)^{1/2}\,, \\
& & \phi=\int {\rm d}t
\left(-\frac{2\dot{H}}{3H^2}
\right)^{1/2}\,.
\end{eqnarray}
Then the tachyon potential giving the power-law expansion, 
$a \propto t^p$, is
\begin{eqnarray}
\label{inpower}
V(\phi)=\frac{2p}{4\pi G} \left(1-\frac{2}{3p}
\right)^{1/2} \phi^{-2}\,.
\end{eqnarray}
In this case the evolution of the tachyon is given by 
$\phi=\sqrt{2/3p}\,t$ (where we set an integration constant to zero).
The above inverse square power-law potential
corresponds to the one in the case of scaling 
solutions \cite{AL,CGST}, as we will see later.
Tachyon potentials which are not steep compared to 
$V(\phi) \propto \phi^{-2}$ lead to an accelerated 
expansion.  In Sec.~\ref{cdynamics} we will consider the  
cosmological evolution for a more general 
inverse power-law potential given by $V(\phi) \propto \phi^{-n}$.
There have been a number of papers written concerning the cosmology of tachyons. 
A fairly comprehensive listing can be seen in Ref.~\cite{tachyonpapers}.

\subsection{Phantom (ghost) field}

Recent observational data indicates
that the equation of state  parameter $w$ lies
in a narrow strip around $w=-1$ and is quite consistent with being below 
this value \cite{Corasaniti:2004sz,ASSS}.
The scalar field models discussed in the previous subsections correspond 
to an equation of state $w \ge -1$. 
The region where the equation of state
is less than $-1$ is typically referred to as a being due to some form of phantom (ghost) dark energy.
Specific models in braneworlds or Brans-Dicke scalar-tensor gravity
can lead to phantom energy \cite{SS,NOphan}. 
Meanwhile the simplest explanation for the phantom dark energy is provided by a
scalar field with a negative kinetic energy \cite{Caldwell02}. 
Such a field may be motivated from $S$-brane constructions
in string theory \cite{Sbranes}.

Historically, phantom fields were first introduced in
Hoyle's version of the steady state theory. 
In adherence to the perfect cosmological principle, 
a creation field (C-field) was introduced by Hoyle
to reconcile the model with the homogeneous density of the universe by the creation of 
new matter in the voids caused by the expansion of the universe  \cite{Hoyle}.
It was further refined and reformulated in the Hoyle and Narlikar theory 
of gravitation \cite{HN} (see also Ref.~\cite{NP} on a similar theme).
The action of the phantom field minimally coupled to gravity 
is given by
\begin{eqnarray}
S=\int {\rm d}^4x \sqrt{-g} \left[ \frac12 
(\nabla \phi)^2-V(\phi) \right]\,, 
\label{lagphantom}
\end{eqnarray}
where the sign of the kinetic term is opposite
compared to the action (\ref{NG}) 
for an ordinary scalar field.
Since the energy density and pressure density are given
by $\rho=-\dot{\phi}^2/2+V(\phi)$ and 
$p=-\dot{\phi}^2/2-V(\phi)$ respectively, 
the equation of state of the field is
\begin{equation}
w_{\phi} ={p \over \rho} 
=\frac{\dot{\phi}^2+2V(\phi)} 
{\dot{\phi}^2-2V(\phi)}\,.
\end{equation}
Then we obtain $w_{\phi}<-1$ for $\dot{\phi}^2/2<V(\phi)$.

As discussed in Sec.~\ref{FRW} the curvature of the universe grows 
toward infinity within a finite time in the universe dominated by a 
phantom fluid. In the case of a phantom scalar field this Big Rip 
singularity may be avoided if the potential has a maximum, e.g., 
\begin{equation}
\label{bell}
V(\phi)=V_0\left[\cosh \left({{\alpha \phi}
\over m_{\rm pl}}\right)\right]^{-1}\,,
\end{equation}
where $\alpha$ is  constant \cite{SSN}.
Due to its peculiar properties, the phantom field evolves towards the top of the potential 
and crosses over to the other side. 
It turns back to execute a period of  damped oscillations 
about the maximum of the potential at $\phi=0$. 
After a certain period of time the motion ceases and the field 
settles at the top of the potential
to mimic the de-Sitter like behavior ($w_{\phi}=-1$). 
This behavior is generic if the potential has a maximum, see 
e.g.,~Ref.~\cite{Carroll03}.
In the case of exponential potentials the system approaches a 
constant equation of state with $w_{\phi}<-1$ \cite{ST04}, 
as we will see in Sec.~\ref{cdynamics}.

Although the above behavior of the phantom field is intriguing as a ``classical cosmological'' field, 
unfortunately phantom fields  are generally plagued by 
severe Ultra-Violet (UV) quantum instabilities.
Since the energy density of a phantom field is 
unbounded from below, the vacuum becomes 
unstable against the production of 
ghosts and normal (positive energy) fields \cite{Carroll03}. 
Even when ghosts are decoupled from matter fields, they couple to gravitons 
which mediate vacuum decay processes of 
the type: vacuum $\rightarrow 2$\, ghosts + $2 \gamma$. 
It was shown by Cline {\it et al.} \cite{Cline} that we require 
an unnatural Lorenz invariance breaking term with cut off of order 
$\sim$ MeV to prevent an 
overproduction of cosmic gamma rays.
Hence the fundamental origin of the phantom field still 
poses an interesting challenge for theoreticians.
See Refs.~\cite{phantompapers} for a selection of papers 
covering various cosmological aspects 
of phantom fields.

 \subsection{Dilatonic dark energy}

We have already mentioned in the previous subsection that the phantom 
field with a negative kinetic term has a problem with quantum instabilities.
Let us consider the stability of perturbations by
decomposing the field $\phi$ into a homogeneous part
$\phi_0$ and a fluctuation $\delta \phi$, as
\begin{eqnarray} \label{decompose}
\phi(t, {\bf x})=
\phi_0(t)+\delta \phi(t, {\bf x}) \,.
\label{phide}
\end{eqnarray}
Since we are concerned with the UV instability of the vacuum,
it is not too restrictive to choose a Minkowski background metric
when studying quantum fluctuations, 
because we are interested in high energy, short distance effects.

Let us start with a general Lagrangian density $p(\phi, X)$.
Expanding $p(X,\phi)$ to second order in $\delta \phi$ it is
straightforward to find the Lagrangian together with
the Hamiltonian for the fluctuations.
The Hamiltonian is given by \cite{PT}
\begin{eqnarray}
{\cal H} &=& \left(p_{,X}
+2 Xp_{,XX}\right)
 \frac{(\delta \dot{\phi})^2}{2} \nonumber \\
& &
+p_{,X}
\frac{(\nabla \delta \phi)^2}{2}
-p_{,\phi \phi}
\frac{(\delta \phi)^2}{2}\,,
\label{Ldelpsi}
\end{eqnarray}
where $p_{,X} \equiv \partial p/\partial X$.
It is positive as long as the following conditions hold
\begin{eqnarray}
\label{xi}
\hspace*{-1.0em}
& &\xi_1\,  \equiv \, 
p_{,X} +2 X p_{,XX}
\, \ge \, 0,~~
\xi_2 \, \equiv \, 
p_{,X}
\, \ge \, 0, \\
\hspace*{-1.0em}
& &\xi_3 \, \equiv \, 
-p_{,\phi \phi}
\, \ge \, 0\,. 
\label{xi2}
\end{eqnarray}

The speed of sound is given by 
\begin{eqnarray}
 \label{sound}
 c_s^2 \equiv \frac{p_{,X}}
 {\rho_{,X}}
 =\frac{\xi_2}{\xi_1}\,,
\end{eqnarray}
which is often used when we discuss the stability of
classical perturbations, since it appears as a coefficient of the
$k^2/a^2$ term ($k$ is a comoving wavenumber). 
Although the classical fluctuations may be
regarded to be stable when $c_s^2>0$, the stability of 
quantum fluctuations requires both $\xi_1>0$ and $\xi_2 \ge 0$.
We note that the instability prevented by the condition (\ref{xi2})
is essentially an Infra-Red (IR) instability which is less dramatic compared to
the instability associated with the violation of the condition (\ref{xi}).
In fact this IR instability appears in the context of density 
perturbations generated in inflationary cosmology.
Hence we shall adopt (\ref{xi}) but not (\ref{xi2}) as the 
fundamental criteria for the consistency of the theory.
These two conditions prevent an instability
related to the presence of negative energy ghost
states which render the vacuum unstable under a
catastrophic production of ghosts and
photons pairs \cite{Cline}.
This is essentially an Ultra-Violet  instability with which
the rate of production from the vacuum
is simply proportional to the phase space integral
on all possible final states.

In the case of a phantom scalar field $\phi$
with a potential $V(\phi)$, i.e., $p=-X-V(\phi)$,
we find that $\xi_1=\xi_2=-1$.
Hence the system is quantum mechanically unstable
even though the speed of sound is positive ($c_s^2>0$).
It was shown in Ref.~\cite{Arkani-hamad} that
a scalar field with a negative kinetic term does not
necessarily lead to inconsistencies, provided that a suitable
structure of higher-order kinetic terms are present in the
effective theory. The simplest model that realizes
this stability is $p=-X+X^2$ \cite{Arkani-hamad}.
In this case one has $\xi_1=-1+6X$ and $\xi_2=-1+2X$.
When $\xi_1>0$ and $\xi_2 \ge 0$, corresponding to $X \ge 1/2$,
the system is completely stable at the quantum level.
In the region of $0 \le X < 1/6$ one has $\xi_1<0$ and $\xi_2<0$ so that
the perturbations are classically stable due to the
positive sign of $c_s^2$. This vacuum state is,
however, generally quantum mechanically unstable.

It is difficult to apply the model $p=-X+X^2$ for
dark energy as it is. This is because the small energy density of
the scalar field relative to the Planck density gives the
condition $|X| \gg X^2$, in which case one can not ensure
the stability of quantum fluctuations.
Instead one may consider the following 
{\it dilatonic ghost condensate} model:
\begin{equation}
\label{ghostcon}
p=-X+ce^{\lambda \phi}X^2\,,
\end{equation}
where $c$ is a positive constant.
This is motivated by dilatonic higher-order corrections
to the tree-level action in low energy effective string theory \cite{PT}.
We assume that the dilaton is effectively decoupled
from gravity in the limit $\phi \to \infty$.
This is the so-called the runaway dilaton 
scenario \cite{runaway} in which the coupling functions
in Eq.~(\ref{staction}) are given by 
\begin{eqnarray}
& &B_g(\phi)=C_g+D_ge^{-\phi}+
{\cal O}(e^{-2\phi})\,, \\
& &B_\phi^{(0)}(\phi)=C_\phi^{(0)}+D_\phi^{(0)}
e^{-\phi}+
{\cal O}(e^{-2\phi})\,.
\end{eqnarray}
In this case $B_g(\phi)$ and $B_\phi^{(0)}(\phi)$ 
approach constant values as $\phi \to \infty$.
Hence the dilaton gradually decouples from gravity 
as the field evolves toward the region $\phi \gg 1$
from the weakly coupled regime.

In the Einstein frame the function $K(\phi)$ given by 
Eq.~(\ref{Kphi}) also approaches
a constant value, whose sign depends upon the coefficients
of $B_g(\phi)$ and $B_\phi^{(0)}(\phi)$.
The dilatonic ghost condensate model corresponds to 
negative $K(\phi)$.
{}From Eq.~(\ref{Lphi}) we find that 
the coefficient in front of the $(\nabla \phi)^4$ term
has a dependence $B_{\phi}^{(1)} \propto e^{\lambda \phi}$
in the dilatonic ghost condensate.
Since the $e^{\lambda \phi}$ term in Eq.~(\ref{ghostcon})
can be large for $\phi \to \infty$, the second term in
Eq.~(\ref{ghostcon}) can stabilize the vacuum
even if $X$ is much smaller than the Planck scale.
The condition for quantum stability is characterized by 
the condition $ce^{\lambda \phi}X \ge 1/2$ from Eq.~(\ref{xi}).
 
It is worth mentioning that the Lagrangian density (\ref{ghostcon})
is transformed to Eq.~(\ref{pX2}) with $f(\phi) \propto 
(\phi-\phi_{0})^{-2}$ by a field redefinition.
In subsection B we showed that this case has a scaling solution 
in the radiation or matter dominating era.
This means that dilatonic ghost condensate model has
scaling solutions. In Sec.~\ref{scalingsec} we will show this 
in a more rigorous way and carry out a detailed analysis in 
Sec.~\ref{cdynamics} about the cosmological 
evolution for the Lagrangian density (\ref{ghostcon}).
The above discussion explicitly tells us that (dilatonic) ghost 
condensate models fall into the category of K-essence.

Gasperini {\it et al.} proposed a runaway dilatonic 
quintessence scenario \cite{runaway} 
in which $K(\phi)$ approaches a {\it positive} constant as 
$\phi \to \infty$. They assumed the presence of an exponential 
potential $V(\phi)=V_0 e^{-\lambda \phi}$ which vanishes
for $\phi \to \infty$. The higher-order kinetic term $X^2$
is neglected in their analysis.
They took into account the coupling between the field $\phi$
and dark matter, since the dilaton is naturally coupled to 
matter fields. This model is also an interesting attempt to 
explain the origin of dark energy using string theory.

 \subsection{Chaplygin gas}
So far we have discussed a number of scalar-field models of dark energy. 
There exist another interesting class of dark energy models involving a fluid known as a Chaplygin 
gas \cite{Kamenshchik:2001cp}. This fluid also leads to the acceleration of the universe
at late times, and in its simplest form has the following specific  equation of state:
\begin{equation}
p=-\frac{A}{\rho}\,,
\label{chapgas}
\end{equation}
where $A$ is a positive constant. 
We recall that $p=-V^2(\phi)/\rho$ for the tachyon 
from Eqs.~(\ref{rhotach}) and (\ref{ptach}).
Hence the Chaplygin gas can be regarded as a special case of a
tachyon with a constant potential.

The equation of state for the Chaplygin gas can be 
derived from the Nambu-Goto action for a D-brane 
moving in the $D+1$ dimensional bulk \cite{bilicthanks,Jackiw:2000mm}.  
For the case of the moving brane (via the Born-Infeld Lagrangian),
the derivation of the Chaplygin gas equation of state was 
first discussed in the context of braneworld cosmologies 
in \cite{Bilic:2002vm}.

With the equation of state (\ref{chapgas}) 
the continuity equation (\ref{conteq}) can be integrated to give
\begin{equation}
\rho=\sqrt{A+\frac{B}{a^6}}\,,
\label{chaprho}
\end{equation}
where $B$ is a constant.
Then we find the following asymptotic behavior:
\begin{eqnarray}
\label{rho1}
&& \rho \sim \frac{\sqrt{B}}{a^3}\,,~~~~a \ll (B/A)^{1/6} \,, \\
\label{rho2}
&&\rho \sim -p \sim \sqrt{A}~~~~a \gg (B/A)^{1/6}\,.
\end{eqnarray}
This is the intriguing result for the Chaplygin gas. At early times when $a$ is small, 
the gas behaves as a pressureless dust.
Meanwhile it behaves as a cosmological constant at late times,
thus leading to an accelerated expansion.

One can obtain a corresponding potential for the Chaplygin gas
by treating it as an ordinary scalar field $\phi$.
Using Eqs.~(\ref{chapgas}) and (\ref{chaprho}) together with 
$\rho=\dot{\phi}^2/2+V(\phi)$ and $p=\dot{\phi}^2/2-V(\phi)$, 
we find 
\begin{eqnarray}
\label{Chapdot}
&& \dot{\phi}^2=\frac{B}{a^6\sqrt{A+B/a^6}}
\,, \\
\label{ChapV}
&& V=\frac12 \left[\sqrt{A+B/a^6}+
\frac{A}{\sqrt{A+B/a^6}} \right]\,.
\end{eqnarray}
We note that this procedure is analogous to the reconstruction 
methods we adopted for the quintessence and tachyon potentials.
Since the Hubble expansion rate is given by 
$H=(8\pi \rho/3m_{\rm pl}^2)^{1/2}$, we can rewrite 
Eq.~(\ref{Chapdot}) in terms of the derivative of $a$:
\begin{eqnarray}
\frac{\kappa}{\sqrt{3}} \frac{{\rm d}\phi}
{{\rm d} a}=\frac{\sqrt{B}}
{a\sqrt{Aa^6+B}}\,.
\end{eqnarray}
This is easily integrated to give
\begin{eqnarray}
a^6=\frac{4B e^{2\sqrt{3}\kappa \phi}}
{A(1-e^{2\sqrt{3}\kappa \phi})^2}\,.
\end{eqnarray}
Substituting this for Eq.~(\ref{ChapV}) we obtain 
the following potential:
\begin{equation}
V(\phi)=\frac{\sqrt{A}}{2}\left(\cosh \sqrt{3}\kappa \phi
+\frac{1}{\cosh \sqrt{3}\kappa \phi} \right)\,.
\end{equation}
Hence, a minimally coupled field with this potential is 
equivalent to the Chaplygin gas model.

Chaplygin gas provides an interesting possibility for the unification of dark energy
and dark matter. However it was shown in Ref.~\cite{AFBC} that the Chaplygin 
gas models are under strong observational pressure from CMB anisotropies
(see also Ref.~\cite{Bilic02,Bento03}).
This comes from the fact that the Jeans instability of perturbations 
in Chaplygin gas models behaves similarly to cold dark matter fluctuations in the 
dust-dominant stage given by (\ref{rho1}) but disappears 
in the acceleration stage given by (\ref{rho2}).
The combined effect of the suppression of perturbations and the 
presence of a non-zero Jeans length gives rise to a strong integrated 
Sachs-Wolfe (ISW) effect, thereby leading to the loss of power 
in CMB anisotropies.
This situation can be alleviated in the generalized 
Chaplygin gas model introduced in Ref.~\cite{Bento:2002ps}
with $p=-A/\rho^{\alpha}$, $0<\alpha<1$.
However, even in this case the parameter $\alpha$ is rather severely 
constrained, i.e., $0 \le \alpha <0.2$
at  the 95\% confidence level \cite{AFBC}.
For further details of the cosmology associated 
with generalized Chaplygin gas models, see  Refs.~\cite{Chappapers}.

\section{Cosmological dynamics of scalar fields
in the presence of a barotropic perfect fluid}
\label{cdynamics}

In order to obtain viable dark energy models, we require 
that the energy density of the scalar field remains
subdominant during the radiation and matter dominating eras, emerging only at late times to
give rise to the current observed acceleration of the universe.
In this section we shall carry out cosmological dynamics 
of a scalar field $\phi$ in the presence of a barotropic fluid
whose equation of state is given by 
$w_m=p_m/\rho_m$.
We denote pressure and energy densities
of the scalar field as $p_{\phi}$ and 
$\rho_{\phi}$ with an equation
of state $w_{\phi}=p_{\phi}/\rho_{\phi}$.
Equations (\ref{HubbleeqI}) and (\ref{dotHeq}) give 
\begin{eqnarray}
\label{Htwo}
& & H^2=\frac{8\pi G}{3} (\rho_{\phi}+\rho_{m})\,, \\
\label{dotHtwo}
& & \dot{H}=-4\pi G (\rho_{\phi}+p_{\phi}+\rho_{m}+p_{m})\,.
\end{eqnarray}
Here the energy densities $\rho_{\phi}$ and $\rho_{m}$
satisfy
\begin{eqnarray}
\label{dotrhotwo}
& & \dot{\rho}_{\phi}+3H(1+w_{\phi})\rho_{\phi}=0\,, \\
& & \dot{\rho}_{m}+3H(1+w_m)\rho_{m}=0\,.
\end{eqnarray}
In what follows we shall assume that $w_{m}$ is 
constant, which means that the fluid energy 
is given by $\rho_{m}=\rho_{0} a^{-3(1+w_{m})}$.
Meanwhile $w_{\phi}$ dynamically changes in general.

Of particular importance in the investigation of cosmological scenarios
are those solutions in which the energy density of the scalar field mimics 
the background fluid energy density. 
Cosmological solutions which satisfy this condition 
are called  ``{\it scaling solutions}'' \cite{CLW}
(see also Refs.~\cite{LS99,van,MP,NM,Ng,RB,Uzan,Chinas}).
Namely scaling solutions are characterized by the relation 
\be
\label{ratio}
\rho_{\phi}/\rho_m=C\,,
\ee
where $C$ is a {\it nonzero} constant.
As we have already mentioned in the previous section, exponential 
potentials give rise to scaling solutions and so can play an important role in quintessence scenarios, 
allowing the field energy density to mimic the 
background being sub-dominant during radiation and matter 
dominating eras. In this case, as long as the scaling solution is the attractor, 
then for any generic initial conditions, the field would sooner 
or later enter the scaling regime, thereby opening up a new line of
attack on the fine tuning problem of dark energy. 

We note that the system needs to exit from the scaling regime 
characterized by Eq.~(\ref{ratio}) in order to give rise to 
an accelerated expansion.
This is realized if the slope of the field potential becomes shallow 
at late times compared to the one corresponding to 
the scaling solution \cite{BCN99,SW99}.
We shall study these models in more details in Sec.~\ref{QK}.
It is worth mentioning that scaling solutions live on the the border between
acceleration and deceleration. Hence the energy density of 
the field catches up to that of the fluid provided that the 
potential is shallow relative to the one corresponding to the scaling solutions. 
In what follows we shall
study the dynamics of scalar fields in great detail 
for a variety of dark energy  models.
First, we explain the property of an autonomous
system before entering the detailed analysis.

\subsection{Autonomous system of scalar-field dark energy models}

A dynamical system which plays an important role in cosmology belongs to
the class of so called autonomous systems \cite{CLW,Derek}. 
We first briefly present some basic definitions related to dynamical systems 
(see also \cite{Szydlowski:2005uq,Szydlowski:2006ma} for a related approach).
For simplicity we shall study the system of two first-order 
differential equations, but
the analysis can be extended to a system of any number of equations. 
Let us consider the following coupled differential equations 
for two variables $x(t)$ and $y(t)$:
\begin{eqnarray}
\dot{x}=f(x,y,t) \,, \quad
\dot{y}=g(x,y,t)\,,
\label{fgxyt}
\label{auto}
\end{eqnarray}
where $f$ and $g$ are the functions in terms of $x, y$ and $t$. 
The system (\ref{auto}) is said to be autonomous
if $f$ and $g$ do not contain explicit time-dependent
terms. The dynamics of the autonomous systems can be
analyzed in the following way.

\subsubsection{Fixed or critical points}

A point $(x_c,y_c)$ is said to be a {\it fixed point} 
or a {\it critical point} of the autonomous system if
\begin{equation}
(f,g)|_{(x_c,y_c)}=0\,.
\end{equation}
A critical point $(x_c,y_c)$ is called an {\it attractor}
when it satisfies the condition
\begin{equation}
\left(x(t),y(t) \right) \to (x_c,y_c)~~{\rm for}~~
t \to \infty\,.
\end{equation}
\subsubsection{Stability around the fixed points}

We can find whether the system approaches one of the  
critical points or not by studying the stability around 
the fixed points.
Let us consider small perturbations $\delta x$ and 
$\delta y$ around the critical point $(x_{c}, y_{c})$, i.e.,
\bea
\label{per}
x=x_{c}+\delta x\,,~~~
y=y_{c}+\delta y\,.
\eea
Then substituting into Eqs.~(\ref{fgxyt}) 
leads to the first-order differential equations:
\begin{eqnarray}
\frac{\rd }{\rd N}
\left(
\begin{array}{c}
\delta x \\
\delta y
\end{array}
\right) = {\cal M} \left(
\begin{array}{c}
\delta x \\
\delta y
\end{array}
\right) \,,
\label{uvdif}
\end{eqnarray}
where $N={\rm ln}\,(a)$ is the number of $e$-foldings
which is convenient to use for the dynamics of dark energy.
The matrix ${\cal M}$ depends upon
$x_c$ and $y_c$, and is given by 
\begin{eqnarray}
 \label{matM}
{\cal M}=\left( \begin{array}{cc}
\frac{\partial f}{\partial x}& \frac{\partial f}{\partial y}\\
\frac{\partial g}{\partial x}& \frac{\partial g}{\partial y}
\end{array} \right)_{(x=x_c,y=y_c)}\,.
\end{eqnarray}

This possesses two eigenvalues $\mu_1$
and $\mu_2$. The general solution for the evolution of 
linear perturbations can be written as
\begin{eqnarray}
&&\delta x=C_1e^{\mu_1 N}+C_2e^{\mu_2 N}\,, \\
&&\delta y=C_3e^{\mu_1 N}+C_4e^{\mu_2 N}\,,
\end{eqnarray}
where $C_1, C_2, C_3, C_4$ are integration constants.
Thus the stability around the fixed points depends upon the
nature of the eigenvalues. 
One generally uses the following 
classification \cite{CLW,GNST}:

\begin{itemize}

\item (i) Stable node: $\mu_1<0$ and $\mu_2<0$.

\item (ii) Unstable node: $\mu_1>0$ and $\mu_2>0$.

\item (iii) Saddle point: $\mu_1<0$ and $\mu_2>0$ (or $\mu_1 >0$ and
$\mu_2<0$).

\item (iv) Stable spiral: The determinant of the matrix ${\cal M}$
is negative and the real parts of $\mu_1$ and $\mu_2$ are negative.

\end {itemize}

A fixed point is an attractor in the cases (i) and (iv), but
it is not so in the cases (ii) and (iii).

\subsection{Quintessence}

Let us consider a minimally coupled scalar field $\phi$
with a potential $V(\phi)$ whose Lagrangian density 
is given by 
\be
\label{quin}
{\cal L}=\frac12 \epsilon\dot{\phi}^2+V(\phi)\,,
\ee
where $\epsilon=+1$ for an ordinary scalar field.
Here we also allow for the possibility of a
phantom ($\epsilon=-1$) as we see in the 
next subsection.
For the above Lagrangian density (\ref{quin}), 
Eqs.~(\ref{Htwo}), (\ref{dotHtwo}) and 
(\ref{dotrhotwo}) read
\bea
& & H^2=\frac{\kappa^2}{3}
\left[\frac12 \epsilon \dot{\phi}^2+V(\phi)
+\rho_{m} \right]\,, \\
& & \dot{H}=-\frac{\kappa^2}{2}
\left[ \epsilon \dot{\phi}^2+(1+w_{m})\rho_{m}
\right]\,, \\
& & \epsilon \ddot{\phi}+3H\dot{\phi}+
\frac{{\rm d}V}{{\rm d}\phi}=0\,.
\eea

Let us introduce the following dimensionless quantities
\bea
& &
x \equiv \frac{\kappa \dot{\phi}}{\sqrt{6}H}\,,~~
y \equiv \frac{\kappa \sqrt{V}}{\sqrt{3}H}\,, 
\nonumber \\
& &
\label{lamGam}
\lambda \equiv -\frac{V_{,\phi}}{\kappa V}\,,~~
\Gamma \equiv \frac{VV_{,\phi \phi}}{V_{,\phi}^2}\,,
\eea
where $V_{,\phi} \equiv {\rm d}V/{\rm d}\phi$.
Then the above equations can be written in the 
following autonomous form \cite{CLW,Ng}:
\bea
\label{autoquin1}
\hspace*{-1.5em}
\frac{\d x}{\d N} &=&
-3x+\frac{\sqrt{6}}{2} \epsilon \lambda y^2 \nonumber \\
& & +\frac32 x
\left[(1-w_m)\epsilon x^2 +(1+w_m)(1-y^2)\right], \\
\label{autoquin2}
\hspace*{-1.5em}
\frac{\d y}{\d N} &=&
-\frac{\sqrt{6}}{2}\lambda xy \nonumber \\
& & +\frac32 y
 \left[(1-w_m)\epsilon x^2 +(1+w_m)(1-y^2)\right], \\
 \label{autoquin3}
 \hspace*{-1.5em}
 \frac{\d \lambda}{\d N} &=&
 -\sqrt{6} \lambda^2 (\Gamma-1)x\,,
\eea
together with a constraint equation
\be
\label{confine}
\epsilon x^2+y^2+\frac{\kappa^2 \rho_{m}}{3H^2}=1\,.
\ee

The equation of state $w_{\phi}$ and the fraction of
the energy density $\Omega_{\phi}$ for the field $\phi$ is
\bea
\label{wphiquin}
& &
w_{\phi} \equiv \frac{p_{\phi}}
{\rho_{\phi}}=\frac{\epsilon x^2-y^2}
{\epsilon x^2+y^2}\,, \\
& &
\label{Omephiquin}
\Omega_{\phi} \equiv 
\frac{\kappa^2 \rho_{\phi}}{3H^2}
=\epsilon x^2+y^2\,.
\eea
We also define the total effective equation of state:
\bea
\label{weffquin}
w_{\rm eff} 
&\equiv& \frac{p_{\phi}+p_m}
{\rho_{\phi}+\rho_m} \nonumber \\
&=& w_m+(1-w_m)\epsilon x^2-(1+w_m)y^2\,.
\eea
An accelerated expansion occurs for $w_{\rm eff}<-1/3$.
In this subsection we shall consider the case of 
quintessence ($\epsilon=+1$).
We define new variables $\gamma_{\phi}$ and $\gamma$ 
as $\gamma_{\phi} \equiv 1+w_{\phi}$ and 
$\gamma \equiv 1+w_{m}$.

\subsubsection{Constant $\lambda$}

{}From Eq.~(\ref{lamGam}) we find that the case of constant $\lambda$
corresponds to an exponential potential \cite{CLW,Ng}:
\bea
\label {exp}
V(\phi)=V_{0}e^{-\kappa \lambda \phi}\,.
\eea
In this case Eq.~(\ref{autoquin3}) is trivially satisfied because $\Gamma =1$.
One can obtain the fixed points by setting $\d x/\d N=0$ and $\d y/\d N=0$
in Eqs.~(\ref{autoquin1}) and (\ref{autoquin2}).
We summarize the fixed points and their stabilities for 
quintessence ($\epsilon=+1$) in TABLE I.

\begin{table*}[t]
\begin{center}
\begin{tabular}{|c|c|c|c|c|c|c|}
Name &  $x$ & $y$ & Existence & Stability & $\Omega_\phi$
 & $\gamma_\phi$ \\
\hline
\hline
(a) & 0 & 0 & All $\lambda$ and $\gamma$ & Saddle point
for $0 < \gamma <
2$ &   0 & -- \\
\hline
(b1) & 1 & 0 & All $\lambda$ and $\gamma$ & Unstable node for $\lambda <
 \sqrt{6}$ & 1 & 2 \\
 & & & & Saddle point for $\lambda > \sqrt{6}$ & &\\
\hline
(b2) & $-1$ & 0 & All $\lambda$ and $\gamma$ & Unstable node for $\lambda >
-\sqrt{6}$ & 1 & 2 \\
 & & & & Saddle point for $\lambda < -\sqrt{6}$ & & \\
\hline
(c) & $\lambda/\sqrt{6}$ & $[1-\lambda^2/6]^{1/2}$ & $\lambda^2 < 6$ &
Stable node for $\lambda^2 < 3\gamma$ & 1 & $\lambda^2/3$ \\
 & & & & Saddle point for $3\gamma < \lambda^2 < 6$ & & \\
\hline
(d) & $(3/2)^{1/2} \, \gamma/\lambda$ & $[3(2-\gamma)\gamma/2\lambda^2]^{1/2}$
& $\lambda^2 > 3\gamma$ & Stable node
for $3\gamma < \lambda^2 <
24 \gamma^2/(9\gamma -2)$ & $3\gamma/\lambda^2$ & $\gamma$ \\
 & & & & Stable spiral for
 $\lambda^2 > 24 \gamma^2/(9\gamma -2)$ & & \\
\hline
\end{tabular}
\end{center}
\caption[crit]{The properties of the critical points
for the quintessence model (\ref{quin}) with $\epsilon=+1$
for the exponential potential given by Eq.~(\ref{exp}).
}
\label{crit} 
\end{table*}

The eigenvalues of the matrix ${\cal M}$
given in Eq.~(\ref{uvdif}) are as follows.

\begin{itemize}

\item Point (a):  
\be
\mu_1=-\frac32 
(2-\gamma)\,,~~
\mu_2=\frac32\gamma\,.
\ee

\item Point (b1): 
\be
\mu_1=3-\frac{\sqrt{6}}{2}\lambda\,,~~~
\mu_2=3(2-\gamma)\,.
\ee

\item Point (b2): 
\be
\mu_1=3+\frac{\sqrt{6}}{2}\lambda\,,~~~
\mu_2=3(2-\gamma)\,.
\ee

\item Point (c): 
\be
\mu_1=\frac12 (\lambda^2-6)\,,~~~
\mu_2= \lambda^2-3\gamma\,.
\ee

\item Point (d): 
\be
\label{musca}
\mu_{1, 2}=-\frac{3(2-\gamma)}{4} \left[
1 \pm \sqrt{1-\frac{8\gamma(\lambda^2-3\gamma)}
{\lambda^2(2-\gamma)}}\right]\,.
\ee

\end {itemize}

In what follows we clarify the properties of the five fixed points 
given in TABLE I. Basically we are interested in a fluid 
with $0<\gamma<2$.
The point (a) corresponds to a fluid dominated solution and 
is a saddle point since $\mu_{1}<0$ and $\mu_{2}>0$.
The points (b1) and (b2) are either an unstable node or
a saddle point depending upon the value of $\lambda$.
The point (c) is a stable node for $\lambda^2<3\gamma$, 
whereas it is a saddle point for $3\gamma<\lambda^2<6$.
Since the effective equation of state is 
$w_{\rm eff}=w_{\phi}=-1+\lambda^2/3$ from 
Eqs.~(\ref{wphiquin}) and (\ref{weffquin}), 
the universe accelerates for $\lambda^2<2$ in this case.
The point (d) corresponds to a scaling solution in which the energy
density of the field $\phi$ decreases proportionally to that of the
barotropic fluid ($\gamma_{\phi}=\gamma$).
Since both $\mu_{1}$ and $\mu_{2}$ are negative for 
$\lambda^2>3\gamma$ from Eq.~(\ref{musca}), the
point (d) is stable in this case. Meanwhile it is a saddle 
point for $\lambda^2<3\gamma$, but this case is not 
realistic because the condition, 
$\Omega_{\phi}\le 1$, is not satisfied.
We note that the point (d) becomes a stable spiral 
for $\lambda^2>24\gamma^2/(9\gamma-2)$.

\begin{figure}
\includegraphics[height=2.0in,width=3.2in]{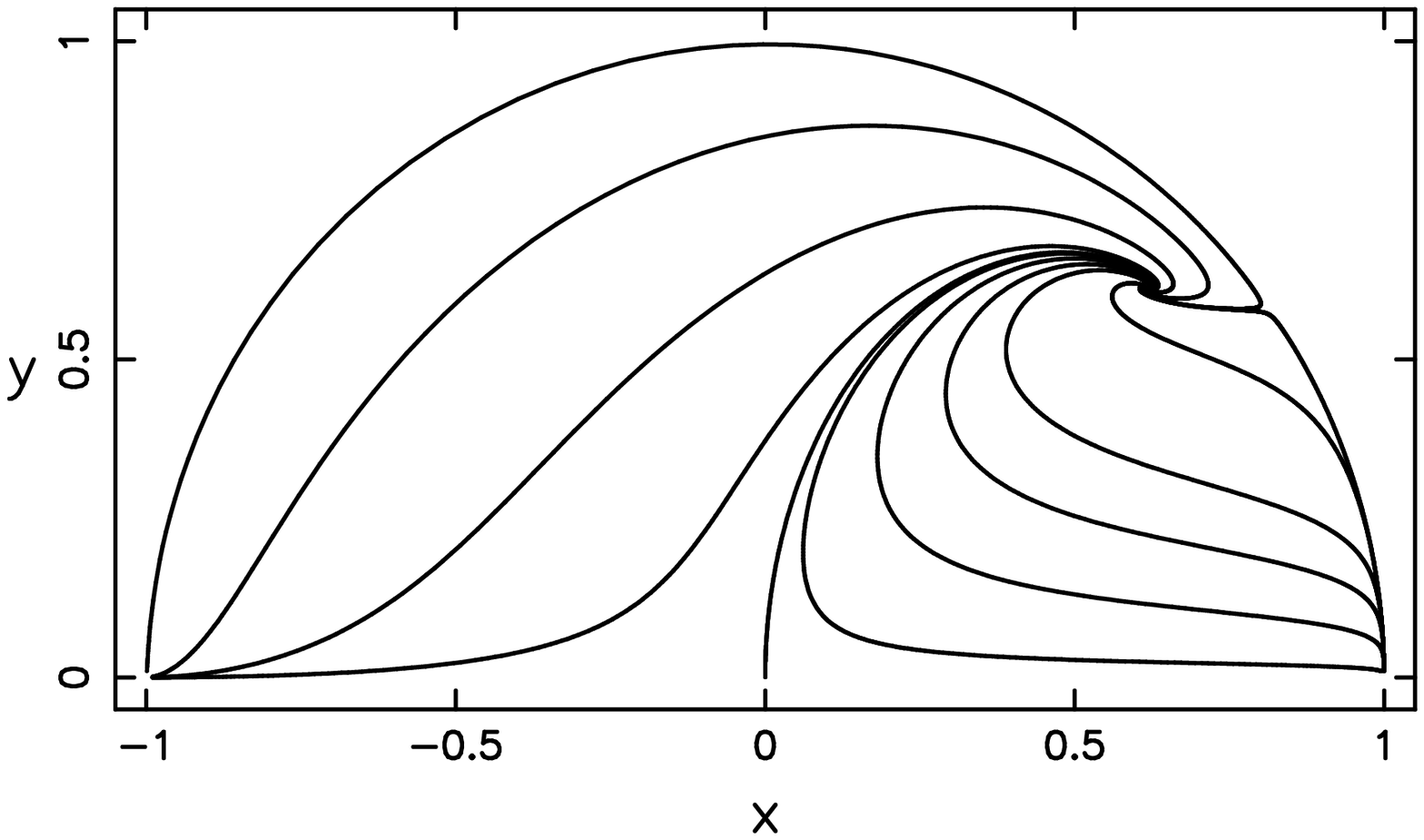}
\caption{The phase plane for $\lambda = 2$ and $\gamma=1$.
The scalar field dominated solution (c) is 
a saddle point at $x = (2/3)^{1/2}$ and 
$y = (1/3)^{1/2}$. 
Since the point (d) is a stable spiral in this case, 
the late-time attractor
is the scaling solution with $x=y=(3/8)^{1/2}$.
{}From Ref.~\cite{CLW}.}
\label{phase} 
\end{figure}

In Fig.~\ref{phase} we show the phase plane plot for 
$\lambda = 2$ and $\gamma=1$. We note that the trajectories are
confined inside the circle given by $x^2+y^2=1$ with $y \ge 0$.
In this case the point (c) is a saddle point, whereas
the point (d) is a stable spiral.
Hence the late-time attractor is the scaling solution (d)
with $x=y=\sqrt{3/8}$.
This behavior is clearly seen in Fig.~\ref{phase}.

The above analysis of the critical points shows that one can obtain
an accelerated expansion provided that the solutions approach
the fixed point (c) with $\lambda^2<2$,
in which case the final state of the universe is the scalar-field
dominated one ($\Omega_{\phi}=1$).
The scaling solution (d) is not viable to explain a late-time
acceleration. However this can be used to provide the 
cosmological evolution in which the energy density of 
the scalar field decreases proportionally to that of 
the background fluid in either a radiation or matter dominated era.
If the slope of the exponential potential becomes shallow enough 
to satisfy $\lambda^2<2$ near to the present, the universe
exits from the scaling regime and
approaches the fixed point (c) giving rise to
an accelerated expansion \cite{BCN99,SW99}. 
This of course requires an effective $\lambda$ which changes 
with time, and we turn to that case in the next subsection. 
However before we do that, we mention that in \cite{LopesFranca:2002ek}, the authors discuss the possibility that the field has not yet reached the
fixed point , and argue that (i) even for  $2 < \lambda^2 < 3$, there
is a non-trivial region of parameter space that can explain the observed values of the cosmological
parameters, such as the equation of state, and (ii) the fine tuning for these
models, is no worse than in other quintessential scenarios.

\subsubsection{Dynamically changing $\lambda$}

Exponential potentials correspond to
constant $\lambda$ and $\Gamma=1$.
Let us consider a potential $V(\phi)$ along which
the field rolls down toward plus infinity with 
$\dot{\phi}>0$.
This means that $x>0$ in Eq.~(\ref{autoquin3}).
Then if the condition,
\bea
\label{track}
\Gamma>1\,,
\eea
is satisfied, $\lambda$ decreases toward 0.
Hence the slope of the potential defined by 
Eq.~(\ref{lamGam}) becomes flat, 
thereby giving rise to
an accelerated expansion at late times.
The condition (\ref{track}) is regarded as
the tracking condition
under which the energy density of $\phi$ eventually
catches up that of the fluid \cite{caldwell98}.
In order to construct viable 
quintessence models, we require that
the potential should satisfy the condition (\ref{track}).
For example, one has $\Gamma=(n+1)/n>1$
for the inverse power-law potential
$V(\phi)=V_0\phi^{-n}$ with $n>0$.
This means that  tracking behaviour occurs for this potential.

When $\Gamma<1$ the quantity $\lambda$ increases
towards infinity. Since the potential is steeper than the one
corresponding to scaling solutions, 
the energy density of the scalar field becomes negligible
compared to that of the fluid.
Then we do not have an accelerated expansion at late times.

In order to obtain dynamical evolution of the system 
we need to solve Eq.~(\ref{autoquin3})
together with Eqs.~(\ref{autoquin1}) and (\ref{autoquin2}).
Although $\lambda$ is a dynamically changing quantity,
one can apply the discussion of constant $\lambda$ to this 
case as well by considering ``instantaneous'' 
critical points \cite{MP,Ng}.
For example, the point (c) in TABLE I dynamically changes 
with time, i.e., $x(N)=\lambda(N)/\sqrt{6}$ and 
$y(N)=[1-\lambda^2(N)/6]^{1/2}$. 
When $\Gamma>1$ this point eventually approaches
$x(N) \to 0$ and $y(N) \to 1$ with an equation of state of a cosmological 
constant ($\gamma_{\phi} \to 0$) as $\lambda(N) \to 0$.
See Refs.~\cite{MP,Ng} for more details.

\subsection{Phantom fields}

The phantom field corresponds to a negative kinetic
sign, i.e, $\epsilon=-1$ in Eq.~(\ref{quin}).
Let us first consider the exponential potential given by
Eq.~(\ref{exp}).
In this case Eq.~(\ref{autoquin3}) is dropped from the dynamical
system. In Table \ref{tablephan} we show fixed points
for the phantom field. 
The points $(x, y)=(\pm 1, 0)$ which exist in the case of
quintessence disappear for the phantom field.
The point (a) corresponds to a saddle point, since the 
eigenvalues of the matrix ${\cal M}$ are
the same as in the quintessence case.

\begin{table*}[t]
\begin{center}
\begin{tabular}{|c|c|c|c|c|c|c|}
Name &  $x$ & $y$ & Existence & Stability &
$\Omega_\phi$ & $\gamma_\phi$ \\
\hline
\hline
(a) & 0 & 0 & No for $0 \le \Omega_\phi \le 1$ & Saddle point
&   0 & -- \\
\hline
(b) & $-\lambda/\sqrt{6}$ & $[1+\lambda^2/6]^{1/2}$ & All values &
Stable node & 1 & $-\lambda^2/3$ \\
\hline
(c) & $(3/2)^{1/2} \, \gamma/\lambda$  &
$[-3(2-\gamma)\gamma/2\lambda^2]^{1/2}$
& $\gamma<0$ & Saddle point for $\lambda^2>-3\gamma$& 
$\frac{-3\gamma}{\lambda^2}$ & $\gamma$ \\
\hline
\end{tabular}
\end{center}
\caption[crit]{The properties of the critical points
for a phantom scalar field ($\epsilon=-1$).}
\label{tablephan}
\end{table*}

The point (b) is a scalar-field dominated solution
whose equation of state is given by 
\bea
\label{phaneq}
w_\phi=-1-\lambda^2/3\,,
\eea
which is less than $-1$.
The eigenvalues of the matrix ${\cal M}$ are
$\mu_{1}=-(\lambda^2+6)/2$ and $\mu_{2}=-\lambda^2-3\gamma$,
which are both negative for $\gamma>0$.
Hence the fixed point (b) is a stable node.
The scaling solution (c) exists only for the phantom fluid
($\gamma<0$).
The eigenvalues of the matrix ${\cal M}$ are
\bea
\mu_{1, 2}=-\frac{3(2-\gamma)}{4} \left[
1 \pm \sqrt{1-\frac{8\gamma(\lambda^2+3\gamma)}
{\lambda^2(2-\gamma)}}\right]\,.
\eea
When $\gamma<0$ the point (c) is a saddle point 
for $\lambda^2>-3\gamma$.

In the presence of a non-relativistic dark matter ($\gamma=1$)
the system approaches the scalar-field dominated
solution (b). Exponential potentials give rise to {\it constant} equation of state
$w_{\phi}$ smaller than $-1$ \cite{ST04}.
Then the universe reaches a Big Rip singularity at which
the Hubble rate and the energy density of the universe diverge.
We recall that the phantom field rolls {\it up} the potential hill,
which leads to the increase of the energy density.

When the potential of the phantom field  is no longer a simple 
exponential, the quantity $\lambda$ can evolve in time. In this case the point (b) 
can be regarded as an instantaneous critical point.
For example, in the case of the bell-type potential 
introduced in Eq.~(\ref{bell}), $\lambda$ decreases to zero 
as the field settles on the top of the potential. 
Hence the equation of state finally approaches $w_{\phi}=-1$.

\subsection{Tachyon fields}

The energy density and the pressure density of a tachyon field
are given by Eqs.~(\ref{rhotach}) and (\ref{ptach}), 
with the tachyon satisfying the equation of motion (\ref{tachevo}).
In the presence of a barotropic fluid whose equation of state is 
$\gamma \equiv 1+w_m=1+p_{m}/\rho_{m}$, 
Equations (\ref{Htwo}) and  (\ref{dotHtwo}) give 
\begin{eqnarray}
& & H^2=\frac{\kappa^2}{3} \left[ \frac{V(\phi)}
{\sqrt{1-\dot{\phi}^2}}+\rho_{m} \right]\,, \\
& & \dot{H}=-\frac{\kappa^2}{2} \left[
\frac{\dot{\phi}^2V(\phi)}
{\sqrt{1-\dot{\phi}^2}}+\gamma \rho_{m} 
\right]\,.
\end{eqnarray}

Let us define the following dimensionless quantities:
\begin{eqnarray}
\label{Dquantity}
x=\dot{\phi}\,,~~~y=\frac{\kappa\sqrt{V(\phi)}}
{\sqrt{3}H}\,.
\end{eqnarray}
Then we obtain the following 
autonomous equations \cite{AL,CGST}
\bea
\label{autotach1}
\frac{\d x}{\d N} &=&
-(1-x^2)(3x-\sqrt{3}\lambda y)\,, \\
\label{autotach2}
\frac{\d y}{\d N} &=&
\frac{y}{2}\left[-\sqrt{3}
\lambda xy-\frac{3(\gamma -x^2)y^2}
{\sqrt{1-x^2}}+3\gamma \right] \,,\\
 \label{autotach3}
 \frac{\d \lambda}{\d N} &=&
 -\sqrt{3}\lambda^2 xy(\Gamma-3/2)\,,
\eea
together with a constraint equation
\begin{eqnarray}
\frac{y^2}{\sqrt{1-x^2}}+\frac{\kappa^2 \rho_{m}}
{3H^2}=1\,.
\end{eqnarray}
Here $\lambda$ and $\Gamma$ are defined by
\begin{eqnarray}
\label{lam}
\lambda \equiv -\frac{V_{,\phi}}{\kappa V^{3/2}}\,,~~~
\Gamma \equiv \frac{VV_{,\phi\phi}}{V_{,\phi}^2}\,.
\label{Gam}
\end{eqnarray}
The equation of state and the fraction of
the energy density in the tachyon field  are given by  
\bea
\gamma_{\phi}=x^2\,,~~~
\Omega_{\phi}=\frac{y^2}{\sqrt{1-x^2}}\,.
\eea
Then the allowed range of
$x$ and $y$ in a phase plane is $0 \le x^2+y^4 \le 1$
from the requirement: $0 \le \Omega_\phi \le 1$.

\subsubsection{Constant $\lambda$}

{}From Eq.~(\ref{autotach3}) we find that
$\lambda$ is constant for $\Gamma=3/2$.
This case corresponds to an inverse square potential
\begin{eqnarray}
V(\phi) =M^2\phi^{-2}\,.
\label{inversesqu}
\end{eqnarray}
As we showed in the previous section, this potential gives 
a power-law expansion, $a \propto t^p$
[see Eq.~(\ref{inpower})].
The fixed points for this potential have been obtained
in Refs.~\cite{AL,CGST}, and are summarized in Table \ref{Tachcrit}.
One can study the stability of the critical points by evaluating
the eigenvalues of the matrix ${\cal M}$.
We do not present all the eigenvalues in this review, but note that they 
are given in Refs.~\cite{AL,CGST}.

\begin{table*}[t]
\begin{center}
\begin{tabular}{|c|c|c|c|c|c|c|}
Name & $x$ & $y$ & Existence & Stability & $\Omega_\phi$
 & $\gamma_\phi$ \\
\hline
\hline
(a) & 0 & 0 & All $\lambda$ and $\gamma$ & Unstable saddle for
$\gamma > 0$  &   0 & 0 \\
& & & & Stable node for $\gamma=0$ & & \\
\hline
(b1) & 1 & 0 & All $\lambda$ and $\gamma$ & Unstable node
& 1 & 1 \\
\hline
(b2) & $-1$ & 0 & All $\lambda$ and $\gamma$ & Unstable node
& 1 & 1 \\
\hline
(c) & $\lambda y_s/\sqrt{3}$ & $y_s$ & All $\lambda$ and $\gamma$ &
Stable node for $\gamma \ge \gamma_s$ & 1 & $\lambda^2 y_s^2/3$ \\
 & & & & Saddle for $\gamma < \gamma_s$ & & \\
\hline
(d1) & $\sqrt{\gamma}$ & $\sqrt{3\gamma}/\lambda$
& $\lambda>0$ and $\gamma<\gamma_s$ & Stable for $\Omega_{\phi}<1$ & 
$\frac{3\gamma}{\lambda^2}
\frac{1}{\sqrt{1-\gamma}}$ & $\gamma$  \\
\hline
(d2) & $-\sqrt{\gamma}$ & $-\sqrt{3\gamma}/\lambda$
& $\lambda<0$ and $\gamma<\gamma_s$ & Stable for $\Omega_{\phi}<1$  & 
$\frac{3\gamma}{\lambda^2}
\frac{1}{\sqrt{1-\gamma}}$ & $\gamma$ \\
\hline
\end{tabular}
\end{center}
\caption[crit]{\label{Tachcrit}
The critical points for the inverse square potential (\ref{inversesqu})
in the case of tachyon.
$\gamma_{s}$ is defined in Eq.~(\ref{gam}).
}
\end{table*}

The fixed points (a), (b1) and (b2) are not stable, so 
they are not a late-time attractor.
The point (c) is a scalar-field dominated solution
($\Omega_\phi=1$) with eigenvalues 
$\mu_1=-3+\lambda^2(\sqrt{\lambda^4+36}-\lambda^2)/12$
and $\mu_2=-3\gamma+\lambda^2(\sqrt{\lambda^4+36}-\lambda^2)/6$.
Hence this point is stable for
\begin{eqnarray}
\gamma \ge \gamma_s \equiv
\frac{\lambda^2}{18}
(\sqrt{\lambda^4+36}-\lambda^2)\,.
\label{gam}
\end{eqnarray}
In TABLE \ref{Tachcrit} the quantity $y_{s}$
is given by 
\begin{eqnarray}
y_s=\left(\frac{\sqrt{\lambda^4+36}-\lambda^2}{6}
\right)^{1/2}\,.
\label{ys}
\end{eqnarray}
Since $\gamma_{\phi}=\lambda^2y_{s}^2/3$ for the 
point (c), an accelerated expansion occurs for 
$\lambda^2 y_s^2<2$.
This translates into the condition 
$\lambda^2<2\sqrt{3}$ \cite{GNST}.

The point (d) is a scaling solution which exists only
for $\gamma<1$, since $\Omega_{\phi}$ is given by 
$\Omega_{\phi}=3\gamma/\lambda^2 \sqrt{1-\gamma}$.
{}From the condition $\Omega_{\phi} \le 1$ we obtain 
\begin{eqnarray}
\gamma \le \gamma_s =
\frac{\lambda^2}{18}
(\sqrt{\lambda^4+36}-\lambda^2)\,.
\label{gam2}
\end{eqnarray}
The eigenvalues of the matrix ${\cal M}$ are
\begin{eqnarray}
\label{mu1mu2}
\mu_{1, 2}= \frac34 \biggl[\gamma-2
\pm \sqrt{17\gamma^2-20\gamma+4+
\frac{48}{\lambda^2}\gamma^2\sqrt{1-\gamma}}
\biggr]. \nonumber \\
\end{eqnarray}
The real parts of $\mu_1$ and $\mu_2$ are both
negative when the condition (\ref{gam2}) is satisfied. 
When the square root in Eq.~(\ref{mu1mu2}) is positive,
the fixed point is a stable node.
The fixed point is a stable spiral  when
the square root in Eq.~(\ref{mu1mu2}) is negative.
In any case the scaling solution is always stable for 
$\Omega_{\phi}<1$, but 
this is not a realistic solution in applying to dark energy 
because of the condition $\gamma<1$.

The above discussion shows that the only viable late-time attractor 
is the scalar-field dominated solution (c).
When the solution approaches the fixed point (c), the
accelerated expansion occurs for $\lambda^2<2\sqrt{3}$.
Since $\lambda$ is given by $\lambda=2M_{\rm pl}/M$, 
the condition for an accelerated expansion gives 
an energy scale which is close to a Planck mass, i.e., 
$M \gtrsim 1.1M_{\rm pl}\simeq 2.6 \times 10^{18}\,{\rm GeV}$. 
The mass scale $M$ becomes smaller 
for the inverse power-law potential
$V(\phi)=M^{4-n}\phi^{-n}$, as we will see below.

\subsubsection{Dynamically changing $\lambda$}

When the potential is different from the inverse square potential
given by Eq.~(\ref{inversesqu}), $\lambda$ is a dynamically
changing quantity. As we have seen in the case
of quintessence, there are basically two cases: 
(i) $\lambda$ evolves toward zero, or
(ii) $|\lambda|$ increases toward infinity.
The case (i) is regarded as the tracking solution in which the energy
density of the tachyon eventually dominates over that of the fluid.
This situation is realized when the following condition
is satisfied \cite{Chitrack}
\begin{eqnarray}
\Gamma>3/2\,,
\end{eqnarray}
which is derived from Eq.~(\ref{autotach3}).
When $\Gamma<3/2$ the energy density
of the scalar field becomes negligible compared 
to that of the fluid.

As an example let us consider the inverse power-law potential 
\begin{eqnarray}
V(\phi)=M^{4-n}\phi^{-n}\,,~~~~
n>0\,.
\end{eqnarray}
Since $\Gamma=(n+1)/n$ in this case, 
the scalar-field energy density dominates at late-times
for $n<2$. The system approaches the ``instantaneous''
critical point (c) for $\gamma \ge 1$.
In the limit $\lambda \to 0$ one has 
$x \to 0$ and $y \to 1$ for the point (c), 
which means that slow-roll approximations
can be used at late-times.
The slow-roll parameter for the tachyon is 
given by \cite{tacinflation}.
\begin{eqnarray}
\epsilon  &\equiv& -\frac{\dot{H}}{H^2} \simeq 
\frac{M_{\rm pl}^2}{2} \left(\frac{V_\phi}{V}
\right)^2 \frac{1}{V} =
\frac{n^2}{2} \left(\frac{M_{\rm pl}}{M}\right)^2
\frac{1}{(\phi M)^{2-n}}\,. \nonumber \\
\end{eqnarray}
We find that $\epsilon$ decreases for $n<2$ 
as the field evolves toward large values.
The condition for the accelerated expansion corresponds
to $\epsilon<1$, which gives
\begin{eqnarray}
\label{phiM}
\phi M>\left(\frac{n}{\sqrt{2}} 
\frac{M_{\rm pl}}{M}\right)^{2/(2-n)}\,.
\end{eqnarray}

\begin{figure}
\includegraphics[height=2.5in,width=3.2in]{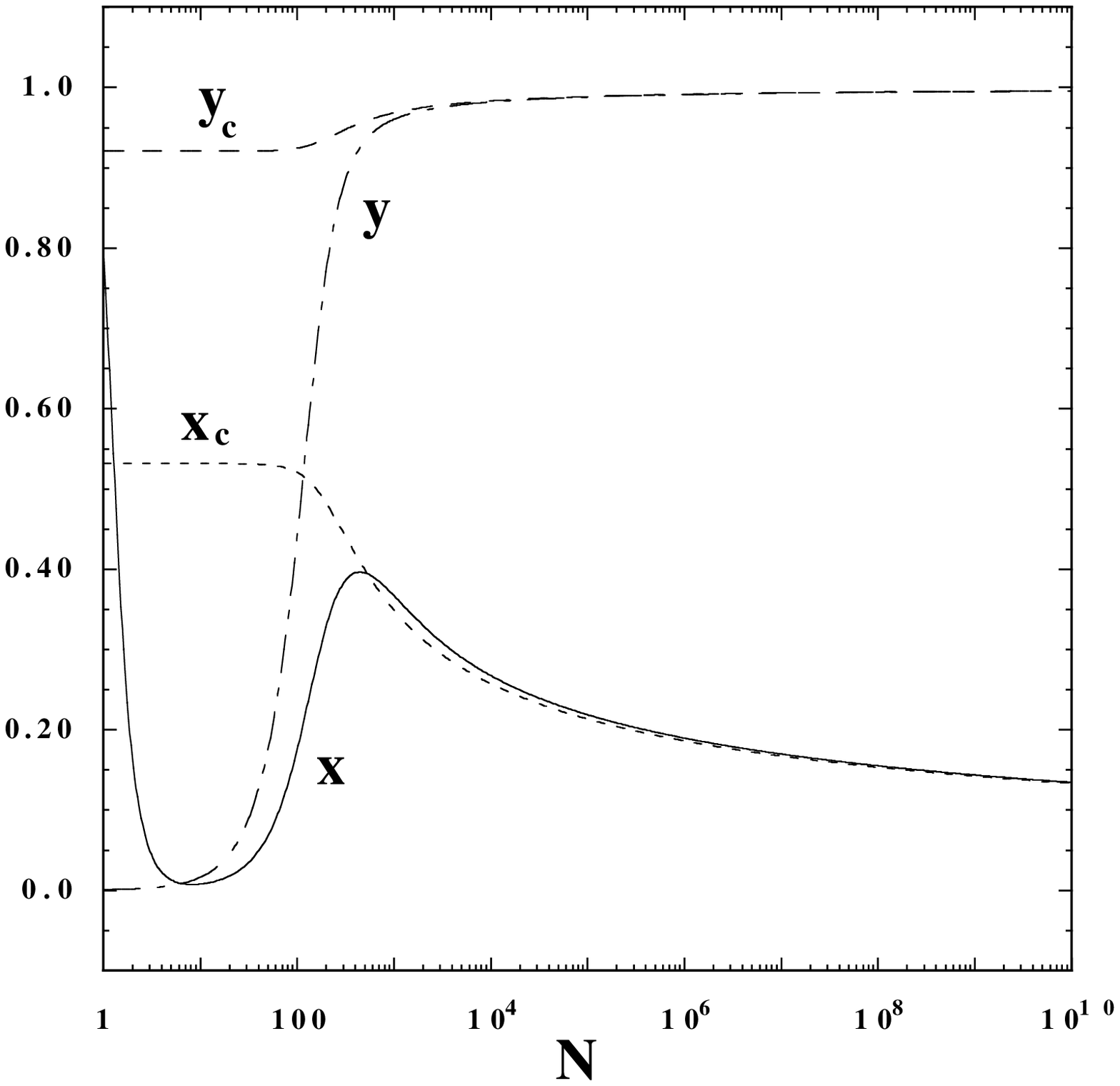}
\caption{Evolution of the parameters $x$ and $y$ 
together with the critical points $x_{c}$ and $y_{c}$
for tachyon with potential $V(\phi)=M^3\phi^{-1}$ 
and a barotropic fluid with $\gamma=1$.
We choose initial conditions 
$x_i=0.8$, $y_i=5.0 \times 10^{-4}$
and $\lambda_i=1.0$. The solution approaches instantaneous critical 
points whose asymptotic values are $x_{c}=0$ and $y_{c}=1$.
{}From Ref.~\cite{CGST}.
}
\label{tachatt} 
\end{figure}

The present potential energy is approximated as
$ \rho_c^{(0)} \simeq V(\phi_0)=M^4/(\phi_0 M)^n \simeq 
10^{-47}\,{\rm GeV}^4$.
Combining this relation with Eq.~(\ref{phiM}) we get
\begin{eqnarray}
\label{Mcons}
\frac{M}{M_{\rm pl}}>\left[\left(\frac{\rho_c^{(0)}}
{M_{\rm pl}^4}\right)
^{1-n/2}\left(\frac{n}{\sqrt{2}}\right)^n\right]^{1/(4-n)}\,.
\end {eqnarray}
While $M$ is close to the Planck scale for $n=2$, 
this problem is alleviated for smaller $n$.
For example one has $M/M_{\rm pl} \gtrsim 10^{-20}$ for $n=1$.
We note that the solutions approach instantaneous
critical points:
$(x_c, y_c)=(\lambda(N) y_s(N)/\sqrt{3}, y_s(N))$
with $y_s(N)=[(\sqrt{\lambda(N)^4+36}-\lambda(N)^2)/6]^{1/2}$.
This behavior is clearly seen in the numerical simulations 
in Fig.~\ref{tachatt}.
Thus the discussion of constant $\lambda$ can be applied to the case
of varying $\lambda$ after the system approaches the stable attractor
solutions.

There exists another tachyon potential in which the quantity $\lambda$
decreases toward zero with oscillations \cite{CGST}.
One example is provided by the potential
\begin{eqnarray}
V(\phi)=V_0 e^{\frac12 M^2\phi^2}\,,
\end {eqnarray}
which, for example, appears as an excitation of the massive state
on the anti D-brane \cite{GST}.
In this case the scalar field approaches the potential
minimum at $\phi=0$ with oscillations, after which
the field stabilizes there.
Since the potential energy $V_0$ remains
at $\phi=0$, this works as a cosmological constant
at late-times.

There are a number of potentials which exhibit the behavior
$|\lambda| \to \infty$ asymptotically.
For example $V(\phi)=M^{4-n}\phi^{-n}$ with $n>2$
and $V(\phi)=V_0 e^{-\mu \phi}$ with $\mu>0$.
In the latter case one has $\Gamma=1$ and
$\rd \lambda/\rd N=(\sqrt{3}/2)\lambda^2xy$, thereby leading
to the growth of $\lambda$ for $x>0$.
In the limit $\lambda \to \infty$ the instantaneous
critical point (c) approaches
$x_c(N) \to 1$ and $y_c(N) \to 0$ with $\gamma_{\phi} \to 1$,
which means the absence of an accelerated expansion.
Although the accelerated expansion does not occur at late-times
in this scenario, it is possible to have a temporal acceleration for
$\lambda \lesssim 1$ and have a deceleration 
for $\lambda \gg 1$ \cite{CGST}.
If this temporal acceleration corresponds to the one
at present, the universe will eventually enter the
non-accelerating regime in which the tachyon
field behaves as a pressureless dust.

\subsection{Dilatonic ghost condensate}

Let us consider the dilatonic ghost condensate model given
by Eq.~(\ref{ghostcon}).
In this case the pressure density and 
the energy density of the field are given by 
$p_{\phi}=p=-X+ce^{\lambda \phi} X^2$ and 
$\rho_{\phi}=2X\partial p_{\phi}/\partial X-p_{\phi}
=-X+3ce^{\lambda \phi}X^2$ with $X=\dot{\phi}^2/2$.
Then Eqs.~(\ref{Htwo}), (\ref{dotHtwo}) and 
(\ref{dotrhotwo}) read
\begin{eqnarray}
& & 3H^2=-\frac12 \dot{\phi}^2+\frac34 ce^{\lambda \phi}
\dot{\phi}^4+\rho_{m}\,, \\
& & 2\dot{H}=\dot{\phi}^2-ce^{\lambda \phi} \dot{\phi}^4
-(1+w_{m})\rho_{m}\,, \\
& & \ddot{\phi} (3ce^{\lambda \phi} \dot{\phi}^2-1)
+3H\dot{\phi} (ce^{\lambda \phi} \dot{\phi}^2-1)
+\frac34 c\lambda e^{\lambda \phi} \dot{\phi}^4=0\,, \nonumber \\
\end {eqnarray}
where we set $\kappa^2=1$.

Introducing the following quantities
\begin{equation}
\label{quan}
x \equiv \frac{\dot{\phi}}{\sqrt{6}H}\,,~~~
y \equiv \frac{e^{-\lambda \phi/2}}{\sqrt{3}H}\,,
\end{equation}
the above equations can written in an autonomous form
\bea 
\label{dxgho}
\frac{\d x}{\d N} &=&   
\frac32 x \left[1+w_m+(1-w_m) x^2(-1+cY)
-2c w_m x^2Y \right] \nonumber \\
& & +\frac{1}{1-6cY} \left[3(-1+2cY)x+
\frac{3\sqrt{6}}{2}\lambda c x^2 Y
\right]\,, \\
\label{dygho}
\frac{\d y}{\d N} &=&   
-\frac{\sqrt{6}}{2}\lambda xy+\frac32 y
[1+w_m \nonumber \\
& &+(1-w_m)x^2
(-1+cY)-2cw_m x^2 Y]\,,
\eea 
where 
\begin{equation}
Y \equiv \frac{x^2}{y^2}=Xe^{\lambda \phi}\,.
\label{Ytach}
\end{equation}
The equation of state and the fraction of the energy 
density for the field can now be written as  
\bea 
\label{wphidi}
& & w_{\phi}=\frac{1-cY}{1-3cY}\,, \\
\label{Omephidi}
& & \Omega_{\phi}=
-x^2+3c\frac{x^4}{y^2}\,.
\eea 
%

\begin{table*}[t]
\begin{center}
\begin{tabular}{|c|c|c|c|c|}
Name & $x$ & $cY$ & $\Omega_\phi$ & $w_\phi$ \\
\hline
\hline
(a) & 0 & $\infty$ & $\frac{3(w_m+1)}{3w_m-1}$ &  1/3 
\\
\hline
(b) & $-\frac{\sqrt{6}\lambda f_+(\lambda)}{4}$ &
$\frac12+\frac{\lambda^2 f_- (\lambda)}{16}$
& 1 & $\frac{-8+\lambda^2 f_-(\lambda)}
{8+3 \lambda^2 f_- (\lambda)}$ 
\\
\hline
(c) & $-\frac{\sqrt{6}\lambda f_-(\lambda)}{4}$ &
$\frac12+\frac{\lambda^2 f_+(\lambda)}{16}$
& 1 & $\frac{-8+\lambda^2 f_+(\lambda)}{8+3 \lambda^2 f_+ (\lambda)}$
\\
\hline
(d) & $\frac{\sqrt{6}(1+w_m)}{2\lambda}$ &
$\frac{1-w_m}{1-3w_m}$
& $\frac{3(1+w_m)^2}{\lambda^2(1-3w_m)}$
& $w_{m}$
\\
\hline
\end{tabular}
\end{center}
\caption[critghost]{\label{critghost} 
The critical points for the dilatonic ghost condensate
model given by (\ref{ghostcon}). 
Here $Y$ and $f_{\pm}(\lambda)$ are 
defined in Eqs.~(\ref{Ytach}) and (\ref{func}).
}
\end{table*}

In Table~\ref{critghost} we present the fixed points for
the system of the dilatonic ghost condensate.
The point (a) is not realistic, since
we require a phantom fluid ($w_m \le -1$)
to satisfy $0 \le \Omega_\phi \le 1$.

The points (b) and (c) correspond to
the dark-energy dominated universe with $\Omega_\phi=1$.
The functions $f_{\pm} (\lambda)$ are defined by
\bea
\label{func}
f_{\pm} (\lambda) \equiv 1 \pm \sqrt{1+16/(3\lambda^2)}\,.
\eea
The condition (\ref{xi}) for the stability of quantum fluctuations
corresponds to $cY \ge 1/2$.
{}From Eq.~(\ref{wphidi}) one has 
$w_\phi<-1$ for $cY<1/2$ and $w_\phi>-1$ otherwise.
The parameter range of $Y$ for the point (b) is
$1/3<cY<1/2$, which means that
the field $\phi$ behaves as a phantom.
The point (c) belongs to the parameter range given by
$1/2<cY<\infty$, which means that the stability of 
quantum fluctuations is ensured.
An accelerated expansion occurs for $w_{\phi}<-1/3$, 
i.e., $cY<2/3$. This corresponds to the condition 
$\lambda^2 f_+(\lambda)<8/3$, i.e., 
$\lambda <\sqrt{6}/3$.
In the limit $\lambda \to 0$ we have $cY \to 1/2$, $\Omega_\phi \to 1$
and $w_\phi \to -1$ for both points (b) and (c).
The $\lambda=0$ case is the original ghost condensate
scenario proposed in Ref.~\cite{Arkani-hamad}, i.e., $p=-X+X^2$.
The point (d) corresponds to a scaling solution
characterized by $w_{\phi}=w_{m}$, in which case we do not 
have an accelerated expansion unless $w_{m}<-1/3$.

We shall study the stability of the fixed points in 
the case of non-relativistic dark matter ($w_{m}=0$)
with $c=1$.
Numerically the eigenvalues of the matrix ${\cal M}$ were evaluated 
in Ref.~\cite{GNST} and it was shown that the determinant of 
the matrix ${\cal M}$ for the point (b) 
is negative with negative real parts of 
$\mu_1$ and $\mu_2$.
Hence the phantom fixed point (b) is a stable spiral.
The point (c) is a stable node for $0<\lambda<\sqrt{3}$, whereas
it is a saddle point for $\lambda>\sqrt{3}$.
This critical value $\lambda_{*}=\sqrt{3}$ 
is computed by setting the determinant of ${\cal M}$ to be zero.
The point (d) is physically meaningful for $\lambda > \sqrt{3}$
because of the condition $\Omega_{\phi} < 1$, 
and it is a stable node \cite{GNST}.
Hence the point (d) is stable when the point (c) is unstable
and vice versa. It was shown in Ref.~\cite{Tsuji06} that this property 
holds for all scalar-field models which possess scaling solutions.
We recall that the point (b) is not stable at the quantum level.
The above discussion shows that the only viable attractor 
which satisfies the conditions
of an accelerated expansion and the quantum stability
is the point (c). 

\begin{figure}
\includegraphics[height=2.5in,width=3.2in]{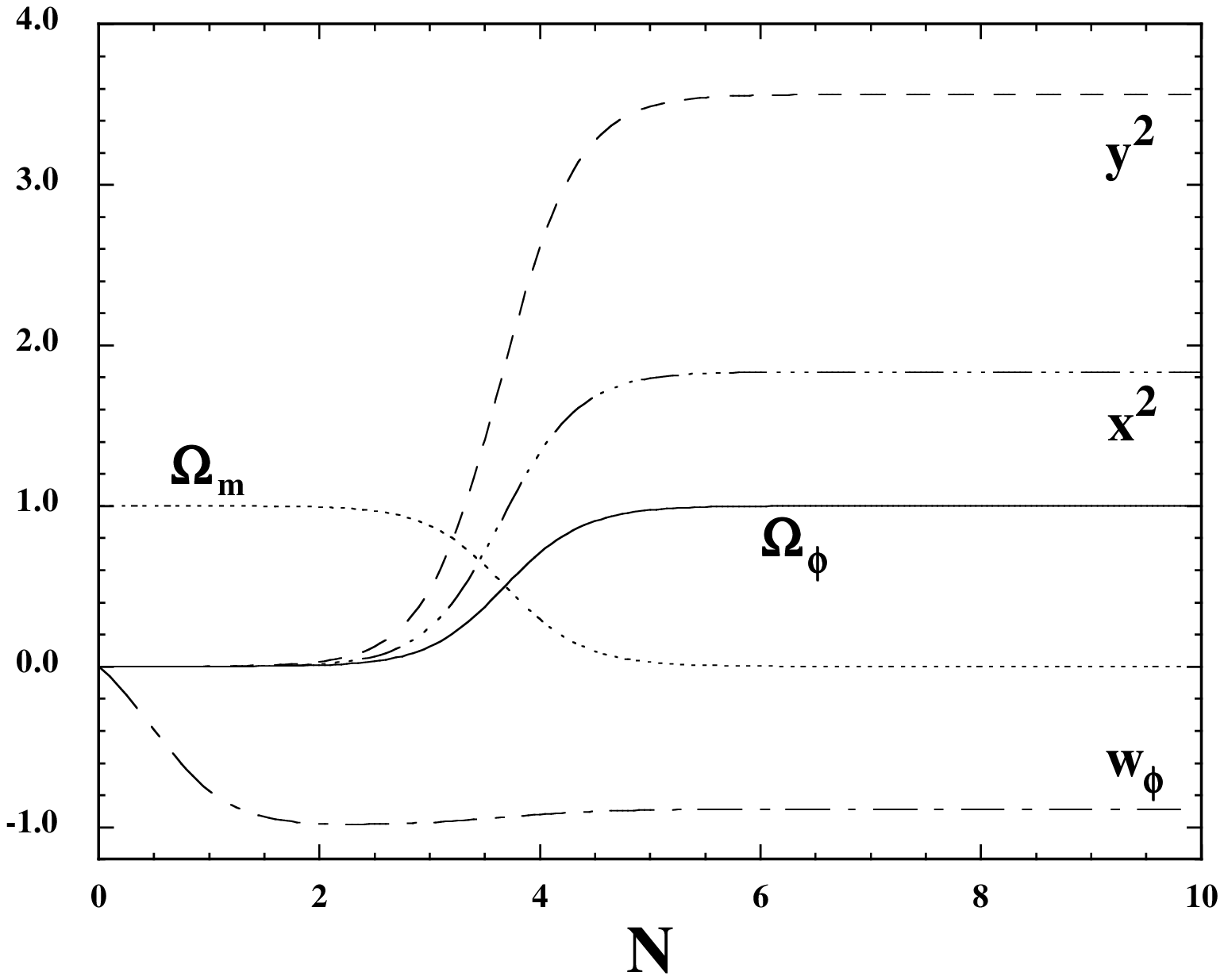}
\caption{Evolution of $\Omega_\phi$, 
$\Omega_m$, $w_\phi$, $x^2$ and
$y^2$ for $c=1$, $w_m=0$ and $\lambda=0.1$
with initial conditions $x_i = 0.0085$ and $y_i =0.0085$.
The solution approaches the scalar-field dominated fixed point (c)
with $x^2 \simeq 1.834$, $y^2=3.561$ and $Y \simeq 0.515$, 
$\Omega_\phi=1$ and $\Omega_m=0$. 
The equation of state in the attractor regime is given by 
$w_\phi=-0.889$.
{}From Ref.~\cite{PT}.
}
\label{evon} 
\end{figure}

In Fig.~\ref{evon} we plot the variation of 
$\Omega_\phi$, $\Omega_m$, $w_\phi$, $x^2$ and
$y^2$ for $c=1$, $w_m=0$ and $\lambda=0.1$.
We find that analytic values of the attractor points agree very
well with our numerical results.

\section{Scaling solutions in a general Cosmological background}
\label{scalingsec}

In the previous section we have seen that there exist
scaling solutions in certain classes of dark energy models.
It is convenient to know the existence of scaling solutions, 
since they give the border of acceleration and deceleration. 
This allows the field energy density to mimic the
background whilst remaining sub-dominant during both the radiation and matter eras. 
Although one does not have an acceleration of the universe
at late-times in this case, it is possible to obtain 
an accelerated expansion if a field $\phi$ (dark energy) 
is coupled to a background fluid 
(dark matter) \cite{coupled1} (see also Ref.~\cite{Luca}).
In this section we implement the coupling $Q$
between the field and the barotropic fluid
and derive a general form of the Lagrangian \cite{PT}
for the existence of scaling solutions. 
We note that this includes uncoupled
dark energy scenarios discussed in the previous section 
by taking the limit $Q \to 0$.

The existence of scaling solutions has been extensively
studied in a number of cosmological scenarios--including standard
General Relativity (GR), braneworlds
[Randall-Sundrum (RS) and Gauss-Bonnet (GB)], tachyon 
and Cardassian scenarios \cite{Maeda00}-\cite{Freese:2002sq,Chinajapan}. 
In what follows we present a unified framework to investigate
scaling solutions in a general cosmological
background characterized by 
$H^2 \propto \rho_{\rm T}^n$, where
$\rho_{\rm T}$ is the total energy density.
The GR, RS, GB and Cardassian cases correspond to 
$n=1$, $n=2$, $n=2/3$ and $n=1/3$,
respectively. Our formalism provides a very generic 
method to study these solutions for all known 
scalar-field dark energy models \cite{TS04}. 

\subsection{General Lagrangian for the existence of scaling solution}

We start with the following general 4-dimensional action
\begin{eqnarray}
\label{action}
S=\int {\rm d}^4 x \sqrt{-g} \left[\frac{M_{{\rm pl}}^2}{2}\, R
+ p(X, \varphi)\right]+S_m (\varphi)\,,
\end{eqnarray}
where $X =-g^{\mu\nu} \partial_\mu \varphi
\partial_\nu \varphi /2$ is a kinetic term of 
a scalar field $\vp$. $S_m$ is an action for a matter fluid
which is generally dependent on $\vp$.
In what follows we set the reduced Planck mass 
$M_{\rm pl}$ to be unity.

Let us consider the following effective Friedmann equation 
in a flat FRW background:
\be
\label{Hubble}
H^2=\beta_n^2 \rho_{\rm T}^n\,,
\ee
where $\beta_n$ and $n$ are constants, and
$\rho_{\rm T}$ is a total energy density of the universe.
We note that a more general analysis can also be undertaken for the case 
where $H^2 \propto L^2(\rho_T)$, where $L(\rho_T)$ is a general 
function of $\rho_T$ \cite{Copeland:2004qe}.
We consider a cosmological scenario in which the universe
is filled by the scalar field $\varphi$ with equation of state 
$w_{\vp}=p_{\vp}/\rho_{\vp}$
and by one type of barotropic perfect fluid with equation of state 
$w_m=p_m/\rho_m$.
Here the pressure density and 
the energy density of the field are given by 
$p_{\vp}=p$ and 
$\rho_{\vp}=2X\partial p_{\vp}/\partial X-p_{\vp}$.

We introduce a scalar charge $\sigma$ corresponding to 
the coupling between the field $\vp$ and matter, 
which is defined by the relation 
$\sigma=-(1/\sqrt{-g})\delta S_{m}/\delta \vp$.
Then the continuity equation for the field $\vp$
is given by 
\begin{eqnarray}
\label{geneeq1}
\frac{\rd \rho_{\vp}}{\rd N}+3(1+w_\vp)\rho_{\vp}
=-Q\rho_m \frac{\rd \vp}{\rd N}\,,
\end{eqnarray}
where $N \equiv {\rm ln}\,a$ and $Q \equiv \sigma/\rho_{m}$.
The energy density $\rho_m$ of the fluid satisfies
\begin{eqnarray}
\label{geneeq2}
 \frac{\rd \rho_m}{\rd N} + 3(1+w_m) \rho_m =  Q \rho_m
 \frac{\rd \vp}{\rd N}\,.
\end{eqnarray}
We define the fractional densities of $\rho_{\vp}$ and $\rho_m$ as
\begin{eqnarray}
\label{Omedef}
\Omega_\varphi \equiv \frac{\rho_{\vp}}{(H/\beta_n)^{2/n}}\,,~~~~
\Omega_m \equiv \frac{\rho_m}{(H/\beta_n)^{2/n}}\,,
\end{eqnarray}
which satisfy $\Omega_\vp+ \Omega_m =1$ 
from Eq.~(\ref{Hubble}).

We are interested in asymptotic scaling solutions which
satisfy the relation (\ref{ratio}), in which case
the fractional density  $\Omega_\vp$ is constant.
We also assume that $w_{\vp}$ and $Q$ are constants
in the scaling regime.
Since Eq.~(\ref{ratio}) is equivalent to the condition
${\rm d} \log \rho_{\vp}/{\rm d}N = {\rm d}\log \rho_m/{\rm d}N $,
we obtain the following relation from
Eqs.~(\ref{geneeq1}) and (\ref{geneeq2}):
\begin{equation}
\label{dphi}
\frac{\rd \vp}{\rd N} = \frac{3\Omega_\vp}{Q}
(w_m - w_\vp) = {\rm const.}
\end{equation}
Then this gives the scaling behavior of 
$\rho_{\vp}$ and $\rho_m$:
\begin{equation}
\label{sca}
\frac{\rd {\rm ln} \rho_{\vp}}{\rd N}=
\frac{\rd {\rm ln} \rho_m}{\rd N}=
-3(1+w_{\rm eff})\,,
\end{equation}
where the effective equation of state is
\begin{equation}
\label{ws}
w_{\rm eff} \equiv \frac{w_{\phi}\rho_{\phi}+w_m \rho_m}
{\rho_{\phi}+\rho_{m}}=
w_m+\Omega_\vp (w_\vp-w_m)\,.
\end{equation}
This expression of $w_{\rm eff}$ is valid 
irrespective of the fact that scaling solutions 
exist or not. The condition for an accelerated
expansion corresponds to $w_{\rm eff}<-1/3$.

{}From the definition of $X$ we obtain
\begin{equation}
\label{Xdef}
2 X = H^2 \left(\frac{\rd \vp}{\rd N}\right)^2 \propto \, H^2
\propto \rho_{\rm T}^n\,.
\end{equation}
This means that the scaling property of $X$ 
is the same as $\rho_{\vp}^n$ and $\rho_m^n$. 
Then we find
\begin{equation}
\label{X2eq}
\frac{\rd {\rm ln} X}{\rd N}=-3n(1+w_{\rm eff})\,.
\end{equation}
Since $p_{\vp}=w_\vp \rho_{\vp}$ scales in the same way 
as $\rho_{\vp}$,
one has $\rd {\rm ln}\,p_{\vp}/\rd N=-3(1+w_{\rm eff})$.
Hence we obtain the following relation
by using Eqs.~(\ref{dphi}) and (\ref{X2eq}):
\begin{equation}
\label{pform}
n\frac{\partial \ln p_{\vp}}{\partial \ln X} -
\frac{1}{\lambda} 
\frac{\partial \ln p_{\vp}}{\partial \vp} = 1\,,
\end{equation}
where
\begin{equation}
\label{lam2}
\lambda\,  \equiv\,
Q \frac{1+w_m - \Omega_\vp (w_m - w_\vp)}
{\Omega_\vp (w_m-w_\vp)}\,.
\end{equation}

Equation (\ref{pform}) gives a constraint on the functional
form of $p(X, \vp)$ for the existence of scaling solutions:
\begin{equation}
\label{scap}
p(X, \vp) = X^{1/n}\,
g\left(X e^{n\lambda \vp}\right)\,,
\end{equation}
where $g$ is any function in terms of $ Y \equiv
X e^{n\lambda \vp}$. 
This expression was first derived
in the GR case ($n=1$) in Ref.~\cite{PT}
and was extended to the case of general $n$
in Ref.~\cite{TS04}.
One can easily show that $Y$ is constant
along the scaling solution:
\begin{equation}
X e^{n\lambda \vp} = Y_0
={\rm const}\,.
\end{equation}
This property tells us that $p$ is proportional to $X^{1/n}$
by Eq.~(\ref{scap}). This could be a defining property of scaling solutions
which means that the Lagrangian or the pressure density
depends upon the kinetic energy alone in the scaling regime.
For an ordinary scalar field it leads to a constant ratio of the 
kinetic to potential energy which is
often taken to be a definition of scaling solutions.

In deriving Eq.~(\ref{scap})
we assumed that the coupling $Q$ is a constant in the scaling regime.
One can also obtain a scaling Lagrangian even when the coupling is 
a free function of the field $\vp$, see Ref.~\cite{AQTW}.
It was also shown that we get the Lagrangian (\ref{scap}) by 
appropriate field redefinitions. This means that one can always work 
with the Lagrangian (\ref{scap}), no matter what kind of coupling 
one has in mind.

\subsection{General properties of scaling solutions}

Combining Eq.~(\ref{ws}) with Eq.~(\ref{lam2}) we find that 
the effective equation of state for scaling solutions is 
given by 
\begin{equation}
\label{weffgene}
w_{\rm eff}=\frac{w_{m}\lambda -Q}{Q+\lambda}\,.
\end{equation}
This property holds irrespective of the form of the
function $g(Y)$.
In the case of nonrelativistic dark matter ($w_m=0$)
we have $w_{\rm eff}=0$ for $Q=0$ and $w_{\rm eff} \rightarrow -1$
in the limit $Q \gg \lambda >0$.

{}From the pressure density (\ref{scap}) we obtain
the energy density $\rho_{\vp}$ as $\rho_{\vp}
=X^{1/n}(2/n-1+2Yg'/g)g$,
where a prime denotes a derivative
in terms of $Y$. 
Then the equation of state 
$w_\vp=p_{\vp}/\rho_{\vp}$ reads
\begin{equation}
\label{wp}
w_\vp=\left(\frac{2}{n}-1+2\alpha\right)^{-1}\,,
\end{equation}
where
\begin{equation}
\label{alp}
\alpha \equiv \, \left. \frac{{\rm d} \log g(Y)}
{{\rm d} \log Y}
\right|_{Y = Y_0}\,.
\end{equation}
Using Eqs.~(\ref{dphi}), (\ref{Xdef}) and (\ref{lam2}), 
we obtain the following relation for the scaling solutions:
\begin{equation}
\label{scare}
3H^2=\frac{2(Q+\lambda)^2}{3(1+w_m)^2}X\,.
\end{equation}
Then the fractional density (\ref{Omedef}) of
the field $\vp$ is given by 
\begin{equation}
\label{Omevp}
\Omega_\vp=\left[\frac{9\beta_n^2(1+w_m)^2}
{2(Q+\lambda)^2}\right]^{1/n}\frac{g(Y_0)}{w_\vp}\,.
\end{equation}
By combining  Eq.~(\ref{dphi}) with Eq.~(\ref{Omedef})
together with the relation $w_\vp=p_{\vp}/\rho_{\vp}$,
we find that $g$ in Eq.~(\ref{scap}) can be written as
\begin{equation}
\label{gY0}
g(Y_0)=-Q\left(\frac{2}{9\beta_n^2}\right)^{1/n}
\frac{w_\vp}{w_\vp- w_m}\left(\frac{1+w_m}
{Q+\lambda}\right)^{(n-2)/n}\,.
\end{equation}
Then Eq.~(\ref{Omevp}) yields
\begin{equation}
\label{Omephi}
\Omega_\vp=\frac{Q}{Q+\lambda}
\frac{1+w_m}{w_m-w_\vp}\,.
\end{equation}
Once the functional form of $g(Y)$ is known, the equation of
state $w_\vp$ is determined by Eq.~(\ref{wp})
with Eq.~(\ref{alp}).
We can then derive the fractional density $\Omega_\vp$
from Eq.~(\ref{Omephi}).

For scaling solutions we can define the acceleration parameter 
by 
\begin{equation}
\label{q}
-q \equiv \frac{\ddot{a}a}{\dot{a}^2}=
1-\frac{3n(1+w_m)\lambda}{2(\lambda+Q)}\,.
\end{equation}
When $Q=0$ the condition $-q>0$ corresponds to 
$w_m<2/(3n)-1$. For example $w_m<-1/3$ for $n=1$.
In the case of non-relativistic dark matter ($w_m=0$), 
an accelerated expansion
occurs only for $n<2/3$ (see for example \cite{Freese:2002sq},  
for the case of Cardassian cosmology, 
and \cite{Szydlowski:2004np} for a discussion of  
a class of Cardassian scenarios in terms of dynamical systems).
If we account for the coupling $Q$, it is possible to
get an acceleration even for $n \ge 2/3$.
The condition for acceleration is then
\begin{equation}
\label{acc}
\frac{Q}{\lambda}>\frac{3n(1+w_m)-2}{2}\,.
\end{equation}
One has $Q/\lambda>1/2$ for $w_m=0$ and $n=1$. 
We shall review coupled dark energy scenarios in detail
in Sec.~\ref{cdenergy}.

\subsection{Effective potential corresponding to scaling solutions}

By using the results obtained in previous subsections we can obtain the 
effective potentials corresponding to scaling solutions.

\subsubsection{Ordinary scalar fields}

We first study the case in which the Lagrangian 
density $p$ is written in the form:
\begin{equation}
\label{sum}
p(X, \vp) = f(X)-V(\vp)\,.
\end{equation}
By using Eq.~(\ref{pform}) we find that the functions
$f(X)$ and $V(\vp)$ satisfy
\begin{equation}
\label{sumre}
nX \frac{\rd f}{\rd X}-f(X)=
-\frac{1}{\lambda}\frac{\rd V}{\rd \vp}-V
\equiv C \,,
\end{equation}
where $C$ is a constant.
Hence we obtain $f=c_1X^{1/n}-C$ and
$V=c_2e^{-\lambda \vp}-C$ with $c_{1}$ and $c_{2}$ being constants.
Then the Lagrangian density is given by 
\begin{equation}
\label{lagsca}
p=c_1X^{1/n}-c_2e^{-\lambda \vp}\,.
\end{equation}

This shows that when $n=1$ (GR)
an exponential potential corresponds to the one for 
scaling solutions.
In other cases ($n \ne 1$) the Lagrangian density (\ref{lagsca})
does not have a standard kinetic term, but one can perform
a transformation so that the kinetic term becomes 
a canonical one.
By introducing a new variable $\phi \equiv e^{\beta \lambda \vp}$,
we find  $Y_0=\tilde{X}\phi^{(n-2\beta)/\beta}/
\beta^2\lambda^2={\rm const}$, where $\tilde{X} \equiv
-g^{\mu\nu} \partial_\mu \phi \partial_\nu \phi /2$.
Hence the Lagrangian density (\ref{scap}) can be rewritten as
\begin{equation}
\label{lagor}
p=\frac{Y_0^{1/n}}{\phi^{1/\beta}}g(Y_0)
=Y_0^{1/n} \left(\frac{\tX}{\beta^2 \lambda^2
Y_0}\right)^{1/(n-2\beta)}g(Y_0)\,.
\end{equation}
Since $p$ is proportional to $\tX^{1/(n-2\beta)}$, the transformation that
gives $p \propto \tX$ corresponds to $\beta=(n-1)/2$, i.e.,
$\phi=e^{(n-1)\lambda \vp/2}$.
Then we have $p \propto \phi^{-2/(n-1)}$ from Eq.\,(\ref{lagor}),
which means that the potential of the field $\phi$ corresponding to
scaling solutions is
\begin{equation}
\label{poten}
V(\phi)=V_0\phi^{-2/(n-1)}\,,
\end{equation}
where $V_0$ is constant.
In the case of the RS braneworld ($n=2$) one obtains
an inverse square potential $V(\phi)=V_0\phi^{-2}$ \cite{MLC}.
The Gauss-Bonnet braneworld ($n=2/3$) gives the potential
$V(\phi)=V_0\phi^6$, as shown in Ref.~\cite{SST}.
The Cardassian cosmology ($n=1/3$) corresponds to 
the potential $V(\phi)=V_0\phi^3$.

\subsubsection{Tachyon}

At first glance the tachyon Lagrangian (\ref{ptach})
does not seem to satisfy the condition for the existence of
scaling solutions given in Eq.~(\ref{scap}).
However we can rewrite the Lagrangian (\ref{scap}) by
introducing a new field
$\phi=e^{\beta \lambda \vp}/(\beta \lambda)$.
Since the quantity $Y$ is written as
$Y=\tX (\beta \lambda \phi)^{n/\beta-2}$ with
$\tilde{X} \equiv -g^{\mu\nu} \partial_\mu \phi
\partial_\nu \phi /2$, one has $Y=\tX$ for $\beta=n/2$.
Hence the Lagrangian density (\ref{scap}) yields
\begin{equation}
\label{tachlag2}
p=\left(\frac{n \lambda \phi}{2}\right)^{-2/n}
\tX^{1/n}g(\tX)\,,
\end{equation}
which corresponds to a system $p(\tX, \phi)=V(\phi)f(\tX)$
with potential
\begin{equation}
\label{potacy}
V(\phi)=V_0\phi^{-2/n}\,,
\end{equation}
and $f(\tX)=\tX^{1/n}g(\tX)$.
We note that the tachyon Lagrangian density (\ref{ptach})
is obtained by choosing
\begin{equation}
\label{gYtach}
g(Y)=-cY^{-1/n} \sqrt{1-2Y}\,.
\end{equation}

When $n=1$ (GR), Eq.~(\ref{potacy}) gives the
inverse square potential $V(\phi)=V_0\phi^{-2}$. 
We have earlier studied the dynamics of this system in
Sec.~\ref{cdynamics}.
We also have $V(\phi)=V_0\phi^{-1}$ for $n=2$ (RS),
$V(\phi)=V_0\phi^{-3}$ for $n=2/3$ (GB), and 
$V(\phi)=V_0\phi^{-6}$ for $n=1/3$ (Cardassian cosmology).

\subsubsection{Dilatonic ghost condensate}

The dilatonic ghost condensate model (\ref{ghostcon}) does not have 
a potential.  Let us consider the GR case ($n=1$) in this model.
The Lagrangian density (\ref{ghostcon}) is derived by choosing 
$g(Y)=-1+cY$ in Eq.~(\ref{scap}).
Then by using the relations obtained in 
subsections A and B, we find
\begin{equation}
cY_0=-\frac{2Q(Q+\lambda)-3(1-w_m^2)}
{3(1+w_m)(1-3w_m)}\,,
\end{equation}
and
\begin{eqnarray}
\label{ghostw1}
w_\vp &=& \frac{-3(1+w_m)w_m+Q(Q+\lambda)}
{-3(1+w_m)+3Q(Q+\lambda)}\,, \\
\Omega_\vp &=& \frac{3(1+w_m)
\left[1+w_m-Q(Q+\lambda)
\right]}{(Q+\lambda)^2(1-3w_m)}\,.
\label{ghostw2}
\end{eqnarray}

The condition for an accelerated expansion (\ref{acc})
gives $Q/\lambda>(1+3w_m)/2$ for $n=1$. 
The stability of quantum fluctuations discussed in  
Sec.~\ref{cdynamics} requires $cY_0 \ge 1/2$, 
which translates into the condition
$Q(Q+\lambda) \le 3(1+w_m)^2/4$.
One can obtain viable scaling solutions
if the coupling $Q$ satisfies both conditions.

\subsection{Autonomous system in Einstein gravity}

In this subsection we shall derive autonomous equations 
for the Lagrangian density (\ref{scap}) with $n=1$ (GR), i.e.,
$p=Xg(Xe^{\lambda \vp})$.
In this case the energy density of the field $\vp$
is given by $\rho_{\vp}=p(1+2Yg'/g)$.
We introduce two quantities $x\equiv \dot{\vp}/(\sqrt{6}H)$ and 
$y\equiv e^{-\lambda \vp /2}/(\sqrt{3}H)$.
Using Eqs.~(\ref{Hubble}), (\ref{geneeq1}) and (\ref{geneeq2}),
we obtain the following autonomous
equations \cite{GNST,ATS,Tsuji06}
\begin{eqnarray}
\label{dxsc}
\hspace*{-2.0em}
\frac{{\rm d}x}{{\rm d}N} &=&
\frac{3x}{2} \left[1+gx^2-w_{m} (\Omega_{\vp}
-1)-\frac{\sqrt{6}}{3} \lambda x \right] \nonumber \\
& &+\frac{\sqrt{6}A}{2} \left[ (Q+\lambda)
\Omega_{\vp}-Q-\sqrt{6}(g+Yg')x \right], \\
\label{dysc}
\hspace*{-2.0em}
\frac{{\rm d}y}{{\rm d}N} &=&
\frac{3y}{2} \left[1+gx^2-w_{m} (\Omega_{\vp}
-1)-\frac{\sqrt{6}}{3} \lambda x \right],
\end{eqnarray}
where $A\equiv (g+5Yg'+2Y^2g'')^{-1}$. 
We note that $\Omega_{\vp}$ and $w_{\vp}$ are given by 
\begin{eqnarray}
\label{Omerelation}
\Omega_{\vp}=x^2 (g+2Yg')\,,~~~
w_{\vp}=\frac{g}{g+2Yg'}\,.
\end{eqnarray}
Since $\partial p/\partial X=g+Yg'$, we find 
\begin{eqnarray}
\label{wphige}
w_{\vp}=-1+\frac{2x^2}{\Omega_{\vp}}
\frac{\partial p}{\partial X}\,.
\end{eqnarray}
Eq.~(\ref{wphige}) shows that the field behaves 
as a phantom ($w_{\vp}<-1$) for 
$\partial p/\partial X<0$.

Equation (\ref{scare}) means that there exists the 
following scaling solution for \textit{any} form
of the function $g(Y)$: 
\begin{eqnarray}
x=\frac{\sqrt{6}(1+w_m)}{2(Q+\lambda)}\,.
\label{scalingx}
\end{eqnarray}
In fact it is straightforward to show that this is one of the
critical points
for the autonomous system given by Eqs.~(\ref{dxsc}) and (\ref{dysc}).
We recall that the effective equation of state $w_{\rm eff}$ 
is also independent of $g(Y)$, see Eq.~(\ref{weffgene}).
While $x$ and $w_{\rm eff}$ do not depend on the form of 
$g(Y)$, $\Omega_{\vp}$ and $w_{\vp}$ remain undetermined 
unless we specify the Lagrangian.

\section{The details of quintessence}
\label{QK}

In this section we shall discuss various aspects of quintessence
such as the nucleosynthesis constraint, tracking behavior, 
assisted quintessence, particle physics models and 
quintessential inflation.

\subsection{Nucleosynthesis constraint}

The tightest constraint on the energy density of dark energy 
during a radiation dominated era comes from primarily nucleosynthesis.
The introduction of an extra degree of freedom (on top
of those already present in the standard model of particle physics) 
like a light scalar field affects the abundance of 
light elements in the radiation dominated epoch.
The presence of a quintessence scalar field changes
the expansion rate of the universe at a given temperature.
This effect becomes crucial at the nucleosynthesis epoch with temperature
around $ 1\,{\rm MeV}$ when the weak interactions 
(which keep neutrons and protons in equilibrium) freeze-out. 

The observationally allowed range of the expansion
rate at this temperature leads to a bound on the energy 
density of the scalar field \cite{Ferreira97}
\begin{equation}
{\Omega_{\phi}}({T \sim 1 {\rm MeV}})
<\frac{7\Delta N_{\rm eff}/4}
{10.75+7\Delta N_{\rm eff}/4}\,,
\label{nucconst1}
\end{equation}
where $10.75$ is the effective number of standard model degrees of 
freedom and $\Delta N_{\rm eff}$ is the additional relativistic 
degrees of freedom.
A conservative bound on the additional degrees of freedom 
used in the literature is 
$\Delta N_{\rm eff}\simeq 1.5$ \cite{Sarkar}, whereas a typical one is given 
by $\Delta N_{\rm eff} \simeq 0.9$ \cite{Copi}. 
Taking a conservative one, we obtain the following bound
\begin{equation}
{\Omega_{\phi}}
({T \sim 1 {\rm MeV}}) < 0.2\,.
\label{nucconst2}
\end{equation}
Any quintessence models need to satisfy 
this constraint at the epoch of nucleosynthesis.
We note that Bean {\it et al.} \cite{Bean01} obtained a tighter 
constraint ${\Omega_{\phi}} ({T \sim 1 {\rm MeV}}) < 0.045$
with the use of the observed abundances of primordial nuclides.

As we have already seen in Sec.~\ref{cdynamics}, the exponential
potential (\ref{exp}) possesses the following two attractor solutions
in the presence of a background fluid:

(1) $\lambda^2 > 3\gamma$ : the
scalar field mimics the evolution of the barotropic fluid
 with $\gamma_{\phi} = \gamma$, and the relation
$\Omega_{\phi} = 3\gamma/\lambda^2$ holds.

(2) $\lambda^2 < 3\gamma$: the late time attractor is the scalar field dominated
solution ($\Omega_\phi =1$) with $\gamma_{\phi} 
= \lambda^2/3$.

The case (1) corresponds to a scaling solution in which 
the field energy density mimics that of the
background during radiation or matter dominated era, thus
alleviating the problem of a cosmological constant.
If this scaling solution exists by the epoch of nucleosynthesis
($\gamma=4/3$), the constraint (\ref{nucconst2}) gives 
\begin{equation}
\Omega_{\phi}=\frac{4}{\lambda^2}<0.2~~~ 
\to~~~\lambda^2>20\,.
\label{lamcondition}
\end{equation}
In this case, however, one can not have an
accelerated expansion at late times, since the equation of 
state of the field is the same as that of the background.
In order to lead to a late-time acceleration, the scaling solution (1)
needs to exit to the scalar-field dominated solution (2)
near to the present.
In the next subsection we shall explain quintessence models
which provide this transition.

\subsection{Exit from a scaling regime}

In order to realize the exit from the scaling regime explained above, 
let us consider the following double exponential 
potential \cite{BCN99,doubleexp}
\begin{equation}
\label{2exp}
V(\phi) = V_0 \left( e^{-\lambda \kappa \phi} 
+ e^{-\mu \kappa \phi} \right)\,,
\end{equation}
where $\lambda$ and $\mu$ are positive.
Such potentials are expected to arise as 
a result of compactifications in superstring models, 
hence are well motivated 
(although there remains an issue over how easy it is 
to obtain the required values of $\mu$ and $\lambda$). 
We require that $\lambda$ satisfies the condition (\ref{lamcondition})
under which the energy density of the field mimics the background 
energy density during radiation and matter dominated eras.
When $\mu^2<3$ 
the solution exits from the scaling regime and approaches 
the scalar-field dominated solution (2) with $\Omega_{\phi}=1$.
The accelerated expansion is realized at late times
if $\mu^2 <2$.

\begin{figure}
\includegraphics[height=2.7in,width=3.3in]{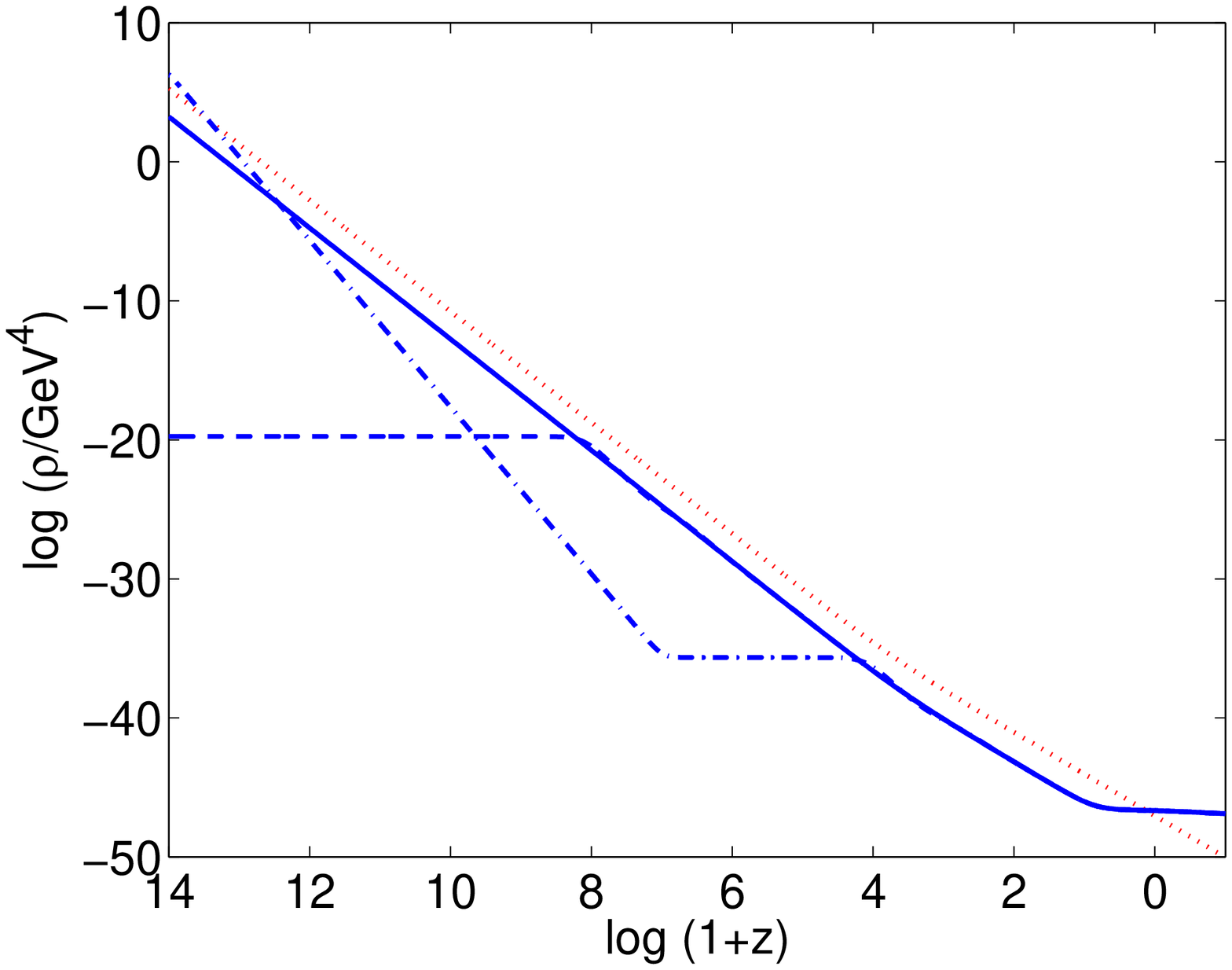}
\caption{
Evolution of the energy density $\rho_{\phi}$
for $\lambda=20$ and $\mu=0.5$.
The background energy density 
$\rho_{\rm matter}+\rho_{\rm radiation}$ is plotted
as a dotted line. Even when $\rho_{\phi}$ is larger than 
$\rho_{\rm matter}+\rho_{\rm radiation}$ at the initial stage,
the solutions approach the scaling regime in which the 
scalar field energy density is subdominant and tracks 
the background fluid.
We thank Nelson J.~Nunes for providing us this figure.
}
\label{trackfig} 
\end{figure}

There is an important advantage to the above double 
exponential potential. For a wide range of initial conditions 
the solutions first enter the scaling regime, which is followed by 
an accelerated expansion of the universe once the potential 
becomes shallow. This behavior is clearly seen in Fig.~\ref{trackfig}.
Interestingly it is acceptable to start with the energy density
of the field $\phi$ larger than that of radiation and then 
approach a subdominant scaling attractor.

Another model which is related to (\ref{2exp}) was suggested by 
Sahni and Wang \cite{SW99}:
\begin{equation}
\label{coshpot}
V(\phi) = V_0 \left[ \cosh( \kappa \lambda \phi) -1 \right]^n \,.
\end{equation}
This potential has following asymptotic forms:
\ba
& & V(\phi)  \simeq \left\{\begin{array}{ll}
\wt{V}_{0} e^{-n\kappa \lambda \phi}
\quad (|\lambda \phi| \gg 1,~\phi<0)\,, & \\
\label{sawa}
\wt{V}_{0} (\kappa \lambda \phi)^{2n}
\quad (|\lambda \phi| \ll 1)\,,
\end{array} \right. 
\ea
where $\wt{V}_{0}=V_{0}/2^n$.
Then the field energy density proportionally decreases  
to that of radiation and matter for $|\lambda \phi| \gg 1$,
in which $\Omega_{\phi}$ is given by 
$\Omega_{\phi}=3\gamma/n^2 \lambda^2$.
As the field approaches the potential minimum at $\phi=0$, 
the system exists from the scaling regime.
During the oscillatory phase in which the potential is given by 
(\ref{sawa}), the virial theorem gives the time-averaged relation
$\langle  \dot{\phi}^2/2 \rangle=n \langle V(\phi) \rangle$.
Then the average equation of state for the field $\phi$ is 
\ba
\langle w_{\phi} \rangle=\frac{n-1}{n+1}\,.
\ea

When $n<1/2$ the field can satisfy the condition 
for an accelerated expansion ($\langle w_{\phi} \rangle<-1/3$).
In fact it was shown in Ref.~\cite{SW99} that tracking solutions which 
give the present-day values $\Omega_\phi \simeq 0.7$ and 
$\Omega_m \simeq 0.3$ can be obtained for a 
wide range of initial conditions.
The field behaves as non-relativistic matter 
($\langle w_{\phi} \rangle=0$) for $n=1$.
This scalar field can give rise to a tracking ``scalar cold dark matter''
if the mass of dark matter is $m_{\rm CDM} \sim 
10^{-26}\,{\rm GeV}$ \cite{MatosLopez}. An interesting attempt of unified description of dark matter and dark energy
with a real scalar field is made in Ref.\cite{AAF}.

Albrecht and Skordis \cite{albrecht} have developed an
interesting model which can be derived from
string theory, in that they claim the parameters are all of order
one in the underlying string theory. The potential has a local
minimum which can be adjusted to have today's critical energy
density value (this is where the fine tuning is to be found as in all Quintessence models). 
The actual potential is a combination of exponential and
power-law terms: 
\begin{equation}
V(\phi) = V_0 e^{- \kappa \lambda \phi} \left[
 A + (\kappa \phi - B)^2 \right] \,.
\end{equation}
For early times the exponential term dominates the dynamics, 
with the energy density
of $\phi$ scaling as radiation and matter.
For suitable choices of the
parameters the field gets trapped in the local minimum because the kinetic
energy during a scaling regime is small.
The field then enters a regime of damped oscillations
leading to $w_\phi \to -1$ and an accelerating universe.

\subsection{Assisted quintessence}

So far we have discussed the case of single-field quintessence.
In early universe inflation it is known that 
multiple scalar fields with exponential potentials lead to 
the phenomenon of {\it assisted inflation} \cite{LMS} whereby they 
collectively drive inflation even if each field has too steep 
a potential to do so on its own.
This property also holds in the context of quintessence 
with steep exponential potentials \cite{BCN99,CvdH,KLT}
(see also Ref.~\cite{Blais}).
Here we shall briefly discuss the dynamics of 
assisted quintessence.

We consider two fields $\phi_1$ and $\phi_2$ each 
with a separate exponential potential 
\begin{eqnarray}
V(\phi_1,\phi_2) =A e^{-\kappa \lambda_1\phi_1} + B 
e^{-\kappa \lambda_2\phi_2}\,,
\end{eqnarray}
where we do not implement  interactions between fields.
Note that such multi-field models may have a link to 
time-dependent compactifications of supergravity 
on symmetric (or twisted product) spaces, see e.g., 
Refs.~\cite{Hyper}.
The original assisted inflation scenario of Liddle {\it et
al.}~\cite{LMS}
corresponds to the case in which no matter is present, which 
gives an effective coupling
\begin{equation}
\label{lameff}
\frac{1}{\lambda_{{\rm eff}}^2} = 
\frac{1}{\lambda_1^2} + \frac{1}{\lambda_2^2} \,.
\end{equation}
Since the effective equation of state is given by 
$\gamma_{{\rm eff}}=\lambda_{\rm eff}^2/3$,
the scale factor evolves as $a \propto t^p$, 
where $p=2/\lambda_{\rm eff}^2$.
Hence an accelerated expansion occurs for $\lambda_{\rm eff}<\sqrt{2}$
even when both $\lambda_1$ and $\lambda_2$ are larger than $\sqrt{2}$.

Lets us take into account a barotropic fluid with an EOS 
given by $\gamma=1+w_{m}$.
In the single field case the fixed points (c) and (d) in Table I 
are stable depending on the values of $\lambda$ and $\gamma$.
By replacing $\lambda$ to $\lambda_1$ and $\lambda_2$, we can 
obtain corresponding fixed points in the multi-field 
case \cite{CvdH,KLT}.
Once a second field is added, the new degrees of freedom 
always render those solutions unstable. The 
late-time attractors instead become either the assisted scalar-field
dominated 
solution with $\gamma_{\phi}=\lambda_{\rm eff}^2/3$ and
$\Omega_\phi=1$
(stable for $\lambda_{\rm eff}^2<3\gamma)$
or the assisted scaling solution 
with $\gamma_{\phi}=\gamma$ and 
$\Omega_\phi=3\gamma/\lambda_{\rm eff}^2$
(stable for $\lambda_{\rm eff}^2>3\gamma)$.

If there are a large number of exponential potentials with different 
initial conditions, more and more fields would join the 
assisted quintessence attractor, which reduces $\lambda_{{\rm eff}}$.
Eventually the attractor can switch from the scaling regime 
$\lambda_{{\rm eff}}^2 > 3 \gamma$ into the regime of 
 scalar field dominance $\lambda_{{\rm eff}}^2 
< 3 \gamma$ \cite{CvdH,KLT}. 
This realizes an accelerated expansion at late-times, 
but we still have a fine-tuning problem to obtain a sufficiently 
negative value of EOS satisfying the current observational 
constraint ($w_{\phi} \lesssim -0.8$).

In Ref.~\cite{Tsuji06} a general analysis was given for 
scalar-field models which possess scaling solutions.
Let us consider $n$ scalar fields 
($\phi_{1}, \phi_{2}, \cdots, \phi_n$) with 
the Lagrangian density:
\begin{eqnarray}
p=\sum_{i=1}^n X_i g(X_i e^{\lambda_i \phi_i})\,,
\end{eqnarray}
where $X_{i} =-g^{\mu\nu} \partial_\mu \phi_i 
\partial_\nu \phi_i/2$ and
$g$ is an arbitrary function.
It was shown that the scalar fominant fixed point
exists for this system with an equation of state:
\begin{eqnarray}
w_{\phi}=-1+\frac{\lambda_{\rm eff}^2}{3p_{,X}}\,,
\end{eqnarray}
where $\lambda_{\rm eff}^2$ is given by \cite{Tsuji06}
\begin{eqnarray}
\label{efflambda}    
\frac{1}{\lambda_{\rm eff}^2}=
\sum_{i=1}^n \frac{1}{\lambda_i^2}\,.
\end{eqnarray}
Here $p_{,X}=g(Y)+Yg'(Y)$ where 
$Y \equiv Y_1=Y_2= \cdots=Y_i=\cdots=Y_n$
with $Y_i \equiv X_i e^{\lambda_i \phi_i}$.

The presence of multiple scalar fields leads to 
the decrease of  the effective $\lambda_{\rm eff}^2$ 
relative to the single-field case. 
Since the quantity $p_{,X}$ is not affected by introducing more
scalar fields \cite{Tsuji06}, the presence of many scalar fields
works to shift the equation of state toward $w_{\phi}=-1$.
Thus for a non-phantom scalar field ($p_{X}>0$)
assisted acceleration always occurs for all 
scalar-field models which have scaling solutions.

\subsection{Particle physics models of Quintessence}
\subsubsection{Supergravity inspired models}

We turn our attention to the thorny issue of finding examples 
of Quintessence in particle physics. Recall, one of the constraints 
we need to satisfy is that the Quintessence potential remains 
flat enough so that we can have slow roll inflation today, 
or some mechanism to trap the scalar field today. 
One of the most interesting approaches is to be found in 
Refs.~\cite{Townsend:2001ea,Townsend:2003fx} (see also \cite{Cornalba:2002fi,Cornalba:2003kd}). 
In  Ref.~\cite{Townsend:2001ea}, Townsend considered the 
possibility of Quintessence arising in M-theory. 
He demonstrated that there is a version of $N=8, D=4$ 
supergravity that has a positive exponential potential, 
obtainable from a ``non-compactification'' of M-theory, 
and this potential can lead to an accelerating cosmological 
solution that realizes ``Eternal Quintessence''. 

There is a problem that such models need to be able to address. 
To derive a Quintessence model from string/M-theory, we would 
expect that any $D=4$ dimensional solution should be a solution 
of $D=11$ supergravity or IIB $D=10$ supergravity. 
Unfortunately this is not so straightforward. 
There exists a no-go theorem due to Gibbons \cite{gibbons-nogo} 
(more recently extended by Maldacena and 
Nu\~nez \cite{maldacena-nunez}), 
which states that when the six or seven dimensional ``internal'' space 
is a time-independent non-singular compact manifold without
boundary there can not be a scalar field with a positive potential, 
hence ruling out the possibility of late-time acceleration 
in any effective $D=4$ supergravity model based on an embedding 
in $D=11$ or $D=10$ supergravity. 

The basic problem concerns the strong energy condition in going 
from $D$ spacetime dimensions to $d<D$ spacetime dimensions 
under a general warped compactification on a compact 
non-singular manifold of dimension $n=D-d$. 
If the non-singular $D$-dimensional metric can be written in the form 
\begin{equation}
\label{assume}
{\rm d}s^2_D = f(y) {\rm d}s^2_d(x) + {\rm d}s^2_n(y)\,,
\end{equation}
where $y$ is the compact dimension, then the positivity 
of $R_{00}$ (the Ricci tensor in $D$ dimensions) implies 
positivity of $r_{00}$ (the Ricci tensor in $d$ dimensions).  
Hence for such compactifications, the strong energy 
condition in $D$ spacetime dimensions implies that
the strong energy condition holds in spacetime dimension $d<D$. 
{}From Einstein equations this then implies  $|g_{00}|V(\phi) \le (d-2)\dot\phi^2$, and
hence the scalar field potential must satisfy $V(\phi ) \le 0$ 
if initial conditions can be chosen such that
$\dot\phi=0$. This fact that the $d$-dimensional strong energy 
condition forbids an accelerating $d$-dimensional universe 
was emphasised in Refs.~\cite{hellerman01,fischler01}, 
in showing how difficult it is to embed accelerating cosmologies 
into string/M-theory, where  the strong energy condition is
satisfied by both $D=11$ supergravity and IIB $D=10$ supergravity.

There exist a number of ways of avoiding the no-go theorem and 
these have been exploited to come up with Quintessence scenarios 
within string/M-theory. One route is to have a ``compactifying'' space 
that is actually  non-compact \cite{Hull:2001ii}. 
In Ref.~\cite{Townsend:2001ea}, Townsend adopted this approach and 
showed that a particular non-compact gauged
$N=8, D=4$ supergravity, obtainable from a warped 
``non-compactification'' of M-theory, has a positive exponential 
potential leading to an accelerating universe, with an equation 
of state $p = -(7/9)\, \rho$. 
Of course, it leaves open the question of how realistic are 
these classes of non-compactified theories, a question 
we will not address here. 

Another way round the problem is 
to allow the compact dimension to be time-dependent. 
A number of authors have adopted this approach, 
but in Ref.~\cite{Townsend:2003fx} the authors pointed out 
that in order to have a transient period of acceleration 
in the Einstein frame in $D=4$ dimensions, what is required is 
a hyperbolic compact internal space evolving in time 
(because the analogous solution of the vacuum Einstein equations 
for an internal manifold of positive curvature does not allow acceleration). 
The advantages of such compactifications have been discussed 
in Refs.~\cite{Kaloper00,starkman01}. 
Of course, as with all Quintessence models to date, all is not 
rosy for this class of time dependent Hyperbolic space solutions. 
In \cite{Emparan:2003gg} the authors developed a four-dimensional 
interpretation of the solutions with a transient accelerating 
phase obtained from compactification in hyperbolic manifolds. 
The solutions correspond to bouncing the radion field off 
its exponential potential, with acceleration occuring at the turning point, 
when the radion stops and the potential energy momentarily dominates. 
There is a degree of fine tuning involved in establishing the radion field 
is close enough to the turning point for sufficient inflation to occur. 
Moreover, in this interpretation in terms of the four dimensional 
effective theory, the precursor of the inflationary phase is a period 
of kinetic domination, whereas we believe the Universe was 
matter dominated before it became dominated by the 
Quintessence field.  Another problem has been highlighted 
in \cite{LindeT} where the authors studied the time 
evolution of the corresponding effective 4d cosmological model 
this time including cold dark matter. 
They concluded that even though it is marginally possible to 
describe the observational data for the late-time cosmic acceleration 
in this model, during the compactification of 11d $\to$ 4d 
the Compton wavelengths of the Kaluza Klein modes in 
this model are of the same order as the size of the observable 
part of the universe. This problem has yet to be resolved. Even so, 
assuming there is a resolution it is encouraging that it is possible 
to obtain late time inflationary solutions in M-theory. 
Another serious problem associated with the approach 
under consideration is related to the fact
that masses of KK-modes in this class of models are 
of the order of the present value of Hubble 
parameter \cite{LindeT}.

An interesting and possibly one of the most promising approaches 
to addressing the origin of dark energy in particle physics is 
due to Burgess and Quevedo along with their 
collaborators \cite{burgess-6d,Burgess:2003mk,Aghababaie:2003ar,Burgess:2004kd,Burgess:2004yq,Burgess:2004dh,Ghilencea:2005vm,Matias:2005gg,Callin:2005wi,Tolley:2005nu}. 
The key idea is that the presence of two large extra dimensions 
(they consider 6-dimensional supersymmetric models) provides 
a natural mechanism to generate the small size of the observed 
dark energy today. Of particular note in the approach is the fact 
that the authors not only ask the question why has the dark energy 
the value it has today, but also why is that value stable to integrating 
over higher energy contributions? 
This question of {\it technical naturalness} is a vital one in particle physics, 
and is at the heart of our understanding the hierarchy problem. 
Therefore it makes sense to adopt it in determining the nature and 
value of the dark energy today, which after all is a very small amount 
compared to the natural scale we would expect it to take. 
Apart from the case of dark energy arising out of axion models 
(which we will come to shortly), the majority of quintessence 
scenarios do not possess such a protection mechanism for 
the mass of the field \cite{lythkolda}. 

To be a bit more specific, following the nice review 
in \cite{Burgess:2005wu}, consider the case where a parameter
$\lambda$ is small when measured in an experiment
performed at an energy scale $\mu$. We wish to understand
this in terms of a microscopic theory of physics which is defined
at energies $\Lambda \gg \mu$. The prediction
for $\lambda$ is given by
\begin{equation}
\lambda(\mu) = \lambda(\Lambda) + 
\delta \lambda(\mu,\Lambda)\,,
\end{equation}
where $\lambda(\Lambda)$ represents the direct contribution to
$\lambda$ due to the parameters in the microscopic theory, and
$\delta\lambda$ represents the contributions to $\lambda$ which
are obtained as we integrate out all of the physics in the energy
range $\mu < E < \Lambda$. For $\lambda$ to be small, barring some 
miraculous cancellation we require that both $ \lambda(\Lambda) $ 
and $ \delta \lambda(\mu,\Lambda)$ are both equally small, 
for any chosen value of $\Lambda$. 
Most models of dark energy to date can generate $\lambda(\mu)$ 
as being small, but can not guarantee the smallness of 
$ \delta \lambda(\mu,\Lambda)$ \cite{Weinberg:1988cp}. 
As we have seen the vacuum's energy density today is 
$\rho \sim (10^{-3}$ eV$)^4$, but for a particle of mass $m$ it typically
contributes an amount $\delta \rho \sim m^4$ when it is integrated
out. It follows that such a small value for $\rho$ can only be understood in a
technically natural way if $\Lambda \sim 10^{-3}$ eV or less.
However, the majority of  the elementary particles 
(including the electron, for which $m_e \sim 5 \times 10^5$ eV)
have $m \gg 10^{-3}$ eV, which violates 
the requirement on $\Lambda$. 

In order to overcome this four dimensional problem, Burgess argues 
that we need to modify the response of gravity to physics at scales 
$E > \mu \sim 10^{-3}$ eV, whilst maintaining that the modification 
does not ruin the excellent agreement with the 
non-gravitational experiments which have been performed covering 
the energy range $\mu < E < \Lambda$, with $\Lambda \sim 10^{11}$ eV. 
The approach adopted to achieve this goal makes use of the framework 
of Large Extra Dimensions \cite{LED}. The observed particles (except the graviton) are constrained to 
live on  a (3+1) dimensional surface, within an extra dimensional 
space, with only gravity probing the extra dimensions. (For the case of a non-supersymmetric string model that can realize the extra dimensions see \cite{Cremades:2002dh,Kokorelis:2002qi}).

The present upper limit for the size of the extra
dimensions is $r < 100 \; \mu$m, or $1/r > 10^{-3}$ eV
which is very close to the scale $\mu$ just described and where a natural 
understanding of the vacuum energy breaks down. Also, it turns out that 
there must be two of these extra dimensions if they are to be this large, 
and if the fundamental scale, $M_g$, of the
extra-dimensional physics is around 10 TeV, due to the relation
$M_{\rm pl} = M_g^2 \, r$ which relates $M_g$ and $r$ to the observed
Planck mass: $M_{\rm pl}= 
(8 \pi G)^{-1/2} \sim 10^{27}$ eV \cite{Burgess:2005wu}. 
With this observation in mind, the idea of Supersymmetric Large
Extra Dimensions (SLED) was introduced \cite{burgess-6d}, 
which has at its heart the existence of  two large (i.e., $r \sim 10$ $\mu$m) 
extra dimensions, within a supersymmetric theory, the supersymmetry 
allowing for the cancellation between bosons and fermions 
which appear in the vacuum energy.

Gravitational physics is effectively
6-dimensional for any energies above the scale, 
$1/r \sim 10^{-2}$ eV, and so the cosmological constant problem 
has to be discussed in 6 dimensions. This means integrating out the
degrees of freedom between the scales $M_g \sim 10$ TeV and $1/r
\sim 10^{-2}$ eV. Once this is done, the cosmological constant within the
effective 4D theory is obtained, describing gravitational physics 
on scales much larger than $r$. The basic procedure undertaken in 
integrating over modes having energies $1/r < E < M_g$ is 
as follows \cite{Burgess:2005wu}. First integrate out (exactly) all of 
the degrees of freedom on the branes, giving the low-energy brane 
dependence on the massless 4D graviton mode. 
The effect of this is to obtain a large effective brane tension, $T \sim M_g^4$ for each
of the 3-branes which might be present, which includes the vacuum
energies of all of the presently-observed elementary particles. 
Following this, there is a classical part of the integration over the
bulk degrees of freedom. This is achieved by solving the classical
supergravity equations to determine how the extra dimensions curve
in response to the brane sources which are scattered throughout
the extra dimensions. The key result from this part of the calculation is that  this classical
response cancels the potentially large contributions from the branes obtained
in the first step. Finally, the quantum part of the integration over
the bulk degrees of freedom is performed. It is this contribution which is
responsible for the fact that the present-day dark energy density
is nonzero, in other words it is a quantum feature! 

Moreover for specific cases they find the small size of the 4D vacuum energy is 
attributed to the very small size with which supersymmetry breaks
in the bulk relative to the scale with which it breaks on the branes. 
We will not go into details of the calculations here, the interested 
reader is referred to \cite{Burgess:2005wu} and references therein. 
There are a few points worth highlighting though.  
In the third part, where the quantum part of the integration is performed, 
in the SLED model for a class of 6D supergravities, these quantum 
corrections lift the flat directions of the classical approximation, and those
loops involving bulk fields do so by an amount leading 
to a potential \cite{Ghilencea:2005vm}
\begin{equation} 
\label{eq:Vofr}
V(r) \sim \frac{1}{r^4} 
\Bigl( a + b \log r \Bigr)\,,
\end{equation}
where $a$ and $b$ are calculable constants and the logarithmic
corrections generically arise due to the renormalization of UV
divergences in even dimensions. 

Such a potential is similar to the Quintessence form introduced 
by Albrecht and Skordis \cite{albrecht}, and it predicts that
scalar-potential domination occurs when $\log (M_{\rm pl} r)$ is of order
$a/b$, which can be obtained  given a modest
hierarchy amongst the coefficients, $a/b \sim 70$. 
The SLED proposal requiring the world to become six-dimensional 
at sub-eV energies is falsifiable (a useful attribute for a model!). 
This is because it has a number of knock on consequences 
for phenomenology that implies particle physics may soon rule it out. 
There is a deviation from the inverse square law for gravity 
around $r/2 \pi \sim 1$ $\mu$m \cite{Callin:2005wi}; distinctive 
missing-energy signals in collider experiments
at the LHC due to the emission of particles into the extra
dimensions \cite{new-paticles} and potential astrophysical 
signals arising from loss of energy into the extra dimensions 
by stars and supernovae. 
It is a an interesting proposal which takes seriously the issue 
of technical naturalness and has a possible resolution of it 
in the context of Quintessence arising in six 
dimensional supersymmetric models.  
Before finishing this section on large extra dimensions 
we should mention the recent work of Sorkin \cite{Sorkin:2005fu}, 
in which he argues that the true quantum gravity scale cannot 
be much larger than the Planck length. 
If it were then the quantum gravity-induced fluctuations 
in $\Lambda$ would be insufficient to produce the observed dark energy. 

In Sec.~\ref{scalarmodel} we discussed dilatonic dark energy models
based upon the low-energy effective string action.
Another approach to supergravity inspired models of quintessence 
makes use of the  inverse power-law potentials which arise in supersymmetric 
gauge theories due to non-perturbative effects, 
see Refs.~\cite{QCD} for details.
In a toy model, taking into account a supergravity correction to globally supersymmetric
theories, Brax and Martin \cite{brax} constructed a quintessence potential  
which possesses a minimum.
The F-term in the scalar potential in general is given by 
\begin{eqnarray}
V &=& e^{\kappa^2K} [( W_i +\kappa^2 W K_i ) K^{j^*i} 
( W_j + \kappa^2 W K_j )^*  \nonumber \\
& &- 3\kappa^2|W|^2]\,,
\label{fterm}
\end{eqnarray}
where $W$ and $K$ are the superpotential and the K\"ahler potential, 
respectively. The subscript $i$ represents the derivative with respect to
the $i$-th field.

With the choice of  a superpotential $W=\Lambda^{3+\alpha} \phi^{-\alpha}$ 
and a flat K\"ahler potential, $K = \phi\phi^*$, we obtain the following potential
\begin{equation}
V(\phi) = e^{\frac{\kappa^2}{2}\phi^2}  
\frac{\Lambda^{4+\beta}}{\phi^{\beta}} \,
\left[\frac{(\beta -2)^2}{4} - (\beta +1) \frac{\kappa^2}{2}\phi^2 + 
\frac{\kappa^4}{4}\phi^4 \right]\,,
\label{flatpot}
\end{equation}
where $\beta = 2\alpha +2$.
This means that the potential can be negative in the presence of 
supergravity corrections for $\phi \sim m_{\rm pl}$.
In order to avoid this problem, Brax and Martin imposed 
the condition that the expectation value of the superpotential
vanishes, i.e., $\langle W \rangle =0$.
In this case the potential (\ref{flatpot}) takes the form 
\begin{equation}
V(\phi) \ =\ \frac{\Lambda^{4+\alpha}}{\phi^{\alpha}}\,  
e^{\frac{\kappa^2}{2}\phi^2}\,.
\label{bmpo}
\end{equation}
Although setting $\langle W \rangle =0$ is restrictive, Brax and Martin argued that 
this can be realized in the presence of matter fields 
in addition to the quintessence field \cite{brax}.

The potential (\ref{bmpo}) has a minimum at 
$\phi=\phi_*\equiv \sqrt{\alpha}/\kappa$.
If $V(\phi_*)$ is of order the present critical density
$\rho_c^{(0)} \sim 10^{-47}\,{\rm GeV}^4$,
it is possible to explain the current acceleration of 
the universe.
{}From Eq.~(\ref{bmpo})  the mass squared
at the potential minimum is given by 
$m^2 \equiv \dfrac{\rd^2 V}{\rd \phi^2}=2\kappa^2 V(\phi_*)$.
Since $3H_0^2 \simeq \kappa^2 V(\phi_*)$, we find 
\begin{equation}
m^2 \simeq 6H_{0}^2\,.
\end{equation}
This is a very small mass scale of order $m \sim H_{0} \sim 
10^{-33}\,{\rm eV}$.
Such a tiny mass is very difficult to reconcile with 
fifth force experiments, 
unless there is a mechanism to prevent $\phi$ 
from having interactions with the other matter fields.
As mentioned this is a problem facing many quintessence models.

One can choose more general K\"ahler potentials when studying 
supergravity corrections.
Lets us consider a theory with superpotential
$W = \Lambda^{3+\alpha}\, \vp^{-\alpha}$
and a K\"ahler $K = -\ln (\kappa \vp + \kappa \vp^*)/\kappa^2$, 
which appears at tree-level in string theory \cite{rosati1}.
Then the potential for a canonically normalized field, 
$\phi = (\ln \kappa \vp)/\sqrt{2} \kappa$, is
\begin{equation}
V(\phi) = M^4 e^{-\sqrt{2} \kappa \beta \phi}\,,
\end{equation}
where $M^4 = \Lambda^{5+\beta}\,\kappa^{1+\beta}~(\beta^2 -3)/2 $ and 
$\beta = 2\alpha +1 $.
We note that $\beta$ needs to be larger than $\sqrt{3}$
to allow for positivity of the potential.
Thus we can obtain an exponential potential giving rise to scaling 
solutions in the context of supergravity.

Kolda and Lyth \cite{lythkolda} argued that supergravity
inspired models suffer from the fact that loop corrections 
always couple the quintessence field to other sources of matter so
as to lift the potential thereby breaking the flatness criteria
required for quintessence today. 
We now go on to discuss a class of models where 
this problem can be avoided.
In the context of $N \ge 2$ extended supergravity models
the mass squared of any ultra-light scalar fields is quantized 
in unit of the Hubble constant $H_{0}$, i.e., 
$m^2=nH_0^2$, where $n$ are of order unity \cite{kalloshlinde,Fre}.

To be concrete let us consider the potential $V(\phi)=\Lambda+
(1/2) m^2 \phi^2$ around the extremum at $\phi=0$.
In extended supergravity theories the mass $m$ is related to 
$\Lambda~(>0)$ via the relation $m^2=n \Lambda/3M_{\rm pl}^2$, where 
$n$ are integers. 
Since $H_0^2=\Lambda/3M_{\rm pl}^2$ in de Sitter 
space, this gives $m^2=nH_{0}^2$.
In the context of $N=2$ gauged supergravity we have $m^2=6H_{0}^2$
for a stable de Sitter vacuum \cite{Fre}, which gives  
\be 
V(\phi) =3H^2_0M_{\rm pl}^2
\left[1 +\left(\phi/M_{\rm pl}\right)^2
\right]\,.
\ee
The $N=8$ supergravity theories give the negative mass squared,
$m^2=-6H_{0}^2$ \cite{kalloshlinde}, in which case we have 
\be 
V(\phi) =3H^2_0M_{\rm pl}^2
\left[1 -\left(\phi/M_{\rm pl}\right)^2\right]\,.
\ee

We note that the constant $\Lambda$ determines the energy scale of 
supersymmetry breaking.
In order to explain the present acceleration we require 
$\Lambda \sim m^2M_{\rm pl}^2 \sim H_{0}^2M_{\rm pl}^2
\sim 10^{-47}\,{\rm GeV}^4$.
This energy scale is so small that quantum corrections to 
$\Lambda$ and $m$ are suppressed.
Hence we naturally obtain ultra-light scalars which are stable 
against quantum corrections.

\subsubsection{Pseudo-Nambu-Goldstone models}

Another approach to dark energy which avoids the serious problem posed 
by Kolda and Lyth \cite{lythkolda} is to consider models in which 
the light mass of the Quintessence field can be protected by an 
underlying symmetry. Such a situation arises in cases where we have 
a pseudo Nambu Goldstone boson acting as the Quintessence field. 
This idea was first introduced by Frieman {\it et al.} \cite{Frieman} 
(see also \cite{Choi}), in response to the first tentative suggestions 
that the universe may actually be dominated by a cosmological constant. 
These axion dark energy models based on $N=1$ supergravity have 
similar properties to the extended supergravity models discussed above. 

The axion potential is 
\be 
V(\phi) =\Lambda \left[C+\cos(\phi/f) \right]\,,
\ee
where the model given by Frieman {\it et al.} \cite{Frieman}
corresponds to $C=1$.
The model with $C=0$ can be obtained by using the superpotential and 
K\"ahler potential motivated from M/string theory \cite{Choi}.
The mass of the field $\phi$ at the potential maximum is 
$m^2=-\Lambda/f^2$. If this energy at potential maximum is responsible
for the current accelerated expansion, we have $3H_{0}^2 \simeq 
\Lambda/M_{\rm pl}^2$. 
Then when $f$ is of order $M_{\rm pl}$, we get 
\be 
m^2=-\Lambda/M_{\rm pl}^2 \simeq -3H_0^2\,.
\ee

The field is frozen at the potential maximum when $|m^2|$ is 
smaller than $H^2$, but begins to roll down 
around present ($|m^2| \sim H_0^2$).
Since the energy at the potential minimum ($\phi=\pi f$) is negative
for $C=0$, the universe collapses in the future within
the next 10-20 billion years \cite{kalloshlinde}.

The possibility of there being an approximate global symmetry being 
present to suppress the natural couplings of the Quintessence 
field to matter, which generally result in long range forces, 
was investigated by Carroll in Ref.~\cite{Carroll:1998zi}. 
He also showed how such a symmetry could allow a coupling 
of $\phi$ to the pseudoscalar $F_{\mu \nu}\tilde{F}^{\mu \nu}$ 
of electromagnetism, the effect being to rotate the polarisation state 
of radiation from distant source. 
Such an effect, although well constrained today, could conceivably 
be used as a way of detecting a cosmolgical scalar field. 

More recently the possibility of the axion providing the dark energy 
has been further developed by Kim and Nilles \cite{Kim:2002tq}, 
as well as Hall and collaborators \cite{Chacko:2004ky,Hall:2005xb,Barbieri:2005gj} 
and Hung \cite{Hung:2005ft}. 
In Ref.~\cite{Kim:2002tq}, the authors consider the model independent axion 
present in string theory, which has a decay constant of order the Planck scale. 
They propose the ``quintaxion'' as the dark energy candidate field, 
the field being made of a linear combination of two axions through 
the hidden sector supergravity breaking. The light cold dark matter axion 
solves the strong CP problem with decay constant determined through 
a hidden sector squark condensation ($F_a \sim 10^{12}$ GeV), and 
the quintaxion with a decay const  as expected for model independent 
axion of string theory ($F_q \sim 10^{18}$ GeV). 
For suitable ranges of couplings, they argue that the potential for 
the quintaxion is responsible for the observed vacuum energy of (0.003 eV)$^4$, 
which remains very flat, because of the smallness of the hidden sector 
quark masses. Hence it is ideal for Quintessence with the Quintessence mass 
protected through the existence of the global symmetry associated with the 
pseudo Nambu-Goldstone boson. 

In Ref.~\cite{Hall:2005xb}, the authors consider an axion model 
which leads to a time dependent equation of state parameter $w(z)$ 
for the Quintessence field. 
As before, the small mass scale is protected against radiative corrections. 
The novel feature they introduce is the seesaw mechanism, 
which allows for two natural scales to play a vital role in determining 
all the other fundamental scales. These are the weak scale, $v$, 
and the Planck scale, $M_{\rm pl}$. 
For example, the dark energy density $\rho_{\rm DE}^{1/4} 
\propto v^2/M_{\rm pl}$,  and the radiatively stable mass 
$m_\phi \propto v^4/M_{\rm pl}^3$. 
Adopting a cosine quintessence potential  they construct an explicit 
hidden axion model, and find a distinctive form for the equation of state $w(z)$. 
The dark energy resides in the potential of the axion field which is generated 
by a new QCD-like force that gets strong at the scale 
$\Lambda \approx v^2/M_{\rm pl} \approx \rho_{\rm DE}^{1/4}$. 
The evolution rate is given by a second seesaw 
that leads to the axion mass, $m_\phi \approx \Lambda^2/f$, 
with $f \approx M_{\rm pl}$.

Many particle physicists believe that the best route to find quintessence 
will be through the axion, and so we can expect much more progress 
in this area over the next few years. 

\subsection{Quintessential inflation}

We now turn our attention  to the case of quintessential
inflation, first developed by Peebles and Vilenkin \cite{Peebles:1998qn} 
(see also Ref.~\cite{Spokoiny:1993kt} for an early example 
which includes some of the features). 
One of the major drawbacks often used to attack models of quintessence
is that it introduces yet another weakly interacting scalar field.
Why can't we use one of those scalars already ``existing'' in
cosmology, to act as the quintessence field? 

This is precisely what Peebles and Vilenkin set about doing (see also 
Ref.~\cite{quininf2}). They
introduced a potential for the field $\phi$ which allowed it to
play the role of the inflaton in the early Universe and later to
play the role of the quintessence field. To do this it is
important that the potential does not have a minimum in which the
inflaton field would completely decay at the end of the initial
period of inflation. They proposed the following potential 
\begin{eqnarray}
V(\phi) &=& \lambda (\phi^4 + M^4)~~~~{\rm for}~\phi<0\,, 
\nonumber \\
&=& \frac{\lambda M^4}{1 + (\phi/M)^\alpha}~~~~
{\rm for}~\phi \ge 0\,.
\end{eqnarray} 

For $\phi <0$ we have ordinary chaotic inflation. 
Much later on, for $\phi > 0$ the universe once again begins to
inflate but this time at the lower energy scale associated with
quintessence. Needless to say quintessential inflation also
requires a degree of fine tuning, in fact perhaps even more than
before as there are no tracker solutions we can rely on for the
initial conditions. The initial period of inflation must produce 
the observed density perturbations, which constrains the coupling to be 
of order $\lambda \sim 10^{-13}$ \cite{LLbook}. 
Demanding that $\Omega_{\phi}^{(0)} \sim 0.7$, we can constrain 
the parameter space of ($\alpha$, $M$). For example,
for $\alpha =4$, we have 
$M \sim 10^{5}$ {\rm GeV} \cite{Peebles:1998qn}.  
Reheating after inflation should have proceeded via gravitational particle 
production (see \cite{Spokoiny:1993kt} for an early example of 
its effect in ending inflation) because of the absence of 
the potential minimum, but 
this mechanism is very inefficient .
However this problem may be alleviated \cite{FKL1} in the instant 
preheating scenario \cite{FKL2}
in the presence of an interaction $(1/2)g^2\phi^2 \chi^2$ between 
the inflaton $\phi$ and another field $\chi$. Of note 
in the quintessential inflation model is that one gets 
a kinetic phase (driven by the kinetic energy of the field) 
before entering the radiation phase. 
This has the effect of  changing the density of 
primordial gravitational waves \cite{Peebles:1998qn,Riazuelo:2000fc}.

An interesting proposal making use of the protected axion as the quintessence 
field has been made in Ref.~\cite{Rosenfeld:2005mt}. One of the problems 
facing the models just described is that the potentials are simply 
constructed to solve the problem at hand, namely to give two periods 
of inflation at early and late times. As such, they are generally 
non-renormalisable. The authors of Ref.~\cite{Rosenfeld:2005mt} 
introduce a renormalizable complex scalar field potential as 
the Quintessential inflation field. They suggest using a complex 
scalar field  with a global $U(1)_{PQ}$ symmetry which is 
spontaneously broken at a high energy scale. This then generates 
a flat potential for the imaginary part of the field (``axion"), 
which is then lifted (explicitly broken) by small instanton 
effects at a much lower energy. In this sense it combines the 
original idea of Natural Inflation \cite{NatInf} and the 
more recent idea of using a pseudo-Nambu Goldstone 
boson for the Quintessence field \cite{Frieman,Hill-Quint}.  
The result is that the model can give both early universe inflation 
(real part of scalar field) and late time inflation 
(imaginary part of the scalar field). 
We also note that there is an interesting 
quintessential inflation model by Dimopoulos \cite{Dimoqu} 
that allows inflation to occur at
lower energy scales than GUT in the context of the curvaton 
mechanism.

Complex scalar fields have also been introduced in the context 
of quintessence models called ``spintessence" \cite{spin1,spin2}. 
In these models the usefulness of the complex nature manifests 
itself in that the model allows for a unified description of 
both  dark matter and dynamical dark energy. 
The field $\phi$ is spinning in a $U(1)$-symmetric potential 
$V(\phi)=V(|\phi|)$, such that as the Universe expands, 
the field spirals slowly toward the origin. 
It has  internal angular momentum which helps drive
the cosmic evolution and fluctuations of the field. 
Depending on the nature of the spin, and the form of the potential, 
the net equation of state for the system can model either 
that of an evolving  dark energy component or 
self-interacting, fuzzy cold dark 
matter \cite{spin2} (see also Ref.~\cite{Kasuya}). 

One of the main difficulties for the realistic construction of quintessential 
inflation is that we need a flat potential during inflation but also 
require a steep potential during the radiation and matter dominated periods. 
The above mentioned axionic models provide one way to guarantee 
that can happen. The possibility that a pseudo-Nambu-Goldstone boson 
could arise in the bulk in a higher dimensional theory was 
investigated in Ref.~\cite{Burgess:2003bi}.  
Another route is through quintessential inflation \cite{steep} 
in braneworld scenarios \cite{Randall}.
Because of the modification of the Friedmann equation in 
braneworlds \cite{Df00,Shiromizu}
($H^2 \propto \rho^2$), it is possible to obtain inflationary solutions
even in the case of a steep exponential potential.
Although the ratio of tensor perturbations to scalar perturbations is large 
and the exponential potential is outside the $2\sigma$ observational 
bound \cite{branecon}, the model can be allowed if
a Gauss-Bonnet term is present in the five dimensional bulk \cite{TSM04}. 
We finish this section with the observation that in Ref.~\cite{DeFelice:2002ir}, 
the authors proposed a mechanism to generate the baryon asymmetry 
of the universe in a class of quintessential inflation models.

\section{Coupled dark energy}
\label{cdenergy}

The  possibility of a scalar field $\phi$ coupled to a matter and 
its cosmological consequences were discussed in Refs.~\cite{Ellis}.
Amendola later proposed a quintessence scenario coupled with dark matter 
\cite{coupled1} as an extension of nonminimally coupled 
theories \cite{Luca}. A related approach in which the 
dark matter and dark energy interact with each other 
exchanging energy has been proposed by Szydlowski 
in \cite{Szydlowski:2005ph} and a method of testing for it 
has been developed in \cite{Szydlowski:2005kv}. 
He is able to show that the cubic correction to the Hubble law, 
as measured by distant supernovae type Ia, can probe this 
interaction, and by considering flat decaying $\Lambda (t)$ 
FRW cosmologies, he argues for the possibility of measuring 
the energy transfer through determination of the cubic and 
higher corrections to Hubble's law.

An interesting aspect of the coupled dark energy scenario \cite{coupled1} 
is that the system can approach scaling solutions (characterized by 
$\Omega_{\phi} \simeq 0.7$) with an associated accelerated expansion.

Earlier in Sec.~\ref{scalingsec} we presented a coupling 
$Q$ between dark energy 
and a barotropic fluid.
This is actually the same coupling studied
in Refs.~\cite{coupled1,Luca}, and in order to show an example of this, 
let us consider the following 4-dimensional 
Lagrangian density with a scalar field $\vp$
and a barotropic perfect fluid:
\ba
\label{stlag}
\tilde{{\cal L}}=\frac12  F(\vp) \tilde{R}-
\frac{1}{2}\zeta(\vp) (\tilde{\nabla} 
\vp)^2-U(\vp)-\tilde{{\cal L}}_m\,,
\ea
where $F(\vp)$, $\zeta(\vp)$ and $U(\vp)$ are 
the functions of $\vp$. 
This includes a wide variety of gravity models--such 
as Brans-Dicke theories, non-minimally coupled 
scalar fields and dilaton gravity.
In fact a number of authors have studied quintessence 
scenarios with a nonminimally coupled scalar 
field \cite{Uzan,nonqui}.
This is related to coupled quintessence 
scenario as we will see below.
We set $\kappa^2=1$ in this section.

After a conformal transformation $g_{\mu\nu}=
F(\vp)\tilde{g}_{\mu\nu}$, the above action 
reduces to that of the Einstein frame:
\ba 
\label{elag}
{\cal L}=\frac{1}{2}R-
\frac12(\nabla \phi)^2-V(\phi)
-{\cal L}_{m}(\phi)\,,
\ea
where 
\ba 
\label{phire}
\phi \equiv \int G(\vp)\rd \vp\,,~~~
G(\vp) \equiv \sqrt{\frac32
\left(\frac{F_{,\vp}}{F}\right)^2+\frac{\zeta}{F}}\,,
\ea
and $F_{,\vp} \equiv {\rm d}F/{\rm d}\vp$.
We note that several quantities in the Einstein frame are
related to those in the Jordan frame via
$a=\sqrt{F}\tilde{a}$, ${\rm d}t=\sqrt{F}
{\rm d} \tilde{t}$, $\rho_m=\tilde{\rho}_m/F^2$, 
$p_m=\tilde{p}_m/F^2$ and $V=U/F^2$. 

In the Jordan frame the energy density $\tilde{\rho}_m$
obeys the continuity equation 
${\rm d}\tilde{\rho}_{m}/{\rm d} \tilde{t}+
3\tilde{H}(\tilde{\rho}_{m}+\tilde{p}_{m})=0$.
By rewriting this equation in terms of the quantities 
in the Einstein frame, we find 
\ba 
\dot{\rho}_m+3H(\rho_m+p_m)=
-\frac{F_{,\vp}}{2FG}(\rho_m-3p_m)\dot{\phi}\,.
\ea
In the case of cold dark matter ($p_m=0$) this
corresponds to Eq.~(\ref{geneeq2}) with a coupling
\ba 
\label{coup}
Q(\vp)=-\frac{F_{,\vp}}{2FG}=
-\frac{F_\vp}{2F} \left[\frac32
\left(\frac{F_\vp}{F}\right)^2+
\frac{\zeta}{F}\right]^{-1/2}\,.
\ea
For example a nonminimally coupled scalar field with a 
coupling $\xi$ corresponds to
$F(\vp)=1-\xi \vp^2$ and $\zeta(\vp)=1$. This gives
\ba 
\label{Qxi}
Q(\vp)=\frac{\xi \vp}{[1-\xi \vp^2(1-6\xi)]^{1/2}}\,.
\ea
In the limit $|\xi| \to \infty$ the coupling approaches 
a constant value $Q(\vp) \to \pm 1/\sqrt{6}$.

Thus a nonminimally coupled scalar field natually leads
to the coupling between dark energy and a barotropic fluid.
In what follows we will derive critical points and study 
their stabilities in a coupled Quintessence scenario
based on the coupling which appears on the RHS of 
Eqs.~(\ref{geneeq1}) and (\ref{geneeq2}). 

Before we investigate in detail the nature of coupled 
dark energy, it is worth mentioning a couple of  
important points that have been emphasised 
in \cite{Das:2005yj} and \cite{Manera:2005ct}. 
In \cite{Das:2005yj} it is pointed out that if 
there is an interaction between dark matter and 
dark energy then this will generically result 
in an effective dark energy equation of state 
of $w<-1$, arising because the interaction 
alters the redshift-dependence of the matter density. 
Therefore an observer who fits the data treating the 
dark matter as non-interacting will infer an effective 
dark energy fluid with $w<-1$. 
The authors go on to argue that the coupled 
dark energy model is consistent with all current 
observations, the tightest constraint coming 
from estimates of the matter density at different 
redshifts. 

In \cite{Manera:2005ct} it is shown that cluster 
number counts can be used to test dark energy 
models where the dark energy candidates 
are coupled to dark matter. Increasing the
coupling reduces significantly the cluster 
number counts, whereas dark energy inhomogeneities 
increase cluster abundances. 
Of possible significance is the fact that wiggles 
in cluster number counts are shown to be a 
specific signature of coupled dark energy models. 
Such oscillations could possibly be detected in future 
experiments, allowing us to discriminate among 
the different dark energy models.

\subsection{Critical points for coupled Quintessence}

We now consider a coupled Quintessence scenario
in Einstein gravity with an exponential potential i.e., 
\ba 
p(X, \phi)=\epsilon X-ce^{-\lambda \phi}\,.
\ea
Here we allow the possibility of a phantom field ($\epsilon<0$).
As we have already shown, exponential potentials possess scaling 
solutions. In fact the above Lagrangian density corresponds 
to the choice $g(Y)=\epsilon- c/Y$ and $n=1$
in Eq.~(\ref{scap}).

The autonomous equations for a general function of $g(Y)$
are given by Eqs.~(\ref{dxsc}) and (\ref{dysc}). 
Substituting $g(Y)=\epsilon- c/Y$ with $Y=x^2/y^2$
for Eqs.~(\ref{dxsc}) and (\ref{dysc}), we obtain the 
following differential equations for $x=\dot{\phi}/(\sqrt{6}H)$
and $y=e^{-\lambda \phi}/(\sqrt{3}H)$:
\ba 
\label{dxnor}
\frac{\d x}{\d N} &=&   
-3x+\frac{\sqrt{6}}{2} \epsilon \lambda c y^2+\frac32
x [(1-w_m)\epsilon x^2 \nonumber \\
& &+(1+w_m)(1-c y^2)] 
-\frac{\sqrt{6}Q}{2}\epsilon (1-\epsilon x^2 -c y^2)\,, 
\nonumber \\
\\
\frac{\d y}{\d N} &=&   
-\frac{\sqrt{6}}{2}\lambda xy+\frac32 y
 [(1-w_m)\epsilon x^2 \nonumber \\
& & +(1+w_m)(1-c y^2)]\,.
 \label{dynor} 
\ea
When $Q=0$ and $c=1$ these equations coincide with 
Eqs.~(\ref{autoquin1}) and (\ref{autoquin2}).
We note that $w_{\phi}$, $\Omega_{\phi}$ and $w_{\rm eff}$
are derived by changing $y^2$ to $cy^2$
in Eqs.~(\ref{wphiquin}), (\ref{Omephiquin}) 
and (\ref{weffquin}). 

The fixed points for the above system can be obtained 
by setting $\d x/\d N=0$ and $\d y/\d N=0$.
We present the fixed points in Table \ref{critquin}.

\begin{table*}[t]
\begin{center}
\begin{tabular}{|c|c|c|c|c|c|}
Name & $x$ & $y$ & $\Omega_\phi$ & 
$w_\phi$ & $w_{\rm eff}$  \\
\hline
\hline
(a) & $-\frac{\sqrt{6}Q}{3\epsilon (1-w_m)}$ & 0 & 
$\frac{2Q^2}{3\epsilon (1-w_m)}$ &  1 &
$w_m+\frac{2Q^2}{3\epsilon (1-w_m)}$  \\
\hline
(b1) & $\frac{1}{\sqrt{\epsilon}}$ & 0 & 1 & 1 &
1 \\
\hline
(b2) & $-\frac{1}{\sqrt{\epsilon}}$ & 0 & 1 & 1 & 
1 \\
\hline
(c) & $\frac{\epsilon \lambda}{\sqrt{6}}$ & 
$[\frac{1}{c}(1-\frac{\epsilon \lambda^2}{6})]^{1/2}$
 & 1 & $-1+\frac{\epsilon \lambda^2}{3}$ &
 $-1+\frac{\epsilon \lambda^2}{3}$ \\
\hline
(d) & $\frac{\sqrt{6}(1+w_m)}{2(\lambda+Q)}$ & 
$[\frac{2Q(\lambda+Q)+3\epsilon (1-w_m^2)}
{2c(\lambda+Q)^2}]^{1/2}$ & 
$\frac{Q(\lambda+Q)+3\epsilon (1+w_m)}{(\lambda+Q)^2}$ & 
$\frac{-Q(\lambda+Q)+3\epsilon w_m (1+w_m)}
{Q(\lambda+Q)+3\epsilon (1+w_m)}$ &
$\frac{w_m \lambda -Q}{\lambda +Q}$ \\
\hline
\end{tabular}
\end{center}
\caption[critquin]
{\label{critquin} The critical points for the ordinary 
(phantom) scalar field with an exponential potential 
in the presence of the coupling $Q$.
The points (b1) and (b2) do not exist 
for the phantom field.}
\end{table*}

\begin{itemize}

\item (i) Ordinary field ($\epsilon=+1$)

The point (a) gives some fraction of the field energy density
for $Q \neq 0$. However this does not provide 
an accelerated expansion, 
since the effective equation of state $w_{\rm eff}$ is positive 
for $0 \le w_m<1$. 
The points (b1) and (b2) are kinetically driven
solutions with $\Omega_\phi=1$ and do not satisfy the 
condition $w_{\rm eff}<-1/3$.
The point (c) is a scalar-field dominating solution 
($\Omega_\phi=1$), which gives an acceleration of the universe
for $\lambda^2<2$. The point (d) corresponds to the cosmological 
scaling solution, which satisfies $w_\phi=w_m$ for $Q=0$. 
When $Q \neq 0$
the accelerated expansion occurs for $Q>\lambda (1+3w_m)/2$.
The points (b1), (b2) and (c) exist irrespective of 
the presence of the coupling $Q$.

\item (ii) Phantom field ($\epsilon=-1$)

The point (a) corresponds to an unrealistic situation
because of the condition $\Omega_\phi<0$ for $0 \le w_m<1$.
The critical points (b1) and (b2) do not exist for the phantom field.
Since $w_{\rm eff}=-1-\lambda^2/3<-1$ for the point (c), the universe
accelerates independent of the values of $\lambda$ and $Q$.
The point (d) gives an accelerated expansion for 
$Q>\lambda (1+3w_m)/2$, and is similar to the case of  
a normal field.

\end{itemize}

\subsection{Stability of critical points}

We shall study the stability around the fixed points.
The eigenvalues of the matrix ${\cal M}$ for the perturbations $\delta x$
and $\delta y$ in Eq.~(\ref{matM}) are \cite{GNST}

\begin{itemize}
    
\item Point (a):  
\ba
\hspace*{-0.5em}& &\mu_1=-\frac32 
(1-w_m)+\frac{Q^2}{\epsilon(1-w _m)}\,, \nonumber \\
\hspace*{-0.5em}& &\mu_2=\frac{1}{\epsilon (1-w_m)}
\left[ Q(\lambda+Q)+\frac{3\epsilon}{2}
(1-w_m^2) \right].
\ea

\item Point (b1): 
\ba
\mu_1=3-\frac{\sqrt{6}}{2}\lambda\,,~~~
\mu_2=3(1-w_m)+\sqrt{6}Q\,.
\ea

\item Point (b2): 
\ba
\mu_1=3+\frac{\sqrt{6}}{2}\lambda\,,~~~
\mu_2=3(1-w_m)-\sqrt{6}Q\,.
\ea

\item Point (c): 
\ba
\mu_1=\frac12 (\epsilon \lambda^2-6)\,,~~
\mu_2= \epsilon \lambda (\lambda+Q) -3(1+w_m).
\ea

\item Point (d): 
\ba
\mu_{1, 2}=-\frac{3\{\lambda(1-w_m)+2Q\}}
{4(\lambda+Q)} [1 \pm \sqrt{1+f(\lambda, Q)}]\,,
\ea
where
\ba
f=\frac{8[3(1+w_m)-\epsilon
\lambda (\lambda+Q)][3\epsilon (1-w_m^2)+2Q(\lambda+Q)]}
{3\{\lambda (1-w_m)+2Q\}^2}. \nonumber \\
\ea

\end {itemize}

\subsubsection{Ordinary field $(\epsilon=+1)$}

We first study the dynamics of an ordinary scalar field
in the presence of a fluid with 
an equation of state: $0 \le w_m<1$.
We shall consider the case of $Q>0$ and 
$\lambda>0$ for simplicity, but it is easy to 
extend the analysis to other cases.

\begin{table*}[t]
\begin{center}
\begin{tabular}{|c|c|c|c|}
Name & Stability & Acceleration & Existence \\
\hline
\hline
(a) & Saddle point for $Q < (3/2)^{1/2}(1-w_m)$  & No 
& $Q<(3/2)^{1/2}(1-w_m)^{1/2}$ 
\\
&  Unstable node for $Q > (3/2)^{1/2}(1-w_m)$ & &  \\
\hline
(b1) & Saddle point for $\lambda >\sqrt{6}$  & No &
All values  \\
&  Unstable node for $\lambda < \sqrt{6}$ & &\\
\hline
(b2) & Saddle point for $Q > (3/2)^{1/2}(1-w_m)$  & No & 
All values \\
&  Unstable node for $Q < (3/2)^{1/2} (1-w_m)$ & &\\
\hline
(c) & Saddle point for $([Q^2+12(1+w_m)]^{1/2}-Q)/2<\lambda<\sqrt{6}$  
& $\lambda<\sqrt{2}$ & $\lambda<\sqrt{6}$    \\
&  Stable node for $\lambda<([Q^2+12(1+w_m)]^{1/2}-Q)/2$ & &\\
\hline
(d) & Stable node for $3(1+w_m)/\lambda-\lambda<Q<Q_*$  
& $Q>\lambda (1+3w_m)/2$ & $Q>3(1+w_m)/\lambda-\lambda$  \\
&  Stable spiral for $Q>Q_*$ & &
\\
\hline
\end{tabular}
\end{center}
\caption[crit2]{\label{crit2} The conditions for stability,  acceleration
and existence for an ordinary scalar field ($\epsilon=+1$).
We consider the situation with positive values of $Q$ and $\lambda$.
Here $Q_*$ is the solution of Eq.~(\ref{Qstar}).
}
\end{table*}

\begin{itemize}

\item Point (a): 

For the point (a) $\mu_1$ is negative
if $Q < \sqrt{3/2}(1-w_m)$ and positive otherwise. 
Meanwhile $\mu_2$ is positive for any value of $Q$ and 
$\lambda$. Therefore this  
is a saddle point for $Q < \sqrt{3/2}(1-w_m)$
and an unstable node for $Q > \sqrt{3/2}(1-w_m)$.
We obtain the condition $Q<\sqrt{(3/2)(1-w_m)}$
from the requirement  $\Omega_\phi<1$.
Hence the point (a) is a saddle point for $w_m=0$
under this condition.

\item Point (b1): 

While $\mu_2$ is always positive, $\mu_1$ is 
negative if $\lambda>\sqrt{6}$ and positive otherwise.
Then (b1) is a saddle point for $\lambda>\sqrt{6}$ and 
an unstable node for $\lambda<\sqrt{6}$.

\item Point (b2): 

Since $\mu_1$ is always positive and $\mu_2$
is negative for $Q > (3/2)^{1/2}(1-w_m)$ and positive otherwise, 
the point (c) is either a saddle point or an
unstable node.

\item Point (c): 

The requirement of the existence of the point (c)
gives $\lambda<\sqrt{6}$, which means that $\mu_1$
is always negative. The eigenvalue $\mu_2$ is negative
for $\lambda<(\sqrt{Q^2+12(1+w_m)}-Q)/2$ and positive otherwise.
Hence the point (c) presents a stable node for $\lambda 
<(\sqrt{Q^2+12(1+w_m)}-Q)/2$, whereas 
it is a saddle point for $(\sqrt{Q^2+12(1+w_m)}-Q)/2<\lambda<\sqrt{6}$.

\item Point (d): 

We first find that $-3\{\lambda(1-w_m)+2Q\}/4(\lambda+Q)<0$ 
in the expression of $\mu_1$ and $\mu_2$.
Secondly we obtain $\lambda(\lambda+Q)>3(1+w_m)$
from the condition, $\Omega_\phi<1$.
Then the point (d) corresponds to a stable node for 
$3(1+w_m)/\lambda-\lambda<Q<Q_*$ and is a stable spiral 
for $Q>Q_*$, where $Q_*$ satisfies the following relation
\begin{eqnarray}
\label{Qstar}
& &8\left[\lambda (\lambda+Q_*)-3(1+w_m)\right]
\left[2Q_*(\lambda+Q_*)+3(1-w_m^2) \right]  \nonumber \\
& &=3[\lambda (1-w_{m})+2Q_{*}]^2\,.
\end{eqnarray}
For example $Q_*=0.868$ for $\lambda=1.5$ and $w_m=0$.

\end {itemize}

The stability around the fixed points and the condition for an 
acceleration are summarized in Table \ref{crit2}.
The scaling solution (d) is always stable provided that 
$\Omega_\phi<1$, whereas the stability of the point (c) 
is dependent on the values of $\lambda$ and $Q$.
It is important to note that the eigenvalue $\mu_2$ for the 
point (c) is positive when the condition for the existence 
of the point (d) is satisfied, i.e., $\lambda(\lambda+Q)>3(1+w_m)$.
Therefore the point (c) is unstable for the parameter range of 
$Q$ and $\lambda$ in which the scaling solution 
(d) exists \cite{Tsuji06}.

\begin{figure}
\includegraphics[height=5.5in,width=4.3in]{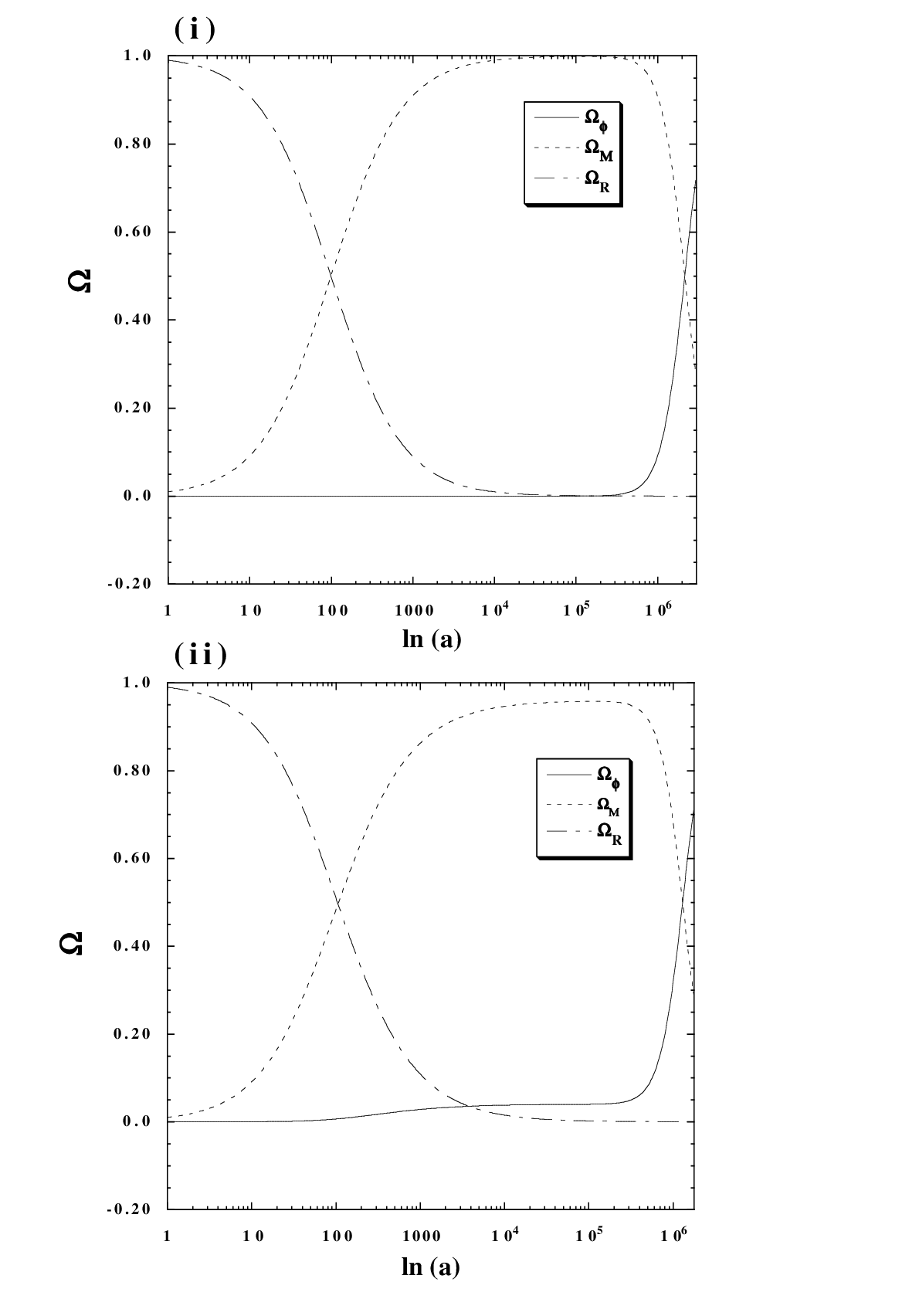}
\caption{Evolution of $\Omega_{R}$, $\Omega_M$ and $\Omega_\phi$
in a coupled quintessence scenario for 
(i) $\lambda=0.1$, $Q=0$ and (ii) $\lambda=0.1$, $Q=0.245$.
In these cases the late-time attractor is the 
scalar-field dominated fixed point (d) in Table \ref{crit2}.
In the case (ii) there exists a transient fixed point (a)
characterized by $\Omega_\phi=2Q^2/3 \simeq 0.04$
during the matter dominated era, 
whereas this behavior 
is absent in the case (i).}
\label{quincoupled} 
\end{figure}

Amendola \cite{coupled1} implemented radiation together with 
cold dark matter and the scalar field $\phi$. Unsurprisingly 
there exist more critical points in this case, but we can use the analysis 
we have just presented to describe the dynamics of 
the system once radiation becomes dynamically unimportant.
In Fig.~\ref{quincoupled} we show the evolution of
the fractional energy densities: $\Omega_R$ (radiation: $w_{m}=1/3$), 
$\Omega_{M}$ (CDM: $w_m=0$), $\Omega_{\phi}$ (dark energy)
for (i) $\lambda=0.1$, $Q=0$ and (ii) $\lambda=0.1$, $Q=0.245$.
In the absence of the coupling $Q$, $\Omega_{\phi}$
is negligibly small compared to $\Omega_M$ during the 
matter dominated era.
Meanwhile when the coupling $Q$ is present
there are some portions of the energy 
density of $\phi$ in the matter dominated era.
This corresponds to the critical point (a) characterized by 
$\Omega_\phi=2Q^2/3$ and $w_{{\rm eff}}=2Q^2/3$.
The presence of this phase (``$\phi$MDE'' \cite{coupled1}) 
can provide a distinguishable feature
for matter density perturbations, as we will see in the next section.
Since the critical point (a) is not stable, the final attractor
is either the scalar-field dominated solution (c) or 
the scaling solution (d).
Figure \ref{quincoupled} corresponds to the case in which 
the system approaches the fixed point (c).

\begin{figure}
\includegraphics[height=4.4in,width=3.3in]{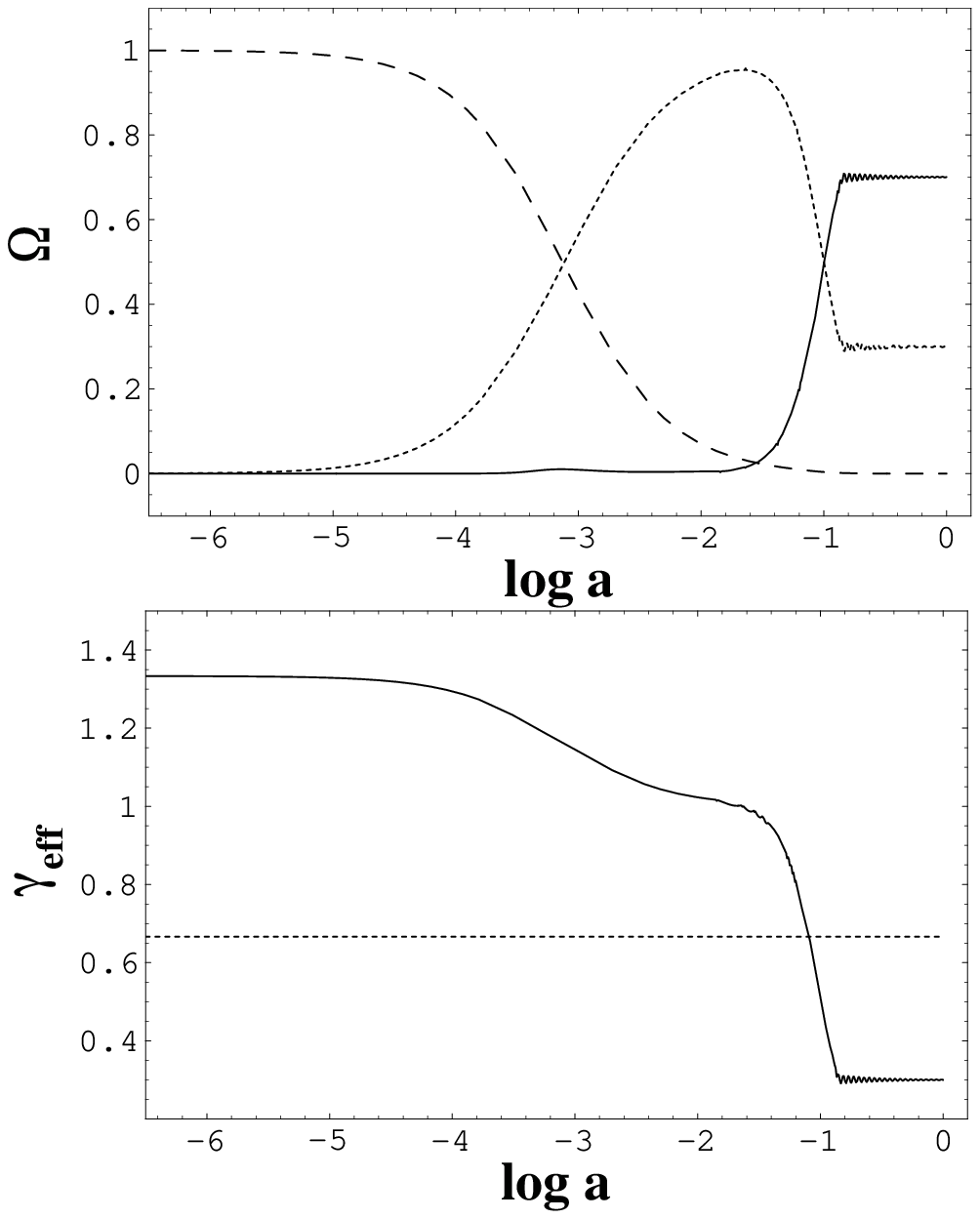}
\caption{Top panel shows the evolution of $\Omega_{R}$ (dashed), 
$\Omega_M$ (dotted) and $\Omega_\phi$ (solid) with $\lambda=30$
in the case where the coupling $Q$ changes rapidly 
from $Q_1=0$ to $Q_2=57.15$.
Bottom panel shows the evolution of the effective equation of state
$\gamma_{\rm eff}=w_{\rm eff}+1$.
It first equals $1/3$, then goes down to $1$, and approaches 
$-0.7$ with an accelerated expansion.
{}From Ref.~\cite{coupled3}.
}
\label{stationary} 
\end{figure}

The system approaches the scaling solution (d) with constant 
$\Omega_{\phi}$ provided that 
the coupling satisfies the condition $Q>3/\lambda-\lambda$.
In addition we have an accelerated expansion for $Q>\lambda/2$.
Then one can consider an interesting situation in which 
the present universe is a global attractor with $\Omega_{\phi} \simeq 0.7$.
However it was pointed out in Ref.~\cite{coupled1}
the universe soon enters the attractor phase  after the radiation 
dominated era for the coupling $Q$ satisfying the condition 
for an accelerated expansion ($Q/\lambda>(1+3w_m)/2$).
This means the absence of a matter dominated era, which is
problematic for structure formation.
It comes from the fact that the coupling $Q$ required for 
acceleration is too large to keep $\Omega_{\phi}=2Q^2/3$ small 
during the matter dominant era.

This problem is overcome if we consider a non-linear 
coupling that changes between a small $Q_1$ 
to a large $Q_2$ \cite{coupled3,tva}.
The authors in Ref.~\cite{coupled3}
adopted the following coupling: 
\ba
\label{nonlinear}
Q (\phi)=\frac{1}{2}\left[ (Q _{2}-Q_{1})\tanh \left( 
\frac{ \phi _{1}-\phi }{\Delta }\right) 
+Q_{2}+Q _{1}\right]. \nonumber \\
\ea
In order to keep $\Omega_{\phi}$ small during the matter
dominated era but to get $\Omega_{\phi} \simeq 0.7$ 
with an accelerated 
expansion, we need to impose the condition 
$Q_{1} \ll \lambda \ll Q_{2}$.
In Fig.~\ref{stationary} we plot the evolution of 
$\Omega_{R}$, $\Omega_M$ and $\Omega_\phi$ 
together with an effective equation of state 
$\gamma_{\rm eff}=w_{\rm eff}+1$
for $\lambda=30$, $Q_1=0$ and $Q_2=57.15$.
We find that there exists a matter dominated era with a small 
value of $\Omega_{\phi}$, 
which allows large-scale structure to grow.
The solution eventually approaches a stationary global attractor 
characterized by $\Omega_{\phi} \simeq 0.7$
with an accelerated expansion.

\subsubsection{Phantom field $(\epsilon=-1)$}
%

\begin{table*}[t]
\begin{center}
\begin{tabular}{|c|c|c|c|}
Name & Stability & Acceleration  & Existence \\
\hline
\hline
(a) & Saddle point for $Q(Q+\lambda) < (3/2)(1-w_m^2)$  
& $Q^2>(1-w_m)(1+3w_m)/2$ & No if the condition 
$0 \le \Omega_\vp \le 1$ 
\\
&  Stable node for $Q(Q+\lambda) > (3/2)(1-w_m^2)$ & &
is imposed
\\
\hline
(c) & Stable node & All values & All values  \\
\hline
(d) & Saddle & Acceleration for $Q>\lambda (1+3w_m)/2$ & 
$Q(Q+\lambda)>(3/2)(1-w_m^2)$ 
\\
\hline
\end{tabular}
\end{center}
\caption[crit3]{\label{crit3} 
The conditions for stability \& acceleration
\& existence for a phantom scalar field ($\epsilon=-1$).
We consider the situation with positive values of $Q$ and $\lambda$.
}
\end{table*}

The fixed points (b) and (c) do not exist for the phantom field.

\begin{itemize}

\item Point (a): 

In this case $ \mu _1$ is always negative, 
whereas $\mu _2$ can be either positive
or negative depending on the values of $Q$ and $\lambda$.
Then this point is a saddle for 
$Q(Q+\lambda) < (3/2)(1-w_m^2)$ and a stable node 
for $Q(Q+\lambda) >(3/2)(1-w_m^2)$.
However, since $\Omega _\phi=-2Q^2/3(1-w_m)<0$ for 
$0 \le w_m<1$, 
the fixed point (a) is not realistic.

\item Point (c): 

Since both $\mu_1$ and $\mu_2$ are negative 
independent of the values of $\lambda$ and $Q$, 
the point (c) is a stable node.

\item Point (d): 

{}From the condition $y^2>0$,
we require that $2Q(Q+\lambda)>3(1-w_m^2)$ for the 
existence of the critical point (d).
Under this condition we find that $\mu_1<0$
and $\mu_2>0$. 
Therefore the point (d) corresponds to a saddle point.

\end{itemize}

The properties of critical points are summarized 
in Table \ref{crit3}.
The scaling solution always becomes unstable for phantom fields.
Therefore one can not construct a coupled dark energy 
scenario in which the present value of $\Omega_\phi$ ($\simeq 0.7$)
is a late-time attractor. 
This property is different from the case of an ordinary field
in which scaling solutions can be stable fixed points.
The only viable stable attractor for phantom fields is 
the fixed point (c), giving the dark energy dominated universe 
($\Omega_\phi=1$) with an equation of state 
$w_\phi=-1-\lambda^2/3<-1$.

\subsection{General properties of fixed points}

In the previous subsections we have considered the case of 
a minimally coupled scalar field with an exponential 
potential. This coupled quintessence scenario can be applied
to other scalar-field dark energy models such as tachyon and 
dilatonic ghost condensate.
For the dark energy models that possess scaling solutions, the 
procedure to derive fixed points is very simple.
The functional form $g(Y)$ is determined by specifying the model.
Then plugging this into Eqs.~(\ref{dxsc}) and (\ref{dysc}), 
we obtain the fixed points for the system.
The stability of fixed points is known by evaluating the 
eigenvalues of the matrix ${\cal M}$.

In fact we can study the stability of fixed points relevant to 
dark energy for the scalar-field models which possess
scaling solutions without specifying any form of $g(Y)$.
{}From Eqs.~(\ref{dxsc}) and (\ref{dysc}) 
the fixed points we are interested in ($y \neq 0$)
satisfy the following equations:
\begin{eqnarray}
\label{basic1}    
& & \sqrt{6} \lambda x=
3\left[1+gx^2-w_{m} (\Omega_{\vp}-1)\right]\,, \\
\label{basic2}  
& & \sqrt{6}(g+Yg')x=(Q+\lambda)
\Omega_{\vp}-Q\,.
\end{eqnarray}
Since $g+Yg'=\Omega_{\vp}(1+w_{\vp})/2x^2$
from Eq.~(\ref{Omerelation}), 
Eqs.~(\ref{basic1}) and (\ref{basic2})
can be written in the form:
\begin{eqnarray}
\label{xexpress}    
x&=& \frac{\sqrt{6}[1+(w_\vp- w_{m})\Omega_{\vp}
+w_{m}]}{2\lambda} \\
&=& \frac{\sqrt{6}(1+w_{\vp})\Omega_{\vp}}
{2[(Q+\lambda)\Omega_{\vp}-Q]}\,,
\end{eqnarray}
This leads to 
\begin{eqnarray}
(\Omega_{\vp}-1) \left[ (w_\vp- w_{m})
(Q+\lambda)\Omega_{\vp}+Q(1+w_{m})\right]=0.
\nonumber \\
\end{eqnarray}

Hence we obtain 
\begin{itemize}
\item (i)
A scalar-field dominant solution with 
\begin{eqnarray}
\label{fixedi}    
\Omega_{\vp}=1\,.
\end{eqnarray}
\item (ii)
A scaling solution with 
\begin{eqnarray}
\label{scaome}    
\Omega_{\vp}=
\frac{(1+w_m)Q}{(w_{m}-w_{\vp})(Q+\lambda)}\,.
\end{eqnarray}
\end{itemize}

In the case (i) Eqs.~(\ref{Omerelation}) and (\ref{basic1}) give
\begin{eqnarray}
\label{eqw}    
w_{\vp}=-1+\frac{\sqrt{6}\lambda x}{3}
=-1+\frac{\lambda^2}{3p_{,X}}\,.
\end{eqnarray}
In the last equality we used the relation
$x=\lambda/\sqrt{6}p_{,X}$ which is 
derived from Eq.~(\ref{basic2}).
Since the scalar-field dominates the dynamics, the 
effective equation is given by $w_{\rm eff}=w_{\phi}$.

Subsitutting Eq.~(\ref{scaome}) for Eq.~(\ref{xexpress}), 
we obtain the value of $x$ given by 
Eq.~(\ref{scalingx}). Hence the fixed point (ii) is actually 
the scaling solution.

In Ref.~\cite{Tsuji06} the stability of the fixed points
relevant to dark energy was studied without specifying 
the form of $g(Y)$. In the presence of 
non-relativistic dark matter with a non-phantom scalar field, 
the final attractor is either a scaling solution 
with constant $\Omega_\vp$ satisfying $0<\Omega_\vp<1$ or a
scalar-field dominant solution with $\Omega_\vp=1$.
Meanwhile a phantom scalar-field dominant fixed point 
($\Omega_{\vp}=1$ and $p_{,X}<0$) is always
classically stable.
Then the universe is eventually dominated by the energy 
density of a scalar field if phantom is responsible for dark energy.
See Ref.~\cite{Tsuji06} for details about 
the stability of fixed points.

\subsection{Can we have two scaling regimes ?}

In the case of Quintessence with an exponential potential
there exists a ``$\phi$MDE'' fixed point (a) presented 
in Table \ref{critquin}. Since $\Omega_{\phi}=
w_{\rm eff}=2Q^2/3$ for $w_{m}=0$, this also corresponds 
to a scaling solution if $Q$ is a constant.
As we mentioned in subsection B, it is not possible to have 
a sequence of the ``$\phi$MDE'' (a) and the accelerated scaling 
attractor (d) with $\Omega_{\phi} \simeq 0.7$.
Then a question arises. Can we have two scaling regimes
for a general class of coupled scalar field Lagrangians?

Let us consider the scaling Lagrangian (\ref{scap}) 
in Einstein gravity ($n=1$).
The $\phi$MDE is a kinetic solution which corresponds 
to $y=0$ in Eq.~(\ref{dynor}).
This point exists only if $g=g(x^{2}/y^{2})$ is non-singular, i.e., only if
one can expand $g$ in positive powers of $y^{2}/x^{2}$,
\begin{eqnarray}
    g=c_{0}+\sum_{n>0}c_{n}\left(\frac{y^{2}}{x^{2}}\right)^{n}\,.
    \label{eq:poly2}
\end{eqnarray}
In this case Eq.~(\ref{dxnor}) is given by
\begin{eqnarray}
\frac{{\textrm{d}}x}{{\textrm{d}}N}=\frac{1}{2}\left(3c_{0}x+\sqrt{6}Q\right)
\left(x^{2}-\frac{1}{c_{0}}\right)=0\,.
\end{eqnarray}
For $c_{0}=0$ this equation gives no real solutions. 
For $c_{0}\ne0$ we get the $\phi$MDE point 
$x=-\sqrt{6}Q/3c_{0}$ together with pure kinematic 
solutions $x=\pm 1/\sqrt{c_{0}}$ (which exists for 
$c_{0}>0$).

In Ref.~\cite{AQTW} it was shown that a sequence of the 
$\phi$MDE and the scaling attractor is not realized for the model
(\ref{eq:poly2}) if $n$ are {\it integers}.
The main reason is that there exist two singularities at $x=0$ and 
$A^{-1}=0$, where $A=g+5Yg'+2Y^2g''=\rho_{,X}^{-1}$ is 
related to a sound speed via $c_{s}^2=Ap_{,X}$.
For the fractional Lagrangian 
\begin{eqnarray}
\label{fract}
g(Y)=c_{0}-cY^{-u}\,,
\end{eqnarray}
we have 
\begin{equation}\label{eq:xsingularity}
    \bigg|\frac{\textrm{d}y/\textrm{d}N}
    {\textrm{d}x/\textrm{d}N}\bigg|_{x\rightarrow0}
    \rightarrow \infty\,, \quad (u \neq 1).
\end{equation}
Thus the solutions can not pass the line $x=0$
except for $u=1$ (the case of Quintessence with an 
exponential potential already excluded).
When $c_0>0$ this singularity is inevitable to be hit when the 
solutions move from the $\phi$MDE to the accelerated 
scaling solution \cite{AQTW}.
When $c_0<0$ one needs to cross either the singularity at 
$A^{-1}=0$ or at $A=0$ ($x=0$), but both cases are 
forbidden.

There is an interesting case in which a sequence of 
the nearly matter dominated era and the accelerated scaling 
solution can be realized. This is the model (\ref{fract})
with $0<u<1$. Although the phase space is separated into
positive and negative abscissa subspaces because of the 
condition (\ref{eq:xsingularity}), one can  have a matter-dominated 
era in the region $x>0$ followed 
by the scaling attractor with $x>0$ (when $Q$ is positive).
It would be of interest to investigate whether this special 
case satisfies observational constraints.

\subsection{Varying mass neutrino scenario}

There is an interesting model called mass-varying neutrinos 
(MaVaNs) in which neutrinos are coupled to dark 
energy \cite{Hung,Fardon}.
This makes use of the fact that the scale of neutrino
mass-squared differences (0.01 {\rm eV})$^2$
is similar to the scale of dark energy ($10^{-3}$\,{\rm eV})$^4$.
According to this scenario the neutrino mass depends upon 
a scalar field called {\it acceleron}, ${\cal A}$, 
which has an instantaneous minimum that varies slowly 
as a function of the density of neutrinos.
The mass of the acceleron can be heavy relative to 
the Hubble rate unlike the case of a slowly rolling 
light scalar field (Quintessence).

The energy density of non-relativistic neutrinos is given by 
$\rho_{\nu}=n_{\nu}m_{\nu}$, where $n_{\nu}$ and 
$m_{\nu}$ are the number density and the mass 
of neutrinos, respectively.
When the acceleron field has a potential $V_0 ({\cal A})$, 
the total effetcive potential for MaVaNs is 
\begin{eqnarray}
V=n_{\nu}m_{\nu}({\cal A})+V_{0}({\cal A})\,.
\end{eqnarray}
Even if the potential $V_{0}({\cal A})$ does not have a minimum, 
the presence of the $n_{\nu}m_{\nu}({\cal A})$ term induces
an instantaneous minimum that varies with time.
Since $\partial V/\partial {\cal A}=0$ at the potential minimum, 
we obtain 
\begin{eqnarray}
n_{\nu}=-\frac{\partial V_0}{\partial m_\nu}\,,
\end{eqnarray}
if $\partial m_\nu/\partial {\cal A} \neq 0$.

Neglecting the contribution of the kinetic energy of the acceleron
field, the energy density and the pressure density of the system is 
given by $\rho \simeq n_{\nu}m_{\nu}+V_{0}$ and 
$p \simeq -V_0$. Hence the equation of state for the 
neutrino/acceleron system is 
\begin{eqnarray}
w=\frac{p}{\rho}=-1+\frac{n_{\nu}m_{\nu}}{V}\,.
\end{eqnarray}
Then we have $w \simeq -1$ provided that the energy density 
of neutrinos is negligible relative to that of the acceleron field
($n_{\nu}m_{\nu} \ll V_0$).

The cosmological consequesnces of this scenario have been studied
by a host of authors, see, e.g., Refs.~\cite{massvary,Brook}.
Since neutrinos are coupled to dark energy, it is expected that we 
may find similar cosmological evolution to the one 
obtained in previous subsections of coupled dark energy.
One difference from the discussions in previous 
subsections is that neutrinos should be described by 
a distribution function $f(x^i, p^i, t)$ in phase space 
instead of being treated by a fluid \cite{Brook}.
When neutrinos are collisionless, 
$f$ is not dependent on time $t$.
Solving a Boltzman equation we obtain 
the energy density of neutrinos, as 
\begin{eqnarray}
\label{rhonu}
\rho_{\nu}=\frac{1}{a^4} \int q^2 {\rm d}q
{\rm d}\Omega \epsilon f_0(q)\,,
\end{eqnarray}
together with the pressure density
\begin{eqnarray}
\label{pnu}
p_{\nu}=\frac{1}{3a^4} \int q^2 {\rm d}q
{\rm d}\Omega \epsilon f_0(q) \frac{q^2}
{\epsilon}\,,
\end{eqnarray}
where $f_{0}$ is a background neutrino Fermi-Dirac 
distribution function.
Here $\epsilon$ is defined by $\epsilon^2=q^2+m_{\nu}^2 a^2$,
where $q$ is the comoving momentum.

If neutrinos decouple while they are still relativistic, the phase-space
density is the function of $q$ only. When the mass of neutrinos
depends on a field $\phi$, Eqs.~(\ref{rhonu}) 
and (\ref{pnu}) give \cite{Brook}
\ba 
\label{neu}
\dot{\rho}_\nu+3H(\rho_\nu+p_\nu)=
\frac{\partial\,{\rm ln}\,m_\nu}{\partial \phi}
(\rho_\nu-3p_\nu)\dot{\phi}\,.
\ea
For non-relativistic neutrinos ($p_\nu=0$), 
comparison of Eq.~(\ref{neu}) with Eq.~(\ref{geneeq2})
shows that the coupling between neutrinos and dark energy is 
given by $Q(\phi)=\partial {\rm ln}\, m_\nu/\partial \phi$.
Hence we can apply the results of cosmological evolution 
in previous subsections to the neutrinos coupled to dark energy.
In Refs.~\cite{Brook} the effect of mass-varying neutrinos on CMB 
and LSS was studied as well as cosmological background evolution
in the case where a light scalar field (Quintessence) is coupled to 
massive neutrinos. This is somewhat different from the 
original MaVaNs scenario in which the mass of the acceleron 
field is much larger than the Hubble rate.
See Refs.~\cite{coupledpapers} for other models of 
coupled dark energy.

\subsection{Dark energy through brane-bulk energy exchange}

In \cite{Kiritsis:2002zf} the authors investigate the brane cosmological 
evolution involving a different method of energy exchange, 
this one being between the brane and the bulk, in the context 
of a non-factorizable background geometry with vanishing 
effective cosmological constant on the brane. 
A number of brane cosmologies are obtained, 
depending on the mechanism underlying the energy 
transfer, the equation of state of brane-matter and 
the spatial topology. Of particular note in their analysis is 
that accelerating eras are generic  features of their solutions. 
The driving force behind the observed cosmic acceleration 
is due to the flow of matter from the bulk to the brane.  

The observational constraints on these type of models in which 
the bulk is not empty has been explored in \cite{Umezu:2005dw}. 
Allowing for the fact that the effect of this energy exchange is 
to modify the evolution of matter fields for an observer on the 
brane the authors determine the constraints from various 
cosmological observations on the flow of matter from the 
bulk into the brane. Intriguingly they claim that a $\Lambda = 0$ 
cosmology to an observer in the brane is allowed which satisfies 
standard cosmological constraints including the CMB temperature 
fluctuations, Type Ia supernovae at high redshift, and the matter 
power spectrum. Moreover it can account for the observed 
suppression of the CMB power spectrum at low multipoles. 
The  cosmology associated with these solutions predicts that the 
present dark-matter content of the universe may be significantly 
larger than that of a $\Lambda$CDM cosmology, its influence, 
being counterbalanced by the dark-radiation term. 

This is an interesting approach to dark energy, is well motivated 
by brane dynamics and has generated quite a bit of interest over 
the past few years \cite{bulk-brane-exchange,kofinas06} and 
for a review see \cite{Kiritsis:2004vy}.

\section{Dark energy and varying alpha}
\label{valpha}

In this section we investigate a possible way to distinguish 
dynamical dark energy models from a cosmological 
constant-- through temporal variation of the effective fine
structure constant $\alpha$. This is just one aspect of the 
more general approach of allowing for the variation of 
constants in general (such as for example the dilaton, Newton's 
constant and possibly the speed of light). For a detailed overview 
of the Fundamental constants and their variation see the 
excellent review of Uzan \cite{Uzan:2002vq}\footnote{See also Ref.\cite{Yego} on the related theme.}. 

In 2001, Webb {\em et al.} \cite{Webb01} reported observational evidence 
for the change of $\alpha$ over a cosmological time between 
$z \simeq 0.5$ and $z=0$. Now, there are a number of existing constraints 
on the allowed variation of $\alpha$. For example, 
the Oklo natural fission reactor \cite{oklo} found the variation 
$\Delta \alpha/\alpha \equiv (\alpha- \alpha_0)/\alpha_0$ is constrained by 
$-0.9\times 10^{-7}< \Delta \alpha/\alpha <1.2\times 10^{-7}$
at a redshift $z \sim 0.16$ \cite{oklo} (here $\alpha_{0}$ is the 
present value of the fine structure constant).
The absorption line spectra of distance 
quasars \cite{savedov,wolf,murphy} provides another route. 
In Ref.~\cite{dzuba} it is claimed that 
$\Delta \alpha/\alpha =(-0.574\pm 0.102) \times 10^{-5}$ for $0.2<z<3.7$ .
The recent detailed analysis of high quality
quasar spectra \cite{chand} gives the lower variation 
$\Delta \alpha/\alpha=(-0.06\pm0.06)\times 10^{-5}$ over the
redshift range $0.4<z<2.3$.
Although there still remains the possibility of systematic errors
\cite{Murphy:2002ve}, should the variation of $\alpha$ hold 
up to closer scrutiny, it will have important
implications for cosmology. 

One powerful conclusion that follows from a varying $\alpha$
is the existence of massless or nearly massless fields coupled to 
gauge fields. 
Quite independently there is a need for a light scalar field
to explain the origin of dark energy.
Thus it is natural to consider that quintessence or another type of scalar field model could be
responsible for the time variation of $\alpha$.
In fact many authors have studied the change of $\alpha$ based on quintessence 
by assuming specific forms for the interaction between a field $\phi$ 
and an electromagnetic field $F_{\mu \nu}$ 
\cite{valphapapers,OP,SBM,CNP}.
Now in general the inclusion of a non-renormalizable interaction of the form 
$ B_F(\phi)F_{\mu \nu}F^{\mu \nu}$ at the quantum level requires the existence of a 
momentum cut-off $\Lambda_{{\rm UV}}$. 
Unfortunately, any particle physics 
motivated choice of $\Lambda_{{\rm UV}}$ destabilizes the 
potential of quintessence, i.e., it could induce a mass term much 
larger than the required one of order $H_0$. 
However, because the nature of this fine-tuning is similar to the 
one required for the smallness of the cosmological constant, 
we are open to proceed hoping that 
both problems could be resolved simultaneously in future.

Originally Bekenstein \cite{Bek} introduced the exponential form 
for the coupling of the scalar field to the electromagnetic field which 
in practice can always be taken in the linear form 
$B_F(\phi)= 1-\zeta \kappa \phi$.
{}From the tests of the equivalence principle the coupling is constrained 
to be $|\zeta|<10^{-3}$ \cite{OP}.
Although the existence of the coupling $\zeta$ alone is 
sufficient to lead to the variation of $\alpha$, the 
resulting change of $\alpha$ was found to be of order 
$10^{-10}$-$10^{-9}$ \cite{Bek}, 
which is too small to be compatible with observations.
This situation is improved by including a potential for the 
field $\phi$ or by introducing a coupling of order 1 between the 
field and dark matter \cite{OP,SBM}. 

In the next subsection we shall study the time variation of $\alpha$
for a minimal Bekenstein-like coupling in the presence of  
a Quintessence potential.
We then discuss the case of a Dirac-Born-Infeld dark energy model
in which the tachyon is naturally coupled to electromagnetic 
fields \cite{GST05va}. In this case we do not need an ad-hoc
assumption for the form of the coupling.

\subsection{Varying alpha from quintessence}

Let us consider an interaction between a Quintessence field $\phi$
and an electromagnetic field $F_{\mu \nu}$, whose Lagrangian 
density is given by 
\begin{equation}
{\cal L}_F (\phi) = - \frac{1}{4} B_F(\phi) 
F_{\mu\nu}F^{\mu\nu}.
\label{bclag}
\end{equation}
We shall assume a linear dependence of the coupling $B_F(\phi)$:
\begin{equation}
\label{BFphi}
B_F(\phi) = 1-\zeta\kappa (\phi-\phi_0)\,,
\end{equation}
where the subscript 0
represents the present value of the quantity.
We note that this is just one example for the form of the coupling.
The fine structure ``constant'' $\alpha$ is inversely 
proportional to $B_F(\phi)$, which can be expressed by 
$\alpha=\alpha_0/B_F(\phi)$.
Then the variation of $\alpha$ is given by 
\begin{equation}
\label{defDelta}
\frac{\Delta \alpha}{\alpha} \equiv 
\frac{\alpha-\alpha_0}{\alpha_0} =
\zeta \kappa (\phi - \phi_0)\,.
\end{equation}

If one uses the information of quasar absorption lines, 
$\Delta \alpha/\alpha \simeq -10^{-5}$, 
around $z=3$ \cite{Webb01}, 
the value of the coupling $\zeta$ can be evaluated as
\begin{equation}
\label{zetavalue}
\zeta \simeq - \frac{10^{-5}}{\kappa \phi(z=3) - 
\kappa \phi(z=0)} \,.
\end{equation}
The bound of atomic clocks is given by 
$|\dot{\alpha}/\alpha|<4.2 \times 10^{-15}\,{\rm yr}^{-1}$
at $z=0$ \cite{Marion}.
In our model the ratio of the variation of $\alpha$ 
around the present can be evaluated as
\begin{equation}
\label{dotalpha}
\frac{\dot{\alpha}}{\alpha} 
\simeq \zeta \kappa \dot{\phi}
\simeq
- \zeta \frac{\rd (\kappa \phi)}{\rd (1+z)}~ H_0\,.
\end{equation}

As an example let us consider an exponential potential 
$V(\phi)=V_0 e^{-\kappa \lambda \phi}$
in the presence of the coupling $Q$ between dark energy 
and dark matter.
The universe can reach the scaling attractor (d) in 
Table \ref{critquin} with an 
accelerated expansion.
Since $x=\sqrt{6}(1+w_{m})/2(Q+\lambda)$ in the scaling regime, 
we find
\begin{equation}
\label{kphi}
\kappa (\phi-\phi_0) = 
- \frac{3 \gamma}{\lambda+Q} \ln (1+z) \,,
\end{equation}
where $\gamma=1+w_m$ and $\phi_{0}$ is the present value
of the field.
Substituting this for Eq.~(\ref{defDelta}), one obtains 
\begin{equation}
\frac{\Delta \alpha}{\alpha}=
-\frac{3\zeta \gamma}{\lambda+Q}
{\rm ln} (1+z)\,.
\end{equation}
{}From Eq.~(\ref{zetavalue}) the coupling $\zeta$ consistent 
with quasar absorption lines is 
\begin{equation}
\label{zevalue}
\zeta = \frac{10^{-5}}{\ln(4)} 
\frac{\lambda+Q}{3 \gamma} \,.
\end{equation}
Substituting Eqs.~(\ref{kphi}) and (\ref{zevalue}) for 
Eq.~(\ref{dotalpha}), we obtain
\begin{equation}
\frac{\dot{\alpha}}{\alpha}=
\frac{10^{-5}}{{\rm ln}\,4}H_{0}
\simeq 4.8 \times 10^{-16}\,{\rm yr}^{-1}\,.
\end{equation}
This satisfies the constraint of  atomic clocks.

The constraint coming from the test of the equivalence principle 
corresponds to $|\zeta|<10^{-3}$.
When $Q=0$ this gives the upper 
bound for $\lambda$ from Eq.~(\ref{zevalue}).
If we also use the constraint (\ref{lamcondition}) coming
from nucleosynthesis, we can restrict the value of $\lambda$
to be $4.5<\lambda<415$.
Of course the uncoupled case is not viable to explain the 
accelerated expansion at late times.
If there exists another source for dark energy, this induces  
errors in the estimation of the evolution of $\alpha$ and the 
coupling $\zeta$.

The acceleration of the universe is realized in the presence 
of the coupling $Q$.
In Sec.~\ref{cdenergy} we showed that 
a matter dominated era does not last sufficiently long for 
large-scale structure to grow if the field $\phi$
is coupled to all dark matter and drives an accelerated 
expansion with a scaling attractor $\Omega_\phi \simeq 0.7$.
In order to avoid this problem
we shall assume the existence of two components 
of dark matter in which one 
is coupled to the scalar field and another is not. 
This is an alternative approach to introducing 
a non-linear coupling given in Eq.~(\ref{nonlinear}).

\begin{figure}
\includegraphics[height=2.5in,width=3.3in]{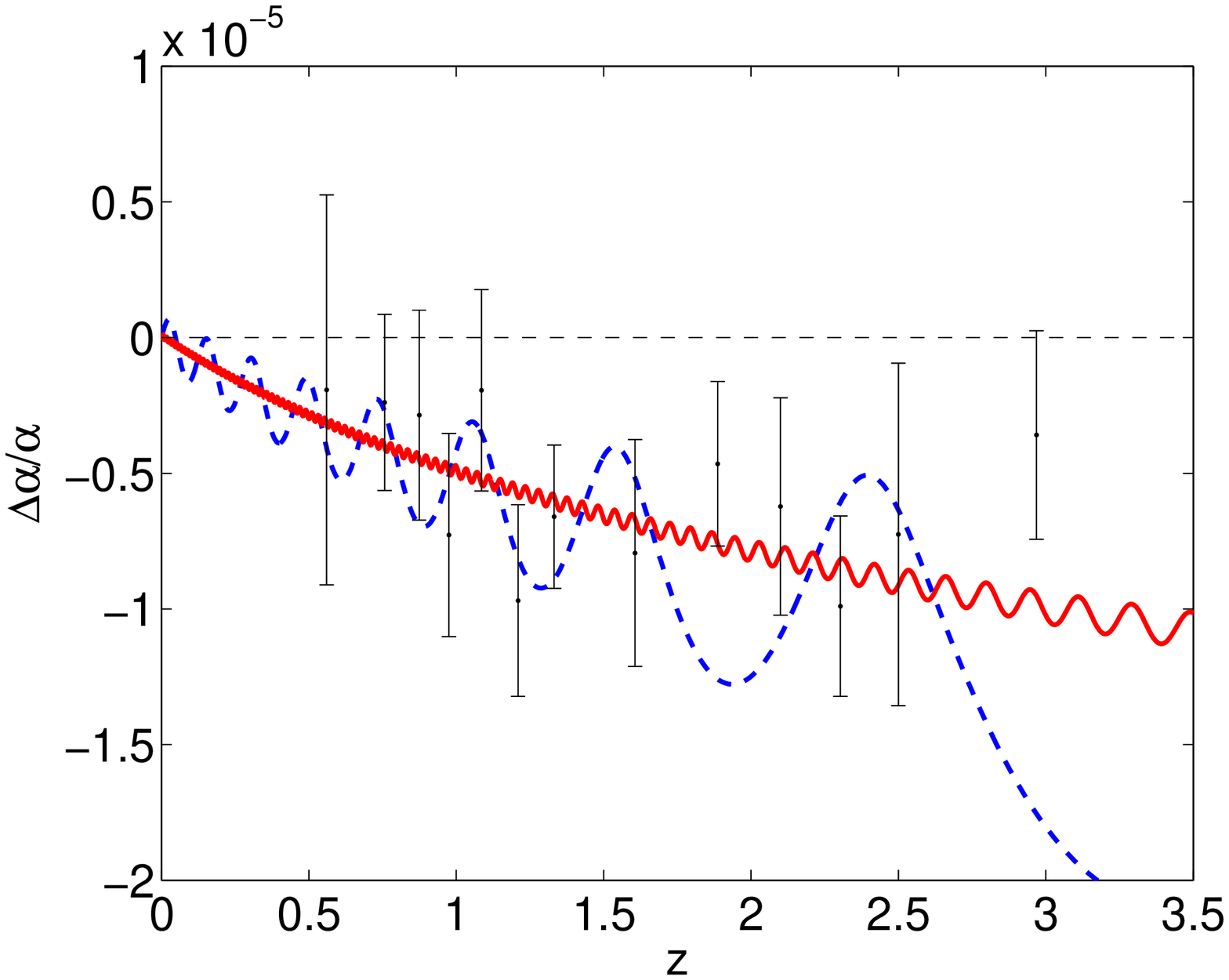}
\caption{
Evolution of $\Delta \alpha/\alpha$ for a coupled quintessence 
model with contributions $\Omega_{m,c}=0.05$
and $\Omega_{\phi}=0.7$ today. The solid and dashed curves
correspond to $\lambda=100$ and $\lambda=10$, respectively.
We also show observational data with error bars.
We thank Nelson J.~Nunes for providing us this figure.
}
\label{varyquins} 
\end{figure}

In Fig.~\ref{varyquins} we plot the evolution of 
$\Delta \alpha/\alpha$ for two different values of 
$\lambda$ when the coupled component of 
dark matter is $\Omega_{m,c}=0.05$ today.
The coupling $Q$ is determined by the condition that 
the scaling attractor corresponds to  $\Omega_{m,c}=0.05$ 
and $\Omega_{\phi}=0.7$ \cite{CNP}.
The oscillation of $\Delta \alpha/\alpha$ in Fig.~\ref{varyquins}
comes from the fact that the solution actually approaches
a scaling attractor.
For smaller values of $\lambda$ we find that the 
attractor is reached at a later stage, which leads to 
the heavy oscillation of $\Delta \alpha/\alpha$.

{}From Eq.~(\ref{zevalue}) we expect that the presence of 
the coupling $Q$ gives larger values of $\zeta$
compared to the uncoupled case.
This is actually the case even when a part of dark matter 
is coupled to the scalar field.
The bound of  the equivalence principle $|\zeta|<10^{-3}$
can be satisfied provided that we choose smaller 
values of $\lambda$ \cite{CNP}.

In Ref.~\cite{CNP} the evolution of $\Delta \alpha/\alpha$
was obtained for a number of other quintessence potentials, 
in which they can in principle be consistent with observations 
if we fine-tune model parameters.
We caution that there is a freedom to choose the coupling $B_F(\phi)$
other than the one given in Eq.~(\ref{BFphi}) and the evolution 
of $\Delta \alpha/\alpha$ crucially depends upon the choice 
of this coupling.
In Ref.~\cite{Nelsonalpha} the possibility of reconstructing dark 
energy equation of state from varying $\alpha$ was studied
for the coupling given by Eq.~(\ref{BFphi});
see also Ref.~\cite{Nelsonalphanew} on the similar theme.

\subsection{Varying alpha from tachyon fields}

The change of $\alpha$ may be explained in 
other types of scalar-field dark energy models
such as those originating from tachyon fields.
In fact a Dirac-Born-Infeld type effective 4-dimensional 
action given below naturally leads to a coupling between 
a tachyon field $\vp$ and a Maxwell tensor 
$F_{\mu \nu}$ \cite{GST05va}:
\begin{equation}
\label{actiontva}
S= -\int \rd^4x\,\wt{V}(\varphi)
\sqrt{-\det(g_{\mu\nu}+
\partial_{\mu}\varphi\partial_{\nu}
\varphi+2\pi\alpha'F_{\mu \nu})}\,,
\end{equation}
where $\wt{V}(\varphi)$ is the potential of the field.
Let us consider a situation in which a brane is located in a
ten-dimensional spacetime with a warped metric \cite{GST}
\begin{eqnarray}
\label{wmetric}
\rd s_{10}^2=\beta g_{\mu \nu}(x) \rd x^{\mu} \rd x^{\nu}+
\beta^{-1}\tilde{g}_{mn}(y) \rd y^m \rd y^n\,,
\end{eqnarray}
where $\beta$ is a warp factor. 

For this metric the action (\ref{actiontva}) is written 
in the form
\begin{eqnarray}
\label{actionmo}    
S= -\int \rd^4x\,V(\phi)
 \sqrt{-\det(g_{\mu\nu}+
 \partial_{\mu}\phi\partial_{\nu}
\phi+2\pi\alpha' \beta^{-1} F_{\mu \nu})},
\nonumber \\
\end{eqnarray}
where 
\begin{eqnarray}
\label{trans}
\phi=\varphi/\sqrt{\beta}\,,~~~
V(\phi)=\beta^2 \wt{V}(\sqrt{\beta}\phi)\,.
\end{eqnarray}
The warped metric (\ref{wmetric}) changes the mass scale on the brane
from the string mass scale $M_s=1/\sqrt{\alpha'}$
to an effective mass which is $m_{\rm eff}=\sqrt{\beta}M_s$.
The expansion of the action (\ref{actionmo}) to second order in
the gauge field, for an arbitrary metric, becomes
\begin{eqnarray}
S & \simeq & \int \rd^4x\Big[-V(\phi)\sqrt{-\det(g_{\mu \nu} +\prt_\mu
\phi\prt_\nu \phi)} \nonumber \\
& & +\frac{(2\pi\alpha')^2V(\phi)}{4\beta^2}\sqrt{-g}
\,{\rm Tr}(g^{-1}Fg^{-1}F)\Big]\,. 
\eea
We have dropped other second order terms that 
involve the derivative of the field $\phi$.
This should be justified provided that the kinetic energy of 
the field is relatively small compared to the potential energy 
of it (as it happens in the context of dark energy).

Comparing the above action with the standard Yang-Mills action,
one finds that the effective fine-structure constant $\alpha$ 
is given by 
\begin{eqnarray}
\alpha \equiv g_{\rm YM}^2
=\frac{\beta^2M_s^4}{2\pi V(\phi)}\,,
\end{eqnarray}
which depends on the field $\phi$.
The variation of $\alpha$ compared to the present value 
$\alpha_0$ is given as
\begin{eqnarray}
\label{delal}
\frac{\Delta \alpha}{\alpha} 
=\frac{V(\phi_0)}{V(\phi)}-1\,.
\end{eqnarray}
For the exponential potential 
$V(\phi)=V_0e^{-\mu\phi}$, we get
\begin{eqnarray}
\label{expdelal}
\frac{\Delta \alpha}{\alpha}
=e^{\mu (\phi-\phi_0)}-1\,,
\end{eqnarray}
and for the massive rolling scalar potential
$V(\phi)=V_0e^{\frac{1}{2}M^2\phi^2}$ considered in
Ref.~\cite{GST}, we have
\begin{eqnarray}
\frac{\Delta \alpha}{\alpha} =
e^{-\frac12 M^2 (\phi^2-\phi_0^2)}-1\,.
\label{varialp}
\end{eqnarray}

We recall that the present value of fine structure constant is 
$\alpha_{0}=1/137$. 
Since the potential energy of $\phi$ at present is estimated as
$3H_0^2 \simeq 8\pi V(\phi_{0})/m_{\rm pl}^2$, one finds
the expression for the warp factor:
\bea
\label{betas}
\beta^2 \simeq \frac{3}{548}\left(\frac{H_0}{M_s}\right)^2
\left(\frac{m_{\rm pl}}{M_s}\right)^2\,.
\eea
When $M_s \sim m_{\rm pl}$ we have $\beta \sim 10^{-62}$.

The model parameters in the tachyon potentials are related to 
the string scale $M_s$ and the brane tension $T_{3}$
if they are motivated by string theory.
The exponential potential $V(\phi)=V_0e^{-\mu\phi}$
introduced above appears in the context of 
the D3 and $\bar{D}$3 branes \cite{KKLMMT}. 
The tachyon potential for the coincident D3-$\bar{D}$3 branes
is twice the potential  for  the non-BPS D3-brane \cite{Senreza}. 
The latter  is given by
$V(\phi)=2\beta^2T_3/\cosh(\sqrt{\beta}M_s\phi)$ \cite{Lambert03}, 
where $\beta$ is a warp factor at the position of the D3-$\bar{D}$3
in the internal compact space and $T_3$ is the tension of the 3 branes. 
Then the potential behaves as $V(\phi)=\beta^2 T_3
e^{-\sqrt{\beta}M_s \phi}$ for large $\phi$, 
which has a correspondence
\begin{eqnarray}
\label{muexp}
V_0 = \beta^2 T_3\,,~~~
\mu = \sqrt{\beta}M_s\,.
\end{eqnarray}
By using the equations (\ref{betas}) and (\ref{muexp}), it was 
shown in Ref.~\cite{GST05va} that the resulting value of 
$\Delta \alpha/\alpha$ evaluated by Eq.~(\ref{expdelal})
is $|\Delta \alpha/\alpha| \gg 1$ for $z={\cal O}(1)$, which 
contradicts the observational bounds, and 
hence implies that these particular string motivated models 
do not work as sources of dark energy. 

Meanwhile for a rolling massive scalar potential 
$V(\phi)=V_{0}e^{\frac12 M^2\phi^2}$ with parameters 
constrained by string theory ($V_{0} \sim \beta^2 T_{3}$ and 
$M \sim \sqrt{\beta}M_s$), it is possible to explain the observed 
values of $\Delta \alpha/\alpha$ at $z={\cal O}(1)$.
In this case the field oscillates around the potential minimum 
at $\phi=0$ and is given by $\phi \simeq \Phi \cos (Mt)$
for $\dot{\phi}^2 \ll 1$, where $\Phi$ is the amplitude of oscillation.
The condition of an accelerated expansion for the tachyon case  
is $\dot{\phi}^2<2/3$. Taking the time average of
$\dot{\phi}^2$, we find that $M^2\phi^2<4/3$.
This then gives
\begin{eqnarray}
\left| \frac{\Delta \alpha}{\alpha} \right| \simeq 
\frac12 M^2 |\phi^2-\phi_{0}^2| 
\lesssim \frac12 M^2 \Phi^2\,.
\end{eqnarray}
It is possible to have $|\Delta \alpha/\alpha|=10^{-6}$-$10^{-5}$
if $|M \Phi|$ is of order $10^{-3}$-$10^{-2}$.
When $M_s \sim m_{\rm pl}$, we have $\beta \sim 10^{-62}$,
in which case the mass $M=\sqrt{\beta}M_s$
is much larger $H_{0} \sim 10^{-42}\,{\rm GeV}$.
Hence the field oscillates for many times while the universe evolves
from $z={\cal O}(1)$ to present, which also leads to the 
oscillation of $\Delta \alpha/\alpha$.

In Ref.~\cite{GST05va} it was found that inverse power-law 
potentials $V(\phi)=M^{4-n}\phi^{-n}$
are not compatible with the observational data of 
$\Delta \alpha/\alpha$  if one uses the mass scale obtained in
the context of string theory.
Thus a varying $\alpha$ provides a powerful tool with which to constrain
tachyon dark energy models.

\section{Perturbations in a universe with dark energy}
\label{perturbations}

In order to confront models of dark energy with observations of 
say the Cosmic Microwave Background (CMB)  and large-scale structure (LSS), 
it is important to study the evolution of density perturbations 
in a universe containing dark energy (see 
e.g., Refs. \cite{Huey99,Boome,Bacci,Bartolo04}). 
Its presence can give rise to features such as the Integrated 
Sachs-Wolfe (ISW) effect, which alters the CMB power spectrum.
In this section we provide the perturbation equations 
in a dark energy dominated universe with a barotropic fluid. 
The system we study covers most of scalar-field 
dark energy models including scalar-tensor theories.
We shall also consider perturbations in coupled dark energy 
scenarios and derive analytic expressions for the solution of 
matter perturbations.

\subsection{Perturbation equations}

A perturbed metric about a FRW background has the following form 
for scalar perturbations in an arbitrary gauge \cite{Kodama}:
\begin{eqnarray}
\hspace*{-0.2em}\rd s^2 &=& - (1+2A)\rd t^2 +
2a\partial_iB \rd x^i\rd t
\nonumber\\
\hspace*{-0.2em}&& +a^2\left[ (1+2\psi)\delta_{ij} + 2\partial_{ij}E 
\right] \rd x^i \rd x^j\,,
\end{eqnarray}
where $\partial_i$ represents the spatial partial derivative
$\partial/\partial x^i$. We will use lower case latin indices to
run over the 3 spatial coordinates. 
We do not consider tensor and vector parts of perturbations.

The model we study is described by the following very 
general action  
\begin{eqnarray}
\label{peraction}
S &=& \int {\rm d}^4 x \sqrt{-g}
\left[ \frac{F(\phi)}{2}R+p(\phi, X)+{\cal L}_{m} \right] \nonumber \\
&\equiv& \int {\rm d}^4 x \sqrt{-g} 
\left[\frac12 f(R, \phi, X)+{\cal L}_{m} \right]\,,
\end{eqnarray}
where $F(\phi)$ is a function of a scalar field $\phi$, 
$p(\phi, X)$ is a function of $\phi$ and 
$X=-(1/2)(\nabla \phi)^2$, and ${\cal L}_m$ 
is the Lagrangian density for a barotropic perfect fluid.
The action (\ref{peraction}) includes a wide variety of 
gravity theories such as Einstein gravity, scalar-tensor 
theories and low-energy effective string theories.

The background equations for this system are given by 
\begin{eqnarray}
& &H^2=\frac{1}{3F} ( 2Xp_{,X}-p-3H\dot{F}+\rho_{m}),\\
\label{dotHgene}
& &\dot{H}= -\frac{1}{2F} ( 2Xp_{,X}+\ddot{F}
-H\dot{F}+\rho_{m}+p_{m} ), \\
& &\frac{1}{a^3} (a^3\dot{\phi}p_{,X})^{\bullet}-p_{,\phi}
-\frac12 F_{,\phi}R=0, \\
& & \dot{\rho}_{m}+3H(\rho_m+p_m)=0\,.
\end{eqnarray}
We define the equation of state for the field $\phi$, as
\begin{eqnarray}
\label{eqge}
w_{\phi}=\frac{p+\ddot{F}+2H\dot{F}}
{2Xp_{,X}-p-3H\dot{F}}\,.
\end{eqnarray}
We have not implemented the coupling between 
the field and the fluid. The case 
of coupled dark energy will be discussed later.
Note that there is another way of defining $w_{\phi}$
when we confront it with observations, see e.g., 
Ref.~\cite{Torres}.

We define several gauge-invariant variables of 
cosmological pertubations.
Under a gauge transformation: $t \to t+\delta t$ and 
$x^i \to x^i+\delta^{ij} \partial_j \delta x$, 
the scalar perturbations transform as \cite{Kodama}
\begin{eqnarray}
& &A \to  A-\dot{\delta t} \,, \quad
B \to B - a^{-1} \delta t + a\dot{\delta x}  
\,, \nonumber \\
& &\psi \to \psi - H\delta t \,, \quad
E \to E - \delta x \,,    
\end{eqnarray}
together with the transformation of the field perturbation:
\begin{eqnarray}
\delta \phi \to \delta \phi -\dot{\phi} \delta t\,.
\end{eqnarray}

The uniform-field gauge corresponds to a
a gauge-transformation  to a frame such that $\delta \phi = 0$, leaving 
the following gauge-invariant variable:
\begin{eqnarray}
{\cal R} \equiv \psi-\frac{H}{\dot{\phi}}\delta \phi\,.
\end{eqnarray}
This is so-called comoving curvature perturbation
first introduced by Lukash \cite{Lukash}.
Meanwhile the longitudinal gauge corresponds to a 
transformation  to a frame such that 
$B=E=0$, giving the gauge-invariant variables:
\begin{eqnarray}
\label{defPhi2}
\Phi &\equiv& A - \frac{{\rm d}}
{{\rm d}t} \left[ a^2(\dot{E}+B/a)\right] \,,\\
\label{defPsi2}
\Psi &\equiv& -\psi + a^2 H (\dot{E}+B/a) \,.
\end{eqnarray}
The above two gauges are often used when we discuss 
cosmological perturbations.
One can construct other gauge invariant variables, see 
e.g., \cite{Hwang05}.

The energy-momentum tensor can 
be decomposed as 
\begin{eqnarray}
& & T_0^0=-(\rho+\delta \rho)\,,\quad 
T^0_{\alpha}=-(\rho+p) v_{,\alpha}\,, \nonumber \\
& & T^{\alpha}_{\beta}=(p+\delta p) 
\delta^{\alpha}_{\beta}+\Pi^{\alpha}_{\beta}\,,
\end{eqnarray}
where $\Pi^{\alpha}_{\beta}$ is a tracefree anisotropic stress.
Note that $\rho$, $\delta \rho$ e.t.c. can be written 
by the sum of the contribution of field and fluid, 
i.e., $\rho=\rho_{\phi}+\rho_{m}$ and 
$\delta \rho=\delta \rho_{\phi}+\delta \rho_{m}$.

We define the following new variables: 
\begin{eqnarray}
\label{defPhi}
\chi \equiv a(B+a\dot{E})\,,\quad 
\xi \equiv 3(HA-\dot{\psi})-\frac{\Delta}{a^2}\chi\,.
\end{eqnarray}
Considering perturbed Einstein equations at linear order 
for the model (\ref{peraction}), 
we obtain \cite{Hwang05} (see also Ref.~\cite{Hwangtachyon})
\begin{eqnarray}
\label{pereq1}
& & \frac{\Delta}{a^2}\psi+H\xi=-4\pi G \delta \rho\,, \\
\label{pereq2}
& & HA-\dot{\psi}=4\pi G a(\rho+p)v\,, \\
\label{pereq3}
& & \dot{\chi}+H\chi-A-\psi=8\pi G \Pi\,,\\
\label{pereq4}
& &  \dot{\xi}+2H\xi+\left( 3\dot{H}+\frac{\Delta}{a^2} 
\right) A=4\pi G (\delta \rho+3\delta p), \\
\label{pereq5}
& & \delta \dot{\rho}_m+3H (\delta \rho_{m}+\delta p_m) \nonumber \\
& &=(\rho_{m}+p_{m}) \left(\xi-3 H A+\frac{\Delta}{a}v_{m}
\right)\,, \\
\label{pereq6}
& &\frac{[a^4(\rho_{m}+p_{m})v_{m}]^{\bullet}}{a^4(\rho_{m}+p_{m})}
=\frac{1}{a} \left( A+\frac{\delta p_{m}}{\rho_{m}+p_{m}}
\right)\,,
\end{eqnarray}
where 
\begin{eqnarray}
\label{delrhoper}    
8\pi G \delta \rho &=& \frac{1}{F}
\biggl[-\frac12 (f_{,\phi}\delta \phi+f_{,X}\delta X)+
\frac12 \dot{\phi}^2 (f_{,X\phi}\delta \phi \nonumber \\
& & +f_{,XX} \delta X)+f_{,X}\dot{\phi} 
\delta \dot{\phi}-3H\delta \dot{F} \nonumber \\
& &+\biggl(3\dot{H}+3H^2
+\frac{\Delta}{a^2}\biggr)\delta F +\dot{F} \xi  \nonumber \\
& &+(3H\dot{F}-f_{,X}\dot{\phi}^2)A \nonumber \\
&& +\delta \rho_m-\frac{\delta F}{F}\rho_{m} \biggr], \\
8\pi G \delta p &=& \frac{1}{F}
\biggl[ \frac12 (f_{,\phi}\delta \phi+f_{,X}\delta X)+
\delta \ddot{F}+2H\delta \dot{F} \nonumber \\
& & -\biggl(\dot{H}+3H^2
+\frac23 \frac{\Delta}{a^2}\biggr) \delta F-\frac23 \dot{F} \xi
-\dot{F}\dot{A} \nonumber \\
& & -2(\ddot{F}+H\dot{F})A+
\delta p_m-\frac{\delta F}{F} p_{m} \biggr], \\
\label{delprhoper} 
8\pi G (\rho+p) v &=& -\frac{1}{aF}
\biggl[-\frac12 f_{,X}\dot{\phi}\delta \phi-\delta \dot{F} 
\nonumber \\
& &+H \delta F+\dot{F} A-\frac{a}{k}(\rho_{m}+p_{m})v_m
\biggr], \\
\label{delpiper}  
8\pi G \Pi&=&\frac{1}{F} (\delta F-\dot{F}\chi).
\end{eqnarray}
Here we have $X=\dot{\phi}^2/2$ and 
$\delta X=\dot{\phi} \delta \phi - \dot{\phi}^2 A$.
Note that the definition of the sign of $X$ is opposite 
compared to the one given in Ref.~\cite{Hwang05}.

Equations (\ref{pereq1})-(\ref{pereq6}) are written 
without fixing any gauge conditions
(so called ``gauge-ready'' form \cite{Hwang02}).
This allows one to choose a temporal gauge condition 
depending upon a situation one is considering.
Readers may be discouraged by rather complicated 
expressions (\ref{delrhoper})-(\ref{delpiper}),
but in subsequent discussions we expect that readers 
will be impressed by beauty of cosmological perturbation theory!

\subsection{Single-field system without a fluid}

Let us first discuss the case in which the barotropic perfect 
fluid is absent (${\cal L}_m=0$).
For the perturbation system given above it is convinient to 
choose the uniform-field gauge ($\delta \phi=0$) and 
derive the equation for the curvature perturbation ${\cal R}$.
Since $\delta F=0$ in this case, Eq.~(\ref{pereq2}) gives 
\begin{eqnarray}
\label{Are}
A=\frac{\dot{\cal R}}{H+\dot{F}/2F}\,.
\end{eqnarray}
{}From Eq.~(\ref{pereq1}) together with the use of 
Eq.~(\ref{Are}), we obtain
\begin{eqnarray}
\label{xieq}
\xi &=&-\frac{1}{H+\dot{F}/2F}
\biggl[\frac{\Delta}{a^2}{\cal R} \nonumber \\
& &+\frac{3H\dot{F}-Xf_{,X}-2X^2f_{,XX}}
{2F(H+\dot{F}/2F)} \dot{\cal R} \biggr]. 
\end{eqnarray}
Substituting Eq.~(\ref{dotHgene}) for Eq.~(\ref{pereq4}), 
we find 
\begin{eqnarray}
\label{ditxieq}
& &\dot{\xi}+\left( 2H+\frac{\dot{F}}{F} \right) \xi+
\frac{3\dot{F}}{2F} \dot{A} \nonumber \\ 
& &+\left[\frac{3\ddot{F}+6H\dot{F}+Xf_{,X}+2X^2f_{,XX}}
{2F}+\frac{\Delta}{a^2}\right]A=0. \nonumber \\
\end{eqnarray}

Plugging Eqs.~(\ref{Are}) and (\ref{xieq}) into 
Eq.~(\ref{ditxieq}), we finally get the following 
differential equation for each Fourier mode of ${\cal R}$:
\begin{eqnarray}
\label{calR}
\ddot{{\cal R}}+\frac{\dot{s}}{s} \dot{R}
+c_{A}^2 \frac{k^2}{a^2} {\cal R}=0\,,
\end{eqnarray}
where 
\begin{eqnarray}
s &\equiv& \frac{a^3(Xf_{,X}+2X^2f_{,XX}+3\dot{F}^2/2F)}
{(H+\dot{F}/2F)^2}\,, \\
\label{cA2}
c_{A}^2 &\equiv& \frac{Xf_{,X}+3\dot{F}^2/2F}
{Xf_{,X}+2X^2f_{,XX}+3\dot{F}^2/2F} \nonumber \\
&=& \frac{p_{,X}+3\dot{F}^2/4FX}
{\rho_{,X}+3\dot{F}^2/4FX}\,.
\end{eqnarray}
Here $\rho_{,X}=p_{,X}+2Xp_{,XX}$.

In the large-scale limit ($c_{A}^2 k^2 \to 0$)
we have the following solution 
\begin{eqnarray}
{\cal R}=C_1+C_2 \int \frac{1}{s} {\rm d}t\,,
\end{eqnarray}
where $C_1$ and $C_2$ are integration constants.
When the field $\phi$ slowly evolves as in the contexts of dark 
energy and inflationary cosmology,
the second-term can be identified as 
a decaying mode \cite{BTW}. 
Then the curvature perturbation is conserved on super-Hubble scales.

On sub-Hubble scales the sign of $c_{A}^2$ is crucially important 
to determine the stability of perturbations. When $c_A^2$ is 
negative, this leads to a violent instability of perturbations.
In Einstein gravity where $F$ is constant, $c_{A}^2$ coincides
with Eq.~(\ref{sound}). In this case $c_{A}^2$ vanishes
for $p_{,X}=0$. Meanwhile $w_{\phi}=-1$ for 
$p_{,X}=0$ from Eq.~(\ref{eqge}).
Hence one has $c_{A}^2=0$ at cosmological constant boundary  
($w_{\phi}=-1$). This suggests the phantom divide crossing is 
typically accompanied by the change of the sign of $c_{A}^2$, 
which leads to the instability of perturbations once the system 
enters the region $w_{\phi}<-1$.
For example in dilatonic ghost condensate model with 
$p=-X+ce^{\lambda \phi} X^2$, we get 
Eq.~(\ref{wphidi}) and $c_{A}^2=\frac{1-2cY}{1-6cY}$
where $Y=e^{\lambda \phi}X$.
The cosmological constant boundary  corresponds to $cY=1/2$.
We find that $c_{A}^2$ is negative for $1/6<cY<1/2$ and 
diverges at $cY=1/6$. The divergence of $c_{A}^2$ occurs 
for $\rho_{,X}=0$.
In scalar-tensor theories ($\dot{F} \neq 0$) the phantom divide 
($w_{\phi}=-1$) does not correspond to the change of the sign of 
$c_{A}^2$. Hence the perturbations can be stable 
even in the region $w_{\phi}<-1$ \cite{Carvalho,Peri}.

It is worth mentioning that one can calculate the spectrum of 
density perturbations generated in inflationary 
cosmology by using the perturbation equation (\ref{calR})
along the line of Ref.~\cite{Hwang05,BTW}.
It is really remarkable that the equation for the curvature 
perturbation reduces to the simple form (\ref{calR})
even for the very general model (\ref{peraction}).

\subsection{Evolution of matter perturbations}

We shall study the evolution of perturbations on sub-Hubble 
scales in the field/fluid system.
In particular we wish to derive the equation for matter pertubations
defined by $\delta_m \equiv \delta \rho_m/\rho_m$.
This is important when we place constraints on dark energy 
from the observation of large-scale galaxy clustering.
We assume that the equation of state $w_{m}$ is constant.

In the lonfitudinal gauge ($B=E=0$),
Eqs.~(\ref{pereq5}) and (\ref{pereq6}) give
the following Fourier-transformed equations:
\begin{eqnarray}
\hspace*{-1.0em}& & \dot{\delta}_{m}=(1+w_m)\left( 3\dot{\Psi}-
\frac{k}{a}v_{m}\right)\,,\\
\hspace*{-1.0em}& & \dot{v}_m +(1-3w_m) Hv_m=
\frac{k}{a} \left( \Phi+\frac{w_m}{1+w_m}
\delta_{m} \right).
\end{eqnarray}
Eliminating the $v_{m}$ term, we obtain 
\begin{eqnarray}
\label{delm}    
& & \ddot{\delta}_{m}+H(2-3w_{m}) \dot{\delta}_{m}
+w_{m} \frac{k^2}{a^2}+(1+w_m) \frac{k^2}{a^2}\Phi
\nonumber \\
& &=3(1+w_{m}) \left[ \ddot{\Psi}+(2-3w_{m})H\dot{\Psi}
\right]\,.
\end{eqnarray}
In what follows we shall study the case of a non-relativistic 
fluid ($w_{m}=0$). On scales which are much smaller 
than the Hubble radius ($k \gg aH$) the contribution of 
metric perturbations on the RHS of Eq.~(\ref{delm}) is neglected, 
which leads to
\begin{equation}
\label{matterper}    
\ddot{\delta}_{m}+2H\dot{\delta}_{m}+\frac{k^2}{a^2}
\Phi \simeq 0\,.
\end{equation}

We shall express $\Phi$ in terms of $\delta_m$ as the next step.
In doing so we use the sub-Horizon approximation 
in which the leading terms correspond to those containing $k^2$
and those with $\delta_{m}$.
Then Eq.~(\ref{pereq1}) gives
\begin{equation}
\frac{k^2}{a^2}\Psi
\simeq \frac{1}{2F} \left( \frac{k^2}{a^2} \delta F
-\delta \rho_{m} \right)\,.
\end{equation}
{}From Eq.~(\ref{pereq3}) one has $\Psi=\Phi+\delta F/F$.
Hence we obtain 
\begin{equation}
\label{dephi1}
\Phi \simeq  -\frac{1}{2F} \frac{a^2}{k^2} \rho_m
\delta_m -\frac{F_{,\phi}}{2F} \delta \phi\,.
\end{equation}

The variation of the scalar-field action in terms of the field
$\phi$ gives the following perturbation 
equation \cite{Hwang05}:
\begin{eqnarray}
\label{delphieq}
& &f_{,X} \biggl[ \delta \ddot{\phi}+\left( 3H+
\frac{\dot{p}_{,X}}{p_{,X}} \right) \delta \dot{\phi}
+\frac{k^2}{a^2}\delta \phi 
-\dot{\phi} (3\dot{\Psi}+\dot{\Phi}) \biggr] \nonumber \\
& &-2f_{,\phi} \Phi +\frac{1}{a^3} (a^3 \dot{\phi}
\delta f_{,X})^{\bullet}-\delta f_{,\phi}=0.
\end{eqnarray}
Note that this equation can be also obtained from 
Eqs.~(\ref{pereq1})-(\ref{pereq4}).
Here $\delta f$ is given by 
\begin{eqnarray}
\delta f =f_{,\phi} \delta \phi +2p_{,X} \delta X
+F_{,\phi}\delta R\,,
\end{eqnarray}
where 
\begin{eqnarray}
\delta R &=& 2 \left[ -\dot{\xi}-4H \xi
+\left( \frac{k^2}{a^2} -3\dot{H} \right) \Phi
-2\frac{k^2}{a^2}\Psi \right] \nonumber \\
&\simeq& 2 \frac{k^2}{a^2} (\Phi-2\Psi)\,.
\end{eqnarray}
Then under the sub-horizon approximation 
Eq.~(\ref{delphieq}) gives
\begin{eqnarray}
\delta \phi \simeq  \frac{F_{,\phi}}{p_{,X}}
(\Psi- 2 \Phi)\,.
\end{eqnarray}
Using the relation $\Psi=\Phi+\delta F/F$ and 
$\delta F/F=(F_{,\phi}/F)\delta \phi$, we find
\begin{eqnarray}
\label{dephi2}
\delta \phi \simeq -\frac{F F_{,\phi}}
{Fp_{,X}+2F_{,\phi}^2} \Phi\,.
\end{eqnarray}

{}From Eqs.~(\ref{dephi1}) and (\ref{dephi2}) the gravitational potential 
can be expressed in terms of $\delta_{m}$, as 
\begin{eqnarray}
\label{Phiexpression}
\Phi \simeq -\frac{a^2}{k^2} \frac{Fp_{,X}+2F_{,\phi}^2}
{F(2Fp_{,X}+3F_{,\phi}^2)} \rho_{m} \delta_{m}\,.
\end{eqnarray}
Substituting this relation for Eq.~(\ref{matterper}), 
we finally obtain the equation for matter perturbations 
on sub-Hubble scales:
\begin{eqnarray}
\label{mpereq}
\ddot{\delta}_m +2H\dot{\delta}_m-4\pi 
G_{\rm eff} \rho_m \delta_m=0\,,
\end{eqnarray}
where 
\begin{eqnarray}
G_{\rm eff}=\frac{Fp_{,X}+2F_{,\phi}^2}
{4\pi F (2Fp_{,X}+3F_{,\phi}^2)}\,.
\end{eqnarray}
For a massless scalar field, $G_{\rm eff}$ corresponds to 
the effective gravitational constant measured by the gravity 
between two test masses.

Equation (\ref{mpereq}) was first derived by Boisseau 
{\it et al.} \cite{BEPS} in the model $p=X-V(\phi)$ to reconstruct 
scalar-tensor theories from the observatios of LSS
(see also Ref.~\cite{EP}).
We have shown that this can be generalized to a more general 
model (\ref{peraction}). 
We note that in Einstein gravity ($F=1/8\pi G$)
the effective gravitational constant reduces to $G$.
Hence we recover the standard form of the equation 
for dust-like matter pertubations.

In Einstein gravity Eqs.~(\ref{Phiexpression}) and 
(\ref{mpereq}) yield
\begin{eqnarray}
\label{Phiper}    
& &\Phi = -\frac{3a^2}{2k^2}H^2\Omega_{m} \delta_{m}, \\
\label{matterperEin}   
& & \frac{\rd^2 \delta _{m}}{\rd N^2} +
\left(2+\frac{1}{H}\frac{\rd H}{\rd N}\right)
\frac{\rd \delta _{m}}{\rd N} -\frac{3}{2}\Omega _{m}
\delta _{m}=0,
\end{eqnarray}
where $N={\rm ln}\,a$.
When $\Omega_m$ is constant, 
the solution for Eq.~(\ref{matterper}) is given by 
\begin{equation}
\label{delmsol}
\delta_{m}=c_{+}a^{n_{+}}+c_{-}a^{n_{-}}\,,
\end{equation}
where $c_{\pm}$ are integration constants and 
\begin{equation}
\label{npm}
n_{\pm}=\frac14 \left[-1 \pm \sqrt{1+24\Omega_{m}} \right]\,.
\end{equation}
In the matter dominated era ($\Omega_{m} \simeq 1$) we find 
that $n_{+}=1$ and $n_-=-3/2$. Hence the perturbations
grow as $\delta_{m} \propto a$, which leads to the 
formation of galaxy clustering.
In this case the gravitational potential is constant,
i.e., $\Phi \propto a^2\rho_{m} \delta_{m} \propto a^0$
from Eq.~(\ref{Phiper}).
In the presence of dark energy $\Omega_{m}$ is smaller 
than 1, which leads to the variation of $\Phi$.
This gives rise to a late-time ISW effect in the 
temperature anisotropies when the 
universe evolves from the matter dominated era 
to a dark energy dominated era. 

We should mention that there exist isocurvature perturbations 
\cite{Abramo01,KMT01,ML02,HwangNoh02,Bartolo04}
in the field/fluid system, which generally leads to the variation 
of the curvature perturbation on super-Hubble 
scales. In order to confront with CMB we need to 
solve the perturbation equations without using the sub-horizon 
approximation. We note that a number of authors showed an interesting 
possibility to explain the suprression of power on largest 
scales observed in the CMB spectrum by accounting for a correalation 
between adiabatic and isocuravture 
perturbations \cite{MT04,Chris04,Chris05}.

\subsection{Perturbations in coupled dark energy}

At the end of this section we shall consider a coupled 
dark energy scenario in which the field is coupled to 
the matter fluid with a coupling $Q$ studied in 
Sec.~\ref{scalingsec}.
The action we study is given by Eq.~(\ref{action}).
The perturbation equations in the presence of the coupling $Q$
are presented in Ref.~\cite{Hwang02} in a gauge-ready form.
Taking a similar procedure as in the uncoupled case, 
we obtain \cite{Amendola2,ATS}
\ba
\label{matterpe2}
\frac{\rd^2 \delta _{m}}{\rd N^2}  &+&
\left(2+\frac{1}{H}\frac{\rd H}{\rd N}+\sqrt{6}Qx\right)
\frac{\rd \delta _{m}}{\rd N}  \nonumber \\
&-& \frac{3}{2}\Omega _{m}
\left(1+2\frac{Q^{2}}{p_{,X}}\right)\delta _{m}=0\,,
\ea
where $x=\dot{\phi}/\sqrt{6}H$ (here we set $\kappa^2=1$).
Note that the gravitaional potential satisfies Eq.~(\ref{Phiper})
in this case as well.

The readers who are interested in the details of the derivation of 
this equation may refer to the references \cite{Amendola2}
(see also \cite{Amendola1}).
When $Q=0$, Eq.~(\ref{matterpe2}) reduces to Eq.~(\ref{matterperEin}).
The presence of the coupling $Q$ leads to 
different evolution of $\delta_{m}$ compared to
the uncoupled case.
In what follows we shall study two cases in which analytic 
solutions can be derived.

\subsubsection{Analytic solutions in scalar-field matter 
dominant stage}

First we apply the perturbation equation 
(\ref{matterper}) to the coupled quintessence scenario 
($p=X-ce^{-\lambda \phi}$)
discussed in Sec.~\ref{cdenergy}.
The characteristic feature of this model is that there 
is a possibility to have an intermediate ``scalar-field matter 
dominated regime ($\phi$MDE)'' \cite{coupled1} characterized 
by $\Omega_{\phi}=2Q^2/3$
before the energy density of dark energy grows rapidly,
see the case (ii) in Fig.~\ref{quincoupled}.
The existence of this stage affects the evolution of 
matter perturbations compared to the case 
without the coupling $Q$.

This transient regime is realized by the fixed point (a)
in Table \ref{critquin}, which corresponds to
\begin{equation}
\label{transient}
x=-\frac{\sqrt{6}}{3}Q, \quad
\Omega_\phi=\frac23 Q^2, \quad 
w_{{\rm eff}}=\frac23 Q^2\,.
\end{equation}
Using the relation 
\begin{equation}
\frac{1}{H} \frac{\rd H}{\rd N}
=-\frac32 \left( 1+w_{\rm eff} \right)\,,
\end{equation}
the perturbation equation (\ref{matterpe2}) 
for the fixed point (\ref{transient}) is given by 
\begin{equation}
\label{delmgene} 
\frac{\rd^2 \delta_{m}}{\rd N^2}+\xi _{1}
\frac{\rd \delta_{m}}{\rd N}+\xi_{2}
\delta _{m}=0\,,
\end{equation}
where 
\begin{eqnarray}
\label{xi12}
\xi_1=\frac12- 3Q^2,~~~
\xi_2=-\frac32 \left(1-\frac32 Q^2 \right)(1+2Q^2)\,.
\end{eqnarray}
Here we used $\Omega_{m} =1-\Omega_\phi$.

For constant $\xi_1$ and $\xi_2$, the general solution for 
Eq.~(\ref{delmgene}) is given by  Eq.~(\ref{delmsol}) 
with indices:
\begin{equation}
n_{\pm}=\frac{1}{2}\left[-\xi _{1}\pm 
\sqrt{\xi _{1}^{2}-4\xi _{2}}\right]\,.
\label{power}
\end{equation}
For the case (\ref{xi12}) we have 
\begin{eqnarray}
\label{nplu}
n_{+}=1+2Q^2\,,\quad
n_{-}=-\frac32+Q^2\,.
\end{eqnarray}
When $Q=0$ this reproduces the result in a matter
dominated era discussed in the previous subsection.
In the presence of the coupling $Q$ the perturbations evolve as
$\delta_{m} \propto a^{1+2Q^2}$, which means that the growth 
rate is higher compared to the uncoupled 
case\footnote{In Ref.~\cite{coupled1} there is an error in the 
sign of $2Q^2$. We thank Luca Amendola for pointing this out.}.
Thus the coupling between dark energy and dark matter makes 
structure formation evolve more quickly.

{}From Eq.~(\ref{Phiper}) the evolution of 
the gravitational potential is given by 
\ba
\label{graev}
\Phi \propto a^{n_{+}-1-3w_{\rm eff}}\,,
\ea
along the fixed point (\ref{transient}).
Then from Eqs.~(\ref{transient}) and (\ref{nplu}) we find that 
$\Phi$ is constant.
Hence there is no ISW effect by the existence of the $\phi$MDE
phase. Meanwhile since the effective equation of state given by 
Eq.~(\ref{transient}) differs from the case of $Q=0$, 
the location of the first acoustic peak is shifted 
because of the change of an angular diameter distance \cite{AGP,coupled1}.
In Ref.~\cite{track} the coupling is constrained 
to be $Q<0.1$ at a $2\sigma$ level by using the first year 
WMAP data.

\subsubsection{Analytic solutions for scaling solutions}

It was shown in Ref.~\cite{ATS} that the perturbation 
equation (\ref{matterpe2}) can be solved analytically 
in the case of scaling solutions.
As we showed in Sec.~\ref{scalingsec}
the existence of scaling solutions restricts the Lagrangian of the form 
$p=Xg(Xe^{\lambda \phi})$ in Einstein gravity, 
where $g$ is an arbitrary function. 
Then there exists the scaling solution given by Eq.~(\ref{scalingx})
for an arbitrary function $g(Y)$.
For this scaling solution we also have the following relation
\begin{equation}
Qx=-\frac{\sqrt{6}w_{\rm eff}}{2}\,, \quad 
p_{,X}=\frac{\Omega _{\phi}+w_{\rm eff}}{2x^2}\,,
\end{equation}
where we have used Eqs.~(\ref{weffgene}), (\ref{wphige}) and (\ref{scalingx}).

Then the equation for matter perturbations (\ref{matterper})
is given by Eq.~(\ref{delmgene}) with coefficients:
\begin{eqnarray}
 &  & \xi _{1}\equiv \frac{1}{2}-\frac{9}{2}w_{\rm eff}\,,\\
 &  & \xi _{2}\equiv -\frac{3}{2}(1-\Omega _{\phi })
 \left(1+\frac{6w_{\rm eff}^{2}}{\Omega _{\phi }+w_{\rm eff}}\right)\, .
\end{eqnarray}
Since $w_{\rm eff}$ and $\Omega _{\phi }$ are constants
in the scaling regime, we obtain the analytic 
solution (\ref{delmsol}) with indices
\ba
\label{npmsca}
n_{\pm} &=& \frac{1}{4}\Biggl[
9w_{\rm eff}-1 \pm \biggl\{ (9w_{\rm eff}-1)^2 \nonumber \\
& & +24(1-\Omega_{\phi}) \left(1+\frac{6w_{\rm eff}^2}
{\Omega_\phi+w_{\rm eff}}\right) \biggr\}^{1/2} \Biggr].
\ea
Remarkably the growth rate of matter perturbations is
determined by two quantities $w_{\rm eff}$ and $\Omega_\phi$ 
only. We stress here that this result holds for any scalar-field
Lagrangian which possesses scaling solutions.
For a non-phantom scalar field characterized by 
$\Omega _{\phi }+w_{\rm eff}\equiv 
\Omega _{\phi }(1+w_{\phi })>0$, 
Eq.~(\ref{npm}) shows that $n_{+}>0$ and $n_{-}<0$.
Hence $\delta _{m}$ (and $\Phi $) grows in the scaling regime, 
i.e., $\delta _{m}\propto a^{n_{+}}$.

In the uncoupled case ($Q=0$) the effective equation of state is
given by $w_{\rm eff}=0$ in Eq.~(\ref{weffgene}).
Then we reproduce the indices given by Eq.~(\ref{npm}).
Since $0 \leq \Omega _{m}\leq 1$
in the scaling regime, the index $n_{+}$ satisfies $n_{+}\leq 1$
for uncoupled scaling solutions. 
Equation (\ref{npmsca}) shows that the index $n_+$ becomes 
larger than 1 for $Q \neq 0$ (i.e., $w_{\rm eff} \neq 0$).
In Fig.~\ref{cplot} we show the contour plot of 
$n_{+}$ as  functions of $\Omega _{\phi }$ and $w_{\rm eff}$. 
The growth rate of perturbations becomes unbounded
as we approach the cosmological constant 
border $\Omega _{\phi}+w_{\rm eff}=0$. 
The large index $n_{+}$ is unacceptable from CMB constraints
because of a strong ISW effect.

\begin{figure}
\includegraphics[height=3.2in,width=3.2in]{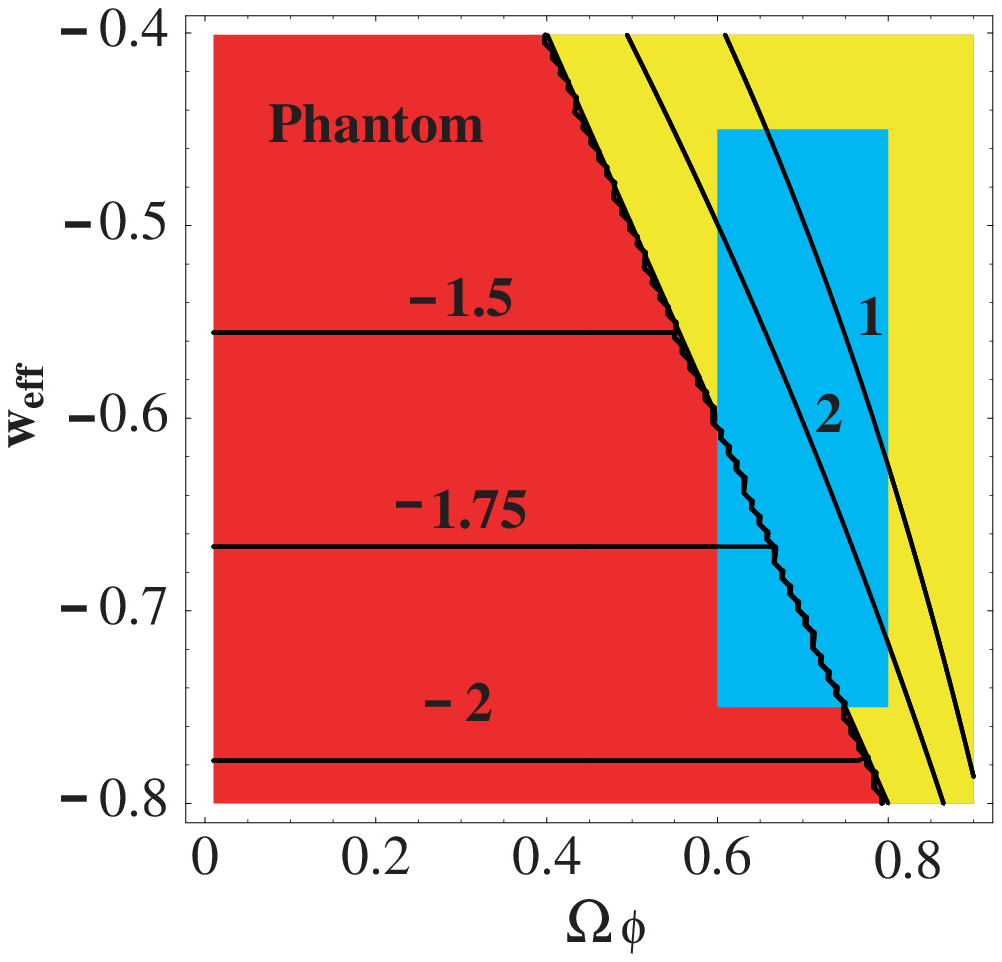}
\caption{Contour plot of the index $n_{+}$ in terms of the functions 
of $\Omega _{\phi }$ and $w_{\rm eff}$. 
The numbers which we show in the figure correspond to
the values $n_{+}$. In the non-phantom region characterized 
by $\Omega _{\phi }+w_{\rm eff}>0$, $n_{+}$ are always positive. 
Meanwhile in the phantom region ($\Omega _{\phi }+w_{\rm eff}<0$)
with $w_{\rm eff}>-1$, $n_{+}$ take complex values with negative real
parts. The real parts of $n_{+}$ are plotted in the phantom region. 
The box (blue in the color version) represents schematically the observational
constraints on $w_{\rm eff},\Omega _{\phi }$ coming from 
the SN Ia data. 
}
\label{cplot} 
\end{figure}

The phantom case corresponds to the parameter range 
$\Omega _{\phi}+w_{\rm eff}<0$. 
In this case we find that $n_{+}$ are either negative real values or complex
values with negative real parts.
Hence the perturbations decay with damped oscillations.
This is understandable, since the repulsive effect
of the phantom coupling dissipates the perturbations.
In Ref.~\cite{ATS} this phenomenon is called 
``phantom damping''.

The evolution of the gravitational potential $\Phi$ is also given by
Eq.~(\ref{graev}). Hence $\Phi$ is constant  
for $n_{+}=3w_{\rm eff}+1$,  which
corresponds to $w_{\rm eff}^{\pm}=[-2 \pm \sqrt{4-3\Omega_{\phi}}]/3$.
Since $0 \le \Omega_{\phi} \le 1$ we find 
$-1/3 \le w_{\rm eff}^+ \le 0$ and $-4/3 \le w_{\rm eff}^- \le -1$.
For example we have 
$w_{\rm eff}^+=-0.207$ and $w_{\rm eff}^-=-1.126$ for 
$\Omega _{\phi }=0.7$.
These values of $w_{\rm eff}$ are currently excluded 
by SN observations, see Fig.~\ref{cplot}.
Nevertheless it is interesting to find that there exist scaling solutions for which the 
gravitational potential is exactly constant.
We should also mention that values
of $w_{\rm eff}$ smaller than $-1$ are allowed if part of 
the dark matter itself is not coupled \cite{AGP}.
Obviously we require further investigations to constrain 
scaling dark energy models using  observations of the 
CMB and large-scale structure.

\section{Reconstruction of dark energy models}
\label{reconstruct}

We now turn our attention to review the reconstruction of 
scalar-field dark energy models from observations.
This reconstruction is in principle simple for a minimally coupled scalar field
with potential $V(\phi)$ \cite{Sta,HT,NC,CN00,GOZ}. In fact one can
reconstruct the potential and the equation of state of the field
by parametrizing the Hubble parameter $H$ in terms of the
redshift $z$ \cite{SRSS}. 
We recall that $H(z)$ is determined by the luminosity 
distance $d_L(z)$ by using the relation (\ref{HdLz}). 
This method was generalized to
scalar-tensor theories \cite{BEPS,EP,Pe}, $f(R)$ gravity
\cite{CCT} and also a dark-energy fluid with viscosity 
terms \cite{CCENO}. 
In scalar-tensor theories a scalar field $\phi$ (the dilaton) is
coupled to a scalar curvature $R$ 
with a coupling $F(\phi)R$. If the evolution
of matter perturbations $\delta_m$ is known observationally,
together with the Hubble parameter $H(z)$, one can even determine
the function $F(\phi)$ together with the potential $V(\phi)$ of
the scalar field \cite{BEPS}.

As we showed in Sec.~\ref{cdenergy} the Lagrangian (\ref{stlag}) 
in the Jordan frame is transformed to the action (\ref{elag}) in 
Einstein frame with a coupling $Q$ between the field 
$\vp$ and a barotropic fluid.
Hence if we carry out a reconstruction procedure for the 
action (\ref{action}) in the Einstein frame,
corresponding reconstruction equations can be derived by 
transforming back to the Jordan frame.
Following Ref.~\cite{Tsuji05} we shall provide the recipe of the 
reconstruction program for the general Lagrangian 
\begin{eqnarray}
\label{actiongeneral}
S=\int {\rm d}^4 x \sqrt{-g} 
\left[\frac{R}{2}+p(X, \phi)\right]+S_m (\phi)\,.
\end{eqnarray}
We consider the same coupling $Q$ as we introduced 
in Sec.~\ref{scalingsec}.
In a flat FRW spacetime the field equations 
for the action (\ref{actiongeneral}) are
\ba \label{basiceq1}
& &3H^2=\rho_m+2Xp_{,X}-p\,, \\
\label{basiceq2}
& & 2\dot{H}=-\rho_m-p_m-2Xp_{,X}\,, \\
\label{basiceq3} & &
\dot{\rho}_m+3H(\rho_m+p_m)
=Q(\phi)\rho_m\dot{\phi}\,.
\ea
In the case of a non-relativistic barotropic fluid
($p_{m}=0$), Eq.~(\ref{basiceq3}) can be written
in an integrated form
\ba 
\label{rhomde}
\rho_m=\rho_m^{(0)} \left(\frac{a_0}{a}\right)^3
I(\phi)\,,
\ea
where 
\ba \label{Idef}
 I(\phi) \equiv \exp \left(\int_{\phi_0}^\phi Q(\phi) \rd
\phi\right)\,. \ea
Here the subscript $0$ represents present values. 
Using Eq.~(\ref{redshift}) together with the relation 
$\rho_m^{(0)}=3H_0^2\Omega_m^{(0)}$, we find that 
Eq.~(\ref{rhomde}) can be written as
\ba \label{rhom}
\rho_m=3\Omega_m^{(0)}H_0^2 (1+z)^3 I(\phi)\,. 
\ea

When $Q=0$, i.e., $I=1$, we can reconstruct the structure of 
theory by using Eqs.~(\ref{basiceq1}), (\ref{basiceq2}) 
and (\ref{rhom}) if the Hubble expansion rate is 
known as a function of $z$.
This was actually carried out for a minimally coupled scalar field 
with a Lagrangian density: 
$p=X-V(\phi)$ \cite{HT,Sta,NC,CN00,GOZ,SRSS}.
In the presence of the coupling $Q$, we require
additional information to determine the strength of the coupling.
We shall make use of the equation of matter density perturbations
for this purpose as in the case of scalar-tensor
theories \cite{BEPS,EP}.

The equation for matter perturbations on sub-Hubble scales
is given by Eq.~(\ref{matterpe2}).
Let us rewrite Eqs.~(\ref{basiceq1}), (\ref{basiceq2}) and
(\ref{matterpe2}) by using a dimensionless quantity 
\ba 
r(z) \equiv H^2(z)/H_0^2\,.
\ea
Then we obtain
\ba 
\label{re1}
& & p=\left[(1+z)r'-3r\right]H_0^2\,, \\
\label{re2} & &
\phi'^2p_{,X}=\frac{r'-3\Omega_m^{(0)}(1+z)^2I}
{r(1+z)}\,,\\
\label{re3} & &
\delta_m''+\left(\frac{r'}{2r}-\frac{1}{1+z}+
\frac{I'}{I} \right) \delta_m'
\nonumber \\ & &-\frac32 \Omega_m^{(0)}
\left(1+\frac{2I'^2}{\phi'^2p_{,X}I^2} \right)
\frac{(1+z)I\delta_m}{r}=0\,,
 \ea
where a prime represents a derivative in terms of $z$.
Eliminating the $\phi'^2p_{,X}$ term from Eqs.~(\ref{re2})
and (\ref{re3}), we obtain
\ba \label{Iprime}
I'=\frac{I}{4r(1+z)A} 
\left[\delta_m' \pm \sqrt{\delta_m'^2-8r(1+z)AB}\right]\,, 
\ea
where
\ba & & A \equiv \frac{3\Omega_m^{(0)}(1+z)\delta_mI}
{2r[r'-3\Omega_m^{(0)}(1+z)^2I]}\,, \\
& &B \equiv [r'-3\Omega_m^{(0)} (1+z)^2I]A \nonumber \\
&&~~~~~~-\delta_m''
-\left(\frac{r'}{2r}-\frac{1}{1+z}
\right)\delta_m'\,.
\ea
We require the condition $\delta_m'^2>8r(1+z)AB$
for the consistency of Eq.~(\ref{Iprime}).

If we know $r$ and $\delta_m$ in terms of $z$
observationally, Eq.~(\ref{Iprime}) is integrated to give the
functional form of $I(z)$.  
Hence the function $I(z)$ 
is determined without specifying the Lagrangian 
density $p(\phi, X)$. 
{}From Eqs.~(\ref{re1}) and
(\ref{re2}), we obtain $p$ and $\phi'^2p_{,X}$ as functions of $z$.
The energy density of the scalar field, $\rho=\dot{\phi}^2 p_{,X}-p$,
is also determined. Equation (\ref{Idef}) gives 
\ba \label{Qevo}
Q=\frac{({\rm ln}\,I)'}{\phi'}\,,
\ea
which means that the coupling $Q$ is obtained once $I$ and $\phi'$ are known.
We have to specify the Lagrangian density $p(\phi, X)$
to find the evolution of $\phi'$ and $Q$.

The equation of state for dark energy, 
$w=p/\rho$, is given by
\ba \label{w1}
w &=& \frac{p}{\dot{\phi}^2p_{,X}-p} \\
\label{w2}
&=& \frac{(1+z)r'-3r}
{3r-3\Omega_m^{(0)}(1+z)^3I}\,.
 \ea
A non-phantom scalar field corresponds to $w>-1$, which translates into
the condition $p_{,X}>0$ by Eq.\,(\ref{w1}). 
{}From Eq.~(\ref{re2})
we find that this condition corresponds to
$r'>3\Omega_m^{(0)}(1+z)^2I$, which can be checked by
Eq.~(\ref{w2}). Meanwhile a phantom field is characterized by the 
condition $p_{,X}<0$ or $r'<3\Omega_m^{(0)}(1+z)^2I$. 
Since $I(z)$ is determined if $r$ and $\delta_m$ are
known observationally, 
the equation of state of dark energy is obtained from
Eq.~(\ref{w2}) without specifying the Lagrangian 
density $p(\phi, X)$. 
In the next section we shall apply our formula to several 
different forms of scalar-field Lagrangians.

\subsection{Application to specific cases}

Most of the proposed scalar-field dark energy models
can be classified into two classes: 
(A) $p=f(X)-V(\phi)$ and (B) $p=f(X)V(\phi)$. 
There are special cases in which 
cosmological scaling solutions exist, which 
corresponds to the Lagrangian density 
(C) $p=Xg(Xe^{\lambda \phi})$ 
[see Eq.~(\ref{scap})].
We will consider these classes of models separately.

\subsubsection{Case of $p=f(X)-V(\phi)$}

This includes quintessence [$f(X)=X$] and 
a phantom field [$f(X)=-X$].
Eq.~(\ref{re2}) gives
\ba 
\label{or1}
\phi'^2 f_{,X}=
\frac{r'-3\Omega_m^{(0)} (1+z)^2I}{r(1+z)}\,.
\ea
If we specify the function $f(X)$, 
the evolution of $\phi'(z)$ and $\phi(z)$ is known 
from $r(z)$ and $I(z)$.
{}From Eq.~(\ref{Qevo}) we can find the coupling $Q$
in terms of $z$ and $\phi$. 
Eq.~(\ref{re1}) gives
\ba 
\label{or2}
V=f+[3r-(1+z)r']H_0^2 \,.
\ea
Now the right hand side is determined as a function of $z$.
Since $z$ is expressed by the field $\phi$, 
one can obtain the potential $V(\phi)$ in
terms of $\phi$. In the case of Quintessence
without a coupling $Q$, this 
was carried out by a number of 
authors \cite{HT,Sta,NC,CN00,GOZ,SRSS}.
We have generalized this to a more general 
Lagrangian density $p=f(X)-V(\phi)$ with 
a coupling $Q$.

\subsubsection{Case of $p=f(X)V(\phi)$}

The Lagrangian density of the form $p=f(X)V(\phi)$ includes 
K-essence \cite{COY,AMS1,AMS2} and tachyon fields \cite{Sen}.
For example the tachyon case corresponds to a choice $f=-\sqrt{1-2X}$.
We obtain the following reconstruction equations
from Eqs.~(\ref{re1}) and (\ref{re2}): 
\ba & & \label{Ke1} 
\phi'^2
\frac{f_{,X}}{f}=\frac{r'-3\Omega_m^{(0)}(1+z)^2I}
{r(1+z)[(1+z)r'-3r]H_0^2}\,, \\
\label{Ke2}
& & V=\frac{[(1+z)r'-3r]H_0^2}{f}\,. 
\ea
Once we specify the form of $f(X)$, one can determine the functions
$\phi'(z)$ and $\phi(z)$ from Eq.~(\ref{Ke1}). Then
we obtain the potential $V(\phi)$ from Eq.~(\ref{Ke2}).

\subsubsection{Scaling solutions}

For the Lagrangian density $p=Xg(Xe^{\lambda \phi})$,
Eqs.~(\ref{re1}) and (\ref{re2}) yield
\ba \label{sca1} & &
Y\frac{g_{,Y}}{g}=\frac{6r-(1+z)r'-3\Omega_m^{(0)}(1+z)^3I}
{2[(1+z)r'-3r]}\,, \\
\label{sca2} & & 
\phi'^2=\frac{2[(1+z)r'-3r]}{r(1+z)^2g}\,, \ea
where $Y=Xe^{\lambda \phi}$. If we specify the functional
form of $g(Y)$, one can determine the function $Y=Y(z)$ from
Eq.~(\ref{sca1}). Then we find $\phi'(z)$ and $\phi(z)$ from
Eq.~(\ref{sca2}). The parameter $\lambda$ is known 
by the relation $Y=(1/2)\dot{\phi}^2 e^{\lambda \phi}$.

As we showed in Sec.~\ref{scalingsec} the quantity $Y$ is
constant along scaling solutions, in which case 
the LHS of Eq.~(\ref{sca1}) is constant. 
Thus the presence of scaling solutions can be directly 
checked by evaluating the RHS of Eq.~(\ref{sca1}) with 
the use of observational data.
We caution however that the above formula needs to be modified
if a part of the dark matter is uncoupled to dark energy.

\subsection{Example of reconstruction}

In order to reconstruct dark energy models
from observations we need to match supernova data to a 
fitting function for $H(z)$.
The fitting functions generally depend upon the models of 
dark energy \cite{NP04}.
Among a number of fitting functions, the following
parametrization for the Hubble parameter
is often used \cite{ASSS}:
\ba 
\label{para}
r(x)=H^2(x)/H_{0}^2
=\Omega_m^{(0)}x^3
+A_0+A_1x+A_2x^2\,,
\ea
where $x \equiv 1+z$ and $A_0 =1-A_1-A_2-\Omega_m^{(0)}$.
The parametrization (\ref{para}) is equivalent to 
the following expansion for dark energy:
\ba 
\rho=\rho_c^{(0)} 
\left(A_0+A_1x+A_2x^2\right)\,,
\ea
where $\rho_c^{(0)} =3H_0^2$.
The $\Lambda$CDM model is included in the above 
parametrization ($A_1=0$, $A_2=0$ and 
$A_0=1-\Omega_m^{(0)}$).

For a prior $\Omega_m^{(0)}=0.3$, the
Gold data set of SN observations gives
$A_1=-4.16 \pm 2.53$ and 
$A_2=1.67 \pm 1.03$ \cite{LNP}. 
We note that the weak energy condition 
for dark energy, $\rho \ge 0$ and $w=p/\rho \ge -1$, 
corresponds to \cite{ASSS}
\ba \label{weak}
A_0+A_1x+A_2x^2 \ge 0\,,
~~~A_1+2A_2x \ge 0\,.
\ea
If we use the best-fit values $A_1=-4.16$ and
$A_2=1.67$, for example, we find that 
the second condition in Eq.~(\ref{weak})
is violated today.
This means that the field
behaves as a phantom ($w<-1$). 
In the case of a non-phantom 
scalar field such as quintessence, 
we need to put a prior $A_1+2A_2 x \ge 0$.

For the moment we have not yet obtained accurate
observational data for the evolution of the matter perturbations
$\delta_m (z)$.
Hence the coupling $Q$ is not well constrained with current 
observations. In what follows we consider the case 
without the coupling $Q$.
Then we only need to use the reconstruction equations
(\ref{re1}) and (\ref{re2}) with $I=1$.
In this case the equation of state $w$ of dark energy 
is determined by Eq.~(\ref{w2}) provided that 
$r=H^2/H_{0}^2$ can be parametrized observationally.
In Ref.~\cite{ASSS} the reconstruction was obtained 
for the parametrization (\ref{para}).

\begin{figure}
\includegraphics[height=3.0in,width=3.2in]{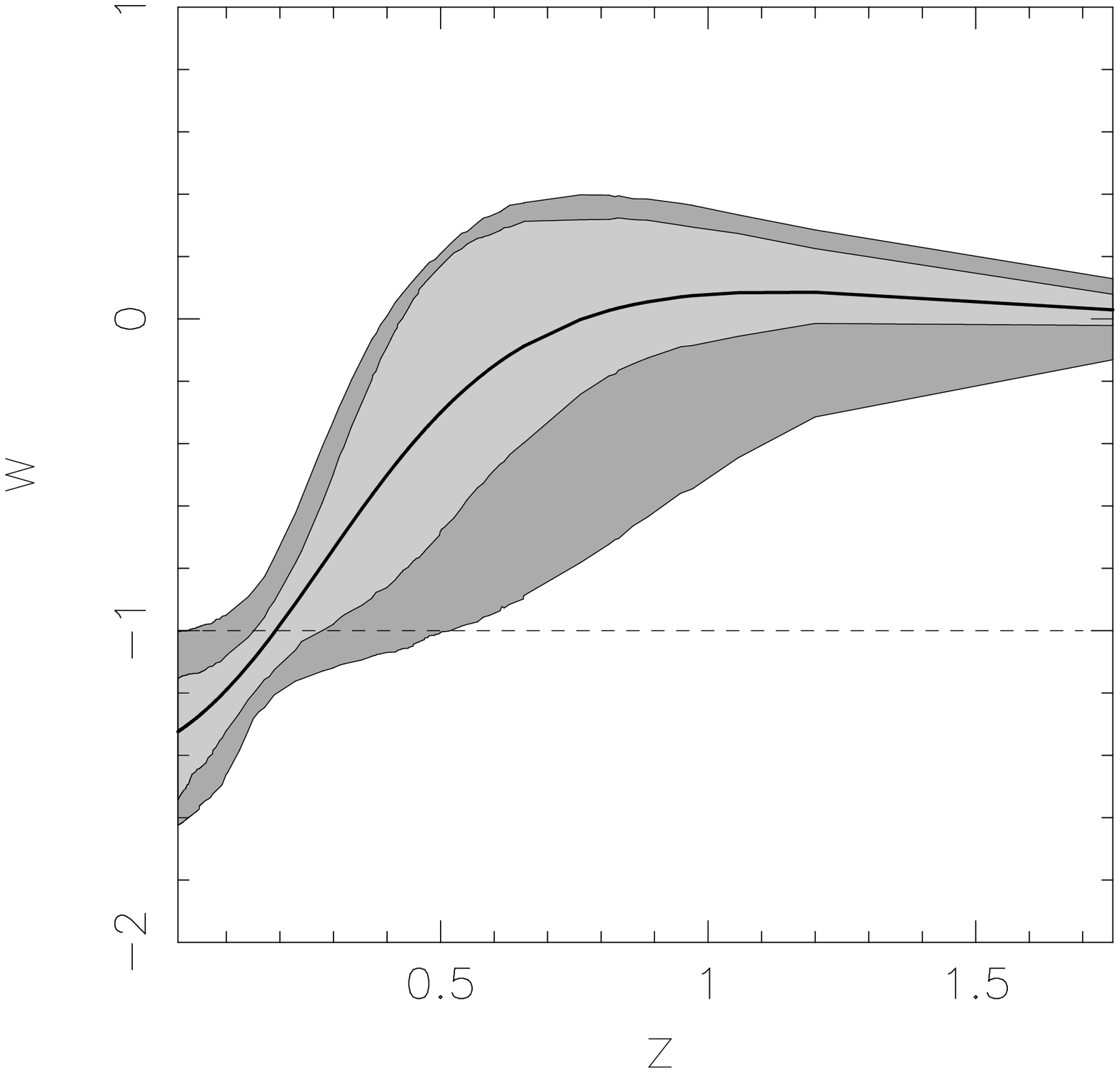}
\caption{Evolution of $w(z)$ versus redshift $z$ 
for $\Omega_m^{(0)}=0.3$ with the parametrization 
given by (\ref{para}). 
Here the thick solid line corresponds to 
the best-fit, the light grey contour represents the $1\sigma$
confidence level, and the dark grey contour represents the $2\sigma$
confidence. The dashed line corresponds to the 
$\Lambda$CDM model.
{}From Ref.~\cite{ASSS}.}
\label{eqstate} 
\end{figure}

We show in Fig.~\ref{eqstate} the evolution of $w$
versus $z$ for $\Omega_m^{(0)}=0.3$.
We find that the equation of state crosses the cosmological 
constant boundary ($w=-1$) for the best-fit parametrization.
Even at the $2\sigma$ confidence level the crossing to 
the phantom region ($w<-1$) is allowed.
In this figure we do not impose any priors for the coefficients 
$A_1$ and $A_2$.
If dark energy originates from an ordinary scalar field 
like quintessence, one needs to put a prior $A_1+2A_2 x \ge 0$.
This case is also consistent with observations, see Ref.~\cite{ASSS}.
In Ref.~\cite{SRSS} the potential of a quintessence field is 
reconstructed by using a parametrization different from 
Eq.~(\ref{para}).
For any parametrization of the Hubble parameter, the potential and 
the kinetic energy of a scalar field can be reconstructed in each 
model of dark energy by using the formula 
(\ref{re1}) and (\ref{re2}).
See Ref.~\cite{Varunstate} for a detailed reconstruction of 
the equation of state of dark energy using higher-order derivatives of $H$
called ``state finder'' \cite{statefinder}, and 
see \cite{Shafieloo:2005nd} where the authors 
have proposed a new non-parametric method of 
smoothing supernova data over redshift using a Gaussian kernel, 
the aim being to reconstruct $H(z)$ and $w(z)$ in a model independent manner.

\subsection{$w=-1$ crossing}

The reconstruction of the equation of state of dark energy
shows that the parametrization of $H(z)$ 
which crosses the cosmological-constant
boundary shows a good fit to recent SN Gold dataset \cite{LNP}, 
but the more recent SNLS dataset favors
$\Lambda$CDM \cite{Nesseris:2005ur}. 
This crossing to the phantom region ($w<-1$) is neither possible 
for an ordinary minimally
coupled scalar field [$p=X-V(\phi)$] nor for a phantom 
field [$p=-X-V(\phi)$]. 
It was shown by Vikman \cite{Vikman} that the $w=-1$ crossing 
is hard to be realized only in the presence of linear terms 
in $X$ in single-field models of dark energy.
We require nonlinear terms in $X$ to realize the $w=-1$ crossing.

This transition is possible for scalar-tensor theories \cite{Peri}, 
multi-field models \cite{Vikman,multifield} 
(called {\it quintom} using phantom and ordinary scalar field), 
coupled dark energy models with specific couplings \cite{NOT} 
and string-inspired models \cite{SS,stringcrossing,kofinas06}
\footnote{We also note that loop quantum 
cosmology \cite{LQC} allows to realize such a possibility \cite{Singh}.}.
A recent interesting result concerning whether in 
scalar-tensor theories of gravity, the equation of state of 
dark energy, $w$, can become smaller than $-1$ without violating 
any energy condition, has been obtained by Martin {\it et al.} \cite{Martin:2005bp}. 
In such models, the value of $w$ today is tied to the level of 
deviations from general relativity which, in turn, is constrained 
by solar system and pulsar timing experiments. 
The authors establish the conditions on these local constraints 
for $w$ to be significantly less than $-1$ and demonstrate that 
this requires the consideration of theories that differ from the 
Jordan-Fierz-Brans-Dicke theory and that involve either a steep 
coupling function or a steep potential. 

In this section we shall present a simple one-field model with 
nonlinear terms in $X$ which realizes
the cosmological-constant boundary crossing and 
perform the reconstruction of such a model. 

Let us consider the following Lagrangian density:
\ba 
\label{gghost}
p=-X+u(\phi)X^2\,,
\ea
where $u(\phi)$ is a function in terms of $\phi$.
Dilatonic ghost condensate models \cite{PT} 
correspond to a choice $u(\phi)=ce^{\lambda \phi}$. 
{}From Eqs.~(\ref{re1}) and (\ref{re2}) we obtain
\ba 
& & \phi'^2
=\frac{12r-3xr'-3\Omega_m^{(0)}x^3I}{rx^2}\,, \\
& & u(\phi)=\frac{2(2xr'-6r+rx^2\phi'^2)}
{H_0^2r^2x^4 \phi'^4}\,.
 \ea

Let us reconstruct the function $u(\phi)$ 
by using the parametrization (\ref{para})
with best-fit values of $A_1$ and $A_2$. 
We caution that this parametrization is not the same as 
the one for the theory (\ref{gghost}), 
but this can approximately describe the fitting of observational 
data which allows the $w=-1$ crossing.

As we see from Fig.~\ref{geghost} 
the crossing of the cosmological-constant boundary 
corresponds to $uX=1/2$, which 
occurs around the redshift $z=0.24$ for the best-fit parametrization. 
The system can enter the phantom region 
($uX<1/2$) without discontinuous behavior of
$u$ and $X$. 

\begin{figure}
\begin{center}
\includegraphics[height=3.2in,width=3.2in]{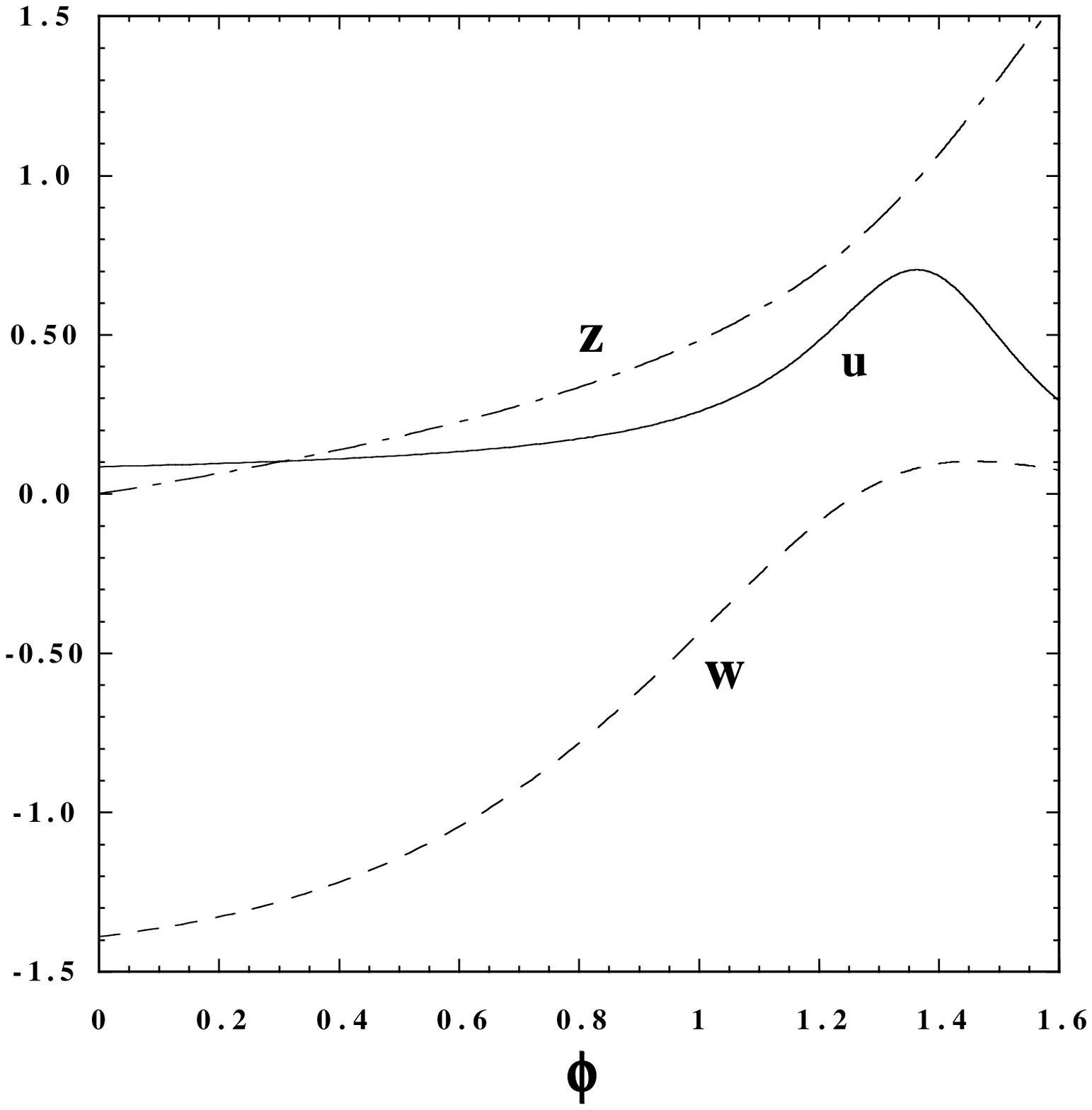}
\caption{\label{geghost} 
Reconstruction of generalized ghost condensate model 
for the parametrization (\ref{para})
with the best-fit parameters $A_1=-4.16$ and
$A_2=1.67$. 
We show $u$, $w$ and $z$ in terms of the function of $\phi$.
This model allows a possibility to cross the 
cosmological-constant boundary ($w=-1$).
}
\end{center}
\end{figure}

We have to caution that the perturbation in $\phi$
is plagued by a quantum instability when the field behaves
as a phantom \cite{PT}. Even at the classical level
the perturbation is  unstable for $1/6<uX<1/2$, 
since $c_{A}^2$ in Eq.~(\ref{cA2}) is negative.
One may avoid this instability
if the phantom behavior is just transient.
In fact transient phantom behavior was found 
in the case of a dilatonic ghost condensate model 
(see, e.g., Fig.\,4 in Ref.\,\cite{PT}). 
In this case the cosmological-constant boundary crossing
occurs again in the future, after which the perturbations 
become stable.
 
We found that the function $u(\phi)$ can be approximated by 
an exponential function $e^{\lambda \phi}$ near to the present, 
although some differences appear for $z \gtrsim 0.2$.
However the current observational data is still not 
sufficient to rule out the dilatonic ghost condensate model.
We hope that future high-precision observations will 
determine the functional form of $u(\phi)$ more
accurately.

\section{Observational constraints on the equation of state of dark energy}
\label{eosobser}

In the previous section we  provided a set of  reconstruction equations
for scalar-field dark energy models. However it is distinctly possible (some would say likely), 
that the origin of dark energy has nothing to do with scalar fields.
Fortunately, even in this case we can express the equation of state $w$ of dark energy 
in terms of $r=H^2/H_{0}^2$.
Let us consider a system of dark energy and  cold dark matter which are not 
directly coupled to each other. Using Eqs.~(\ref{Htwo}) and (\ref{dotHtwo})
with the replacement $\rho_{\phi} \to \rho_{\rm DE}$ 
and $p_{\phi} \to p_{\rm DE}$ 
together with the relation $\rho_m=\rho_m^{(0)}(1+z)^3$, 
we easily find
\ba 
w=\frac{p_{\rm DE}}{\rho_{\rm DE}}=\frac{(1+z)r'-3r}
{3r-3\Omega_m^{(0)}(1+z)^3}\,,
 \ea
which corresponds to $I=1$ in Eq.~(\ref{w2}).
Hence if observational data is accurate enough to express
$r(z)$ in terms of $z$,
we obtain $w_{\rm}(z)$ independent of the model of dark energy.
However the parametrization of $r(z)$ itself 
depends upon dark energy models and current SN Ia observations 
are not sufficiently precise to discriminate which 
parametrizations are favoured.

\subsection{Parametrization of $w_{\rm DE}$}

Instead of expressing the Hubble parameter $H$ in terms of $z$, one can 
parametrize the equation of state of dark energy. 
By using Eq.~(\ref{Htwo}) the Hubble parameter
can be written as
\ba 
\label{Hparaz}
H^2(z)=H_{0}^2 \left[ \Omega_{m}^{(0)} (1+z)^3+
(1- \Omega_{m}^{(0)}) f(z) \right]\,,
 \ea
where 
\ba 
\label{fzob}
f(z) \equiv \frac{\rho_{\rm DE}(z)}{\rho_{\rm DE}^{(0)}}
=\exp \left[3 \int_{0}^z \frac{1+w(\tilde{z})}
{1+\tilde{z}} \rd \tilde{z}\right]\,.
\ea
Hence $H(z)$ is determined once $w(z)$ is parametrized.
Then we can constrain the evolution of $w(z)$ observationally 
by using the relation (\ref{HdLz}).

There are a number of parametrizations of $w(z)$ 
which have been proposed so far, see for example  \cite{BCM04,Johri}. 
Among them, Taylor expansions of $w(z)$ 
are commonly used:
\bea
\label{Taylor}
w(z)=\sum_{n=0}w_n x_n(z)\,,
\eea
where several expansion functions have been considered \cite{BCM04}:
\ba
& &
{\rm (i)\,constant}~w: \quad x_{0}(z)=1;~x_{n}=0\,,~~n \ge 1\,, \\
& &
{\rm (ii)\,redshift:} \quad \quad x_n(z)=z^n\,, \\
& &
{\rm (iii)\,scale~factor:} \quad  x_n(z)
=\left( 1-\frac{a}{a_{0}} \right)^n=
\left(\frac{z}{1+z}\right)^n \,, \nonumber \\ \\
& &
{\rm (iv)\,logarithmic:} \quad  x_n(z)=
\left[{\rm log}(1+z)\right]^n\,.
\ea

Case (i) includes the $\Lambda$CDM model.
Case (ii) was introduced by Huterer and Turner \cite{hutTur}
\& Weller and Albrecht \cite{welleralbrecht} with $n \le 1$, i.e., 
$w_{\rm DE}=w_{0}+w_{1}z$.
In this case Eq.~(\ref{Hparaz}) gives the Hubble parameter
\ba
\hspace*{-1.0em} H^2(z) &=&
H_{0}^2 [ \Omega_{m}^{(0)} (1+z)^3 \nonumber \\
& &+(1- \Omega_{m}^{(0)}) (1+z)^{3(1+w_{0}-w_1)}
e^{3w_1 z}]\,.
\ea
We can then constrain the two parameters $w_{0}$ and $w_{1}$
by using SN Ia data.
Case (iii) was introduced by Chevallier and Polarski \cite{Pola}
\& Linder \cite{Linder}. At linear order we have 
$w(z)=w_{0}+w_{1}\dfrac{z}{z+1}$.
Jassal {\it et al.} \cite{Jassal05} 
extended this to a more general case with 
\ba
\label{Jassalpara}
w(z)=w_{0}+w_{1}\frac{z}{(z+1)^p}\,.
\ea
For example one has $w(\infty)=w_{0}+w_{1}$ for $p=1$
and $w(\infty)=w_{0}$ for $p=2$. Thus the difference appears 
for larger $z$ depending on the values of $p$.
Case (iv) was introduced by Efstathiou \cite{efstathiou}.
Basically the Taylor expansions were taken to linear
order ($n \le 1$) for the cases (ii), (iii) and (iv),
which means that two parameters $w_{0}$ and $w_{1}$
are constrained from observations.

A different approach was proposed by Bassett {\it et al.} \cite{Bruce02} and was further
developed by Corasaniti and Copeland \cite{Cora02}.
It allows for tracker solutions in which there is a rapid evolution in the
equation of state, something that the more conventional 
power-law behavior can not accommodate. 
This has some nice features in that it
allows for a broad class of quintessence models to be accurately
reconstructed and it opens up the possibility of finding evidence 
of quintessence in the CMB both through its contribution to 
the ISW effect \cite{Cora02,Cora03} and  as a way of using 
the normalization of the dark energy power spectrum on cluster 
scales, $\sigma_8$, to discriminate between dynamical models 
of dark energy (Quintessence models) and a conventional cosmological 
constant model \cite{Doran:2001rw,Kunz03}.

This {\it Kink} approach can be
described by a 4-parameter parametrization, which is
\bea
\label{Kink}
w(a)=w_0+(w_m -w_0)
\Gamma(a, a_t, \Delta)\,,
\eea
where $\Gamma$ is the transition function which depends 
upon $a$, $a_{t}$ and $\Delta$.
Here $a_{t}$ is the value of the 
scale factor at a transition point between $w=w_{m}$, 
the value in the matter-dominated era,
and $w=w_{0}$, the value today, with $\Delta$ controlling the width 
of the transition.
The parametrization (\ref{Kink}) is schematically illustrated 
in Fig.~\ref{paramet}.
The transition function used in the 
papers \cite{BCM04,Cora02,Cora04} is of the general form 
\bea
\label{Kink2}
\Gamma(a, a_t, \Delta)=\frac{1+e^{a_{t}/\Delta}}
{1+e^{(a_t-a)/\Delta}}
\frac{1-e^{(1-a)/\Delta}}{1-e^{1/\Delta}}\,.
\eea
Its advantage is that it can cope with rapid evolution of $w$, something which is difficult to 
be realized for the case of the Taylor expansions given above.

\begin{figure}
\begin{center}
\includegraphics[height=2.6in,width=3.2in]{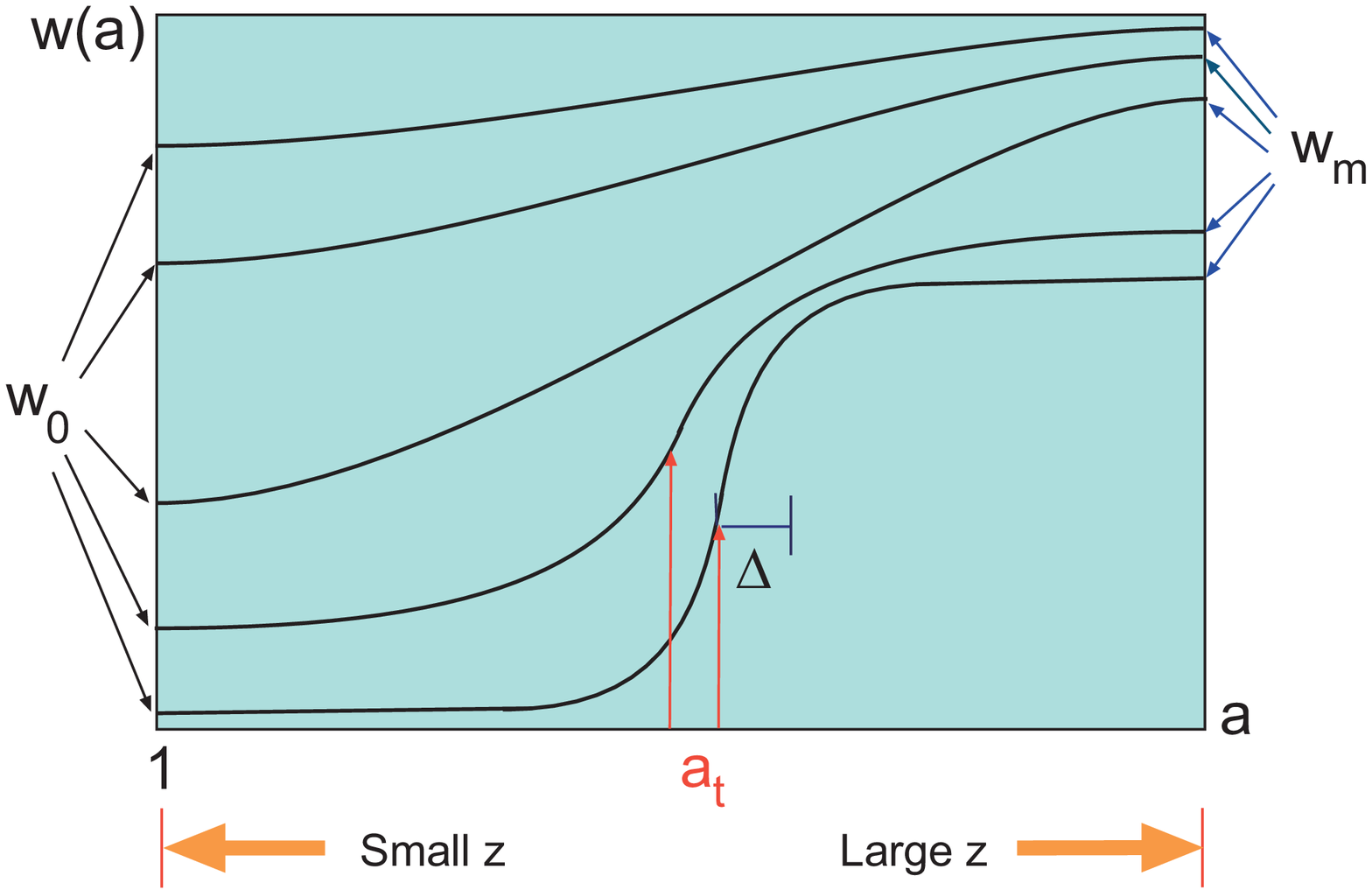}
\caption{
Schemtic illustration of the equation of state of 
dark energy for the kink parametrization (\ref{Kink}).
}
\label{paramet} 
\end{center}
\end{figure}

%
\subsection{Observational constraints from SN Ia data}
%

\begin{table*}[t]
\begin{center}
\begin{tabular}{|c|c|c|}
Parametrization & $w_{0}$ & $w_{1}$  \\
\hline
\hline
Redshift & $-1.30 \pm^{0.43}_{0.52}$  &
$1.57 \pm^{1.68}_{1.41}$    \\
\hline
Scale factor & $-1.48 \pm^{0.57}_{0.64}$  &
$3.11 \pm^{2.98}_{3.12}$    \\
\hline
Logarithmic & $-1.39 \pm^{0.50}_{0.57}$ &
$2.25 \pm^{2.19}_{2.15}$    \\
\hline
\end{tabular}
\end{center}
\caption[bestfit]{Best fits values of $w_{0}$ and 
$w_{1}$ for several different 
Taylor expansions at linear order.
Error bars correspond to the $1\sigma$ 
confidence level.
{}From Ref.~\cite{BCM04}.}
\label{bestfit}
\end{table*}

There has been recent interest in how successfully 
the equation of state of dark energy can be constrained with
SN Ia observations. For the Taylor expansion at linear order ($n \le 1$),
Bassett {\it et al.} \cite{BCM04} found the best fit values
shown in Table \ref{bestfit} by 
running a Markov-Chain Monte Carlo (MCMC) code 
with the Gold data set \cite{riess2}.
Note that these were obtained by minimizing 
$\chi^2=-2{\rm log}\,{\cal L}$,
where ${\cal L}$ is the likelihood value.

Meanwhile the Kink formula (\ref{Kink}) gives 
the best-fit values: $w_0=-2.85$, $w_{m}=-0.41$, 
$a_t=0.94$ and ${\rm ln}(\Delta)=-1.52$
with $\chi^2=172.8$.
This best-fit case corresponds to the equation of state which 
is nearly constant ($w \sim w_m$) for $z>0.1$
and rapidly decreases to $w=w_0$ for $z<0.1$.
This behavior is illustrated in Fig.~\ref{kinkevo}
together with the $2\sigma$ limits of several 
parametrizations.
We find that the best-fit solution passes outside 
the limits of all three Taylor expansions 
for $0.1 \lesssim z \lesssim 0.3$ and $z \lesssim 0.1$.
It suggests that the Taylor expansions at linear order 
are not sufficient to implement the case of such rapid
evolution of $w(z)$. Note the general similarities with the results in 
Ref.~\cite{ASSS} shown in Fig.~\ref{eqstate}.

In Ref.~\cite{BCM04} it was found that the redshift $z_{c}$
at which the universe enters an accelerating stage  
strongly depends upon the parameterizations of $w(z)$.
The $\Lambda$CDM model corresponds to 
$z_{c}=0.66 \pm^{0.11}_{0.11}$,
which is consistent with the estimation (\ref{zccon}).
One has $z_{c}=0.14 \pm^{0.14}_{0.05}$ in redshift parametrization 
[$w(z)=w_{0}+w_{1}z$] and $z_c=0.59\pm^{8.91}_{0.21}$
in scale-factor parametrization
[$w(z)=w_{0}+w_{1}z/(1+z)$].
While these large differences of $z_{c}$ may be used to distinguish the 
cosmological constant from dynamical dark energy models,
this also casts doubt on the use of standard two-parameter 
parametrizations in terms of $w_{0}$ and $w_{1}$.

\begin{figure}
\begin{center}
\includegraphics[height=3.3in,width=3.3in]{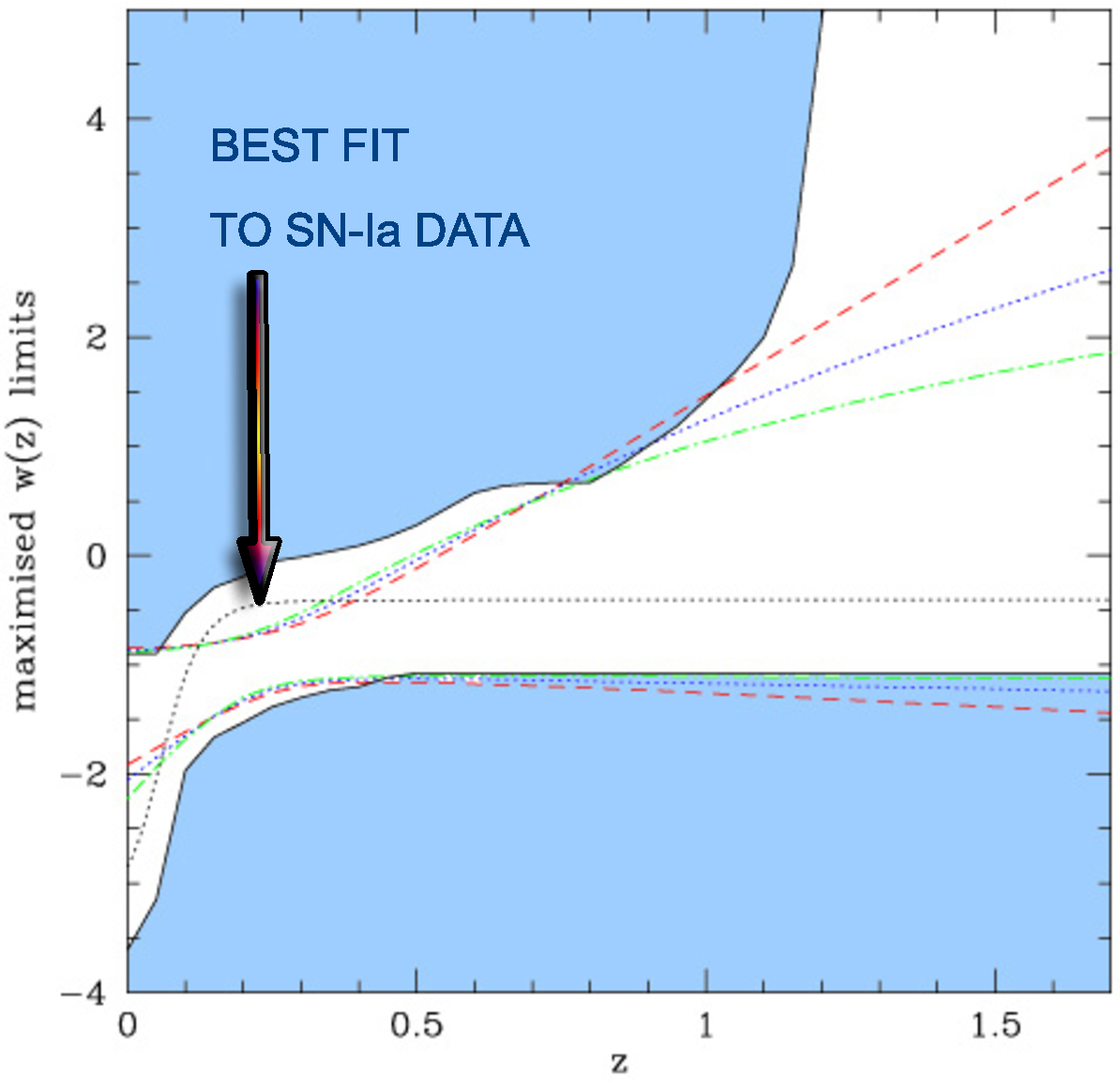}
\caption{Evolution of the equation of state of dark energy  for the best-fit
Kink parametrization. The maximized limits
on $w(z)$ are shown for (a) redshift (red dashed line), 
(b) scale-factor (green dash-dotted), (c)
logarithmic (blue-dotted), and 
(d) Kink (solid black lines) parametrizations. 
{}From Ref.~\cite{BCM04}.
}
\label{kinkevo} 
\end{center}
\end{figure}

If we include higher-order terms ($n \ge 2$) in the 
Taylor expansions (\ref{Taylor}), the above problems 
can be alleviated to some extent.
In this case it was found by Bassett {\it et al.} \cite{BCM04}
that the allowed ranges of $w_{0}$ are shifted toward 
smaller values with a maximum likelihood $w_{0} \sim -4$.
In addition huge values of $w_{1} \sim 50$ and 
$w_{2} \sim -100$ are allowed.

The above results show that observational constraints on
the equation of state of dark energy are sensitive to the parametrization of it 
and that we require at least three parameters to address a 
wide range of the variation of $w(z)$.
In Ref.~\cite{BCM04} it was found that the 
$\chi^2$ for the best-fit Kink 
parametrization is lowest compared to the values in the linear
Taylor expansions. This situation changes if we account for 
the second-order term ($n=2$); then the redshift parametrization 
gives the lowest $\chi^2$.

A question then arises: How many dark energy parameters 
do we need to describe the dark energy dynamics?
This may be addressed by using the Akaike 
information criterion (AIC) 
and Bayesian Information criterion (BIC) \cite{bic,{Liddle:2004nh}}
(see also Refs.~\cite{Linder05}).
These two criteria are defined as:
\begin{eqnarray}
\label{criterion}
{\rm AIC} & = & -2\ln\,{\cal L} + 2k_{p} \,,\\  
{\rm BIC} & = & -2\ln\,{\cal L} + k_{p} \ln \,N \,.
\label{criterion2}
\end{eqnarray}
Here ${\cal L}$ is the maximum value of the likelihood, 
$k_{p}$ is the number of parameters 
and $N$ is the number of data points.  
The optimal model minimizes the AIC or BIC. 
In the limit of large $N$, AIC tends to favour models with more
parameters while BIC more strongly penalizes them 
(since the second term diverges in this limit). 
BIC provides an estimate of the posterior evidence of a model 
assuming no prior information. Hence
BIC is a useful approximation to a full 
evidence calculation when we have no prior 
on the set of models.
Bassett {\it et al.} found that the minimum value of BIC 
corresponds to the $\Lambda$CDM model \cite{BCM04}. 
This general conclusion has been confirmed in the recent work 
of Ref.~\cite{Mukherjee:2005wg}.
It is interesting that the simplest dark energy model with only one 
parameter is preferred over other dynamical dark energy models.
This situation is similar to early universe inflation in which single-field 
models are preferred over multi-field models from two information 
criteria \cite{doubleinflation}. Ending this subsection on a cautionary note, 
Corasaniti has recently emphasised how extinction by intergalactic 
gray dust introduces a magnitude redshift dependent offset in the 
standard-candle relation of SN Ia \cite{Corasaniti:2006cv}.  
It leads to overestimated luminosity distances compared to a 
dust-free universe and understanding this process is crucial 
for an accurate determination of the dark energy parameters.

\subsection{Observational constraints from CMB}

Let us consider observational constraints arising from the CMB.
The temperature anisotropies in CMB are expanded in 
spherical harmonics: $\delta T/T=\Sigma a_{lm}Y_{lm}$.
The CMB spectrum, $C_{l} \equiv \langle |a_{lm}|^2 \rangle$,
is written in the form \cite{LLbook}
\begin{eqnarray}
C_{l}=4\pi \int \frac{\rd k}{k} {\cal P}_{\rm ini}(k)
|\Delta_{l}(k, \eta_{0})|^2\,,
\label{Cl}
\end{eqnarray}
where ${\cal P}_{\rm ini}(k)$ is an initial power spectrum and
$\Delta_{l}(k, \eta_{0})$ is the transfer function 
for the $l$ multipoles of the $k$-th wavenumber
at the present time $\eta_{0}$ (here we use conformal time:
$\eta \equiv \int a^{-1} \rd t$).
The initial power spectrum is nearly scale-invariant, which 
is consistent with the prediction of an inflationary cosmology.

The dynamical evolution of dark energy affects the CMB temperature
anisotropies in at least  two ways. First, the position of the acoustic 
peaks depends on the dark energy dynamics because of the fact that 
an angular diameter distance is related to 
the form of $w(z)$. 
Second, the CMB anisotropies are affected by the ISW effect.

In order to understand the effect of changing the position of 
acoustic peaks, let us start with the constant equation of state $w$ of dark 
energy. The presence of dark energy induces a shift by a linear 
factor $s$ in the $l$-space positions of the acoustic peaks.
This shift is given by \cite{MMOT03}
\begin{eqnarray}
s=\sqrt{\Omega_{m}^{(0)}}\,D\,,
\end{eqnarray}
where $D$ is an angular diameter distance which is written as 
\begin{eqnarray}
D=\int_0^{z_{\rm dec}} 
\frac{\rd z}{\sqrt{\Omega_{m}^{(0)}(1+z)^3
+\Omega_{\rm DE}^{(0)}
(1+z)^{3(1+w)}}},
\nonumber \\
\end{eqnarray}
where $z_{\rm dec}$ is the redshift at decoupling.
The shift of the power spectrum is proportional to $sl$.
In Fig.~\ref{Mel} we show a CMB angular power spectrum 
with the relative denisty in cold dark matter 
$\Omega_{\rm CDM}^{(0)}=0.252$ and that in baryons
$\Omega_{b}^{(0)}=0.046$, 
for various values of $w$.
As we decrease $w$,
the power spectrum is shifted toward smaller 
scales (i.e., larger $l$).

\begin{figure}
\begin{center}
\includegraphics[height=3.2in,width=3.2in]{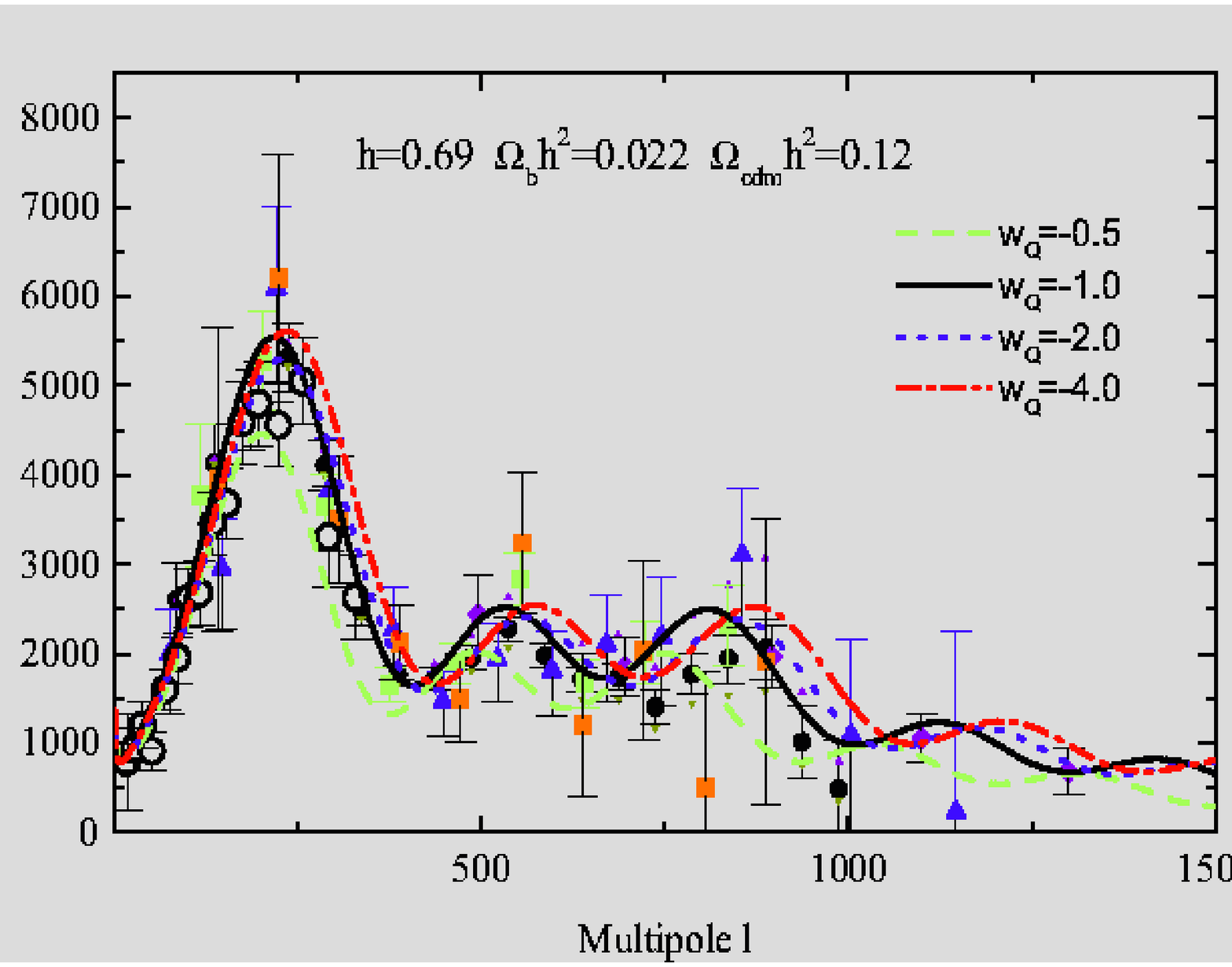}
\caption{The shift of the power spectrum for the equation of state
$w=-0.5, -1.0, -2.0, -4.0$ with $h=0.69$, 
$\Omega_{\rm CDM}^{(0)}=0.252$ and 
$\Omega_{\rm b}^{(0)}=0.046$. The spectrum shifts
toward larger $l$ for smaller $w$.
{}From Ref.~\cite{MMOT03}.}
\label{Mel} 
\end{center}
\end{figure}

The transfer function $\Delta_{l}(k, \eta_{0})$ in 
Eq.~(\ref{Cl}) can be written as the sum of the contribution coming from 
the last scattering surface and the contribution from the 
ISW effect. This ISW contribution is given by \cite{Hu95}
\begin{eqnarray}
\label{delISW}    
\Delta_{l}^{\rm ISW}(k)=2\int \rd \eta e^{-\tau}
\frac{\rd \Phi}{\rd \eta} j_{l}[k(\eta-\eta_{0})]\,,
\end{eqnarray}
where $\tau$ is the optical depth due to scattering of photons, 
$\Phi$ is the gravitational potential, and
$j_{l}$ are the Bessel functions.
As we showed in Sec.~\ref{perturbations}
the gravitational potential is constant in the matter-dominated 
period, which means the absence of the ISW effect.
However the presence of dark energy leads to a variation of 
$\Phi$, which gives rise to the ISW effect.
This is especially important for large-scale perturbations corresponding 
to $l \lesssim 20$.
In particular coupled dark energy models can have a strong 
impact on the CMB spectrum.

There have been a number of papers placing constraints on 
the equation of state of dark energy by combining 
the CMB data sets (WMAP1) together with SN Ia and LSS 
\cite{WA02,WL03,LeeNg,WT04,Upadhye,Hannestad02,Maor,MMOT03,Corasaniti:2004sz,Rape,Pogo,Jassal05}. 
Melchiorri {\it et al.} \cite{MMOT03} studied the case of constant $w$
and found that the combined analysis of CMB (WMAP1), HST, SN Ia and 2dF data 
sets gives $-1.38<w<-0.82$ at the 95 \% confidence level
with a best-fit model $w=-1.05$ and $\Omega_{m}^{(0)}=0.27$.
We note that this result was obtained by neglecting perturbations in 
the dark energy component. 
Pogosian {\it et al.} \cite{Pogo} analysed the way that future measurements 
of the ISW effect could constrain dynamical dark energy 
models as a function of redshift, $w(z)$. Introducing a new 
parameterization of $w$, the mean value of $w(z)$ as an 
explicit parameter, they argue that it allows them to separate 
the information contained in the estimation of the distance to 
the last scattering surface (from the CMB) from the information 
contained in the ISW effect. 

\begin{figure}
\begin{center}
\includegraphics[height=2.5in,width=3.2in]{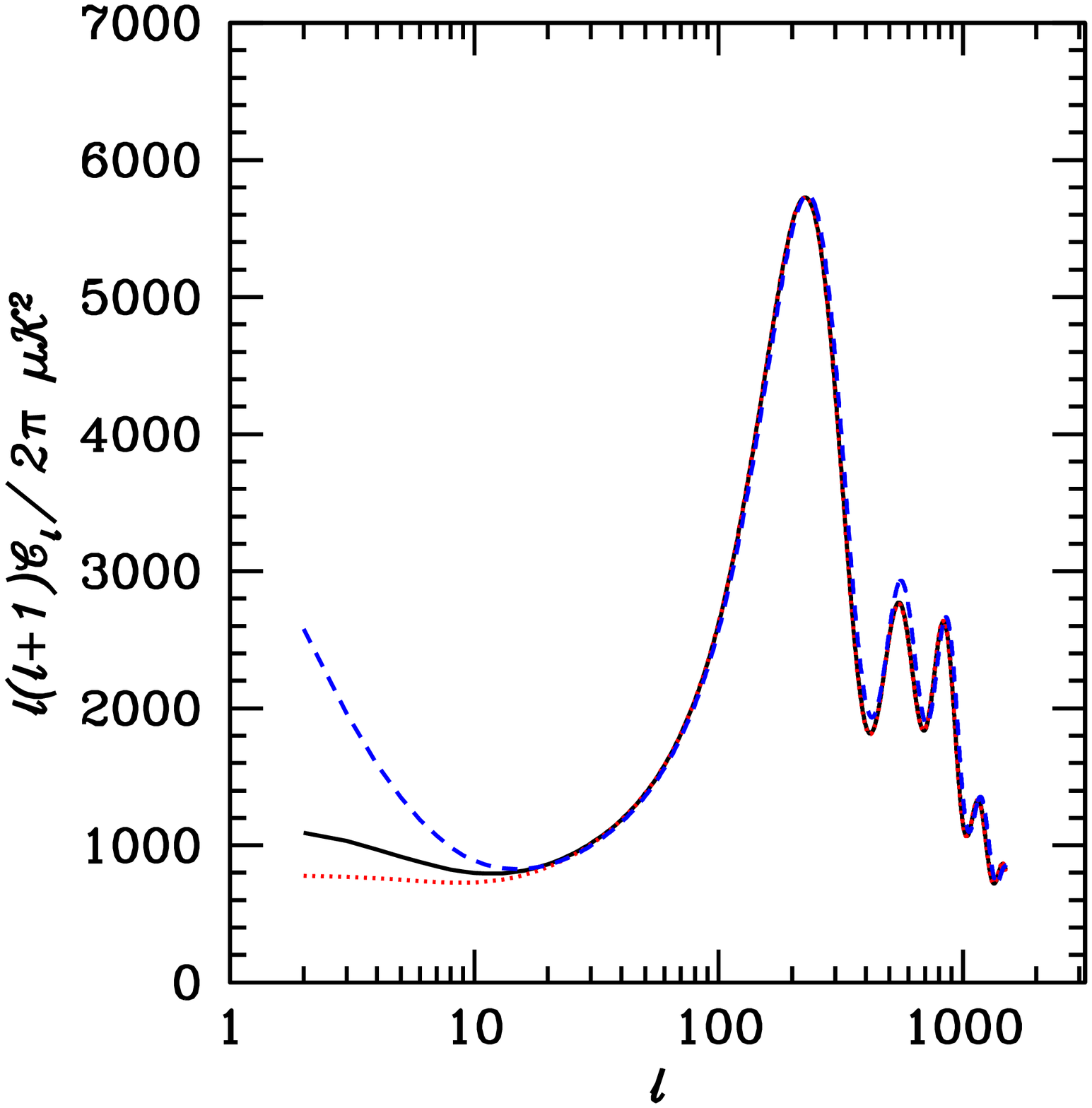}
\caption{CMB angular power spectra for three different 
models without dark energy perturbations.
The solid line, dotted line and dashed line correspond to 
(a) $w=-1$, $\Omega_{m}^{(0)}=0.3$, $\Omega_{b}^{(0)}=0.05$, 
$H_{0}=65\,{\rm km s^{-1} Mpc^{-1}}$,
(b) $w=-0.6$, $\Omega_{m}^{(0)}=0.44$, 
$\Omega_{b}^{(0)}=0.073$, 
$H_{0}=54\,{\rm km s^{-1} Mpc^{-1}}$, and 
(c) $w=-2.0$, $\Omega_{m}^{(0)}=0.17$, 
$\Omega_{b}^{(0)}=0.027$, 
$H_{0}=84\,{\rm km s^{-1} Mpc^{-1}}$,
respectively. When $w=-2$ there is a large contribution 
to low multipoles from the ISW effect.
{}From Ref.~\cite{WL03}.
}
\label{cmbnoper} 
\end{center}
\end{figure}

\begin{figure}
\begin{center}
\includegraphics[height=2.5in,width=3.2in]{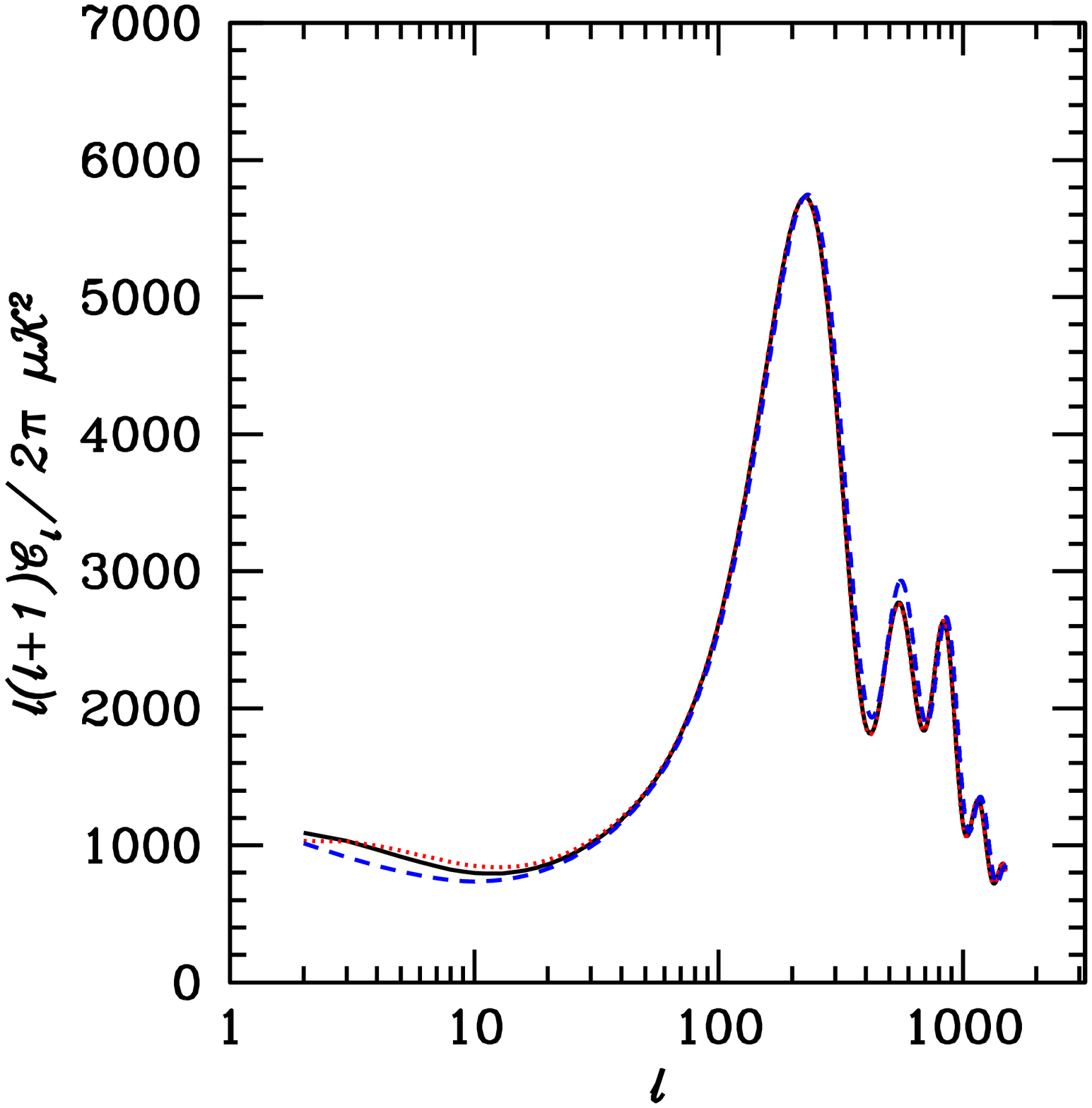}
\caption{CMB angular power spectra for three different 
models with dark energy perturbations.
Each line corresponds to the same model parameters
as in Fig.~\ref{cmbnoper}.
{}From Ref.~\cite{WL03}.
}
\label{cmbper} 
\end{center}
\end{figure}

Weller and Lewis \cite{WL03} studied the contribution of 
dark energy perturbations with constant $w$.
In Fig.~\ref{cmbnoper} the CMB power spectrum is plotted
in the case where dark energy perturbations are neglected
for $w=-1$, $w=-0.6$ and $w=-2$. 
This shows that the ISW effect is siginificant for 
$w<-1$, whereas there is a small contribution 
to the low multipoles for $w>-1$.
The situation changes in the presence of 
dark energy perturbations.
When $w$ is larger than $-1$, the inclusion of dark energy 
perturbations increases the large scale power (see Fig.~\ref{cmbper}).
For $w<-1$ dark energy fluctuations partially cancel the large 
contribution from the different evolution of the background 
via matter perturbations.
As is clearly seen from Figs.~\ref{cmbnoper} and \ref{cmbper}, 
the inclusion of perturbations in the dark energy component 
increases the degeneracies.
In fact the combined analysis of the WMAP1 (first year), ACBAR and CBI data
together with a prior from BBN and HST shows that even the values
$w<-1.5$ are allowed if we 
take into account dark energy fluctuations \cite{WL03}.
When SN Ia data are added in the anlysis, Weller and Lewis \cite{WL03}
obtained a constraint: $w=-1.02 \pm 0.16$ at $1\sigma$ level
for the speed of sound $c_{s}^2=1$.
This result does not change much even allowing for different values 
of the speed of sound, see Fig.~\ref{antony}. 

A similar conclusion is reached in  \cite{Corasaniti:2005pq}, 
where the authors investigate the possibility of constraining dark energy 
with the ISW effect recently detected by cross-correlating
the WMAP1 maps with several LSS surveys, 
concluding that current available data put weak limits on a constant 
dark energy equation of state $w$. In fact they find no constraints 
on the dark energy sound speed $c_{s}^2$. For quintessence-like 
dark energy ($c_{s}^2$=1) they find $w<-0.53$. 
Hopefully, better measurements of the CMB-LSS correlation 
will be possible with the next generation of deep redshift surveys 
and this will provide independent constraints on the dark energy 
which are alternative to those usually inferred from CMB and SN-Ia data.

\begin{figure}
\begin{center}
\includegraphics[height=3.0in,width=3.2in]{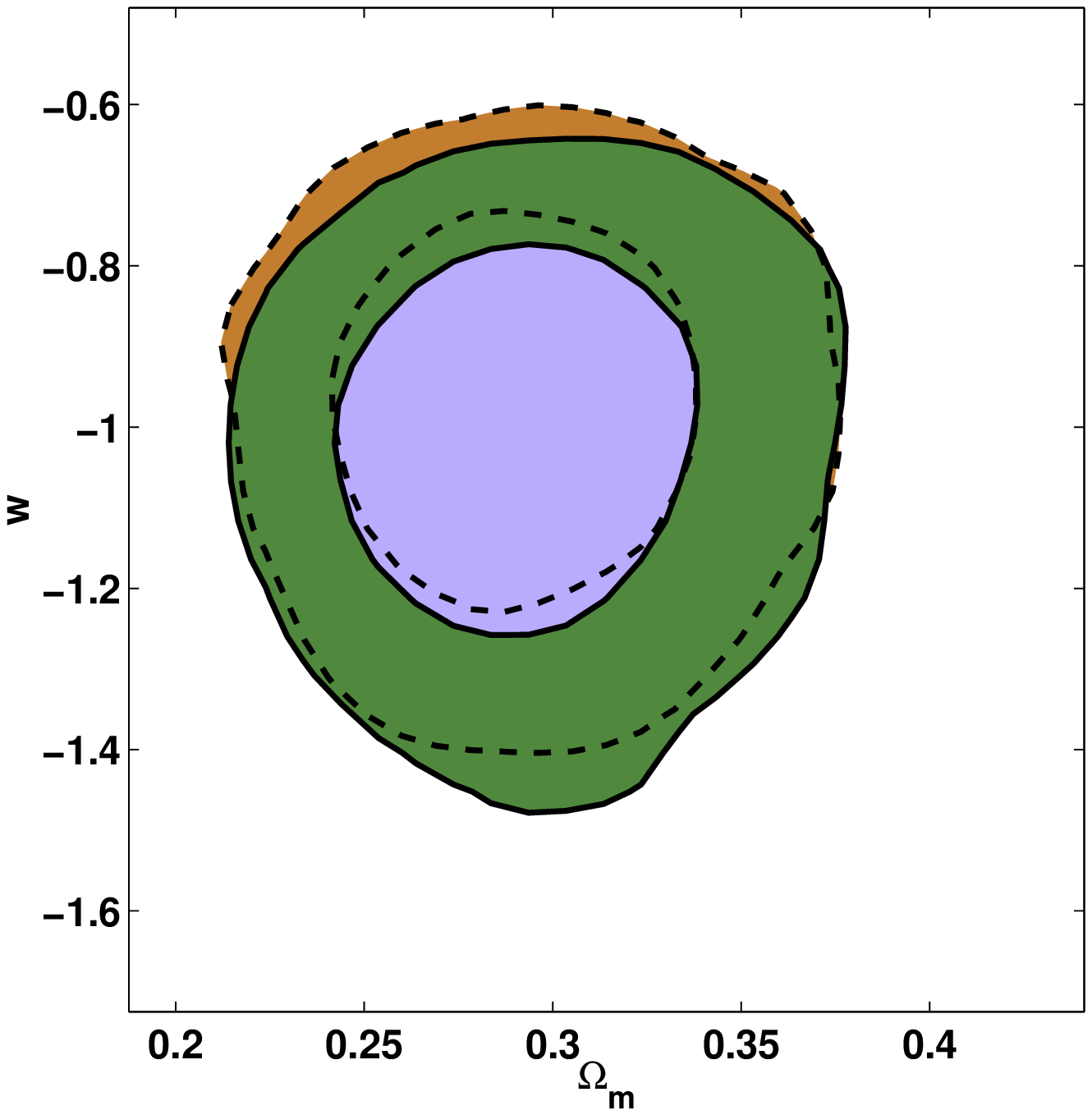}
\caption{Likelihood contours for
$w$ and $\Omega_{m}^{(0)}$ in the case 
of constant equation of state of dark energy at the 68\% 
and 95\%
confidence levels.
This is a combined analysis of CMB (WMAP1), 2dF, SN Ia, HST and BBN
data sets.
The solid line corresponds to $c_{s}^2=1$, whereas the dashed 
line to marginalizing over $c_{s}^2$.
{}From Ref.~\cite{WL03}.
}
\label{antony} 
\end{center}
\end{figure}

Let us next consider the case of a dynamically changing $w$.
If the equation of state changes from $w>-1$ to $w<-1$, 
the perturbations become unstable when the system crosses a 
cosmological constant boundary.
This problem can be alleviated in the presence of non-adiabatic
pressure perturbations.
In fact it was shown in Refs.~\cite{Hu05} that 
the phantom divide crossing can be realized in multiple
scalar field models.

A number of authors \cite{WT04,Corasaniti:2004sz,Jassal05,FB06} placed constraints 
on the dynamical evolution of dark energy by using several parametrizations 
of $w(z)$ or $\rho(z)$.
When the phantom divide crossing occurs, these results should 
be regarded as speculative since the evolution of 
dark energy perturbations around $w=-1$ was not fully addressed.
For a complete analysis we need to take into account non-adiabatic
perturbations which makes dark energy gravitationally stable.

Corasaniti {\it et al.} \cite{Corasaniti:2004sz} 
restricted the models to those with $w(z) \ge -1$
in order to avoid the instability of perturbations.
The authors used the Kink parametrization (\ref{Kink}) with (\ref{Kink2})
and performed the likelihood analysis by varying 
four dark energy parametrs $(w_0, w_m, a_t, \Delta)$ and six cosmological
parametrs $(\Omega_{\rm DE}^{(0)}, 
\Omega_b^{(0)}h^2, h, n_{\rm S}, \tau, A_{\rm S})$. 
The total likelihood is taken to be the product of each data set 
(CMB, SN-Ia and LSS)
\bea
\chi^2_{\rm tot}=\chi^2_{\rm WMAP1}+
\chi^2_{\rm SN Ia}+\chi^2_{2{\rm dF}}\,.
\eea
The 2dF data does not provide strong constraints on dark energy
beyond those obtained using CMB + SN-Ia data set.
The total $\chi^2_{\rm tot}$ of this model is 1602.9,
whereas the best fit $\Lambda$CDM model has $\chi^2_{\rm tot}=1605.8$.
The total number of degrees of freedom is 1514, which shows that none of  
the fits are very good. This is mainly due to the WMAP1 data. 
Corasaniti {\it et al.} evaluated AIC and BIC defined in 
Eqs.~(\ref{criterion}) and (\ref{criterion2}), and found that  
the quintessence models have an AIC of 1622.9 
and the $\Lambda$CDM model of 1617.8.
This means that the $\Lambda$CDM model is favoured over the 
quintessence model. 
This property also holds when the BIC criterion is used.

In the case of quintessence, the best-fit dark energy parameters are 
given by $w_m=-0.13$, $a_t=0.48$, $w_0=-1.00$ and $\Delta=0.06$.
This corresponds to a transition in which $w(z)$ does not vary 
much for $z>2$ ($w(z) \sim w_{m}=-0.13$) and  rapidly changes 
around $z=1$ toward $w_{0}=-1.00$.
Models with $w_m \ge 0$ for $z >1$
with fast transition at $z \leq 1$ are ruled out. 
This is because the
models with transition at $z<10$ with $w_m>-0.1$ lead to 
a non-negligible dark energy 
contribution at decoupling which is strongly constrained by CMB. 
Then perfect tracking behaviour for which $w=0$ during the matter era
with late time fast transition from tracking to acceleration
are disfavoured. 
On the other hand models with approximate tracking behaviour 
slowly varying equation of state with $w_0<-0.8$ and $w_{m}>-0.1$ 
are consistent with data. 
These include quintessence models with inverse power law 
potential \cite{peebles,Paul99}, 
supergravity inspired potentials \cite{brax} 
and off tracking quintessence models \cite{Maco02}. 
We note that models of late-time transitions \cite{Parker99} 
(see also \cite{Sahni:1998at}) have a similar property 
to the best-fit model.

\begin{figure}
\includegraphics[height=3.0in,width=3.2in]{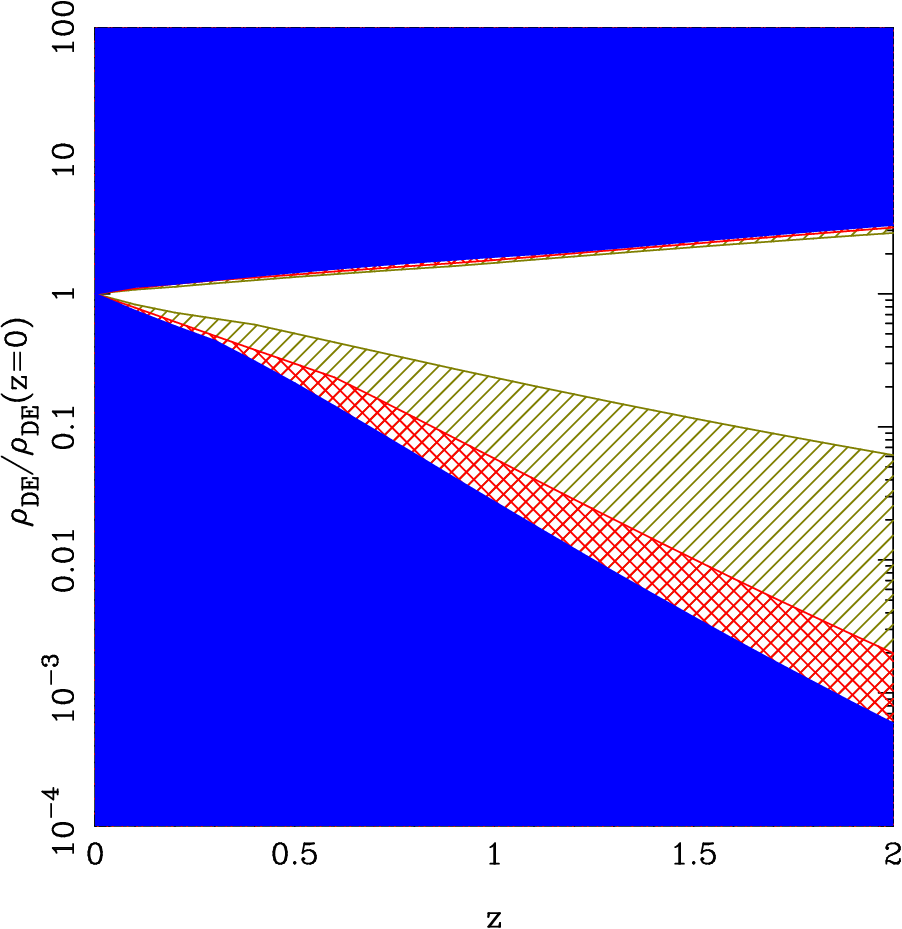}
\caption{The variation of dark energy density is shown as a function of
redshift for the parameterisation (\ref{Jassalpara}) with $p=1$.
This is the combined constraints from WMAP1 and SNLS data. 
The green/hatched region is excluded at $68\%$ confidence limit, 
red/cross-hatched region at $95\%$
confidence level and the blue/solid region at $99\%$ confidence limit. 
The white region shows the allowed range of variation 
of dark energy at $68\%$ confidence limit.  
The phantom like models have 
$\rho_{\rm DE}(z)/\rho_{{\rm DE}}(z=0) <  1$. 
The allowed values of $w_0$ at $95\%$ confidence limit for 
this parameterisation are $-1.89<w_{0}<-0.61$ (with SNLS data),  
$-1.64<w_0<-0.42$ (with WMAP1 data) and
$-1.46<w_0<-0.81$ when we combine the SNLS and WMAP1 data. 
The values for $w_{1}$ for these data sets are constrained to 
lie in the range $-4.82<w_1<3.3$ (SNLS), $-3.09<w_1<1.32$ (WMAP1) 
and $-0.99<w_1<1.04$ (combined). 
This clearly shows that the WMAP1 data is more effective in
constraining the equation of state parameters are compared 
to the supernova data. 
{}From Ref.~\cite{Paddyfigure}.
}
\label{sami_revfig} 
\end{figure}

The recent results published in Ref.~\cite{Jassal:2006gf},  
using a different data set can be seen in Fig~\ref{sami_revfig}.
We note that dark energy perturbations are not taken into 
account in their analysis.
Figure \ref{sami_revfig} shows that everything is 
perfectly consistent with a true non evolving cosmological constant. 
The fact that different data sets have been used 
(e.g., Gold SN  in Ref.~\cite{ASSS} versus SNLS in Ref.~\cite{Jassal:2006gf}), 
as well as different priors, such as  the value of $\Omega_m^{(0)}$ and 
the parametrisation for $w(z)$ could well lead to different conclusions. 
Basically it  is still too early to say whether observations 
prefer varying $w$ or constant $w$ at present. 

An interesting alternative approach to parameterising dark energy 
has been proposed in \cite{Koivisto:2005mm} where they develop 
a phenomenological three parameter fluid description of dark 
energy which allows them to include an imperfect dark energy 
component on the large scale structure. 
In particular in addition to the equation of state and the sound 
speed, they allow a nonzero viscosity parameter for the fluid. 
It means that anisotropic stress perturbations are generated in
the dark energy, something which is not excluded by the present 
day cosmological observations. They also investigate structure 
formation of imperfect fluid dark energy characterized by 
an evolving equation of state, concentrating on unified models 
of dark energy with dark matter, such as the Chaplygin gas or 
the Cardassian expansion, with a shear perturbation included.

\subsection{Cross-correlation Tomography}

An interesting approach for measuring dark energy evolution with 
weak lensing has been proposed by Jain and Taylor \cite{Jain:2003tb}. 
They developed a cross-correlation technique of lensing tomography. 
The key concept they were able to use, was that the variation of the 
weak lensing shear with redshift around massive foreground objects 
such as bright galaxies and clusters depends solely on ratios of 
the angular diameter distances. By using  massive foreground halos 
they can compare relatively high, linear shear values in the same 
part of the sky, allowing them to effectively eliminate the dominant 
source of systematic error in cosmological weak lensing measurements. 

They estimate the constraints that deep lensing surveys with 
photometric redshifts can provide on the $\Omega_{\rm DE}$, 
the equation of state parameter $w$ and  $w' \equiv {\rm d}w/{\rm d}z$. 
They claim that the accuracies on $w$ and $w'$ are: 
$\sigma (w) \simeq  0.02 f_{\rm sky}^{-1/2}$ and 
$\sigma (w') \simeq 0.05 f_{\rm sky}^{-1/2}$, where $f_{\rm sky}$
is the fraction of sky covered by the survey and 
$\sigma (\Omega_{\rm DE})=0.03$ 
is assumed in the marginalization. 
When this cross-correlation method is combined with standard 
lensing tomography, which possess complementary degeneracies, 
Jain and Taylor argue that it will allow measurement of the dark 
energy parameters with significantly better accuracy than 
has previously been obtained \cite{Jain:2003tb}.

In \cite{Schimd:2006pa} constraints on quintessence models 
where the acceleration is driven by a slow-rolling scalar 
field are investigated, focusing on cosmic shear, combined 
with supernovae Ia and CMB data. Based on earlier theoretical 
work developed in \cite{Schimd:2004nq}, the authors combine 
quintessence models with the computation of weak lensing 
observables, and determine several two-point shear statistics 
with data that includes, for the first time, the "gold set" of 
supernovae Ia, the WMAP-1 year data and the VIRMOS-Descart 
and CFHTLS-deep and -wide data for weak lensing. 
In doing so, it is the first analysis of high-energy motivated 
dark energy models that uses weak lensing data, and allows 
for the exploration of larger angular scales, using a synthetic 
realization of the complete CFHTLS-wide survey as well 
as next space-based missions surveys. In other words 
it opens up the possibility of predicting how future 
wide field imagers can be expected to perform.

\subsection{Constraints from baryon oscillations}

In addition to SN Ia, CMB and LSS data, 
the recently observed baryon oscillations in the power spectrum of 
galaxy correlation functions also constrain the nature of 
dark energy \cite{Eis}
(see also Refs~\cite{Bassett:2005kn,Baryonpapers}).
The universe before the decoupling consists of a hot plasma of 
photons, baryons, electrons and dark matter. 
The tight coupling between photons and electrons due to Thompson scattering
leads to oscillations in the hot plasma. As the universe expands 
and cools, electrons and protons combine into atoms  
making the universe neutral.
The acoustic oscillations then cease but
become imprinted on the radiation as well as on the baryons and 
should be seen in the spectrum of galaxy correlations today. 

The detection of imprints of these oscillations in the
galaxy correlation function is difficult as the signal is supressed by the fractional
energy density of baryons which is about $4\%$ of the total cosmic budget. 
Thus a large volume of the universe is required to be surveyed in order 
to detect the signature. Recently, the imprints of baryon oscillations 
were observed by the Sloan Digital Sky Survey \cite{Eis}.
A peak in the correlation
function was found around $100 h^{-1} {\rm Mpc}$ separation. 
With this finding it has been possible 
to measure the ratio of the distances at redshifts $z=0.35$ and 
$z=1089$ to a high accuracy. 

{}From the CMB radiation it is possible to  
constrain the angular diameter distance at a redshift $z=1089$ 
for fixed values of $\Omega_m^{(0)} h^2$ and $\Omega_b^{(0)} h^2$. 
In the case of a flat model with a cosmological constant $\Lambda$, 
this distance depends only on the energy fraction of $\Lambda$.
These mesurements therefore can be used to 
constrain $\Omega_\Lambda^{(0)}$ or $\Omega_m^{(0)}$ to good precision.  
The consideration of the flat model with an unknown equation of
state $w \ne -1$ provides us a  2-dimensional
parameter space [for instance ($\Omega_{\Lambda}^{(0)}, w)$] 
which requires more information in addition to the CMB acoustic scale. 
With  recent detection of baryon oscillations, we have the possibilty 
to accurately constrain one more parameter, say, the equation of state or
$\Omega_K^{(0)}$. In the case of constant $w$, Eisentein {\it et al.} found that 
$w=-0.80 \pm 0.18$ and $\Omega_m^{(0)}=0.326 \pm 0.037$ \cite{Eis}, 
which gives an independent confirmation of dark energy. 
For another approach to dark energy including the input 
of baryon oscillations see \cite{Ichikawa:2005nb}. 

The measurements of baryon oscillations, however,
can say nothing about the dynamics of dark energy at present.
For that, the dynamical equation of state $w(z)$ would 
require additional information coming from LSS such as 
the observation on baryon oscillations at higher values of 
the redshift which is one of the dreams of future missions of LSS 
studies. Finally, a word of caution. Forcing $w$ to be equal 
to a constant can lead to bias, thereby hiding the actual 
dynamics of dark energy. 
Presumably, future surveys of large scale structure 
at other redshifts or perhaps more abitious measurements
of $H(z)$ at different values of $z$ will provide vital information
for establishing the nature of dark energy \cite{Bassett:2005kn}. 
In an interesting approach using the current astronomical data 
and based on the use of the Bayesian information criteria of 
model selection, Szydlowski {\it et al.} have analysed a class of 
models of dynamical dark energy, arriving at their top ten accelerating 
cosmological models \cite{Szydlowski:2006ay}. 
The interested reader, wishing to learn more about the observational 
status of dark energy may want to look at the recent lectures of 
Perivolaropoulos \cite{Perivolaropoulos:2006ce}. 

\begin{figure}
\includegraphics[height=2.5in,width=3.4in]{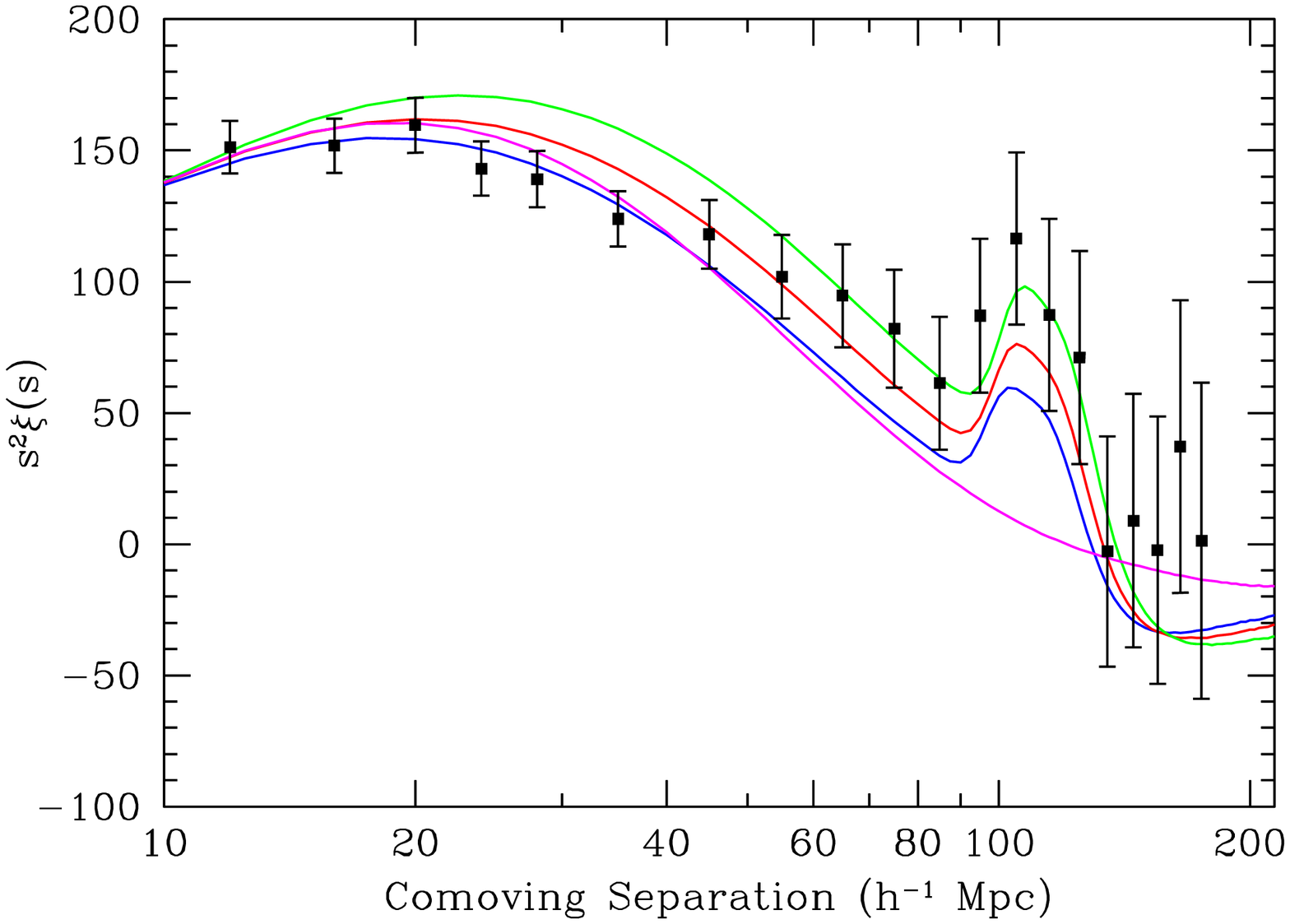}
\caption{Large-scale redshift-space correlation function 
(multiplied by the square of the separation $s$) 
from the SDSS survey. {}From top to bottom the models 
are $\Omega_m^{(0)} h^2=0.12$, $0.13$, $0.14$ and 
$0.105$ with $\Omega_{b}^{(0)}h^2=0.024$.
The bottom one corresponds to a pure cold dark matter model, which 
does not have an acoustic peak.
Meanwhile there exist acoustic peaks around 
$100h^{-1}\,{\rm Mpc}$ in other models. 
{}From Ref.~\cite{Eis}.
}
\label{bao}
\end{figure}

\section{The fate of a dark energy universe--future singularities}
\label{fate}

In this section we shall discuss the future singularities which can 
in principle appear in a dark energy universe.
When the equation of state of dark energy is less than $-1$, 
the universe reaches a Big Rip singularity within a finite time.
In this case the null energy condition 
\ba 
\rho+p \ge 0\,,
\ea
is violated.
Barrow \cite{Barrow} showed that a different type of future 
singularity can appear at a finite time even when the strong 
energy condition 
\ba 
\label{senergy}
\rho+3p \ge 0\,, \quad
\rho+p \ge 0\,,
\ea
is satisfied (see also Refs.~\cite{Barrow2}). 
This sudden future singularity corresponds to the 
one in which the pressure density $p$ diverges at $t=t_{s}$
but the energy density density $\rho$ and the scale factor $a$ 
are finite.

There exist a number of different finite-time singularities
in a dark energy universe. 
The future-singularities can be classified into the following 
five classes \cite{NOT}:
\begin{itemize}
\item  Type I (``Big Rip'') : For $t \to t_s$, $a \to \infty$,
$\rho \to \infty$ and $|p| \to \infty$
\item  Type II (``sudden'') : For $t \to t_s$, $a \to a_s$,
$\rho \to \rho_s$ and $|p| \to \infty$
\item  Type III : For $t \to t_s$, $a \to a_s$,
$\rho \to \infty$ and $|p| \to \infty$
\item  Type IV : For $t \to t_s$, $a \to a_s$,
$\rho \to 0$, $|p| \to 0$ and higher derivatives of $H$ diverge.
\item  Type V : For $t \to t_s$, $a \to \infty$,
$\rho \to \rho_{s}$, $p \to p_s$ and higher 
derivatives of $H$ diverge.
\end{itemize}
Here $t_s$, $a_s$, $\rho_s$ and $p_s$ are constants with
$a_s \neq 0$. The type I  corresponds to the Big Rip 
singularity \cite{CKW}, whereas the type II  corresponds to the sudden future 
singularity mentioned above.
The type III singularity has been discovered in the model
of Ref.~\cite{final} and is different from the sudden future singularity 
in the sense that $\rho$ diverges. 
The type IV is found in Ref.~\cite{NOT} for the model described below.
This also includes the case when $\rho$ ($p$) or both of them tend to 
approach some finite values while higher derivative of $H$ diverge.
The type V is called a ``quiescent singularity'' that appears in 
braneworld models \cite{typeV}.

In what follows we shall describe some concrete models which give rise 
to the above singularities.
Let us consider the equation of state of dark energy
which is given by 
\be
\label{EOS1}
p=-\rho - f(\rho)\,,
\ee
where $f(\rho)$ is a function in terms of $\rho$.
We note that this type of an equation of state may be related 
to bulk viscosity \cite{Barrowvis}.
The function $f(\rho)$ characterizes the deviation from a
$\Lambda$CDM cosmology.
Nojiri and Odintsov \cite{final} proposed the function 
of the form $f(\rho)\propto \rho^\alpha$ and 
this case was studied in detail in Ref.~\cite{Stefancic}. 
For the equation of state (\ref{EOS1}) with $f(\rho) \neq 0$
the continuity equation 
(\ref{dotrhotwo}) is written in an integrated form as:
\be
\label{EOS4}
a=a_0\exp \left(\frac{1}{3}\int \frac{\d\rho}{f(\rho)}\right)\,,
\ee
where $a_0$ is constant.
In the absence of any barotropic fluid other than dark energy, 
the Hubble rate satisfies Eq.~(\ref{HubbleeqI}) with $K=0$.
Then we obtain the following relation 
\be
\label{trho}
t=\int \frac{{\rm d} \rho}{\kappa \sqrt{3\rho}f(\rho)}\,.
\ee

In what follows we shall study the properties of future 
singularities for several choices of the function $f(\rho)$
establishing the relation between the singularities and 
the behavior of $f(\rho)$.

\subsection{Type I and III singularities}

Type I and III singularities appear when the function $f(\rho)$ is given by 
\be
\label{funchoice1}
f(\rho)=A\rho^{\alpha}\,,
\ee
where $A$ and $\alpha$ are constants.
Let us consider a situation in which $\rho$ goes to 
infinity with positive $\alpha$.
{}From Eqs.~(\ref{EOS4}) and (\ref{trho}) we have
\be
\label{scaleI}
a =a_0 \exp \left[\frac{\rho^{1-\alpha}}
{3(1-\alpha)A} \right]\,,
\ee
and
\ba
\label{EOS31}
& & t = t_s + \frac{2}{\sqrt{3}\kappa A}
\frac{\rho^{-\alpha+1/2}}{1-2\alpha}\,,~~
{\rm for}~~\alpha \neq \frac12\,, \\
& & t = t_s + \frac{{\rm ln}\,\rho}
{\sqrt{3}\kappa A}\,,~~
{\rm for}~~\alpha=\frac12\,,
\ea
where $t_s$ is constant.

When $\alpha> 1$, the scale factor is finite
even for $\rho \to \infty$.
When $\alpha<1$ we find $a\to \infty$ $(a\to 0)$ as 
$\rho\to \infty$ for $A>0$ $(A<0)$.
If $\alpha>1/2$ the energy density $\rho$ diverges 
in the finite future or past ($t=t_s$).
On the other hand, if $\alpha \le 1/2$,  
$\rho$ diverges in the infinite future or past.

Since the pressure is given by
$p\sim -\rho - A\rho^\alpha$,
$p$ always diverges when $\rho$ becomes infinite.
The equation of state of dark energy is 
\be
w=\frac{p}{\rho}=-1-A\rho^{\alpha-1}\,.
\ee
When $\alpha>1$ one has $w \to +\infty$ ($w \to -\infty$)
as $\rho \to \infty$ for $A<0$ ($A>0$).
Meanwhile when $\alpha<1$, $w \to -1+0\,(-1-0)$ 
for $A<0$ ($A>0$).

{}From the above argument, one can classify 
the singularities as follows:

\noindent
\begin{enumerate}
\item $\alpha> 1$:

There exists a type III singularity.
$w \to +\infty$ ($-\infty$) if $A<0$ ($A>0$).

\item $1/2<\alpha <1$:

There is a type I  future singularity for $A>0$.
When $A<0$, one has $a \to 0$ as $\rho \to \infty$.
Hence if the singularity exists in the past (future), we may 
call it Big Bang (Big Crunch) singularity.
$w \to -1+0$ ($-1-0$) if $A<0$ ($A>0$).

\item $0< \alpha \leq {1}/{2}$:

There is no finite future singularity.
\end{enumerate}

When $\alpha<0$, it was shown in Ref.~\cite{Stefancic} that
the type II singularity appears when $\rho$ approaches 0.
In the next subsection we shall generalize this to a 
more general case.

\subsection{Type II singularity}

Let us consider the function 
\be
\label{EOS33}
f(\rho)=C (\rho_0 - \rho)^{-\gamma}\,,
\ee
where $C$, $\rho_{0}$ and $\gamma$ are constants with $\gamma>0$.
We study the case in which  $\rho$ is smaller than $\rho_0$.
In the limit $\rho \to \rho_0$, the pressure $p$ becomes infinite
because of the divergence of $f(\rho)$.
The scalar curvature $R$ also diverges
since $R=2\kappa^2\left(\rho - 3p\right)$.
The equation of state of dark energy is
\be
w=-1 - \frac{C}{\rho (\rho_0-\rho)^\gamma}\,.
\ee
Hence, $w \to -\infty$ for $C>0$ and
$w \to \infty$ for $C<0$ as $\rho \to \rho_0$.

{}From Eq.~(\ref{EOS4}) the scale factor is given by 
\be
\label{EOS34}
a=a_0 \exp \left[ - \frac{(\rho_0 - \rho)^{\gamma+1}}
{3C (\gamma + 1)} \right]\,,
\ee
which means that $a$ is finite for $\rho=\rho_0$.
Since the Hubble rate $H \propto \sqrt{\rho}$ is nonsingular,
$\dot{a}$ remains finite.
On the other hand Eq.~(\ref{acceleq}) shows that 
$\ddot{a}$ diverges for $\rho \to \rho_0$.
By using Eq.~(\ref{trho}) we find the following relation
around $\rho \sim \rho_0$:
\be
\label{EOS35}
t \simeq t_s - \frac{(\rho_0-\rho)^{\gamma+1}}
{\kappa C \sqrt{3\rho_0} (\gamma+1)}\,,
\ee
where $t_s$ is an integration constant.
Then we have $t=t_s$ for $\rho=\rho_0$.
The above discussion shows that the function $f(\rho)$
in Eq.~(\ref{EOS33}) gives rise to the type II
singularity. We note that the strong energy  condition (\ref{senergy})
is satisfied for $C<0$ around $\rho=\rho_{0}$.
This means that the sudden singularity appears even in 
the case of a non-phantom dark energy ($w>-1$).

This type II singularity always appears when the denominator
of $f(\rho)$ vanishes at a finite value of $\rho$.
The model (\ref{funchoice1}) with negative $\alpha$ 
is a special case of the model (\ref{EOS33}) with $\rho_0=0$.

\subsection{Type IV singularity}

In Ref.~\cite{NOT} it was shown that the type IV singularity can 
appear in the model given by 
\be
\label{EOS15}
f(\rho)= \frac{AB \rho^{\alpha+\beta}}
{A\rho^\alpha + B \rho^\beta}\,,
\ee
where $A$, $B$, $\alpha$ and $\beta$ are constants.
We note that this model also gives rise to the 
type I, II, III singularities \cite{NOT}.

In what follows we shall study the case with $\alpha=2\beta -1$.
Then Eqs.~(\ref{EOS4}) and (\ref{trho}) are integrated to give
\be
\label{EOS17}
a=a_0 \exp \left\{-\frac{1}{3}\left[\frac{\rho^{-\alpha + 1}}{(\alpha - 1)A}
+ \frac{\rho^{-\beta + 1}}{(\beta - 1)B}\right] \right\}\,,
\ee
and
\ba
\label{EOS19}
& & \frac{2}{4\beta - 3} \rho^{-\frac{4\beta - 3}{2}}
+ \frac{2A}{(2\beta - 1)B} \rho^{-\frac{2\beta - 1}{2}} \nn
& &=- \sqrt{3}\kappa A (t-t_s) \equiv \tau\,,
\ea
where $t_s$ is an integration constant.
Equation (\ref{EOS19}) is valid for $\beta \neq 1$, $\beta \neq 3/4$, 
and $\beta \neq 1/2$.

Let us consider the case with $0<\beta<1/2$.
In this case the pressure density behaves as 
\ba
\label{plim}
p \sim -\rho - B \rho^{\beta}, \quad
{\rm when} \quad \rho \sim 0\,.
\ea
Equation (\ref{EOS19}) shows that $t \to t_s$ as $\rho \to \rho_{0}$.
Then from Eq.~(\ref{plim}) one has $p \to 0$ and $\rho \to 0$
as $t \to t_s$. By using Eqs.~(\ref{EOS17}) and (\ref{EOS19}) 
we obtain the following relation around $t=t_s$:
\ba
\label{appa}
{\rm ln}\,(a/a_{0}) \propto \tau^s, \quad
s=1-\frac{1}{2\beta-1}\,.
\ea
Hence the scale factor is finite ($a=a_{0}$) at $t=t_s$.

{}From Eq.~(\ref{appa}) we find that $s>2$ for $0<\beta<1/2$, 
which means that $H$ and $\dot{H}$ are finite.
However $\d^n H /\d t^n$ diverges for $n>-1/(2\beta -1)$
as long as $s$ is not an integer.
This corresponds to the type IV singularity in which higher-order derivatives of
$H$ exhibit divergence even if $a$, $\rho$ and $p$ are finite
as $t \to t_s$. In this case $w \to +\infty$ ($-\infty$) for $B<0$ ($B>0$).

\vspace{0.5cm}

Thus we have shown that the equation of state given by Eq.~(\ref{EOS1})
has a rich structure giving rise to four types of sudden singularities.
We note that there are other types of equation of state which lead to the 
singularities mentioned above, see Refs.~\cite{Barrow,Barrow2}.
In the presence of a bulk viscosity $\zeta$ 
the effective pressure density is given by 
$p_{\rm eff}=p-3\zeta H$ \cite{Brevik}.
This was generalized to a more general inhomogeneous dark energy 
universe in Ref.~\cite{NOvis}.
Such inhomogeneous effects
can change the type of singularities discussed in this section.
See Refs.~\cite{singularitypapers} for other interesting aspects
of future singularities.

Finally, we should mention that the model studied in 
Ref.~\cite{Anth4} provides an alternative mechanism for
the emergence of future singularities, 
see Ref.~\cite{AnthLC} for details.

\section{Dark energy with higher-order curvature corrections}
\label{hcorrections}

In the previous section, we saw that a dark energy universe 
with singularities is typically associated with the growth of 
the curvature of the universe. For the models we have been 
considering, the type I, II, III singularities lead to the divergence
of the Ricci scalar $R$ at finite time. 
In such circumstances we expect that the effect of higher-order
curvature terms can be important around 
singularities \cite{final,quantum,ENO04,NOT,NOS05,STTT,CTS,Tsujian,Neupane05}. 
This may moderate or even remove the singularities. 

Let us consider the following action with 
a correction term ${\cal L}_c^{(\phi)}$:
\ba\label{actcur}
S &=& \int {\rm d}^D x \sqrt{-g}\Biggl[\frac12 
f(\phi,R)-\frac12 \zeta(\phi)(\nabla\phi)^2
-V(\phi) \nonumber \\
& &+\xi(\phi){\cal L}_c
+{\cal L}_\rho^{(\phi)}\Biggr]\,,
\ea
where $f$ is a generic function of a 
scalar field $\phi$ and the Ricci scalar $R$.
$\zeta$, $\xi$ and $V$ are functions of $\phi$. 
${\cal L}_\rho^{(\phi)}$ is the Lagrangian of a 
perfect fluid with energy density $\rho$ and 
pressure density $p$. 
The barotropic index, $w\equiv p/\rho$, is assumed to be 
constant. In general the fluid can couple to the scalar field $\phi$.
We note that the action (\ref{actcur}) 
includes a wide variety of gravity theories such as Einstein gravity,
scalar-tensor theories and low-energy effective string theories.
In what follows we shall consider two types of higher-order correction
terms and investigate the effects on the future singularities.

\subsection{Quantum effects from a conformal anomaly}

Let us first study the effect of quantum effects in four dimensions 
by taking into account the contribution of the conformal anomaly
as a backreaction.
We shall consider the case of a fixed scalar field without a potential 
in which the barotropic fluid ${\cal L}_\rho$ is responsible 
for dark energy, i.e., $f=R$, $\zeta=0$, $V=0$ and 
$\xi=1$ in Eq.~(\ref{actcur}). 

The conformal anomaly $T_A$ takes the following 
form \cite{final,quantum,NOT}:
\be
\label{OVII}
T_A=b_1\left(F+{2 \over 3}\Box R\right) + 
b_2 G + b_3\Box R\,,
\ee
where $F$ is the square of a 4-dimensional Weyl tensor,
$G$ is a Gauss-Bonnet curvature invariant, 
which are given by
\bea
\label{GF}
F&=& (1/3)R^2 -2 R_{ij}R^{ij}+ R_{ijkl}R^{ijkl}\,, \\
G&=&R^2 -4 R_{ij}R^{ij}+ R_{ijkl}R^{ijkl}\,.
\eea
With $N$ scalar, $N_{1/2}$ spinor, $N_1$ vector fields, $N_2$ ($=0$ or $1$)
gravitons and $N_{\rm HD}$ higher derivative conformal scalars,
the coefficients $b_1$ and $b_2$ are given by
\bea
\label{bs}
\hspace*{-2.2em} &&
b_1={N +6N_{1/2}+12N_1 + 611 N_2 - 8N_{\rm HD} 
\over 120(4\pi)^2}\,, \\
\hspace*{-2.2em} &&
b_2=-{N+11N_{1/2}+62N_1 + 1411 N_2 -28 
N_{\rm HD} \over 360(4\pi)^2}\,.
\eea
We have $b_1>0$ and $b_2<0$ for the usual matter except for
higher derivative conformal scalars.
We note that $b_2$ can be shifted by a finite renormalization of the
local counterterm $R^2$, so $b_2$ can be arbitrary.

The conformal anomaly is given by $T_A=-\rho_A + 3p_A$
in terms of the corresponding energy density $\rho_A$ and the
pressure density $p_A$.
Using the continuity equation
\be
\label{CB1}
\dot{\rho}_A+3 H\left(\rho_A + p_A\right)=0\,,
\ee
$T_{A}$ can be expressed as
\be
\label{CB2}
T_A=-4\rho_A -\dot{\rho}_A/H \,.
\ee
This then gives the following expression for $\rho_A$:
\bea
\label{CB3}
\hspace*{-0.4em} \rho_A&=& -\frac{1}{a^4} \int \d t\, a^4 H T_A \nn
\hspace*{-0.4em}&=&  -\frac{1}{a^4} \int \d t\, a^4 H 
\Bigl[-12b_1 \dot{H}^2
+ 24b_2 (-\dot{H}^2 + H^2 \dot{H} + H^4)  \nn
\hspace*{-0.5em}& &- (4b_1+ 6b_3)\left(\dddot{H}
+ 7 H \ddot{H} + 4\dot{H}^2 +
12 H^2 \dot{H} \right) \Bigr]\,.
\eea
In Ref.~\cite{NOqc} a different form of $\rho_A$ was obtained 
by requiring that the quantum corrected energy momentum tensor
$T_{A\,\mu\nu}$ has the form as $T_{A\,\mu\nu}=(T_A/4)g_{\mu\nu}$
in the conformal metric case rather than
assuming the conservation law (\ref{CB1}).

Now, we are considering a universe with a dark energy fluid 
and quantum corrections.
Then the Friedmann equation is given by 
\be
\label{EOSq1}
3H^2=\kappa^2 \left(\rho + \rho_A \right)\,.
\ee
Since the curvature is large around the singularity, 
we may assume $(3/\kappa^2)H^2\ll \left| \rho_A\right|$.
This gives $\rho_{A} \sim -\rho$, which reflects the fact 
that the conformal anomaly can give rise a negative
energy density coming from higher-order curvature terms.
\bea
\label{EOSq2}
\hspace*{-2.2em}&& \dot{\rho}+4H\rho \nn
\hspace*{-2.2em}&& =  H\Bigl[-12b_1 \dot{H}^2
+ 24b_2 (-\dot{H}^2 + H^2 \dot{H} + H^4)  \nn
\hspace*{-2.2em}& &- (4b_1 + 6b_3)\left(\dddot{H}
+ 7 H \ddot{H} + 4\dot{H}^2 +
12 H^2 \dot{H} \right) \Bigr]\,.
\eea

One can understand whether the singularities may be moderated 
or not by using this equation.
We consider a dark energy fluid with an equation of state given by 
Eq.~(\ref{EOS1}). As an example we study the case of 
the type II (sudden) singularity 
for the model (\ref{EOS33}).
{}From Eq.~(\ref{EOS35})  the evolution of the energy density 
$\rho$ around the singularity is described by 
\be
\label{EOSq13}
\rho \sim \rho_0 - \left[ \kappa C\sqrt{3\rho_0}
(\gamma + 1) \left(t_s - t\right)
\right]^{\frac{1}{\gamma + 1}}\,.
\ee
In the absence of quantum corrections the Hubble parameter is 
given by 
\be
\label{EOSq14}
H\sim \kappa\sqrt{\frac{\rho_0}{3}}\left\{1
- \frac{1}{2\rho_0}\left[\kappa C\sqrt{3\rho_0}
(\gamma + 1) \left(t_s - t\right)
\right]^{\frac{1}{\gamma + 1}} 
\right\}\,.
\ee
Although $H$ is finite at $t=t_s$, $\dot H$ diverges there 
because of the condition: $0<1/(\gamma + 1)<1$.

The situation is different if we include quantum corrections.
Let us assume the following form of $\rho$ around $t=t_s$:
\be
\label{EOSq15}
\rho = \rho_0 + \rho_1 \left(t_s-t\right)^\nu\,,
\ee
where $\nu$ is a positive constant.
The continuity equation (\ref{conteq}) gives $H=\dot{\rho}/3f(\rho)$
for the equation of state (\ref{EOS1}), thereby giving 
\be
\label{EOSq16}
H\sim \frac{\nu\rho_1^{1+\gamma}}{3(-C)}
\left(t_s - t \right)^{-1 + \nu\left(1+\gamma\right)}\ .
\ee
Here $\nu\left(1+\gamma\right)$ is positive.
Picking up the most singular terms in Eq.~(\ref{EOSq2}) around the 
singularity, we find 
\be
\label{rhoapp}
\dot{\rho} \sim -6 \left( \frac23 b_1+b_3 \right)
H \dddot{H}\,.
\ee
Then substituting Eqs.~(\ref{EOSq15}) and (\ref{EOSq16})
for Eq.~(\ref{rhoapp}), we obtain
$\nu = 4/(2\gamma + 1)$ and
\bea
& & \rho = \rho_0 + \rho_1 \left(t_s-t\right)^{\frac{4}{2\gamma + 1}}\,, \\
& & H \propto (t_s-t)^{\frac{2\gamma+3}{2\gamma+1}}\,.
\eea

This shows that both $H$ and $\dot H$ are finite 
because $(2\gamma+3)/(2\gamma+1)$ is larger than 1.
Hence quantum effects works to prevent the type II singularity. 
When the quantum correction becomes important, this typically
works to provide a negative energy density $\rho_A$ which nearly
cancels with the energy density $\rho$ of dark energy.
This is the reason why the Hubble rate does not diverge
in such a case.
It was shown in Ref.~\cite{NOT} that the type I and III singularities
can be moderated as well in the presence of quantum corrections.
This property also holds for scalar-field dark energy 
models \cite{quantum}. 
Thus quantum effects can work to make the 
universe less singular or completely nonsingular.

\subsection{String curvature corrections}

We now turn our attention to study the effect of higher-order 
corrections \cite{stringcorre} in low-energy effective string theory
in the presence of a dark energy fluid (see Ref.\cite{Burgeesrev} for cosmological relevance of strings and branes).
In this case the field $\phi$ in the action (\ref{actcur}) 
corresponds to either the dilaton or another modulus.
At tree level the potential of the field $\phi$ vanishes, so  we include the $\alpha'$-order quantum 
corrections of the form:
\be 
\label{Lc}
{\cal L}_c
=a_1R_{ijkl}R^{ijkl}
+a_2R_{ij}R^{ij}
+a_3R^2+a_4(\nabla\phi)^4\,,
\ee
where $a_i$ are coefficients depending on the string model one is 
considering. The Gauss-Bonnet parametrization 
($a_{1}=1$, $a_{2}=-4$ and $a_{3}=1$) corresponds to the ghost-free 
gravitational Lagrangian, which we shall focus on below.
See Ref.~\cite{CTS} for the cosmological dynamics 
in the case of other parametrizations. 
The reader should note that the expansion does not include 
the quantum loop expansion which is governed by the string 
coupling constant, as these have not been fully determined. 
It could well be that these would also play an important role 
in any dynamics, but we have to ignore them for this argument. 

For a massless dilaton field the action (\ref{actcur})
is given by \cite{Gas}
\ba
F = -\zeta =e^{-\phi}, \quad
V = 0, \quad
\xi = \frac{\lambda}{2} e^{-\phi},
\label{tree}
\ea
where $\lambda= 1/4, 1/8$ for the bosonic and heterotic 
string, respectively, whereas $\lambda=0$ in the Type II superstring. 
The choice of $\xi$ corresponds to the tree-level correction.
In general the full contribution of $n$-loop corrections is 
given by $\xi(\phi)=\sum C_n e^{(n-1)\phi}$, 
with coefficients $C_n$. 

Generally moduli fields appear whenever a submanifold of the target 
spacetime is compactified with radii described by the expectation 
values of the moduli themselves. In the case of a single modulus (one common 
characteristic length) and heterotic string ($\lambda=1/8$), 
the four-dimensional action corresponds to \cite{ART}
\ba
\label{modu1}
F = 1, \quad
\zeta = 3/2, \quad a_4 = 0, \quad
\xi = -\frac{\delta}{16}\ln 
[2e^\phi\eta^4(ie^\phi)], \nonumber \\
\label{modu2}
\ea
where $\eta$ is the Dedekind function and $\delta$ is a constant proportional 
to the 4D trace anomaly. $\delta$ depends on the number of chiral, vector, 
and spin-$3/2$ massless supermultiplets of the $N=2$ sector of the theory. 
In general it can be either positive or negative, 
but it is positive for the theories which do not
have too many vector bosons present. 
Again the scalar field corresponds to a flat direction 
in the space of nonequivalent vacua and $V=0$.
At large $\phi$ the last equation can be approximated as 
\ba
\xi_{,\phi} \approx \xi_0 \sin h\phi, \quad
\xi_0 \equiv \frac{1}{24} \pi \delta\,.
\label{ximodu2}
\ea
As shown in Ref.~\cite{Easther96} this is a very good approximation
to the exact expression (\ref{modu1}).

In Ref.~\cite{CTS} cosmological solutions based on the action (\ref{actcur})
without a potential ($V=0$) were discussed in details for three cases--(i) 
fixed scalar field ($\dot{\phi}=0$), (ii)
linear dilaton ($\dot{\phi}={\rm const}$), and (iii) 
logarithmic modulus ($\dot{\phi} \propto 1/t$).
For case (i) we obtain geometrical inflationary solutions only for $D \ne 4$.
Case (ii) leads to pure de-Sitter solutions in the string frame, but 
this corresponds to a contracting universe in the Einstein frame.
These solutions are unrealistic when we apply to dark energy scenarios.
In what follows we shall focus on cosmological solutions 
in case (iii) in four dimensions ($D=4$).
We assume that the dilaton is stabilized 
by some non-perturbative mechanism.

In general the field $\phi$ can be coupled to a barotropic fluid.
We choose the covariant coupling $Q$ introduced in Sec.~\ref{scalingsec}.
Then the energy density $\rho$ of the dark energy fluid satisfies
\ba
\dot{\rho}=\left[-3H(1+w)+Q\dot{\phi}\right]\rho\,.
\ea
We also obtain the equations of motion for the modulus system:
\ba
\label{stdotH}
& & \dot{H}=\frac{2\dot{\rho}+3\dot{\phi}\ddot{\phi}
-48 H^3 \ddot{\xi}}{12H(1+12H \dot{\xi})}\,, \\
\label{stdotphi}
& & \ddot{\phi}=16\frac{{\rm d}\xi}{{\rm d}\phi} 
H^2(H^2+\dot{H})-3H\dot{\phi}
-\frac{2}{3}Q\rho \,.
\ea
Let us search for future asymptotic solutions 
with the following form
\ba
\label{asso1}
\hspace*{-1.0em} H &\sim& \omega_1 t^\beta,\quad
\phi \sim  \phi_0+\omega_2\ln t, \quad
\xi \sim \frac12 \xi_0e^{\phi_0}t^{\omega_2},\\
\label{asso2}
\hspace*{-1.0em} 
\rho &\sim& \rho_0t^{Q\omega_2}\exp\left[-\frac{3(1+w)\omega_1}
{\beta+1}t^{\beta+1}\right],\,\,\,\beta\neq -1,\label{zne1}\\
\label{asso3}
\hspace*{-1.0em}
\rho &\sim& \rho_0t^{\alpha},\,\,\,\beta= -1,\label{asso5}
\ea
where $\omega_1$ and $\omega_2$ are real values of constants, and 
\be\label{Qw}
\alpha \equiv Q\omega_2-3(1+w)\omega_1 \,.
\ee
Substituting Eqs.~(\ref{asso1}), (\ref{asso2}) and (\ref{asso3})
into Eqs.~(\ref{stdotH}) and (\ref{stdotphi}), we can obtain 
a number of asymptotic solutions depending on the regimes 
we are in \cite{CTS}. 
Among them the following two solutions are particularly important. 

\begin{enumerate}
\item \underline{Solution in a low-curvature regime}

The solution which appears in a low-curvature regime
in which the $\xi$ terms 
are subdominant at late times, is characterized by 
\ba \label{grsol}
\beta=-1,\qquad \omega_2<2,
\ea
together with the constraints
\ba
\label{lowome1}
& & \omega_1 = \frac{1}{3}-\frac{2Q \rho_0t^{\alpha+2}}{9\omega_2}, \\
\label{lowome2}
& & 3\omega_2^2 = 12\omega_1^2-4\rho_0t^{\alpha+2}, \\
& &\alpha=Q\omega_2-3(1+w) \leq -2.\label{Qeq}
\ea
This corresponds to the solution `$A_{\infty}$' in Ref.~\cite{ART}
and describes the asymptotic solution of the tree-level system 
($\delta=0$).

\item \underline{Solution in a asymptotically flat-space regime}

This solution appears in a situation where some of the $\xi$ terms 
contribute to the dynamics, and is given by 
\ba
\label{minsol}
& &\beta=-2,\quad \omega_2=5,\quad Q\leq-2/5, \\
& &\omega_1^3
=\frac{1}{24c_1 \xi_{0} e^{\phi_0}}
\left(15-2Q \rho_0t^{5Q+2}\right),
\ea
for a non-vanishing fluid.
We note that this is different from the 
high-curvature solution in which the $\xi$ terms completely 
dominate the dynamics \cite{CTS}.
The solution corresponds to `$C_{\infty}$' in Ref.~\cite{ART}
and describes an asymptotically flat universe with slowly 
expanding (or contracting) scale factor.
In fact an expanding solution is given by $a(t)\sim 
a_0\exp(-\omega_1/t)$, 
which exhibits superinflation as $t \to -0$.

\end{enumerate}

These solutions can be joined to each other
if the coupling constant $\delta$ is negative \cite{ART}.
There exists an exact solution for Eqs.~(\ref{stdotH}) 
and (\ref{stdotphi}), but this is found to 
be unstable in numerical simulations of Ref.~\cite{CTS}.
In the asymptotic future the solutions tend to approach the 
low-curvature one given by Eq.~(\ref{grsol}) rather than 
the others, irrespective of the sign of 
the modulus-to-curvature coupling $\delta$.

Let us consider the case in which a phantom fluid ($w<-1$)
is present together with the modulus string corrections.
Equation (\ref{Qeq}) shows that the condition for the existence 
of the low-curvature solution (\ref{grsol}) is not satisfied 
for $Q=0$ and $w<-1$. However the presence of the coupling 
$Q$ can fulfill this condition. This suggests that the Big Rip
singularity may be avoided when the modulus field $\phi$
is coupled to dark energy.

\begin{figure}[htb]
\begin{center}
\includegraphics[height=3.1in,width=3.2in]{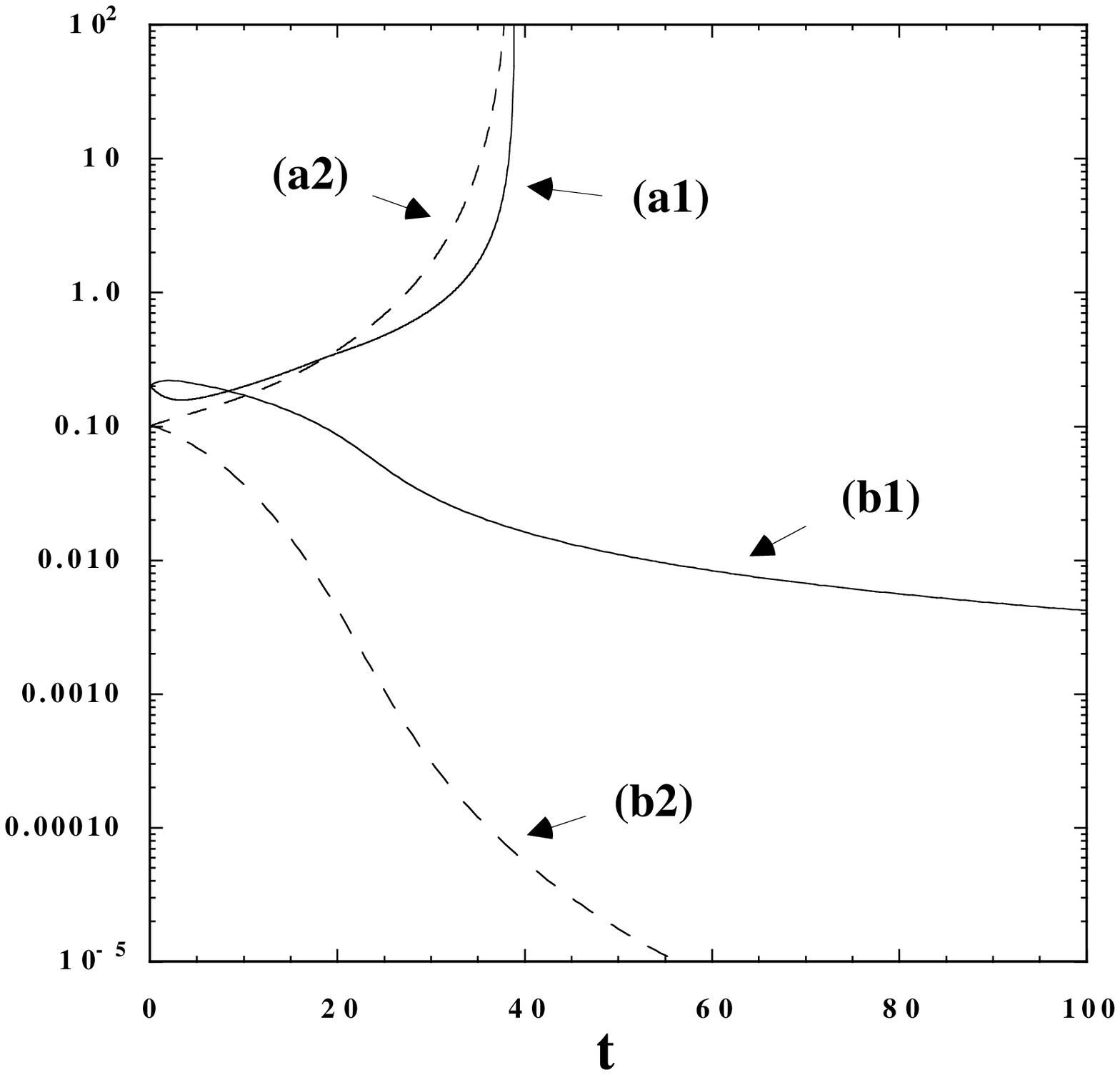}    
\caption{
Evolution of $H$ and $\rho$ with $\xi_{0}=-2$, $w=-1.1$ for (a) 
$Q=0$ and (b) $Q=-5$. We choose initial conditions as $H_{i}=0.2$, 
$\phi_{i}=2.0$ and $\rho_{i}=0.1$.
The curves (a1) and (b1) represent the evolution of $H$ for
$Q=0$ and $Q=-5$, respectively, while the curves (a2) and (b2)
show the evolution of $\rho$ for corresponding $Q$.}
\label{modulus}
\end{center}
\end{figure}

In Fig.~\ref{modulus} we show the evolution of $H$ and $\rho$
with negative $\delta$
for (a) $Q=0$ and (b) $Q=-5$ in the presence of the phantom fluid
with $w=-1.1$.
Although the solution approaches a Big Rip singularity for $Q=0$, 
this singularity is actually avoided for $Q=-5$. 
In the latter case the universe approaches the low-curvature
solution given by Eq.~(\ref{grsol}) at late times.
Since the asymptotic values of $\omega_1$ and $\omega_2$
are $\omega_{1}=1/3$ and $\omega_{2}=2/3$
from Eqs.~(\ref{lowome1}) and (\ref{lowome2}), 
the condition (\ref{Qeq}) for the existence of low-curvature 
solution is $Q<3(w-1)/2=-3.15$.
Numerical calculations show that the Big Rip singularity 
can be avoided for a wide range of initial 
conditions \cite{CTS}.

When $Q>0$ the condition (\ref{Qeq}) is not satisfied
for $\omega_{2}>0$. 
However it is numerically found that  
the system approaches the low-energy regime  
characterized by $\omega_{1}=1/3$ and 
$\omega_{2}=-2/3$ \cite{CTS}.
This negative value of $\omega_{2}$ means that 
the Big Rip singularity may be avoided even for positive $Q$.
In fact $H$ and $\rho$ decrease when the condition (\ref{Qeq})
is satisfied in the asymptotic regime.

When $\delta>0$, there is another interesting circumstance in 
which the Hubble rate decreases but the 
energy density of the fluid increases \cite{CTS}. 
This corresponds to the solution
in which the growing energy density $\rho$
can balance with the GB term ($\rho \approx 24H^3\dot{\xi}$ in 
the Friedmann equation). 
Hence the Big Rip singularity does not appear 
even when $w<-1$ and $Q=0$. 

The above discussion shows that for a restricted class of 
modulus-type string corrections there exists the possibility of avoiding 
the Big Rip singularity. We also note that recent development of 
loop quantum cosmology allows us to avoid several future singularities
discussed in Sec.~\ref{fate} \cite{SST06}.

\section{Cosmic acceleration from modified gravity and other alternatives to
dark energy}
\label{modified}

The contribution of the matter content of the universe is represented by the energy
momentum tensor on the right hand side of Einstein equations, 
whereas the left hand side is represented by pure geometry.
There are then two ways to give rise to 
an accelerated expansion: (i) either by supplementing the energy 
momentum tensor by an exotic form of matter such as a cosmological 
constant or scalar field;
 (ii) by modifying the geometry itself. 
The geometrical modifications 
can arise from quantum effects such as 
higher curvature corrections to the Einstein Hilbert action. 
In the previous section we have used such curvature corrections
to avoid future singularities in the presence of a dark energy fluid.
In this section we are interested in whether it is possible to 
obtain an accelerated expansion driven by geometrical terms alone.

It is well known that the quadratic term in $R$ leads to an 
inflationary solution in the early universe \cite{Stainf}.
In this model the effective potential in the Einstein frame vanishes
at a potential minimum, in which case we can not have 
a late time accelerated expansion of the universe.
However, it was pointed out in Refs.~\cite{Capo,Carroll:2003wy}
that late time acceleration can be realized by terms containing inverse powers of
the Ricci scalar added to Einstein Hilbert action\footnote{We note 
that inflationary solutions in such cosmological models were 
already studied in 1993 in Ref.~\cite{CapoLuca}.}.
However the original model (${\cal L} \propto 1/R$)
is not compatible with solar system 
experiments \cite{Chiba03} and possess 
instabilities \cite{DK03,woodard} 
(see Refs.~\cite{odintsovl,woodard1} for recent reviews).
It was argued by Nojiri and Odintsov \cite{NOmo1}
that the situation
could be remedied by adding a counterterm term proportional 
to $R^2$ in the action (see also Refs.~\cite{NOmo2}, however also see \cite{Navarro:2005ux} for a different take on the problem). 

Another interesting approach which can avoid the above mentioned problem is
provided by Palatini formalism \cite{Pala1,Pala2,Pala3,Vollick2,Pala4}. 
The Palatini formalism leads to differential equations of second order
even in presence of non-linear terms in $R$ in the gravitational action
and is free from the problem of instabilities \cite{Pala1,Pala2}.
A variety of different aspects of $f(R)$ gravity and 
associated cosmological dynamics
is discussed in Ref.~\cite{more}. 
An interesting possibily of obtaining
late time acceleration from modified Gauss-Bonnet gravity is discussed
in Ref.~\cite{f(G)}.

The other exiting possibility of obtaining accelerated expansion is provided
by theories with large extra dimensions known as braneworlds. 
Being inspired by string theory, our four dimensional spacetime ({\it brane}) 
is assumed to be embedded in a higher dimensional {\it bulk} spacetime. 
In these scenarios all matter fields are confined 
on the brane whereas gravity being
a true universal interaction can propagate into the anti de Sitter bulk. 
In Randall-Sundrum (RS) braneworld \cite{Randall}
the Einstein equations are modified by high energy corrections 
\cite{Df00,Shiromizu},
but this modification is generally not thought to be important for late-time 
cosmology (However, see 
\cite{steep} for interesting possibilities).
The situation is reversed in the braneworld model of Dvali-Gabadadze-Porrati (DGP) 
 \cite{DGP} in which the brane is 
embedded in a {\it Minkowski} bulk.
They differ from the RS
brane world by a curvature term on the brane 
(see Ref.~\cite{royrev} for review and Refs.~\cite{DGPrelated}
for related works). 
Unlike the RS scenario, in DGP braneworld, gravity remains
four dimensional at short distances but can leak into the bulk 
at large distances leading to infrared modifications to Einstein gravity.
In the DGP model there is a cross-over scale
around which gravity manifests these higher-dimensional properties.
This scenario is a simple one parameter model which can account for the
current acceleration of the universe provided the cross-over scale
is fine tuned to match observations.

In this section we shall briefly describe these 
two approaches for obtaining the current
acceleration of the universe from modified theories of gravity.

\subsection{$f(R)$ gravities}

Let us start with an action \cite{Capo,Carroll:2003wy}
\be
S=\int \rd^4 x\sqrt{-g}\,f(R)\,,
\label{fgravity}
\ee
where $f(R)$ is an arbitrary function in terms of $R$.
By varying the action (\ref{fgravity}) with respect to the metric leads to the following 
field equations
\bea
G_{\mu \nu}&=&\left(\frac{\partial f}{\partial R}\right)^{-1}
\Biggl[ \frac12 g_{\mu \nu} \left(f-\frac{\partial f}{\partial R}R
\right) \nonumber \\
& &+\left\{ \nabla_{\mu} \nabla_{\nu} \frac{\partial f}{\partial R}
-\Box \left( \frac{\partial f}{\partial R}\right) g_{\mu \nu}
\right\} \Biggr]\,,
\label{eqfgr1}
\eea
where $G_{\mu \nu}$ is an Einstein tensor.

Equation (\ref{eqfgr1}) looks complicated but can acquire a simple form 
after a conformal transformation 
\bea
g_{\mu \nu}^{(E)}=e^{2\omega}g_{\mu \nu}\,,
\label{confor}
\eea
where $w$ is a smooth and positive function of space time 
coordinates. Here `$E$' denotes the metric in the Einstein frame.
{}From Eqs.~(\ref{eqfgr1}) and (\ref{confor}) we find that 
the Einstein tensor in the $g_{\mu \nu}^{(E)}$ metric 
can be written as \cite{Maeda89}
\bea
G_{\mu \nu}^{(E)}&=&\left(\frac{\partial f}{\partial R}\right)^{-1}
\Biggl[ \frac12 g_{\mu \nu} \left(f-\frac{\partial f}{\partial R}R
\right) \nonumber \\
& &+\left\{ \nabla_{\mu} \nabla_{\nu} \frac{\partial f}{\partial R}
-\Box \left( \frac{\partial f}{\partial R}\right) g_{\mu \nu}
\right\} \Biggr]\,,
\nonumber \\
& &-2(\nabla_{\mu} \nabla_{\nu} \omega-\Box \omega g_{\mu \nu})
+2 \nabla_{\mu} \omega \nabla_{\nu} \omega+(\nabla \omega)^2
g_{\mu \nu}. \nonumber \\
\label{eqfgr2}
\eea

If we choose the conformal factor of the form 
\bea
2\omega={\rm ln} \left[ 2\kappa^2 
\left| \frac{\partial f}{\partial R} \right| \right]\,,
\eea
we find that the term $(\partial f/\partial R)^{-1}
\nabla_{\mu} \nabla_{\nu} (\partial f/\partial R)$
cancels with the term $-2\nabla_{\mu} \nabla_{\nu} \omega$
in Eq.~(\ref{eqfgr2}).
In this case $\omega$ behaves like a scalar field $\phi$, 
which is defined by  
\bea
\kappa \phi \equiv \sqrt{6}\omega=\frac{\sqrt{6}}{2}
{\rm ln} \left[ 2\kappa^2 
\left| \frac{\partial f}{\partial R} \right| \right]\,.
\label{phirelation}
\eea
Then the action in the Einstein frame is given by 
$S_E=\int \rd^4 x\sqrt{-g_{E}} {\cal L}$ with 
Lagrangian density
\bea
{\cal L}=\frac{1}{2\kappa^2}R(g_{E})-
\frac12 (\nabla_{E} \phi)^2-U(\phi)\,,
\eea
where 
\bea
\label{effpotential}
U(\phi)=({\rm sign}) e^{-\frac{2\sqrt{6}}{3}
\kappa \phi}  \left[ \frac{({\rm sign})}{2\kappa^2}R
e^{\frac{\sqrt{6}}{3} \kappa \phi}-f\right]\,,
\eea
and $({\rm sign})=(\partial f/\partial R)/|\partial f/\partial R|$.

We now consider the modified gravity action given 
by \cite{Capo,Carroll:2003wy}
\bea
f(R)=\frac{1}{2\kappa^2}
\left[R-\mu^{2(n+1)}/R^n
\right] \,,\quad n>0\,,
\label{frn}
\eea
where $\mu$ is a parameter with units of mass.
{}From Eq.~(\ref{effpotential}) the effective potential in 
Einstein frame is 
\bea
U(\phi)=\mu^2 M_{\rm pl}^2 \frac{n+1}{2n} n^{\frac{1}{n+1}}
e^{-\frac{2\sqrt{6}}{3}\kappa \phi}
(e^{\frac{\sqrt{6}}{3}\kappa \phi}-1)^{\frac{n}{n+1}}\,,
\nonumber \\
\eea
where we used the relation (\ref{phirelation}).
This potential has a maximum at $\kappa \phi=2(n+1)/(n+2)$
and has the following asymptotic form:
\bea
\label{expapp}
U(\phi) \propto \exp \left(-\frac{\sqrt{6}\kappa}{3}
\frac{n+2}{n+1} \phi \right)\,,\quad \kappa\phi \gg 1\,.
\eea

Taking note that the potential (\ref{exp}) leads to 
the power-law expansion (\ref{plaw}), we find that 
the evolution of the scale factor in the Einstein frame is given by 
\bea
a_{E} \propto t_{E}^p\,,\quad
p=\frac{3(n+1)^2}{(n+2)^2}\,.
\eea
When $n=1$ one has $p=4/3$, which corresponds to 
an accelerated expansion.
The power-law index $p$ increases for larger $n$
and asymptotically approaches $p \to 3$
as $n \to \infty$.

We note that scale factor $a$ and cosmic time $t$
in the Jordan frame are related to those in Einstein 
frame via the relation $a=e^{-\kappa \phi/\sqrt{6}}a_E$
and $\rd t=e^{-\kappa \phi/\sqrt{6}}\rd t_{E}$.
Since the field $\phi$ is given by 
$\kappa \rd\phi/\rd t_{E}=\sqrt{2p}/t_E$ for the 
potential (\ref{expapp}), we find that the evolution of
scale factor in the Jordan frame is 
\bea
a \propto t^q\,,\quad
q=\frac{(2n+1)(n+1)}{(n+2)}\,.
\eea
{}From Eq.~(\ref{sol2}) this corresponds to the 
effective equation of state:
\bea
w_{\rm DE}=-1+\frac{2(n+2)}{3(2n+1)(n+1)}\,.
\eea
When $n=1$ we have $q=2$ and $w_{\rm DE}=-2/3$.
WMAP in concert with other observations have really 
begun to constrain the current value of the dark energy 
equation of state, although it does depend on the priors. 
For instance, in a flat universe, the combination of WMAP3 
and the Supernova Legacy Survey (SNLS) gives 
$w_{\rm DE}=-0.97^{+0.07}_{-0.09}$, whereas 
even if we do not include a prior of a flat universe, 
then by combining WMAP3 with large scale structure 
and supernova data we obtain $w_{\rm DE}=-1.06^{+0.13}_{-0.08}$ 
at the $2\sigma$ level \cite{WMAP3}. It follows that the $n=1$ 
case is outside of observational bounds.
However the model is compatible with observations
when $n \ge 2$. We note that the $n \to \infty$ limit 
corresponds to the equation of state of cosmological 
constant ($w_{\rm DE}=-1$). 
The effects of modification
should become important only at late times, 
which requires the tuning of the energy scale $\mu$. 

It was pointed out by Chiba \cite{Chiba03} that 
theories of the type (\ref{frn}) give the Brans-Dicke parameter
$\omega_{\rm BD}=0$, which contradicts with the constraint of solar-system 
experiments ($\omega_{\rm BD}>40000$ \cite{Bertotti03}).
This means that the field $\phi$ couples to matter with a comparable
strength as gravity.
Dolgov and Kawasaki \cite{DK03} showed that a non-linear gravitational 
action (\ref{frn}) suffers from serious instabilities 
which lead to a dramatic change of gravitational fields associated 
with any gravitational bodies.
Nojiri and Odintsov \cite{NOmo1} have argued 
that this problem can be alleviated by adding a counter term 
$R^n$ to the modified action with appropriate coefficients. However, 
there is some debate as to whether this can actually work. 
In \cite{Navarro:2005ux}, Navarro and Acoleyen argue that 
for this mechanism to work, it relies on a particular value for 
the background scalar curvature and that if it deviates from 
this background value, as will happen in the Solar System, 
the mass of this scalar field decreases again to a value $m \sim H$, 
hence we would observe corrections to Einstein gravity. 

In Ref.~\cite{APT} the authors show that in all $f(R)$ theories that 
behave as a power of $R$ at large or small $R$ the scale factor 
during the matter dominated stage evolves as $a \propto t^{1/2}$
instead of $a \propto t^{2/3}$, except for Einstein gravity
(see also Ref.~\cite{APT2}). 
This means that these cases are incompatible with cosmological 
observations such as WMAP.
The absence of the standard matter dominated era also holds for 
the model given by (\ref{frn}). It would be of interest 
to find $f(R)$ dark energy models in which a matter
dominated epoch exists before the late-time acceleration.

Another interesting way to tackle the problem
is provided by the so-called Palatini formalism \cite{Pala1,Pala2}.
In this formalism the action is varied with respect to the metric
and connection by treating them as independent field variables.
In the case of the Einstein Hilbert action this method leads to the same
field equations as the one derived from a standard variation principle.
However when the action includes nonlinear functions of the 
Ricci scalar $R$, the two approaches give different field equations.
An important point is that the Palatini formalism provides 
second-order field equations, which are free from the instability 
problem mentioned above.
It was pointed out by Flanagan \cite{Pala3} that 
even in the Palatini formalism matter fields of the standard model at an 
energy scale of order $10^{-3}$\, eV can have interactions, thus the 
model (\ref{frn}) may be excluded by particle physics 
experiments. This is based on the argument that minimally coupled 
fermions are included in the Jordan frame and that transforming to 
Einstein frame induces additional interactions between matter fields.
Vollick \cite{Vollick2} argued that the equivalence between 
the two frames discussed
by Flanagan is not physical but mathematical.
The physical interpretation of the difference of the frames is 
a thorny subject, which we will not enter in detail.
Setting these subtleties aside we shall proceed with the discussion 
of the observational constraints on $f(R)$ gravity theories.

Amarzguioui {\it et al.} \cite{Aobs1}
tried to place constraints on $f(R)$ gravity models 
with the Palatini formalism using
several observational data sets (see also Ref.~\cite{MSW05}).
They parameterized the gravity Lagrangian of the form 
\bea
f(R)=R \left[ 1+\alpha 
\left(-\frac{R}{H_0^2} 
\right)^{\beta-1} \right]\,,
\eea
where $\alpha$ and $\beta$ are dimensionless constants.
Using the combined analysis of SN Ia, CMB, baryon oscillations and 
LSS data sets, the best fit values of the model parameters were
found to be $(\alpha, \beta)=(-3.6, 0.09)$. 
As is clearly seen in Fig.~\ref{palatini}, 
the $\beta=-1$ case is ruled out observationally. 
The allowed values of $\beta$ exist in the narrow 
range: $|\beta|<0.2$.
Hence $f(R)$ gravity models do not exhibit
any significant observational preference 
compared to the GR case ($\beta=0$).

\begin{figure}
\includegraphics[height=3.0in,width=3.5in]{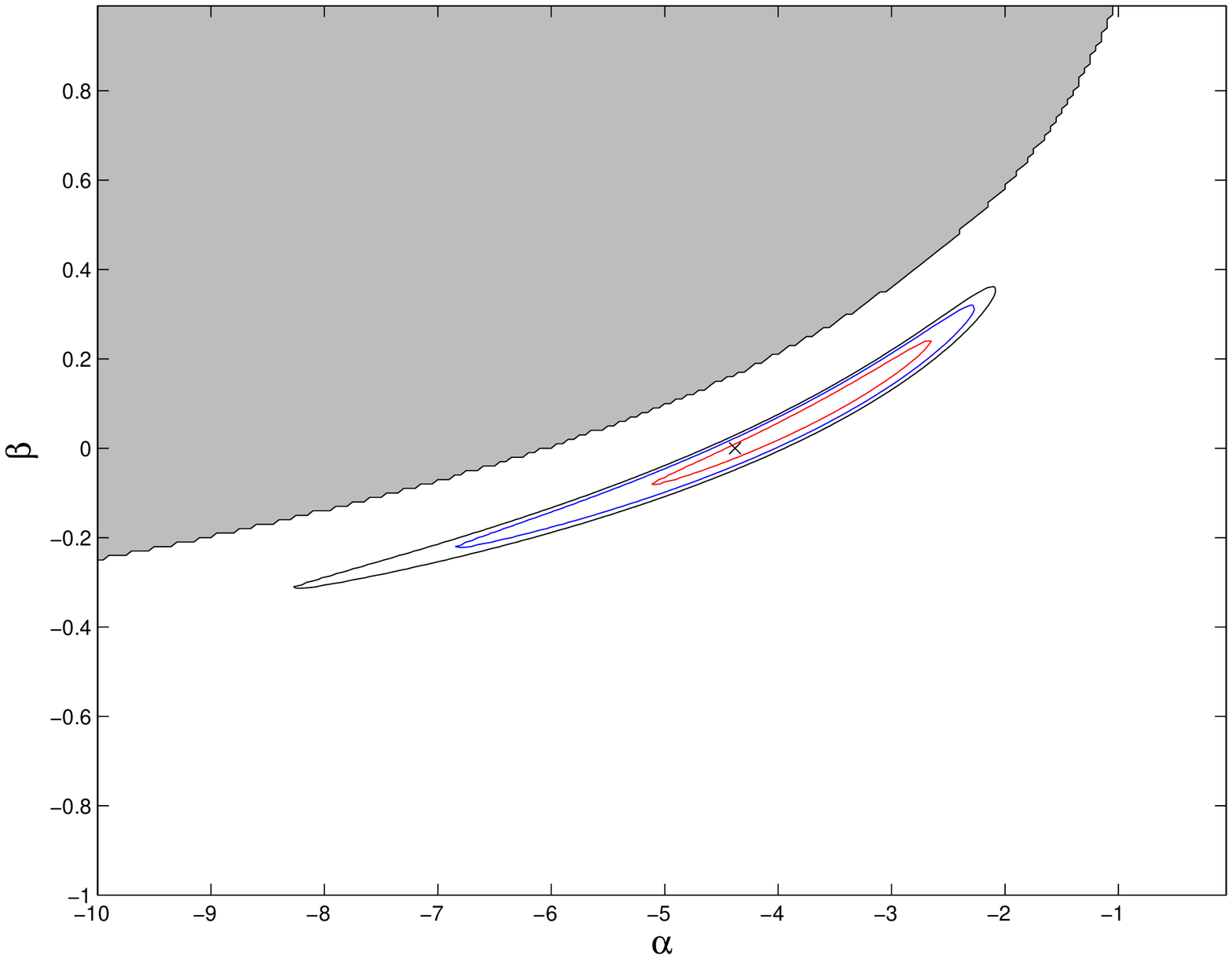}
\caption{The $1\sigma$, 
$2\sigma$ and $3\sigma$ contours drawn using the compilation of 
data sets from SN Ia, CMB, baryon oscillations and LSS observations. 
The $\Lambda$CDM model ($\alpha=-4.38$, $\beta=0$)
is marked with the cross. The grey part represents the region 
which is excluded observationally.
{}From Ref.~\cite{Aobs1}.}
\label{palatini}
\end{figure}

%
\subsection{DGP model}

Let us now discuss a DGP braneworld 
model \cite{DGP,Dvalimore} which 
can also lead to an accelerated expansion.
We consider a brane embedded in a 5-dimensional 
Minkowski bulk described by the action
\bea 
S&=&-\frac{M_{5}^3}{2}\int \rd^5 X 
\sqrt{-g}R_{5}-\frac{M_{\rm pl}^2}{2}
\int \rd^4 x\sqrt{-h}R_{4} \nonumber \\ 
& &+\int \rd^4 x\sqrt{-h} {\cal L}_{m}
+S_{\rm GH}\,,
\label{DGPaction}
\eea
where $ g_{ab}$ is the metric in the bulk and $h_{\mu \nu}$ is the 
induced metric on the brane. ${\cal L}_m$
is the matter Lagrangian confined to the brane. 
The second term containing the 4-dimensional Ricci scalar on the brane
is an extra piece appearing in the DGP model in contrast to the RS scenario.
Such a term can be induced by quantum effects in the matter sector
on the brane. 
The last term $S_{\rm GH}$ is a Gibbons-Hawking boundary term 
necessary for the consistency of the variational procedure and leads to the Israel 
junction conditions.

The ratio of the two scales, namely, the 4-dimensional Planck mass 
$M_{\rm pl}$ and its counter part $M_{5}$ in the 5-dimensional bulk, 
defines a cross over scale
\bea
r_c=\frac{M_{\rm pl}^2}{2M_5^3} \,.
\eea
For characteristic length scales much smaller than $r_c$,
gravity manifests itself as four dimensional theory whereas at 
large distances it leaks into the bulk making the higher dimensional
effects important. Across the crossover
scale $r_c$, the weak-field gravitational potential behaves as
\begin{equation}
\Phi \sim \left\{ \begin{array}{lll} r^{-1} & \mbox{for} & r< r_c\,,
\\ r^{-2} & \mbox{for} & r> r_c\,.
\end{array}\right.
\end{equation}
We are interested in a situation in which the cross over occurs
around the present epoch. In this case $r_{c}$ is the same order as the 
present Hubble radius $H_{0}^{-1}$, which corresponds to 
the choice $M_5=10$-$100$\,MeV.

In the FRW brane characterized by the metric (\ref{frwmet})
we obtain the following modified 
Hubble equation \cite{Deffayet}
\bea
H^2+\frac{K}{a^2}=\left(\sqrt{\frac{\rho}{3M_{\rm pl}^2}
+\frac{1}{4r_c^2}}+ \frac{1}{2r_{c}}\right)^2\,,
\label{DGPHeq}
\eea
where $\rho$ is the total cosmic fluid 
energy density on the brane which satisfies
the standard conservation equation (\ref{conteq}).
For a flat geometry ($K=0$) we find that Eq.~(\ref{DGPHeq})
reduces to 
\bea
\label{mofrw}
H^2-\frac{\epsilon}{r_c}H=\frac{\rho}{3M_{\rm pl}^2}\,,
\eea
where $\epsilon =\pm 1$.
When the Hubble length $H^{-1}$ is much smaller 
than the distance scale $r_{c}$, i.e. $H^{-1} \ll r_{c}$, 
the second term on the left hand side of Eq.~(\ref{mofrw}) is negligible 
relative to the first term, thus giving the Friedmann equation,
$H^2=\rho/3M_{\rm pl}^2$.
The second term in Eq.~(\ref{mofrw}) 
becomes important on scales comparable to the cross-over 
scale ($H^{-1} \gtrsim r_{c}$).
Depending on the sign of $\epsilon$ we have two different regimes
of the DGP model.
When $\epsilon=+1$, Eq.~(\ref{mofrw}) shows that in a 
CDM dominated situation characterized by $\rho \propto a^{-3}$
the universe approaches the de Sitter solution 
\bea
H \to H_{\infty}=\frac{1}{r_{c}}\,.
\eea
Thus we can have an accelerated expansion at late times
without invoking dark energy.
In order to explain acceleration now we require that 
$H_{0}$ is of order $H_{\infty}$, which means that the 
cross-over scale approximately corresponds to the present 
Hubble radius ($r_c \sim H_0^{-1}$).
We stress here that this phenomenon arises in DGP from the 
the gravity leakage at late times. In other words it is not due to the presence of a
negative pressure fluid but rather to the weakening of gravity on
the brane.

When $\epsilon=-1$ and $H^{-1} \gg r_{c}$
the second term in Eq.~(\ref{mofrw}) dominates over 
the first one, which gives 
\bea
H^2=\frac{\rho^2}{36 M_{5}^6}\,.
\eea
This is similar to the modified FRW equations in RS cosmology 
at high energy. However this does not give rise to an accelerated
expansion unless we introduce dark energy on the brane.
Hence in what follows we shall concentrate on the case 
of positive $\epsilon$.

The FRW equation (\ref{mofrw}) can be 
written in the form
\bea
\label{Hparame}
H(z)=H_{0} \left[\sqrt{\Omega_{r_{c}}^{(0)}}+
\sqrt{\Omega_{r_{c}}^{(0)}+\Omega_m^{(0)}
(1+z)^3}\right]\,,
\eea
where $\Omega_m^{(0)}$ is the matter density parameter and 
\bea
\Omega_{r_{c}}^{(0)} \equiv 
\frac{1}{4r_c^2 H_{0}^2}\,.
\eea
Setting $z=0$ in Eq.~(\ref{Hparame}), we get the normalization 
condition 
\bea
\label{conDGP}
\Omega_{r_{c}}^{(0)}=\frac{(1-\Omega_m^{(0)})^2}{4}\,.
\eea

Deffayet {\it et al.} \cite{Deffa02} placed observational 
constraints coming from SN Ia and CMB (WMAP1) data sets.
When only SN Ia data \cite{perlmutter} is used in likelihood analysis, 
the best fit values were found to be 
$\Omega_{m}^{(0)}=0.18^{+0.07}_{-0.06}$ and 
$\Omega_{r_{c}}^{(0)}=0.17^{+0.03}_{-0.02}$.
If we include CMB data sets \cite{TegZa}, it was shown in
Ref.~\cite{Deffa02} that larger values of $\Omega_m^{(0)}$
are allowed. In particular a concordance model with $\Omega_m^{(0)}=0.3$
is consistent with both SN Ia and CMB (WMAP1)data sets.
The cross-over scale was constrained to be $r_{c} \sim 1.4 H_0^{-1}$.
We caution that the analysis in Ref.~\cite{Deffa02}
made use of the observational data before the WMAP1 data 
appeared.
Updated observational constraints on the DGP model have been 
carried out by a number of authors \cite{SS,DGPobser}.

Recently, in Ref.~\cite{Sawicki:2005cc}, Sawicki and Carroll 
looked at the evolution of cosmological perturbations on large
scales in the DGP model. 
They found that at late times, perturbations enter a DGP regime 
with an increase in the effective value of Newton's constant 
because  the background density diminishes. 
This in turn leads to a suppression of the integrated 
Sachs-Wolfe effect, which has the effect of making the 
DGP gravity fit the WMAP1 data better than conventional 
$\Lambda$CDM. This conclusion has been questioned in \cite{KoyamaR} 
where it is argued that the authors of \cite{Sawicki:2005cc} are using 
an inconsistent assumption for the truncation of the 5D perturbations.
More precisely, their ansatz leads to the breakdown of the 4D Bianchi
identity, making their results for the suppressions of the integrated 
Sachs-Wolfe effect as being unreliable.

In \cite{Sawicki:2005cc}, the authors also found a significantly worse 
fit to supernova data and the distance to the last-scattering 
surface in the pure DGP model as compared to the 
$\Lambda$CDM model, concluding that $\Lambda$CDM 
overall provides the best fit. A similar conclusion appears 
to be reached in \cite{Fairbairn} and \cite{AlamS}, 
where the two groups have also tried to constrain the
DGP model using SN Ia data
and the baryon acoustic peak in the Sloan Digital Sky Survey.
In Fig.~\ref{Malcolm} we show observational contour bounds
together with the constraint relation (\ref{conDGP}) in a flat
universe in the DGP model.
This was obtained by using Supernova Legacy Survey (SNLS) data \cite{Astier}
and recent results of baryon oscillations \cite{Eis}, which shows 
that the original DGP model discussed above is ruled out at 3$\sigma$
level \cite{Fairbairn}. The analysis of Ref.~\cite{AlamS}
which is the combined anlysis of SN Ia Gold data \cite{riess2} and baryon 
oscillations \cite{Eis}, for a spatially flat cosmology $(K=0)$,
shows that the model is allowed at the $2\sigma$ level. 
Figure \ref{Royfig} shows the analysis of SN Ia Gold 
data \cite{riess2} and baryon oscillations \cite{Eis} 
with $\Omega_K$ being varied \cite{Roycommun}.
The flat DGP model is marginally on the border of the $2\sigma$
contour bound. Clearly the results are sensitive
to which SN Ia data are used in the analysis. 
Thus SN Ia observations alone are not yet reliable enough to reach
a definite conclusion.

Both the $\Lambda$CDM and DGP models can describe the 
current acceleration of the universe
provided that $\Lambda \sim H^2_0$ and $r_c \sim H^{-1}_0$. 
The degeneracy can be broken using LSS data as the two models
predict different evolution of density perturbations. The comprehensive treatment
of perturbations in DGP model is still an open problem.
A possible solution to this problem and its future 
perspectives were discussed in details in Ref.~\cite{KoyamaR}.
We should also mention that apart from the fine tuning of 
the cross-over scale, the DGP model is plagued with
an instability problem related to ghosts
and strong couplings. 
Thus the model deserves further 
investigations perhaps along the lines suggested in
Ref.~\cite{Koyamaper}.

Finally we should mention that a generalization of the DGP model 
was proposed in Ref.~\cite{SS}. The model contains
additional free parameters but exhibits an intrguing possibility 
of transient phantom phase in the presence of 
a non-zero cosmological constant on the brane \cite{SS,Starkman}.

\begin{figure}
\includegraphics[height=3.3in,width=3.4in]{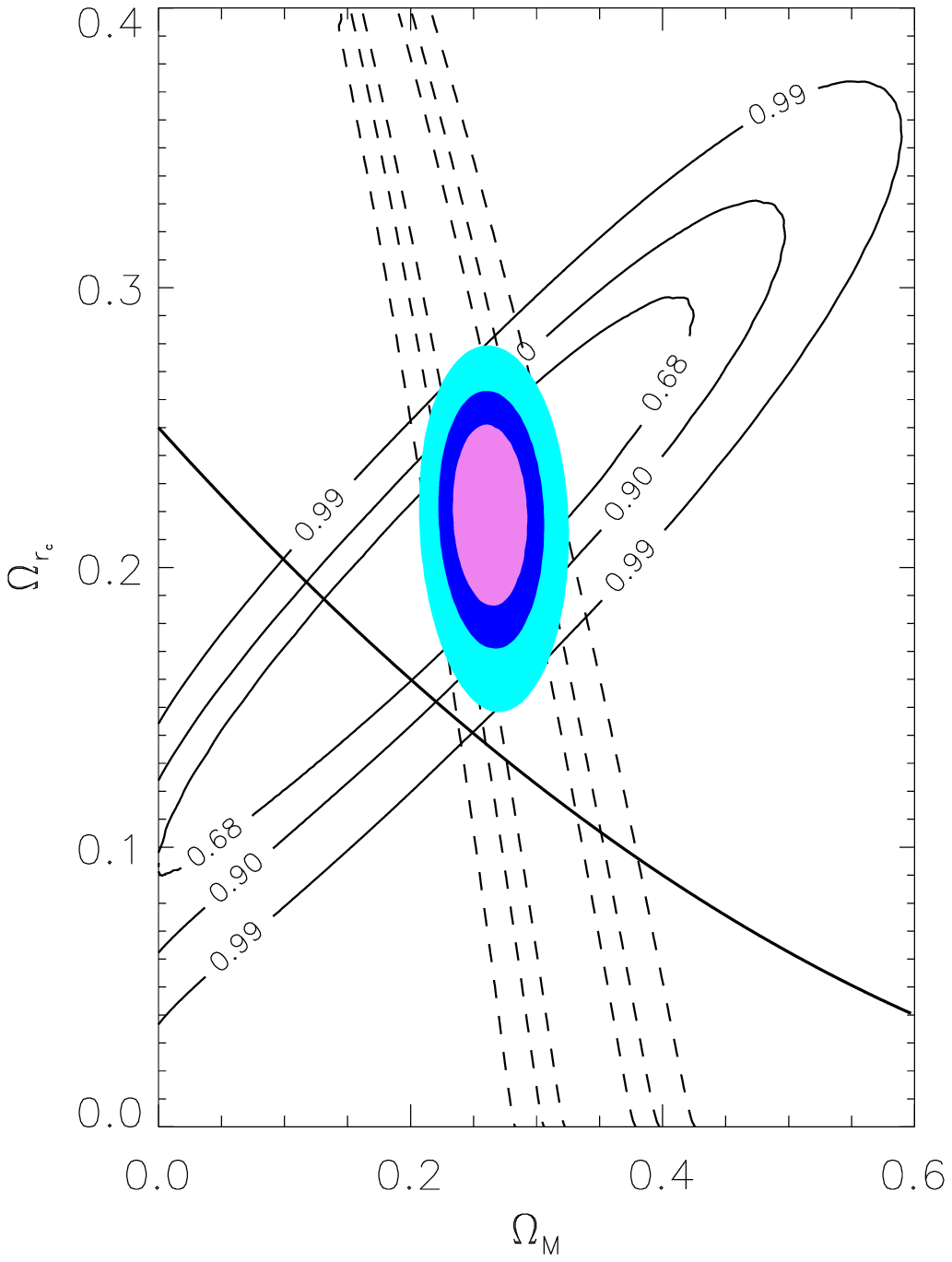}
\caption{The allowed parameter space in the 
$\Omega_m^{(0)}$-$\Omega_{r_c}^{(0)}$ plane
in the DGP braneworld model with $\Omega_K=0$ 
from a combined analysis of 
the first year SNLS data \cite{Astier} and the 
baryon oscillation data \cite{Eis}. 
The thick solid line shows the constraint relation 
(\ref{conDGP}) in a flat universe.
The solid thin contours correspond to the
allowed parameter regions 
at the $1\sigma$, $2\sigma$ and $3\sigma$
confidence levels coming from
the SNLS data.
The dashed lines represent the corresponding regions 
from the baryon oscillation peak. 
The colored contours
show the result of the combination of both data-sets. 
{}From Ref.~\cite{Fairbairn}.
}
\label{Malcolm}
\end{figure}

\begin{figure}
\includegraphics[height=3.3in,width=3.4in]{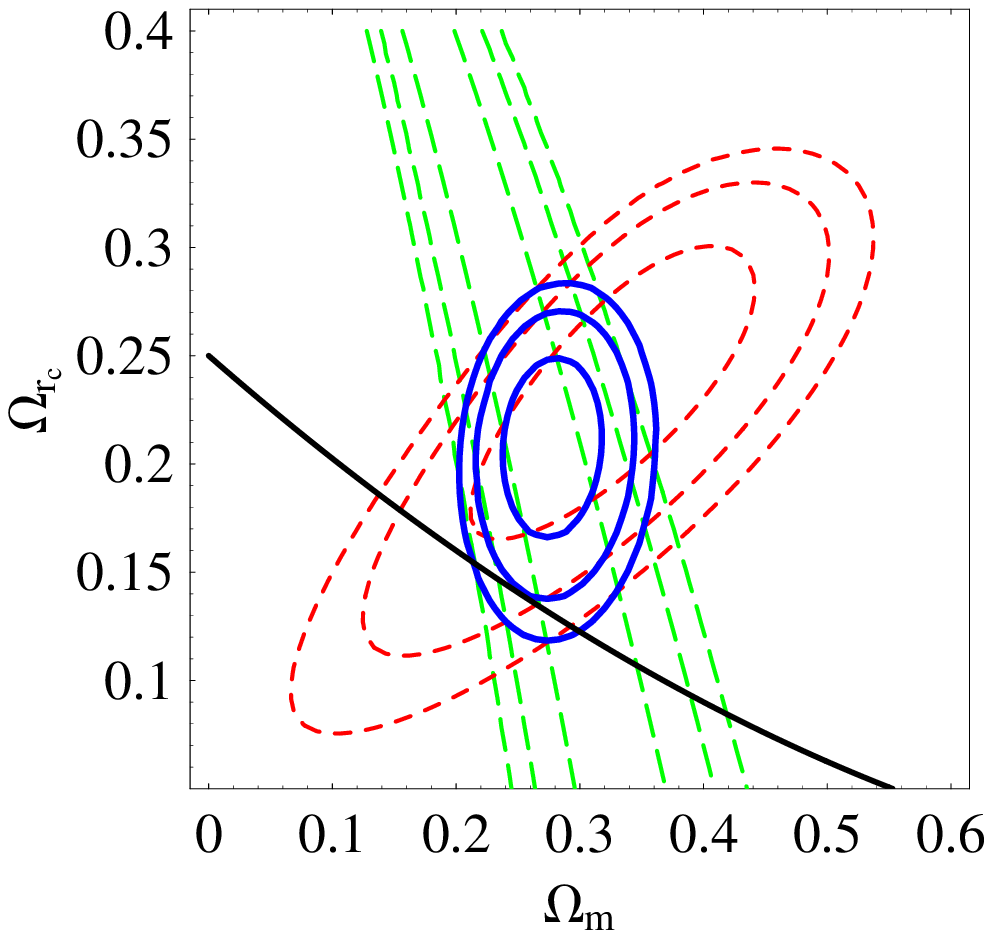}
\caption{The allowed parameter space in the 
$\Omega_m^{(0)}$-$\Omega_{r_c}^{(0)}$ plane
in DGP braneworld model from a combined analysis of 
the SN Ia Gold data set \cite{riess} and the 
baryon oscillation data \cite{Eis}.
In this figure $\Omega_{K}$ is varied in the likelihod analysis.
{}From Ref.~\cite{Roycommun}.} 
\label{Royfig}
\end{figure}

%
\subsection{Dark energy arising from the Trans-Planckian Regime}
A novel approach to addressing the issue of the origin of the 
dark energy is to link it to another unknown, that of 
the transplanckian regime, or what are the observable affects 
of physics occurring in the early Universe on length scales 
below the Planck scale, or energies well above the Planck 
scale? In \cite{Mersini-Houghton:2001su} the authors 
model the transplanckian regime by replacing the usual 
linear dispersion relation $w^2(k) = k^2$ with a one-parameter 
family of smooth non-linear dispersion relations 
which modify the dispersion relation at very short distances. 
In particular motivated by arguments from superstring 
duality (see \cite{Bastero-Gil:2001nu} for a justification 
of the argument), they choose an Epstein function
\be
w^2(k) = k^2 \left( {\epsilon_1 \over 1+e^x} + 
{\epsilon_2 e^x \over 1+e^x} + {\epsilon_3 e^x 
\over (1+e^x)^2} \right) \,,
\label{modify-dispersion}
\ee
where $x=(k/k_c)^{1/\beta}$.  
$\beta$ is the constant determining the rate 
of expansion in the inflating universe given 
by $a(\eta) \propto |\eta|^{-\beta}$ where the scale factor 
is evaluated in conformal time, and $k_c$ is the 
wavenumber where the frequency reaches a maximum. 
The constants satisfy $\epsilon_2 =0$, 
$\epsilon_1/2 + \epsilon_3/4 =1$, giving a one parameter 
(say $\epsilon_1$) family of 
functions \cite{Mersini-Houghton:2001su}. 

A particular feature of the family of dispersion functions 
they choose is the production of ultralow frequencies 
at very high momenta $k> m_{\rm pl}$, and there are 
a range of ultralow energy modes (of very short distances) 
that have frequencies equal or less than the current 
Hubble rate $H_0$, known as the $\it{tail}$ modes. 
These modes are still frozen today due to the expansion 
of the universe. Calculating their energy today, 
the authors argue that the $tail$ provides a strong 
candidate for the {\it dark energy} of the universe. 
In fact during inflation, their energy is about 122-123 
orders of magnitude smaller than the total energy, 
for any random value of the free parameter in the 
modified dispersion relations. The exact solutions 
of the system show that the CMBR spectrum is 
that of a (nearly) black body, and that the adiabatic 
vacuum is the only choice for the initial conditions. 
In a nice follow up paper, Bastero-Gil and Mersini-Houghton 
investigate a more general class of models and show 
how demanding they satisfy both SN1a and CMBR 
data severely constrains the viability of these models, 
the most important constraint coming from the 
CMBR \cite{Bastero-Gil:2001rv}.

\subsection{Acceleration due to the backreaction of cosmological perturbations}

The role of gravitational backreaction in inflating cosmologies 
has a long history \cite{robert}. It was pioneered by in a series 
of papers by Tsamis and Woodard 
\cite{Tsamis:1992sx,Tsamis:1996qm,Tsamis:1996qq,Tsamis:1997rk} 
who investigated the quantum gravitational back-reaction on an initially inflating, 
homogeneous and isotropic universe and showed that the role of long 
wavelength gravitational waves back-reacting on an inflationary background, 
was to slow the rate of inflation. 

In  \cite{Mukhanov:1996ak,Abramo:1997hu} the authors derive the 
effective gauge-invariant energy-momentum tensor for cosmological 
perturbations and use it to study the influence of perturbations on the 
behaviour of the Friedmann background in inflationary Universe scenarios. 
In particular they found that the back reaction of cosmological perturbations 
on the background can become important  at energies below the 
self-reproduction scale. For the cases of scalar metric fluctuations and 
gravitational waves in chaotic inflation, the backreaction resulting  from 
the effective gauge-invariant energy-momentum tensor is such that for long 
wavelength scalar and tensor perturbations, the effective energy density 
is negative and counteracts any pre-existing cosmological constant. 
This then leads the authors to speculate that gravitational back-reaction 
may lead to a dynamical cancellation mechanism for a bare cosmological 
constant, and yield a scaling fixed point in the asymptotic future in which 
the remnant cosmological constant 
satisfies $\Omega_{\Lambda} \sim 1$\cite{Brandenberger:2002sk} . 
  
More recently, in a series of papers, Rasanen \cite{Rasanen}, Barausse {\it et al.} \cite{barausse} 
and Kolb {\it et al.} \cite{Kolb:2005me,Kolb:2005da}, 
have returned to the basic idea of the backreaction being important. 
They have explored the possibility that
the observed acceleration of the universe has nothing to do with
either a new form of dark energy, or a modification of gravity.
Rather it is due to the effect of the backreaction of either super or
sub-horizon cosmological perturbations. By considering the effective
Friedmann equations  describing an inhomogeneous Universe after
smoothing (for a derivation see \cite{Buchert:1999er,Buchert:2001sa}), 
they argued that it is possible to have acceleration in
our local Hubble patch even if the fluid elements themselves do not
individually undergo accelerated expansion. 

The time behavior of the regularized general-relativistic cosmological 
perturbations possesses an instability which occurs in the perturbative 
expansion involving sub-Hubble modes. 
The above authors interpret this as acceleration in our
Hubble patch originating from the backreaction of cosmological
perturbations on observable scales. The conclusion has raised a
considerable amount of interest and criticism
\cite{Hirata,wald} (see also Refs.~\cite{backothers}). 
Ishibashi and Wald \cite{wald} have argued
that it is not plausible for acceleration to arise in general
relativity from a back-reaction effect of inhomogeneities in our
universe, unless there is either a  cosmological constant or some
form of dark energy. Basically the fact our universe is so well
described by a FLRW metric perturbed by Newtonian mechanics implies
the back-reaction of inhomogeneities on the dynamics of the universe
is negligible. Moreover, they argue that the acceleration of the
scale factor may accelerate in these models without there being any
physically observable consequences of this acceleration. 
It has been argued that the no-go theorems due to 
Hirata and Seljak \cite{Hirata} do not hold for the case 
of Refs.~\cite{Mukhanov:1996ak,Abramo:1997hu} where there is 
a large positive bare cosmological constant 
which dominates the dynamics \cite{robert}.

In an interesting recent paper, Buchert {\it et al} \cite{Buchert:2006ya}, have demonstrated there exists a correspondence between the kinematical backreaction and more conventional scalar field cosmologies, with particular potentials for their  'morphon field'. For example, they argue that it is possible reinterpret, say, quintessence scenarios by demonstrating that the physical origin of the scalar field source can be ascribed to inhomogeneities in the Universe. Through such a correspondence they explain the origin of dark energy as emerging from the morphon fields. The averaged cosmology is characterized by a weak decay (quintessence) or growth (phantom quintessence) of kinematical fluctuations, feeded by `curvature energy' that is stored in the averaged 3-Ricci curvature. 

Although the idea of sub-horizon perturbations in a conventional 
cosmology driving the current acceleration may not be flavour of 
the month, in many ways it would be great if this idea was to work
out, it would allow us to live in a universe where gravity is
conventional, there is no negative-pressure fluid out there waiting
to be discovered, and no cosmological constant needed. 
Unfortunately the Universe looks like it has not been so obliging.

\section{Conclusions}

The question of the nature of the dark energy that is driving the observed 
acceleration of the Universe today is without doubt one of the most exciting 
and challenging problems facing physicists and astronomers alike. 
It is at the heart of current astronomical observations and proposals, 
and is driving the way particle theorists are trying to understand 
the nature of the early and late universe. It has led to a remarkable 
explosive surge in  publications over the past few years. 
For example over 900 papers with the words ``Dark Energy'' in the 
title have appeared on the archives since 1998, 
and nearly 800 with the words ``Cosmological constant'' have appeared. 

Writing a review on the subject has been a daunting task, 
it is just impossible to properly do justice to all the avenues 
of investigation that people have ventured down.  Instead we have 
concentrated on a subset of all the work that has gone on, 
trying to link it wherever possible to the other works. 
In particular we have decided to take seriously the prospect 
that the dark energy may be dynamical in origin, and so have 
performed quite a thorough investigation into both the nature 
of the cosmological constant in string theory, as well as the nature
of Quintessence type scenarios. This has allowed us to  compare many 
models which are in the literature and to point out where they are 
generally fine tuned and lacking motivation. Unfortunately it is a problem 
that faces many such scalar field inspired scenarios. 
On the odd occasion where a really promising candidate field 
seems to have emerged, we have said so. 

Alongside the modification due to the presence of new sources 
of energy momentum in Einstein's equations, another route 
we have explored is to allow for the possibility that Einstein's 
equations themselves require some form of modification, 
in other words the geometry part of the calculation needs rethinking. 
Although there is no reason as of yet to believe this is the case, 
it is perfectly possible and so we have spent some time looking 
at alternatives to Einstein gravity as a source of the current acceleration 
today. As we have mentioned, there is more that we have not dealt 
with, than we have. For example we have not addressed the issues 
related to the holographic approach \cite{holdarkenergy}
and other observational aspects about dark energy, such as 
gravitational lensing which can serve as an important probe 
of dark energy \cite{Glensing}.

We should also mention recent developments related to 
Bekenstein's relativistic theory
of modified Newton dynamics (MOND) \cite{MOND1}. 
Bekenstein's theory is a multi-field theory
which necessarily contains a vector and a scalar field 
apart from a spin two field-- so called tensor-vector-scalar theory 
(TeVeS) \cite{MOND2} (see the review of
Sanders and references therein \cite{MOND3}, as well as
the recent detailed work of Skordis \cite{Skordis:2005eu}). 
Since TeVeS contains a scalar field, 
it is natural to ask whether this theory
can account for late-time acceleration and inflation. 
Recently efforts have been made to
capture these two important aspects of cosmological dynamics in the frame 
work of TeVeS \cite{MOND4,Skordis:2005xk,MOND5}. 
However, these investigations are
at the preliminary level at present. 

In the context of modifed gravity models there has recently 
been some interesting work which can be related to 
MOND \cite{Navarro:2005ux,Navarro:2006mq}. 
The authors have proposed a class of actions for the spacetime 
metric that introduce corrections to the Einstein-Hilbert Lagrangian 
depending on the logarithm of some curvature scalars, 
as opposed to power-law corrections discussed earlier in the review. 
For some choices of these invariants the models are ghost 
free and modify Newtonian gravity below a characteristic 
acceleration scale given by $a_0 = c\mu$, where $c$ is 
the speed of light and $\mu$ is a parameter of the model 
that also determines the late-time Hubble constant $H_0 \sim \mu$. 
The model has a massless spin two graviton, but also a scalar 
excitation of the spacetime metric whose mass depends on the 
background curvature. Although almost massless in vacuum, 
the scalar becomes massive and effectively decouples close to 
any source leading to the recovery of an acceptable weak 
field limit at short distances. The classical ``running'' of 
Newton's constant with the distance to the sources and 
gravity is easily enhanced at large distances by a large ratio 
opening up the possibility of building a model with a MOND-like 
Newtonian limit that could explain the rotation curves of 
galaxies without introducing Dark Matter using this kind of actions.  
Perhaps advances in our ability to perform solar and stellar system 
tests of the cosmological constant, will allow us to discriminate 
different models for $\Lambda$. 

On the observational front, to many people's frustration, pretty 
much everything seems perfectly consistent with the true cosmological 
constant being the source of the acceleration, but of course we are 
not really sure (well some of us aren't anyway) why it has the value it 
does have, or why it should be coming to dominate so recently. 
However, there are a number of exciting observational proposals 
on the horizon (including solar and stellar system tests of the 
cosmological constant \cite{Jetzer:2006gn,Sereno:2006re})
which if they come up trumps may well provide us with vital
information about the nature and magnitude of 
the cosmological constant today. 

They include the Dark Energy Survey (DES) \cite{Abbott:2005bi}, 
a proposed optical-near infrared survey of 5000 sq. deg of the South 
Galactic Cap. It will allow for the measurement of the dark energy
and dark matter densities and the dark energy equation of state through: 
galaxy clusters, weak gravitational lensing tomography, galaxy angular 
clustering, and supernova distances. The beauty of this is that the 
methods constrain different combinations of the cosmological 
model parameters and are subject to different systematic errors. 

A second proposed mission which has generated a lot of excitement 
is SNAP \cite{Albert:2005um}. It seeks to place constraints on the 
dark energy using two distinct methods, first through obtaining 
more and deeper Type Ia SN, and the second through weak 
gravitational lensing, which relies on the coherent distortions 
in the shapes of background galaxies by foreground mass 
structures. Once again, the two methods for probing dark 
energy are highly complementary with error contours from 
the two methods that are largely orthogonal. 

A third proposed mission (which is funded!) is the 
Planck CMB satellite which, although probably not having the 
sensitivity to measure any evolution in the dark energy equation 
of state, should be able to tell us whether or not it is a true 
cosmological constant with $w=-1$, or whether $w$ is 
different from that value. Such a result if it proved the case 
would be as dramatic as evidence for evolution in the dark 
energy. What form of matter would be giving us such a result? 

Recently the suggestions that 
Gamma Ray Bursters may actually be excellent standard 
candles have been revisited, with some interesting tentative 
initial conclusions \cite{G-R-B}. 
The significance of such a result, 
if true, is hard to underestimate. GRB's are some of the 
brightest objects in the universe and so can be seen much 
further than Type Ia Supernovae. In principle they could 
be seen out to redshifts of around $z \sim 10$, which 
would allow us to have a much more detailed Hubble 
diagram, and to probe more accurately whether there 
is evidence of evolution in the dark energy equation 
of state. Although the error bars are still large, the initial 
evidence actually suggests that for GRB's out to a redshift 
of 6, the Hubble diagram is best fit with a dynamical equation 
of state, as opposed to a cosmological constant.  
It may not be statistically significant, but what the 
heck its a fun and tantalising way to end this review!

\section*{ACKNOWLEDGEMENTS}

We thank D.~V.~Ahluwalia-Khalilova for inviting us to write 
this review for International Journal of Modern Physics D.
We are grateful to Luca Amendola, Tiago Barreiro, 
Bruce A. Bassett, Gianluca Calcagni, 
Pier Stefano Corasaniti, Naresh Dadhich, M. R.~Garousi, 
Burin Gumjudpai, Soo A Kim, Martin Kunz,
Seung-Joo Lee, Andew R. Liddle, Jim Lidsey, Michael Malquarti, 
Shuntaro Mizuno, Tapan Naskar, Shin'ichi Nojiri, 
Nelson J.~Nunes, Sergei Odintsov, T.~Padmanabhan,
Sudhakar Panda, David Parkinson, Federico Piazza, David Polarski,
M.~Pospelov, Francesca Rosati, Varun Sahni, N. Savchenko, Parampreet Singh, Alexey Toporensky, Peter V. Tretjakov, 
Mark Trodden, Carlo Ungarelli, David Wands, and John Ward
for fruitful collaborations about dark energy.
We thank Bruce A. Bassett, Sean M. Carroll, Naresh Dadhich, Abha Dev,
Deepak Jain, Andrei Linde, Roy Maartens, 
Jayant V. Narlikar, Ishwaree Neupane, Nelson J.~Nunes, 
Sergei Odintsov, T.~Padmanabhan, Leandros Perivolaropoulos, 
David Polarski, Varun Sahni, P. Sharan and Alexei Starobinsky 
for giving us very useful comments. 
We also thank Greg Aldering, Daniel Eisenstein, 
Malcolm Fairbairn, Martin Kunz, Antony Lewis, 
Roy Maartens, Alessandro Melchiorri, David F. Mota, Nelson J. Nunes, 
T.~Padmanabhan, and Varun Sahni
for permission to include figures from their papers. 
We are also very grateful to the 60 people who 
contacted us following the inital submission of the review to the archives.  
Hopefully we have successfully incorporated most of their comments and 
suggestions in this revised version. 

E.\,J.\,C. would like to thank the Aspen Center for Physics,
for their hospitality during the time some of this work was completed. 
M.\,S. thanks IUCAA (Pune) for hospitality where this work was started.
M.\,S. is supported by DST-JSPS grant and 
thanks Gunma National College of Technology (Japan) 
for hospitality.
S.\,T. is supported by JSPS (No.\,30318802). 
M.\,S. and S.\,T. also thank the organisers of the 
third Aegean Summer School where part of the work 
was presented.


\end{document}